\documentclass[12pt]{article}
\usepackage{jheppub,amsmath,amssymb}

\usepackage{tikz}
\usetikzlibrary{arrows}
\usetikzlibrary{shapes.geometric,calc, arrows, positioning,shapes.misc}

\newcommand{\arrowpath}[3]{
	\draw[line width=1pt] #1--#2;
	\draw[->, >=stealth, line width=1pt, black] #1--($#1!#3!#2$);
}

\newcommand{\arrowpathdashed}[3]{
	\draw[line width=1pt,dashed] #1--#2;
	\draw[->, >=stealth, line width=1pt, black,dashed] #1--($#1!#3!#2$);
}

\newcommand{\arrowpathdouble}[3]{
	\draw[double] #1--#2;
	\draw[double,->, >=stealth] #1--($#1!#3!#2$);
}

\newcommand{\paddedline}[3]{
	\draw[white, fill=white] ($#1-#3$)--($#1+#3$)--($#2+#3$)--($#2-#3$)--cycle;
}

\newcommand{\bezierarrowpath}[5]{
	\def\t{#5};
	\def\tminuseps{#5-0.001};
	\coordinate (A) at ($#1!#5!#2$);
	\coordinate (B) at ($#2!#5!#3$);
	\coordinate (C) at ($#3!#5!#4$);

	\coordinate (D) at ($(A)!#5!(B)$);
	\coordinate (E) at ($(B)!#5!(C)$);
	
	\coordinate (P) at ($(D)!#5!(E)$);
	\coordinate (Pme) at ($(D)!\tminuseps!(E)$);
	\draw #1 .. controls #2 and #3 .. #4;
	\draw[->, >=stealth, line width=1pt, black] (Pme)--(P);	
}

\newcommand{\bezierarrowpathdouble}[5]{
	\def\t{#5};
	\def\tminuseps{#5-0.001};
	\coordinate (A) at ($#1!#5!#2$);
	\coordinate (B) at ($#2!#5!#3$);
	\coordinate (C) at ($#3!#5!#4$);

	\coordinate (D) at ($(A)!#5!(B)$);
	\coordinate (E) at ($(B)!#5!(C)$);
	
	\coordinate (P) at ($(D)!#5!(E)$);
	\coordinate (Pme) at ($(D)!\tminuseps!(E)$);
	\draw[double] #1 .. controls #2 and #3 .. #4;
	\draw[double,->, >=stealth, black] (Pme)--(P);	
}

\renewcommand{\ker}{{\rm ker}}

\newcommand{\Hom}{{\rm Hom}}
\newcommand{\ra}{{\rightarrow}}
\newcommand{\cG}{{\mathcal G}}
\newcommand{\cF}{{\mathcal F}}
\newcommand{\ZZ}{{\mathbb Z}}
\newcommand{\RR}{{\mathbb R}}
\newcommand{\CC}{{\mathbb C}}

\newcommand{\Vect}{{\rm Vect}}

\newcommand{\cA}{{\mathcal A}}
\newcommand{\cH}{{\mathcal H}}
\newcommand{\cM}{{\mathcal M}}

\newcommand{\cC}{{\mathcal C}}

\newcommand{\fT}{{\mathfrak T}}
\newcommand{\fB}{{\mathfrak B}}

\newcommand{\cD}{{\mathcal D}}
\newcommand{\cI}{{\mathcal I}}

\newcommand{\Cl}{{\rm Cl}}

\newcommand{\eps}{\epsilon}

\newcommand{\hG}{{\hat G}}

\newcommand{\fA}{{\mathcal A}}
\newcommand{\eE}{{\Lambda}}
\newcommand{\cZ}{{\mathcal Z}}
\newcommand{\Exp}{{\rm Exp}}

\newcommand{\frP}{{\mathfrak P}}

\newcommand{\sA}{{\mathsf A}}
\newcommand{\sB}{{\mathsf B}}

\newcommand{\bbF}{{\mathbb F}}
\newcommand{\bbG}{{\mathbb G}}

\title{State sum constructions of spin-TFTs and string net constructions of fermionic phases of matter}

\author[1]{Lakshya Bhardwaj,}
\author[1]{Davide Gaiotto,} 
\author[2]{Anton Kapustin}

\affiliation[1]{Perimeter Institute for Theoretical Physics, Waterloo, Ontario, Canada N2L 2Y5}
\affiliation[2]{California Institute of Technology, Pasadena, CA 91125, United States}

\abstract{It is possible to describe fermionic phases of matter and spin-topological field theories in $2+1d$ 
in terms of bosonic ``shadow'' theories, which are obtained from the original theory by ``gauging fermionic parity''. 
The fermionic/spin theories are recovered from their shadow by a process of fermionic anyon condensation:
gauging a one-form symmetry generated by quasi-particles with fermionic statistics. 
We apply the formalism to theories which admit gapped boundary conditions. 
We obtain Turaev-Viro-like and Levin-Wen-like constructions of fermionic phases of matter. 
We describe the group structure of fermionic SPT phases protected by $\ZZ_2^f \times G$. 
The quaternion group makes a surprise appearance. }

\begin{document}

\maketitle

\section{Introduction}
The general purpose of this paper is to explore the properties of spin-topological quantum field theories in $2+1$ dimensions \cite{dijkgraaf1990}
and their relation to fermionic gapped phases of matter \cite{Fidkowski:2010aa,chen2010local}. A concrete objective of this paper is to leverage the relation between 
these two notions in order to produce explicit lattice Hamiltonians for new fermionic phases of matter. 

Spin-topological quantum field theories are topological quantum field theories which are defined only on manifolds 
equipped with a spin structure. Fermionic phases of matter are phases defined by a microscopic local Hamiltonian which contains 
fermionic degrees of freedom. The relation between these two notions is most obvious for relativistic theories, 
thanks to the spin-statistics theorem. It is far from obvious for non-relativistic theories or discrete lattice systems \cite{GaiottoKapustin}. 

Standard (unitary) TFTs in $2+1$ dimensions are rather well-understood in terms of properties of 
their line defects, which form a modular tensor category \cite{Kitaev:aa,moore1989}. 
A similar characterization for spin-TFTs is not as well developed. 
The expected relation to fermionic phases of matter suggests the existence of a formulation involving 
some kind of modular super-tensor category, involving vector spaces with non-trivial fermion number grading.
We do not know how to give such a description or how to reconstruct a spin-TFT from this type of data. 

Instead, we follow a different strategy: we encode a spin-TFT $\fT_s$ into the data of a ``shadow'' TFT $\fT_f$, 
a standard TFT equipped with an extra piece of data, a fermionic quasi-particle $\Pi$ which fuses with itself to the identity. 
\footnote{See appendix \ref{app:RCFT} for a simple justification of this statement for TFTs which are associated to 2d RCFTs.}

In the language of \cite{GaiottoKapustin}, the spin-TFT is obtained from its shadow by a procedure of ``fermionic anyon condensation''. 
Conversely, if we pick a spin manifold $M$ and add up the $\fT_s$ partition function over all possible choices of spin structure $\eta$ 
we recover the $\fT_f$ partition function.  

The relation between spin-TFTs and standard TFTs equipped with appropriate fermionic quasi-particles was also discovered in the mathematical literature
\cite{Beliakova:2014aa}. This reference proposes a Reshetikhin-Turaev-like construction of a spin TFT partition function from the data of a 
modular tensor category equipped with invertible fermionic lines. The partition function of the spin TFT summed over the possible choices of spin structure 
reproduces the Reshetikhin-Turaev partition function of the underlying modular tensor category. 

Similarly, the relation between fermionic phases of matter and bosonic phases equipped with a special fermionic quasi-particle 
was proposed in \cite{Gu:2013ab,Cheng:2015aa,Lan:2015aa} as a form of ``gauging fermionic parity''.

It is natural to wonder if all spin TFTs should admit a shadow. We believe that should be the case.
Given a spin TFT and a spin manifold, we can add up the partition function over all possible spin structures
in order to define tentatively the partition function of its shadow. This procedure essentially corresponds to ``gauging fermionic parity''
and assigns to every spin manifold a partition function which does not depend on a choice of spin structure. The key question is if 
one can extend this definition to general manifolds which may not admit a spin structure. In $2+1$ dimensions TFTs can be reconstructed 
from the properties of their quasi-particles, which should be computable from the data of spin manifold partition functions. 

The fermionic anyon condensation procedure computes the partition function of $\fT_s$ on a spin manifold $M$ from partition functions of $\fT_f$ on $M$ 
decorated by collections of fermionic quasi-particles. The calculation involves some crucial signs involving the Gu-Wen Grassmann integral \cite{Gu:2012aa}
and a choice of spin structure on $M$. 

A slightly more physical perspective on the construction can be given as follows. Consider some microscopic bosonic physical system 
$S$ which engineers $\fT_f$ at low energy. Combine $S$ with a system of free massive fermions. The $\Pi$ quasi-particles in $S$ can 
combine with the free fermions $\psi$ to produce a a bosonic composite particle $\psi \Pi$. Condensation of $\psi \Pi$ produces 
a new, fermionic phase of matter which we identify as a physical realization of $\fT_s$.

We will focus in most of the paper on theories which admit a state-sum construction. Concretely, that means that the shadow TFT $\fT_f$ is fully captured by the 
data of a spherical fusion category $\cC_f$, which can be fed into the Turaev-Viro construction  \cite{Turaev:1992hq} of the partition function or the
Levin-Wen construction  \cite{Levin:2004mi} of a bosonic commuting projector Hamiltonian. We will learn how to modify these standard constructions to compute 
the partition function of $\fT_s$ on a spin manifold and a fermionic commuting projector Hamiltonian for $\fT_s$. 
This is an extension of the proposal of \cite{Gu:2010aa}.

As an application of these ideas, we propose an explicit construction for all the fermionic SPT phases 
which are predicted by the spin-cobordism groups \cite{Kapustin:2014dxa}. In particular, this includes phases which lie outside
of the Gu-Wen super-cohomology construction. The classification of such phases has been previously proposed in \cite{Cheng:2015aa}, and our results agree with theirs.  
The novelty here is that we construct explicit state sums and Hamiltonians for all the phases and make explicit their dependence on spin structure.
Furthermore, we give a cohomological description of the classification and determine explicitly the group structure of fermionic SPT phases
under the stacking operation.

While this paper was in gestation, there appeared two papers which address some of the same questions. Lan et al. \cite{LanKongWen} also discuss topological phases of fermions using the theory of spherical fusion categories.  From our point of view, they identify bosonic shadows of fermionic phases. Tarantino and Fidkowski \cite{TarantinoFidkowski} very recently constructed an explicit commuting projector Hamiltonian for nonabelian fermionic SPT phases on a honeycomb lattice. They show that the result depends on a Kasteleyn orientation. This is an alternative way of thinking about spin structures on a lattice. 

\section{Overview}

\subsection{One-form symmetries and their anomalies}
In order to understand the relation between $\fT_s$ and $\fT_f$, it is useful to look at an analogous 
relation between standard ``bosonic'' TFTs. Consider TFTs $\fT_{\ZZ_2}$ endowed with a (non-anomalous) $\ZZ_2$ global symmetry, i.e. TFTs which are defined on manifolds equipped with a flat connection.  The dimension of space-time is arbitrary at this stage. For a mathematical definition TFTs with symmetries in $2+1d$, see e.g. \cite{Turaev:aa,Turaev:2012aa,Turaev:2013aa} and references therein.

Given such a TFT, we can build a new TFT $\fT_{b}$ by coupling the $\ZZ_2$ global symmetry to a dynamical gauge field. The partition function 
for $\fT_{b}$ on a manifold $M$ is computed by summing up the $\fT_{\ZZ_2}$ partition functions over all 
possible inequivalent $\ZZ_2$ flat connections (with the same weight):
\begin{equation}
Z[M;\fT_b] = \frac{1}{|H^0(M,\ZZ_2)|}\sum_{[\alpha_1] \in H^1(M,\ZZ_2)} Z[M;\fT_{\ZZ_2};[\alpha_1]] 
\end{equation}

\begin{figure}
\begin{center}
\includegraphics[width=15cm]{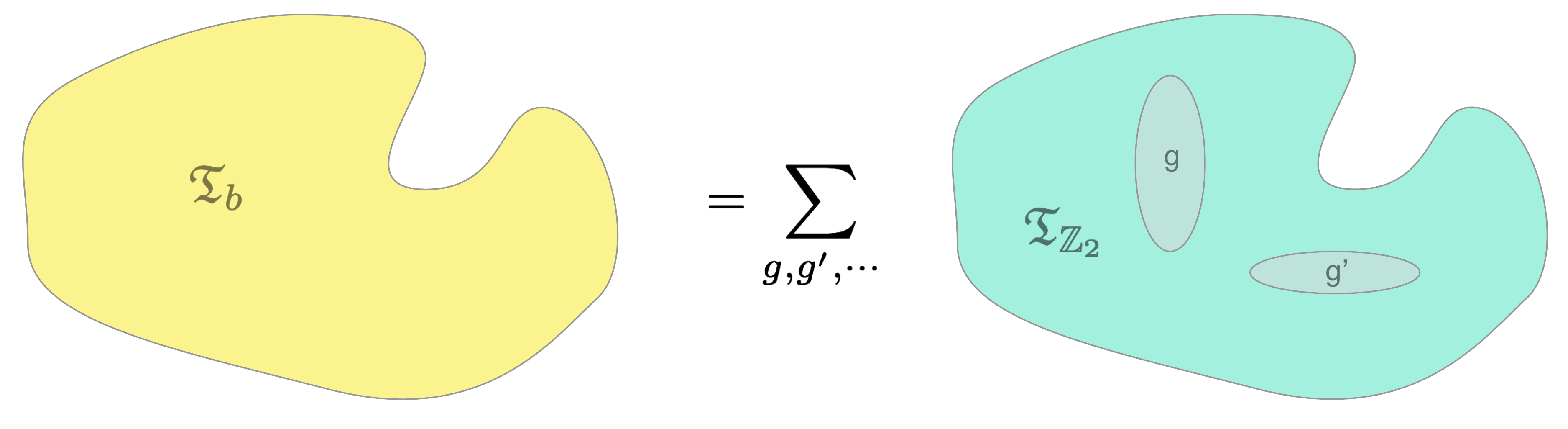}
\end{center}
\caption{A graphical depiction of the map from $\fT_{\ZZ_2}$ to $\fT_{b}$. On the right we have the partition function of $\fT_{\ZZ_2}$ on a three-dimensional manifold, 
equipped with a $\ZZ_2$ flat connection. We represent the connection as a collection of domain walls implementing $\ZZ_2$ symmetry transformations $g$, $g'$, etc. 
On the left we have the partition function of $\fT_{b}$, obtained by summing the $\fT_{\ZZ_2}$ partition function over all possible choices of $\ZZ_2$ flat connection. }
\label{fig:gau}
\end{figure}

The theory $\fT_{b}$ is always equipped with a bosonic quasi-particle $B$, 
the Wilson line defect, which fuses with itself to the identity in a canonical way. 
We can recover $\fT_{\ZZ_2}$ from $\fT_{b}$ by condensing $B$. Intuitively, the insertion of $A=1 \oplus B$ 
along a cycle in $M$ forces the flat connection to be trivial along that cycle (i.e. the partition function vanishes unless the holonomy of the connection is trivial). Adding a sufficient number of $A$'s to $M$ will set the flat connection to zero. 

\begin{figure}
\centering
\begin{tikzpicture}[line width=1pt]
\begin{scope}[every node/.style={sloped,allow upside down}]

\coordinate (lowa1) at (0,0);
\coordinate (lowa2) at ($(lowa1)+(2,0)$);
\coordinate (higha1) at ($(lowa1)+(0,2)$);
\coordinate (higha2) at ($(lowa2)+(0,2)$);

\draw[double] (lowa1) to[bend right] (higha1);
\draw[double] (lowa2) to[bend left] (higha2);

\coordinate (eq1) at ($(lowa2)+(1,1)$);
\node at (eq1) {$=$};

\coordinate (lowb1) at ($(lowa1)+(4,0)$);
\coordinate (lowb2) at ($(lowa2)+(4,0)$);
\coordinate (highb1) at ($(higha1)+(4,0)$);
\coordinate (highb2) at ($(higha2)+(4,0)$);

\draw[double] (lowb1) to[bend left] (lowb2);
\draw[double] (highb1) to[bend right] (highb2);

\coordinate (eq2) at ($(lowb2)+(1,1)$);
\node at (eq2) {$=$};

\coordinate (lowc1) at ($(lowb1)+(4,0)$);
\coordinate (lowc2) at ($(lowb2)+(4,0)$);
\coordinate (highc1) at ($(highb1)+(4,0)$);
\coordinate (highc2) at ($(highb2)+(4,0)$);

\draw[double] (lowc1) to (highc2);
\paddedline{(highc1)}{(lowc2)}{(.1,0)};
\draw[double] (highc1) to (lowc2);

\end{scope}
\end{tikzpicture}
\caption{Wilson lines in $\ZZ_2$ gauge theory have trivial statistics and can be freely recombined. We use a double-line notation for quasi-particles and line defects to indicate a choice of framing, 
but the Wilson loops have no framing dependence, i.e. represent bosonic quasi-particles. In general, these abstract properties characterize the quasi-particle generators $B$ of non-anomalous $\ZZ_2$ 1-form symmetries.}
\label{fig:wilson}
\end{figure}
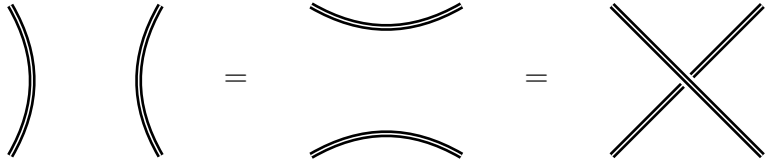

In 2+1 dimensions, it is useful to think about this process as gauging a (non-anomalous) $\ZZ_2$ 1-form symmetry generated by $B$.  
By definition, a global $\ZZ_2$ 1-form symmetry is parameterized by an element of $H^1(M,\ZZ_2)$ \cite{Gaiotto:2014aa}. 
Gauging this symmetry amounts to coupling the theory to a flat $\ZZ_2$-valued 2-form gauge field
\footnote{We will try be be careful and distinguish a 2-cocycle $\beta_2$ from its cohomology class $[\beta_2]$.} 
$[\beta_2]\in H^2(M,\ZZ_2)$. Thus $\fT_{b}$ has more structure than an ordinary TFT: it can associate a partition function to a manifold equipped with a 2-form gauge field $[\beta_2]$. 

\begin{figure}
\begin{center}
\includegraphics[width=15cm]{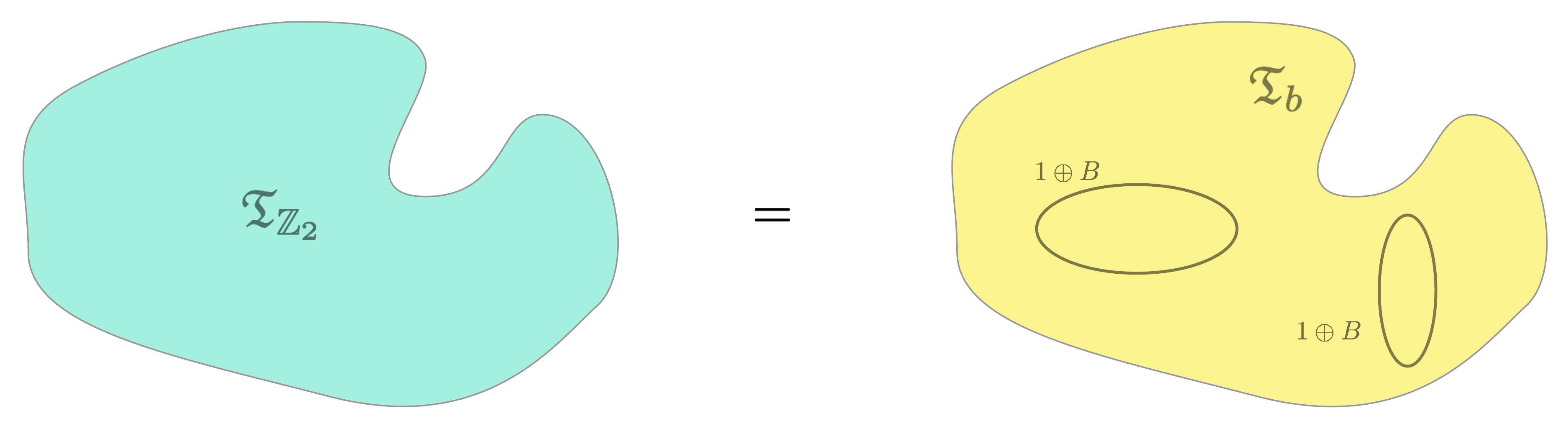}
\end{center}
\caption{A graphical depiction of the map from $\fT_{b}$ to $\fT_{\ZZ_2}$. On the right we have the partition function of $\fT_{b}$ on a three-dimensional manifold, 
possibly decorated with Wilson line operators $B$ along non-trivial cycles, dual to the domain walls of the previous picture. Abstractly, the choice of Wilson lines equips 
the manifold with a flat connection $[\beta_2]$ for a dual $\ZZ_2$ 1-form symmetry of $\fT_{b}$. Summing over all choices gives back the $\fT_{\ZZ_2}$ partition function.}
\label{fig:ungau}
\end{figure}

Concretely, we can triangulate the manifold $M$ and represent $\beta_2$ as a 2-cocycle, an assignment of elements of $\ZZ_2$ to faces of the triangulation such that the sum over faces of each tetrahedron vanishes.
\footnote{An arbitrary $\ZZ_2$-valued function on faces is called a 2-cochain with values in $\ZZ_2$, and the condition that the sum over faces of each tetrahedron vanishes is written as $\delta \beta_2=0$, i.e. the 2-cochain is closed. A 1-form gauge transformation is parameterized by a 1-cochain $\lambda_1$, i.e. a $\ZZ_2$-valued function on the links, and transforms $\beta_2$ to $\beta_2+\delta\lambda_1$.} We can define the partition function $Z[M;\fT_b;\beta_2]$ of $\fT_{b}$ coupled to $\beta_2$ by decorating $M$ with $B$ lines which pass an (even) odd number of times through each face labelled by the (trivial) nontrivial element of $\ZZ_2$.
\footnote{We write concrete elements of $\ZZ_2$ additively. That is, the trivial element will be denoted $0$, while the nontrivial one will be $1$. In particular, when we discuss cochains with values in $\ZZ_2$, we will write the group operation additively.}
\begin{equation}
Z[M;\fT_b;\beta_2] = \frac{1}{|H^0(M,\ZZ_2)|}\sum_{[\alpha_1] \in H^1(M,\ZZ_2)} (-1)^{\int_M \alpha_1 \cup \beta_2} Z[M;\fT_{\ZZ_2};[\alpha_1]] 
\end{equation}

An even number of $B$ lines enter each tetrahedron and can be connected to each other in any way we wish without changing the answer,
thanks to the statistics and fusion properties of $B$. It is relatively straightforward to verify that the partition function does not change if we replace 
$\beta_2$ with a gauge-equivalent cocycle $\beta_2 + \delta \lambda_1$ or if we 
change the triangulation of $M$. In either case, the collection of $B$ lines is deformed or re-organized. Thus
\begin{equation}
Z[M;\fT_b;\beta_2 + \delta \lambda_1] = Z[M;\fT_b;\beta_2] \equiv Z[M;\fT_b;[\beta_2]]
\end{equation}
Summing up this decorated partition function over all possible $\beta_2$ will insert enough $A$'s to project us back to the partition function of $\fT_{\ZZ_2}$:
\begin{equation} \label{eq:gauz2}
Z[M;\fT_{\ZZ_2}] = \frac{|H^0(M,\ZZ_2)|}{|H^1(M,\ZZ_2)|}\sum_{[\beta_2] \in H^2(M,\ZZ_2)} Z[M;\fT_b;[\beta_2]] 
\end{equation}
We can introduce extra signs to select a specific $\ZZ_2$ flat connection $\alpha_1$: 
\begin{equation}\label{eq:gauz22}
Z[M;\fT_{\ZZ_2};\alpha_1] =  \frac{|H^0(M,\ZZ_2)|}{|H^1(M,\ZZ_2)|} \sum_{[\beta_2] \in H^2(M,\ZZ_2)} (-1)^{\int_M \alpha_1 \cup \beta_2} Z[M;\fT_b;[\beta_2]] 
\end{equation}

Vice versa, we can consider a theory $\fT_{b}$ equipped with a non-anomalous $\ZZ_2$ 1-form symmetry generated by some quasi-particle $B$. Gauging the 
$\ZZ_2$ 1-form symmetry with the same formulae \ref{eq:gauz2} and \ref{eq:gauz22} results into a new theory $\fT_{\ZZ_2}$ which is always equipped with 
a $Z_2$ global symmetry generated by Wilson surfaces. 

Now we can go back to $\fT_{f}$. By definition, this theory contains a particle which is a fermion. That is, a particle $\Pi$ which is (1) an abelian anyon (2) generates a $\ZZ_2$ subgroup in the group of abelian anyons and (3) has topological spin $-1$. 
The first two conditions mean that $\fT_{f}$ has a 1-form $\ZZ_2$ symmetry, while the third one implies that this symmetry is anomalous, i.e. there is an obstruction to coupling the theory to a 2-form gauge field in $2+1$ dimensions in a gauge-invariant manner.

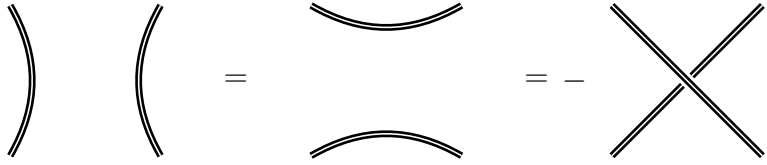
\begin{figure}
\centering
\begin{tikzpicture}[line width=1pt]
\begin{scope}[every node/.style={sloped,allow upside down}]

\coordinate (lowa1) at (0,0);
\coordinate (lowa2) at ($(lowa1)+(2,0)$);
\coordinate (higha1) at ($(lowa1)+(0,2)$);
\coordinate (higha2) at ($(lowa2)+(0,2)$);

\draw[double] (lowa1) to[bend right] (higha1);
\draw[double] (lowa2) to[bend left] (higha2);

\coordinate (eq1) at ($(lowa2)+(1,1)$);
\node at (eq1) {$=$};

\coordinate (lowb1) at ($(lowa1)+(4,0)$);
\coordinate (lowb2) at ($(lowa2)+(4,0)$);
\coordinate (highb1) at ($(higha1)+(4,0)$);
\coordinate (highb2) at ($(higha2)+(4,0)$);

\draw[double] (lowb1) to[bend left] (lowb2);
\draw[double] (highb1) to[bend right] (highb2);

\coordinate (eq2) at ($(lowb2)+(1,1)$);
\node at (eq2) {$=$};

\coordinate (eq3) at ($(eq2)+(.5,0)$);
\node at (eq3) {$-$};

\coordinate (lowc1) at ($(lowb1)+(4,0)$);
\coordinate (lowc2) at ($(lowb2)+(4,0)$);
\coordinate (highc1) at ($(highb1)+(4,0)$);
\coordinate (highc2) at ($(highb2)+(4,0)$);

\draw[double] (lowc1) to (highc2);
\paddedline{(highc1)}{(lowc2)}{(.1,0)};
\draw[double] (highc1) to (lowc2);

\end{scope}
\end{tikzpicture}
\caption{The $\Pi$ lines have fermionic statistics and thus extra signs may occur  as the worldlines are recombined.}
\label{fig:piwilson}
\end{figure}

Concretely, in order to couple $\fT_{f}$ to the 2-cocycle $\beta_2$, we again pick a triangulation of $M$. 
Up to some choices of conventions for how to frame the quasi-particle worldlines, we can populate $M$ with $\Pi$ lines which pass an (even) odd number of times through each face labelled by the (trivial) non-trivial element of $\ZZ_2$, joined together inside each 
tetrahedron. This produces some tentative partition function $Z[M;\fT_f;\beta_2]$. 
The anomalous nature of the $\ZZ_2$ 1-form symmetry implies that the partition function changes by some signs when the 2-cocycle $\beta_2$ is replaced by a cohomologous one, i.e. when the $\Pi$ lines are deformed and recombined. Signs may also arise when one re-triangulates $M$
and, obviously, if we change our conventions of how to connect or frame the collection of $\Pi$ lines representing $\beta_2$. 

It is quite clear that the anomaly we encounter here does not depend on the specific choice of theory.  
If we are given two such TFTs $\fT_1$ and $\fT_2$, then their product $\fT_1 \times \fT_2$ has a 
standard, non-anomalous 1-form symmetry with generator $\Pi_1 \Pi_2$. That means that we can define unambiguously 
the partition function for the product theory coupled to a background $\ZZ_2$ two-form connection, implemented by decorating $M$ by a collection of $\Pi_1 \Pi_2$ defects. 

As we are considering a product theory and products of lines in the two factors, we can factor the partition function as
\begin{equation}
Z[M;\fT_1 \times \fT_2;\beta_2] = Z[M;\fT_1;\beta_2] Z[M;\fT_2;\beta_2] 
\end{equation}
Thus the individual partition functions can only change sign simultaneously under gauge transformations 
of changes of triangulation. 

We would like to argue that we can pick our conventions of how to connect and frame $\Pi$ lines in such a way that 
the gauge and re-triangulation anomalies coincide with the ones which emerged in the study of Gu-Wen fermionic SPT phases \cite{Gu:2012aa} and their relation to spin-TFTs \cite{GaiottoKapustin}. 
The Gu-Wen Grassmann integral combined 
with a spin-structure-dependent sign gives a $\ZZ_2$-valued function $z_{\Pi}(M;\beta_2)$ of a triangulated manifold 
endowed with a cocycle $\beta_2$ and a spin structure. This function changes
in a specific manner as one changes the cocycle by a gauge transformation $ \beta_2 \to  \beta_2 + \delta \lambda_1$ or the triangulation. 
We claim these are the same transformation rules as for $Z[M;\fT;\beta_2]$. 

In particular, the combination $z_{\Pi}( \beta_2)Z[M;\fT_f;\beta_2]$ is well-defined and gives us a spin-TFT 
with a bosonic $\ZZ_2$ one-form symmetry. Gauging that symmetry gives us the spin-TFT $\fT_s$, with a partition function 
\begin{equation}
Z[M;\fT_s]= \frac{|H^0(M,\ZZ_2)|}{|H^1(M,\ZZ_2)|}\sum_{[\beta_2] \in H^2(M,\ZZ_2)} z_{\Pi}(\beta_2)Z[M;\fT_f;\beta_2]
\end{equation}
This is our basic prescription to recover $\fT_s$ from its shadow $\fT_f$.

An alternative way to express the expected anomalous transformation laws of $Z[M;\fT_f;\beta_2]$ is to say that the 1-form $\ZZ_2$ symmetry generated by the $\Pi$ lines 
can only be gauged if we regard the (2+1)-dimensional theory $\fT_{f}$ as living on a boundary of a (3+1)-dimensional TFT containing a 2-form gauge field $\beta_2$. Concretely, 
the action of this (3+1)-dimensional TFT is \cite{GaiottoKapustin}
\begin{equation}\label{oneformanomalyaction}
S_4=i \pi \int_{M_4} \beta_2\cup\beta_2.
\end{equation}
This action is invariant under $\beta_2\ra\beta_2+\delta\lambda_1$ if $M_4$ is closed, but on a general compact manifold it varies by a boundary term
\begin{equation}\label{oneformanomaly}
S_4\ra S_4+i \pi \int_{\partial M_4} \fA(\beta_2,\lambda_1),
\end{equation}
where the $\ZZ_2$-valued 3-cochain $\fA$ is given by
\begin{equation}
\fA(\beta_2,\lambda_1)=\lambda_1\cup \beta_2+\beta_2\cup\lambda_1+\lambda_1\cup\delta\lambda_1.
\end{equation}
Note that one cannot discard the first two terms in parentheses because the cup product is not supercommutative on the cochain level. 

The anomalous nature of the 1-form $\ZZ_2$ symmetry means that when $\fT_{f}$ on $M=\partial M_4$ is coupled to a 2-form gauge field $\beta_2$, its partition function,
with an appropriate choice of conventions for drawing and framing the $\Pi$ lines encoding $\beta_2$, transforms under 1-form gauge symmetry precisely as in (\ref{oneformanomaly}).
\begin{equation}
Z[M;\fT_f;\beta_2 + \delta \lambda_1] = (-1)^{\int_M \fA(\beta_2,\lambda_1)} Z[M;\fT_f;\beta_2] \
\end{equation}

More generally, both gauge transformations and changes of triangulations can be interpreted as triangulated bordisms $M \times [0,1]$
with $\beta_2$ defined over the whole 4-manifold, interpolating between $\beta_2|_0$ and $\beta_2|_1$ at the two ends. 
Then $Z[M;\fT_f;\beta_2]$ changes under such 
manipulations as 
\begin{equation}\label{oneformanomalyaction}
Z[M;\fT_f;\beta_2|_1] = e^{i \pi \int_{M \times [0,1]} \beta_2\cup\beta_2} Z[M;\fT_f;\beta_2|_0].
\end{equation}

\subsection{Shadow of a product theory}
The fermionic sign $z_{\Pi}( \beta_2)$ is almost multiplicative \cite{GaiottoKapustin}: 
\begin{equation}
z_{\Pi}( \beta_2) z_{\Pi}( \beta'_2) = (-1)^{\int_M \beta_2 \cup_1 \beta'_2} z_{\Pi}( \beta_2+ \beta'_2)
\end{equation}

This observation allows us to re-write the product of two spin-TFT partition functions in a suggestive way
\begin{equation}
Z[M;\fT_s] Z[M;\fT'_s]= \frac{|H^0(M,\ZZ_2)|}{|H^1(M,\ZZ_2)|}\sum_{[\beta_2] \in H^2(M,\ZZ_2)}  z_{\Pi}( \beta_2) Z[M;\fT_f \times_\Pi;\beta_2]
\end{equation}
with 
\begin{equation}
Z[M;\fT_f \times_f \fT'_f;\beta_2] \equiv \frac{|H^0(M,\ZZ_2)|}{|H^1(M,\ZZ_2)|}\sum_{[\beta'_2] \in H^2(M,\ZZ_2)} (-1)^{\int_M (\beta_2+ \beta'_2) \cup_1 \beta'_2}Z[M;\fT_f;\beta_2 + \beta'_2] Z[M;\fT'_f;\beta'_2]
\end{equation}
This is a recipe expressing the shadow of the product of two spin-TFTs in terms of the product of the shadows. 

The physical interpretation of this formula is straightforward. The product of shadow theories  $\fT_f$ and $\fT'_f$ is endowed with a bosonic $\ZZ_2$ 
1-form symmetry generated by the product $\Pi \otimes \Pi'$ of the fermionic lines of the two theories. 
Gauging that symmetry leaves us with a new theory with fermionic 1-form symmetry generated by $\Pi_1$, which 
we interpret as the shadow of the product $\fT_s \times \fT'_s$ of the corresponding spin TFTs.
This agrees with the stacking construction proposed in \cite{Lan:2015aa}.

A simple check of this proposal is that the multiplication is associative: the product of three shadow theories  
has a $\ZZ_2 \times \ZZ_2$ bosonic 1-form symmetry with non-trivial generators 
$\Pi \otimes \Pi'$, $\Pi' \otimes \Pi''$, $\Pi \otimes \Pi''$. 

We will use this construction systematically in order to explore the group structure of fermionic SPT phases. 

\subsection{Gu-Wen and beyond}

The starting point of the Gu-Wen construction of femionic SPT phases \cite{Gu:2012aa} is a group super-cohomology element $(\nu_3, n_2)$, i.e. a pair of cochains on $BG$ with values in $\RR/\ZZ$ and 
$\ZZ_2$, respectively, satisfying 
\begin{equation}\label{GuWeneqs}
\delta n_2=0,\quad \delta \nu_3 = \frac12 n_2 \cup n_2. 
\end{equation}
Given a flat $G$-connection on $M$, one can pull back the cochains $\nu_3$ and $n_2$ on $BG$ to cochains 
on $M$ which we can still denote as $\nu_3$ and $n_2$. Then the Gu-Wen Grassmann integral $z_{\Pi}(n_2)$ can be combined with the product of $\nu_3$ over all tetrahedra in $M$
in order to give the partition function of an invertible spin-TFT with a symmetry $G$.

Our strategy to prove that $z_{\Pi}(\beta_2)$ captures the anomaly of fermionic 1-form symmetries 
will be to re-cast the Gu-Wen construction in this form, by defining an appropriate  
bosonic theory $\fT_f[\nu_3, n_2]$ such that the associated partition function reproduces the 
product of $\nu_3$ over all tetrahedra in $M$. 

The construction proceeds as follows. A 2-cocycle $n_2$ gives rise to a central extension
\begin{equation}
0 \to \ZZ_2 \to \hat G \to G \to 0
\end{equation}
Consider a bosonic SPT phase for $\hat G$, labelled by a  $\hat G$-cocycle $\hat \nu_3$ with values in $\RR/\ZZ$ \cite{dijkgraaf1990,Chen:2011aa}. We can gauge the $\ZZ_2$ subgroup and get a bosonic TFT with symmetry $G$. 
The resulting theory is essentially an enriched version of the toric code, where the $G$ symmetry acts on quasi-particles in a way determined by $n_2$ and $\hat \nu_3$. 
If this theory has a bulk line defect $\Pi$ which is a fermion and is acted upon trivially by the $G$ symmetry, it is a candidate for a shadow of a Gu-Wen phase. 

We will determine the condition for the bulk fermion $\Pi$ to exist. The existence of $\Pi$ will restrict $\hat \nu_3$ to 
be a specific combination of $n_2$ and a group cochain $\nu_3$ which satisfies (\ref{GuWeneqs}). We will denote this bosonic TFT $\cG_{\nu_3, n_2}$.
The result is a one-to-one map between Gu-Wen fermionic SPT phases  and bosonic SET phases of the form $\cG_{\nu_3, n_2}$. 

We will compute explicitly $Z[M;\cG_{\nu_3, n_2};\beta_2]$ to find that it is only non-vanishing if 
$\beta_2$ equals the pull-back of $n_2$ to $M$, in which case the partition function is essentially equal
to the product of $\nu_3$ over all tetrahedra in $M$. This will verify that 
$Z[M;\fT_s]$ for these theories coincide with the Gu-Wen partition sum and 
$z_{\Pi}(\beta_2)$ is the correct kernel for fermionic anyon condensation. 

Cobordism theory \cite{Kapustin:2014dxa,Cheng:2015aa} predicts the existence of a more general class of fermionic SPT phases protected by 
fermion number symmetry together with a global symmetry $G$, 
labelled by a triple $(\nu_3,n_2,\pi_1)$, where $\pi_1$ is a $\ZZ_2$-valued $1$-cocycle on $G$, $n_2$ is a $\ZZ_2$-valued 2-cochain on $G$, and $\nu_3$ is an $\RR/\ZZ$-valued 3-cochain on $G$. We will show that $n_2$ is in fact a cocycle, and $\nu_3$ and $n_2$ must again satisfy the Gu-Wen equations  (\ref{GuWeneqs}). Thus the set of fermionic SPT phases with symmetry $G$ can be identified with the product of the set of Gu-Wen phases and the set $H^1(G,\ZZ_2)$ parameterized by $\pi_1$. 

The meaning of $\pi_1$ is a group homomorphism from $G$ to $\ZZ_2$, which is used to pull-back a certain ``root'' $\ZZ_2$ fermionic SPT phase along $\pi_1$.
The ``root'' $\ZZ_2$ phase is expected to be the phase whose shadow is the toric code, enriched by the $\ZZ_2$ 
symmetry which exchanges the $e$ and $m$ quasi-particles. Such a $\ZZ_2$ symmetry is not manifest in the standard formulation of the toric code and only emerges at low energy. 
With a bit of effort, though, one can produce a microscopic description of the toric code with explicit $\ZZ_2$ symmetry \cite{Chang:2014aa}, 
starting from an Ising fusion category. 

We will verify that the $\ZZ_2$-equivariant toric code $\cI$ is indeed the shadow of root fermionic SPT phase with $\ZZ_2$ global symmetry, 
by explicitly computing $\cI \times_f \cI$ and matching it with a Gu-Wen phase. 

The Ising pull-back phases $\cI_{\pi_1}$ can be combined with a standard Gu-Wen phase $\cG_{\nu_3, n_2}$ to give a candidate $\cG_{\nu_3, n_2} \times_f \cI_\pi$ for the shadow of the most general fermionic SPT phase.
We will verify this combination is indeed the most general symmetry-enriched version of the toric code which admits a suitable fermion $\Pi$. 

Finally, we will compute the twisted products of general fermionic SPT phases with the help of a relation 
of the schematic form
\begin{equation}
\cI_{\pi_1} \times_f \cI_{\pi'_1} = \cG_{\nu_3, n_2}\times_f \cI_{\pi_1+ \pi'_1}
\end{equation}
where the Gu-Wen phase $\cG_{\nu_3, n_2}$ is determined canonically from $\pi_1$ and $\pi'_1$. 

This result explicitly realizes the group of fermionic SPT phases as an extension of 
$H^1(G, \ZZ_2)$ by the super-cohomology group of Gu-Wen phases (which itself is an extension of 
$H^2(G, \ZZ_2)$ by $H^3(G, \ZZ_2)$). This extension is nontrivial. That is, while the set of fermionic SPT phases is the product of the group $H^1(G,\ZZ_2)$ and the group of Gu-Wen phases with symmetry $G$, the abelian group structure on this set is not the product structure. 

\subsection{A Hamiltonian perspective}
We would like to describe the relation between a gapped bosonic Hamiltonian which engineers the shadow bosonic TFT $\fT_f$
and a gapped fermionic Hamiltonian which can engineer the related spin TFT $\fT_s$. Again, it is useful to first 
look at a pair of bosonic Hamiltonians for $\fT_b$ and $\fT_{\ZZ_2}$, related by gauging standard or 1-form non-anomalous $\ZZ_2$ symmetries.

The procedure for gauging a standard on-site $\ZZ_2$ global symmetry of some lattice realization of $\fT_{\ZZ_2}$ is well understood. 
One extends the Hilbert space by adding $\ZZ_2$-valued edge variables playing the role of a flat $\ZZ_2$ connection $\alpha_1$. Flatness is imposed locally by 
extra placquette terms in the Hamiltonian enforcing $\delta \alpha_1 =0$. The Hamiltonian for $\fT_{\ZZ_2}$ deformed by the coupling to the flat connection can be denoted as 
$H_{\ZZ_2}[\alpha_1]$ and the Hamiltonian on the enlarged Hilbert space is schematically
\begin{equation}
H'_{\ZZ_2}=H_{\ZZ_2}\left[\frac{1+\hat\sigma^z}{2}\right].
\end{equation}
Here $\hat \sigma^{x,y,z}_e$ are Pauli matrices acting on the $\ZZ_2$ variables at the $e$-th edge. More explicitly, suppose $H_{\ZZ_2}[\alpha]$ is given as a sum of local terms:
\begin{equation}\label{HZ}
H_{\ZZ_2}[\alpha]=\sum_v H_{\ZZ_2}^v[\alpha]
\end{equation}
where $H_{\ZZ_2}^v$ acts nontrivially only on the degrees of freedom in a neighborhood of the vertex $v$. 
We can take $H_{\ZZ_2}^v[\alpha]$ to vanish if the connection is not flat in a neighbourhood of $v$. 

Let $P_f$ be a projector which enforces the flatness of $\ZZ_2$-valued edge variables 
at a face $f$. Concretely, denoting edges and faces as pairs and triples of vertices,
\begin{equation}
P_{012} =\frac12(1+ \hat\sigma^z_{01} \hat\sigma^z_{12} \hat\sigma^z_{20}).
\end{equation}
Then the Hamiltonian in the enlarged Hilbert space is also a sum  of local terms
\begin{equation}\label{HprimeZ}
H'_{\ZZ_2}=\sum_v H_{\ZZ_2}^v + \sum_f (1-P_f)
\end{equation}

The resulting enlarged Hilbert space is then projected to gauge-invariant states by a collection of projectors $U^{\ZZ_2}_{v}$ 
which act by a local $\ZZ_2$ transformation on the degrees of freedom at the lattice site $v$ and shift the connection on the nearby 
edges. Concretely, we can write
\begin{equation}
U^{\ZZ_2}_v = \hat u_v \prod_{v'} \hat\sigma^x_{v v'} 
\end{equation}
Here $\hat u_v$ acts on the local degrees of freedom of the original theory at $v$ as a local $\ZZ_2$ symmetry transformations. 

More generally, one can define operators $U^{\ZZ_2}_{\lambda_0}$ which implement gauge transformations $\alpha_1 \to \alpha_1 + \delta \lambda_0$
with a parameter $\lambda_0$ which is a $\ZZ_2$-valued 0-cochain. Absence of anomalies means that 
\begin{equation}
U^{\ZZ_2}_{\lambda_0}U^{\ZZ_2}_{\lambda'_0} = U^{\ZZ_2}_{\lambda_0+ \lambda'_0}
\end{equation}

Hence the final Hilbert space $\cH[\fT_b]$ is obtained by the projection 
\begin{equation}
U^{\ZZ_2}_{\lambda_0} |\Psi_b \rangle =  |\Psi_b \rangle 
\end{equation} 
Thus we can define a Hamiltonian for $\fT_b$ as 
\begin{equation}
H_{\ZZ_2 \to b} =H'_{\ZZ_2}+ \sum_v \frac{1}{2}(1-U^{\ZZ_2}_v)
\end{equation}

Wilson line quasi-particles can be added at the vertices of the lattice by flipping the sign of the Coulomb branch constraints there. 
For convenience, we will choose a branching structure on the lattice, taken to be triangular, and define the Hilbert space $\cH[\fT_b;\beta_2]$ as
\begin{equation}
U^{\ZZ_2}_{\lambda_0} |\Psi_b;\beta_2 \rangle = (-1)^{\int \lambda_0 \cup \beta_2} |\Psi_b ;\beta_2\rangle 
\end{equation} 
i.e. 
\begin{equation}
H_{\ZZ_2 \to b}[\beta_2] =H'_{\ZZ_2}+ \sum_v \frac{1}{2}\left(1- (-1)^{\int \lambda^v_0 \cup \beta_2} U^{\ZZ_2}_v\right)
\end{equation}
where $\lambda^v_0$ is a delta function at the vertex $v$. 
Concretely, each face $t$ with $\beta_2(t) =1$ will contribute a Wilson loop at its first vertex. 
This makes the $\ZZ_2$ 1-form symmetry of $\fT_b$ manifest ``on-site''. 

The construction can be readily generalized to non-anomalous symmetry realizations which do not act on-site. 
We can introduce a triangular lattice in the system, with a lattice scale which is much larger than the 
scale set by the gap in $\fT_{\ZZ_2}$, and add the $\ZZ_2$ connection to the edges of that lattice. 
Operators $U^{\ZZ_2}_{\lambda_0}$ with the correct properties will still be defined, up to 
exponentially suppressed effects. 

Conversely, starting from a generic theory $\fT_b$ with non-anomalous 1-form symmetry, 
the Hilbert space of $\fT_{\ZZ_2}$ is obtained by first summing the 
Hilbert spaces of $\fT_b$ with one or none insertions of the $B$ quasi-particle  and then projecting 
to the sub-space which is fixed by the action of closed $B$ string operators, 
i.e. closed $B$ lines wrapping non-trivial cycles on the space manifold $\Sigma$. 

We can obtain a more local description by enlarging further the original Hilbert space and the subsequent projector. 
If the theory $\fT_b$ is given in a form which allows a direct coupling to a 2-form connection 
on the lattice by a local Hamiltonian $H_b[\beta_2]$ we just make $\beta_2$ into a collection 
of dynamical $\ZZ_2$ variables attached to the faces of the lattice. 

If not, we introduce a new triangular lattice in the system, with a lattice scale which is much larger than the 
scale set by the gap in $\fT_b$. We can attach a $\ZZ_2$ variable $\beta_2(t)$ to each face $t$ of the lattice
and denote as $\cH[\beta_2]$ the space of ground states of $\fT_b$ with a $B$ quasi-particle inserted 
in the middle of each face with $\beta_2(t)=1$. In particular, $\cH[0]$ is the usual space of ground states of $\fT_b$. 

In either case, we define the enlarged Hilbert space as the direct sum $\cH' = \oplus_{\beta_2} \cH[\beta_2]$ 
over all 2-cocycles $\beta_2$. Concretely, the Hilbert space $\cH[\beta_2]$ is realized as the space of zero energy states of a local Hamiltonian $H_b[\beta_2]$ 
acting on the microscopic Hilbert space. 
We can realize $\cH'$ as the space of zero energy states of a local Hamiltonian 
\begin{equation}\label{Hprime}
H'=H_b\left[\frac{1+\hat\sigma^z}{2}\right].
\end{equation}
Here $\hat \sigma^{x,y,z}_t$ are Pauli matrices acting on the $\ZZ_2$ variables at the $t$-th face.\footnote{Since in two dimensions any 2-cochain is closed, there is no need for projectors in $H'$.}

Due to the properties of the $B$ quasi-particles, we must have unitary transformations 
\begin{equation}
U_{\lambda_1}: \cH[\beta_2] \to \cH[\beta_2 + \delta \lambda_1]
\end{equation}
which move $B$ quasi-particles from one site to another or create or annihilate pairs of $B$ quasi-particles. 
For example, if $\lambda_1$ is concentrated on one edge $e$, the corresponding unitary operator $U_e$ will move, create or annihilate $B$ particles in the two faces adjacent to that edge. 
In particular, it will anti-commute with the $\hat \sigma^z$ variables for these two faces, commute with all others. 

We expect the $U_e$ operator to be an operator which only acts in the neighborhood of the edge $e$, i.e. local at the scale of our lattice. 
There is a certain degree of freedom in defining the $U_{\lambda_1}$. As the $B$ quasi-particles are bosons, it should be possible to use that freedom to ensure that 
different ways to transport the $B$ particles are all equivalent, i.e. 
\begin{equation}
U_{\lambda_1} U_{\lambda'_1} = U_{\lambda_1 + \lambda'_1}
\end{equation}
In other words, $U_{\lambda_1}$ implement the 1-form gauge symmetry of the theory $\fT_b$, which should not be anomalous. In particular, for every edge $e$ we have $U_e^2=1$, and $[U_e,U_e']=0$ for all $e,e'$.  We must also ensure $U_{\delta\mu_0}=1$ for all 0-cochains $\mu_0$. This requirement means that 1-form symmetry transformations with parameters $\lambda_1$ and $\lambda_1+\delta\mu_0$ are physically indistinguishable. 

We want to define the Hilbert space for $\fT_{\ZZ_2}$ as the subspace of the enlarged Hilbert space fixed by the action of these unitary transformations. 
We can define a commuting projector Hamiltonian acting on the enlarged Hilbert space $\cH'$ as 
\begin{equation}
\sum_e \frac{1}{2}(1-U_e).
\end{equation}
It engineers the space of ground states of $\fT_{\ZZ_2}$.This construction makes the $\ZZ_2$ global symmetry of $\fT_{\ZZ_2}$ manifest: it acts on the face variables as $\prod_t \hat\sigma^z_t$ and commutes with the Hamiltonian. 

Note that the $U_{\lambda_1}$ operators for closed 1-cochains, which satisfy $\delta \lambda_1 = 0$,  can be identified with the closed $B$ string operators we discussed originally, 
while the general $U_{\lambda_1}$ operators are open $B$ string operators. We can denote the closed string operators as $U^{cl}_{\lambda_1}$. They map each summand in the 
Hilbert space back to itself. 

Thus we define a microscopic Hamiltonian for $\fT_{\ZZ_2}$ as 
\begin{equation}
H_{b\to \ZZ_2} =H_b\left[\frac{1+\hat\sigma^z}{2}\right]+ \sum_e \frac{1}{2}(1-U_e)
\end{equation}
acting on the tensor product of the microscopic Hilbert space of $\fT_b$ and of the $\ZZ_2$ face degrees of freedom

Now consider the case of a fermionic $\ZZ_2$ 1-form symmetry, i.e. a $\ZZ_2$ 1-form symmetry with an anomaly (\ref{oneformanomaly}). As a warm-up, we can focus on how to define consistently the action of closed $\Pi$ string operators 
on the original Hilbert space of ground states for $\fT_f$. If we triangulate the space manifold and pick a 1-cocycle $\lambda_1$, i.e. a $\ZZ_2$-valued function on edges $\lambda_1$ satisfying $\delta \lambda_1=0$, 
we can draw a collection of non-intersecting $\Pi$ lines which cross each edge $e$ $\lambda_1(e)$ times modulo 2. We can relate different such 
collections for the same $\lambda_1$ without ever braiding the $\Pi$ lines, and thus we should be able to define 
a corresponding composite string operator $V^{cl}_{\lambda_1}$ acting on the space of ground states of $\fT_f$.

If we compose two such closed string operators $V^{cl}_{\lambda_1}$ and $V^{cl}_{\lambda_1'}$, we get a collection of string which may have intersections. 
Resolving each intersection will cost us a $-1$ sign. The total number of intersections modulo $2$ should coincide with $\int \lambda_1 \cup \lambda'_1$.
Thus we expect to be able to consistently define the closed string operators in such a way that 
\begin{equation}
V^{cl}_{\lambda_1} V^{cl}_{\lambda'_1} = (-1)^{\int_\Sigma \lambda_1 \cup \lambda'_1} V^{cl}_{\lambda_1 + \lambda'_1}
\end{equation}
In particular, there is no consistent way for a ground state to be fixed by all $V^{cl}$.

There is a natural way to correct the closed string operators in such a way that a consistent projection becomes possible: 
we can dress them by some extra sign $\sigma_2(\lambda_1)$ which also satisfies 
\begin{equation}
\sigma_2(\lambda_1)\sigma_2(\lambda'_1) = (-1)^{\int_\Sigma \lambda_1 \cup \lambda'_1} \sigma_2(\lambda_1 + \lambda'_1)
\end{equation}

If the space manifold is endowed with a spin structure, we can use the spin structure to define such a sign. 
Moreover, the Gu-Wen grassmann integral in two dimensions combined with a spin structure provides a local definition of
precisely the same sign $\sigma_2(\lambda_1)$ provided we enlarge the Hilbert space with fermionic degrees of freedom living on faces \cite{GaiottoKapustin}. In other words, $\sigma_2(\lambda_1)$ can be written as a product of local fermionic operators situated on the edges $e$ for which $\lambda_1(e)\neq 0$. 

In order to get a fully explicit and local definition of the space of $\fT_s$ ground states, we need to extend this logic to 
open $\Pi$ string operators, or equivalently to $V_{\lambda_1}$ for not necessarily closed 1-cochains $\lambda_1$. 

We can proceed as before and consider the sum of Hilbert spaces $\cH[\beta_2]$ over all 2-cocycles $\beta_2$, where  $\cH[\beta_2]$ is the space of ground states of $\fT_f$ with a $\Pi$ quasi-particle inserted in the middle of each face with $\beta_2(t)=1$. 
We can define as before unitary operators $V_{\lambda_1}$ which re-arrange the location of the $\Pi$ quasi-particles, 
but the fermionic nature of the quasi-particles, or the anomaly of the corresponding 1-form symmetry, indicates that the 
algebra of $V_{\lambda_1}$ will only close up to signs: 
\begin{equation}
V_{\lambda_1} V_{\lambda'_1} = V_{\lambda_1 + \lambda'_1} (-1)^{\omega_\Sigma (\beta_2,\lambda_1, \lambda_1')} .
\end{equation}

Similar considerations as for the partition function show that the anomaly $\omega_\Sigma$ must be universal
for all theories with a fermionic $\ZZ_2$ 1-form symmetry.  We can get a concrete expression for as follows. Consider the 3+1d TFT (\ref{oneformanomalyaction}) on $M_4=\Sigma \times D^2$, coupled to the 2+1d TFT on $\partial M_4=\Sigma \times S^1$. The operator $V_{\lambda_1}$ in the 2+1d  theory which implements the $\ZZ_2$ 1-form symmetry transformations also shifts the 2-form gauge field $\beta_2$ by $\delta\lambda_1$. By continuing $\lambda_1$ into the bulk, we may regard $V_{\lambda_1}$ as a boundary of a codimension-1 defect in the 3+1d TFT. By considering three such defects with parameters $\lambda_1,\lambda'_1$ and $-(\lambda_1+\lambda'_1)$ meeting at the origin of $D^2$, one can see that 
\begin{equation}
V_{\lambda_1} V_{\lambda'_1}  V_{-\lambda_1 - \lambda'_1}=\int_\Sigma \omega(\beta_2,\lambda_1,\lambda'_1).
\end{equation}
The 2-cochain $\omega$ is defined as a solution of the equation
\begin{equation}
\delta \omega(\beta_2,\lambda_1,\lambda_1')=\cA(\beta_2,\lambda_1)+\fA(\beta_2+\delta\lambda_1,\lambda'_1)-\fA(\beta_2,\lambda_1+\lambda'_1).
\end{equation}
Using (\ref{oneformanomaly}), we find
\begin{equation}
\omega(\beta_2,\lambda_1,\lambda_1')=\lambda_1\cup\lambda_1'+\delta\lambda\cup_1\lambda_1',
\end{equation}
where $\cup_1$ is the Steenrod higher product \cite{Higher} (see also Appendix B.1. of \cite{Kapustin:2014aa} for a brief summary). In particular, we see that $\omega_\Sigma=\int_\Sigma \omega$ does not depend on $\beta_2$ in this case. 

Another manifestation of the anomaly is that the operators $V_{\lambda_1}$ are not invariant under $\lambda_1\mapsto\lambda_1+\delta\mu_0$, where $\mu_0$ is an arbitrary $\ZZ_2$-valued 0-cochain. Namely, by considering two defects implementing 1-form gauge transformations with parameters $\lambda_1$ and $\lambda_1+\delta\mu_0$, we find
\begin{equation}\label{Vnoninv}
V_{\lambda_1+\delta\mu_0}V_{-\lambda_1}=(-1)^{\int_\Sigma \left(\mu_0\cup\delta\lambda_1+\beta_2\cup\mu_0+\mu_0\cup\beta_2\right)}. 
\end{equation}

One way to deal with this anomaly would be to couple the system to the Hamiltonian version of Gu-Wen Grassmann integral. The Gu-Wen Grassmann integral on a 
bordism geometry $\Sigma \times [0,1]$ with $\beta_2$ and $\beta_2 + \delta \lambda_1$ at the two ends will provide dressing operators $U^{z_\Pi}_{\lambda_1}$
which should correct the $V_{\lambda_1}$ to a set of commuting projectors. This is somewhat cumbersome, though, and we will 
propose a more direct alternative lattice construction. 

We will promote the face variables $\beta_2(f)$ to occupation numbers $n_f$ for fermionic degrees of freedom.  Thus at each face
we have a pair of fermionic creation and annihilation operators, or equivalently a pair of Majorana fermions $\gamma_f,\gamma'_f$. 
We combine the individual edge operators $V_e$ with Majorana fermions on the two faces $f_L$ and $f_R$ sharing $e$ and define new edge operators 
\begin{equation}
U^f_e =  \pm V_e \gamma_{f_L} \gamma'_{f_R} 
\end{equation}
in such a way as to make the following fermionic Hamiltonian 
well-defined
\begin{equation}
H_{f\to s} = H_f[n_f] + \sum_e \frac{1}{2}(1-U^f_e )
\end{equation}
The sign in the definition of $U^f_e$ is determined by a certain 1-chain $E$ with values in $\ZZ_2$. This chain encodes a choice of spin structure on $\Sigma$. 

If $\fT_f$ admits a Levin-Wen construction, we will show how to incorporate directly the effect of the $\Pi$ particles to get a
string net construction for $\fT_s$.

\subsection{Open questions and future directions}

Classification of fermionic SPT phases can be generalized in several directions. Most obviously, one would like to classify SPT phases protected by $\hat G$ which is a central extension of $G$ by $\ZZ_2^f$. A natural guess is that the corresponding shadow theory must have both ordinary symmetry $G$ and a fermionic one-form symmetry $\ZZ_2$, but with a mixed anomaly between the two. 

The mixed anomaly is determined by the extension class of the short exact sequence $0\ra \ZZ_2\ra \hat G\ra G\ra 0$. Concretely, this means that the shadow theory is described by a $G$-graded fusion category, but the crossing conditions for the fermion are modified by the 2-cocycle $\psi_2$ representing the extension class. 
Physically, intersections of domain walls implementing $G$ symmetry transformations will carry non-trivial fermion number. 

Following the approach of Appendix B, we get  a generalization of the Gu-Wen equations:
\begin{equation}
\delta\nu_3=\frac12 n_2\cup n_2+\frac12 \psi_2\cup n_2,\quad \delta n_2=0.
\end{equation}
It would be interesting to study the group structure on the space of such fermionic SPT phases.  

Another possible generalization is to extend the discussion to unorientable theories. This is important for classifying fermionic SPT and SET phases with anti-unitary symmetries.

It would be very interesting to extend the shadow theory approach to fermionic phases in higher dimensions. For example, it has been proposed in \cite{GaiottoKapustin} that 3+1d fermionic phases are related to bosonic phases with an anomalous 2-form $\ZZ_2$ symmetry, where the 5d anomaly action is
\begin{equation}
\int_{M_5} Sq^2 \,C_3,
\end{equation}
with $C$ being the background 3-form $\ZZ_2$ gauge field and $Sq^2$ denoting a Steenrod square. Gu-Wen equations in 3+1d can be interpreted as describing shadow theories of this sort, and it should be possible to use the methods of this paper to produce more general SPT phases.

Optimistically, one might hope that every fermionic theory in every dimension has a bosonic shadow. Recent results of Brundan and Ellis \cite{Brundan:2016aa} indicate that this is true in 2+1d. In particular, it would be very interesting to understand shadows of general spin-TFTs in 2+1d which have framing anomalies. This would require developing the theory of "super modular tensor categories.''

Finally, we hope that the study of shadows of fermionic theories could shed light on the fermion doubling problem in lattice field theory.

\section{Spherical fusion categories and fermions}
The bosonic theory $\fT_f[\nu_3, n_2]$ we will associate to the Gu-Wen fermionic SPT phases 
belongs to the special class of TFTs which admit a Turaev-Viro state sum construction of the partition function \cite{Turaev:1992hq}
and a Levine-Wen string net construction of a microscopic lattice Hamiltonian \cite{Levin:2004mi}. 

The Turaev-Viro construction allows one to define a large class of three-dimensional topological field theories. 
The mathematical input for the construction is a spherical fusion category $\cC$. The output is the partition function 
of a topological field theory, whose quasi-particles are described by the modular tensor category $Z[\cC]$, the 
Drinfeld center of $\cC$. 

The physical meaning of the mathematical input becomes manifest through the following observation:
the Turaev-Viro construction produces topological field theories which admit a canonical topological boundary condition,
which in turns supports topological line defects labelled by the objects in $\cC$ \cite{2010arXiv1004.1533K}. 

\begin{figure}
\begin{center}
\includegraphics[width=13cm]{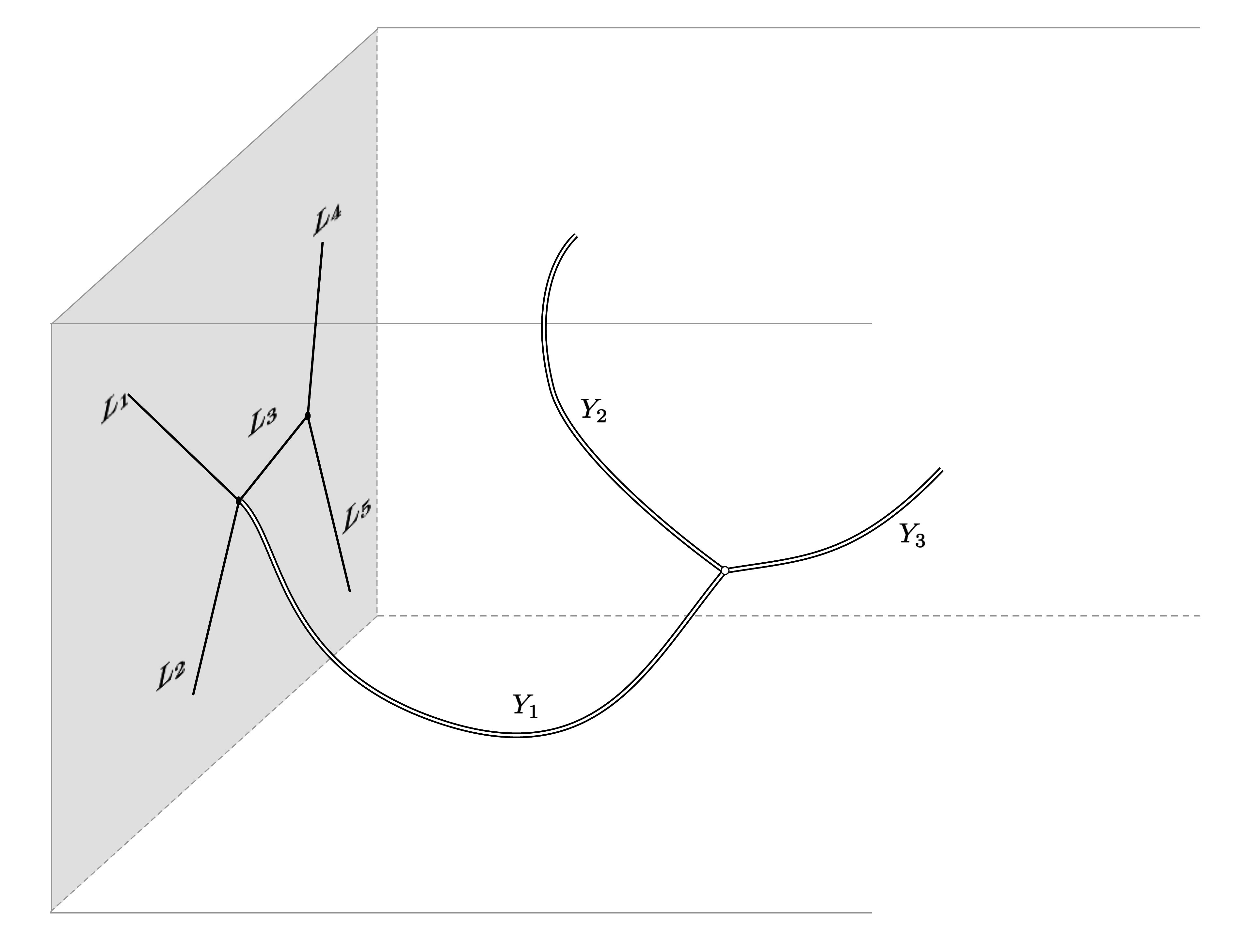}
\end{center}
\caption{A topological field theory with a gapped boundary condition. Boundary lines are labelled by objects $L_i$ in a spherical fusion category $\cC$
which controls their topological fusion and junctions. Bulk lines are labelled by objects $Y_a$ in a modular tensor category which can be recovered as the Drinfeld center 
$Z[\cC]$ of the boundary lines. Junctions of lines are labelled by choices of local operators, i.e. elements in certain morphism spaces. 
We use a double-line notation to indicate the dependence of bulk lines on a choice of framing. The partition function can be computed 
by a Turaev-Viro state sum.}
\label{fig:bougap}
\end{figure}

This suggests the following physical statement: any (irreducible, unitary) three-dimensional topological field theory $\fT$ equipped with a 
topological boundary condition $\fB$ will admit a Turaev-Viro construction based on the category of topological line defects 
supported on $\fB$. 

At first sight, it may appear surprising that the whole bulk topological field theory could be reconstructed from the properties 
of a single boundary condition. This is related to the cobordism hypothesis \cite{Lurie:2009aa}. 
There is a simple ``swiss cheese'' argument which demonstrates this fact in $2+1d$ and motivates the 
structure of the Turaev-Viro state sum model, which we review in a later section \ref{sec:swiss}. 

The same argument justifies the observation that several properties and enrichments of the bulk theory can be encoded in terms of the 
spherical fusion category $\cC$. For example, if $\fT$ has a non-anomalous ($0$-form) symmetry group $G$ then 
$\cC$ will admit an extension to a $G$-graded category $\cC_G$, which can be used to extend the Turaev-Viro
construction to manifolds endowed with a $G$-valued flat connection \cite{Turaev:2012aa}. 

With this motivation in mind, we can review some useful facts about spherical fusion categories ant their physical interpretation. 

\subsection{Categories of boundary line defects}

In the following we use the term topological field theory to denote the low energy/large distance effective field theory 
description of a gapped unitary quantum field theory. Similarly, a topological boundary condition is 
simply the low energy description of a gapped boundary condition. 

The mathematical description of topological field theories 
involves a variety of operations which have an intuitive interpretation as a ``fusion'' of local operators or defects. 
The precise physical interpretation is that the local operators or defects to be fused are brought to relative distances 
which are still much larger than the gap, but smaller than the scale of the low energy effective field theory. 
This allows one to replace them by a single effective local operator or defect. 

A gapped system may have multiple vacua, either due to spontaneous breaking of a symmetry 
or to accidental degeneracy. In the bulk theory, the presence of multiple vacua manifests itself 
in the existence of non-trivial local operators, whose expectation value labels different vacua. 
Mathematically, the local operators which survive at very low energy form a ring under the fusion operation 
described above (because of cluster decomposition).  
The identity operator can be decomposed into a sum of idempotents which project the system to a specific vacuum:
\begin{equation}
1 = \sum_v 1_v \qquad \qquad \qquad 1_v 1_{v'} = \delta_{v,v'} 1_v
\end{equation}

The same idea applies to defects of lower co-dimension. As an example consider line defects, which could be the effective description of a quasi-particle 
or of a microscopic line defect. Line defects can be fused with each other and may support non-trivial local operators, including 
local operators which interpolate between two or more lines. Again, the existence of a local multiplicity of vacua for a line defect manifests itself 
in the existence of non-trivial idempotent local operators.

Mathematically, line defects can be organized into a fusion category. The objects in the category are the line defects themselves, and 
the morphisms are the local operators interpolating between two line defects. The physical fusion operation is encoded into a tensor product operation 
and accidental degeneracies into a sum operation. Line defects with no accidental degeneracy map to ``simple'' objects in the category. 

\begin{figure}
\begin{tikzpicture}[line width=1pt]
\begin{scope}[every node/.style={sloped,allow upside down}]
\coordinate (lowA) at (0,0);
\coordinate (highA) at (0,4);
\arrowpath{(lowA)}{(highA)}{0.5};
\node[below] at (lowA) {\scriptsize{$L$}};
\node[below] at ($(lowA) - (0, 0.5)$) {\scriptsize{(a)}};

\coordinate (lowB) at ($(lowA)+(4,0)$);
\coordinate (highB) at ($(lowB) + (0,4)$);
\coordinate (midB) at ($(lowB)!0.5!(highB)$);
\arrowpath{(lowB)}{(midB)}{0.5};
\arrowpath{(midB)}{(highB)}{0.5};
\draw[fill=black] (midB) circle(2pt);
\node[right] at (midB) {\scriptsize{$m$}};
\node[below] at (lowB) {\scriptsize{$L_1$}};
\node[above] at (highB) {\scriptsize{$L_2$}};
\node[below] at ($(lowB) - (0, 0.5)$) {\scriptsize{(b)}};

\coordinate (lowC) at ($(lowB)+(4,0)$);
\coordinate (highC) at ($(lowC) + (0,4)$);
\coordinate (midC) at ($(lowC)!0.5!(highC)$);
\arrowpath{(lowC)}{(midC)}{0.5};
\arrowpath{(midC)}{(highC)}{0.5};
\coordinate (halfbox) at (0.25, 0.2);
\draw[fill=white] ($(midC) - (halfbox)$) rectangle ($(midC) + (halfbox)$);
\node at (midC) {\scriptsize{$m$}};
\node[below] at (lowC) {\scriptsize{$L_1$}};
\node[above] at (highC) {\scriptsize{$L_2$}};
\node[below] at ($(lowC) - (0, 0.5)$) {\scriptsize{(c)}};

\coordinate (lowD) at ($(lowC)+(4,0)$);
\coordinate (highD) at ($(lowD) + (0,4)$);
\coordinate (midD) at ($(lowD)!0.5!(highD)$);
\arrowpath{(lowD)}{(midD)}{0.5};
\arrowpath{(midD)}{(highD)}{0.5};
\coordinate (halfbox) at (0.25, 0.2);
\draw[fill=white] ($(midD) - (halfbox)$) rectangle ($(midD) + (halfbox)$);
\node at (midD) {\scriptsize{$\pi_1$}};
\node[below] at (lowD) {\scriptsize{$L_1 \oplus L_2$}};
\node[above] at (highD) {\scriptsize{$L_1$}};
\node[below] at ($(lowD) - (0, 0.5)$) {\scriptsize{(d)}};
\end{scope}
\end{tikzpicture}
\caption{Data of a category $\mathcal C$: (a) A line defect (shown here with its orientation) is an object in $\mathcal C$. (b) A local operator between two lines is a morphism in $\mathcal C$. (c) Another representation of the previous figure (common in mathematics literature) in which morphisms are denoted by boxes. (d) The direct sum $L_1 \oplus L_2$ of two line defects can be projected to an individual summand by a local operator}\label{fig:lines1}
\end{figure}
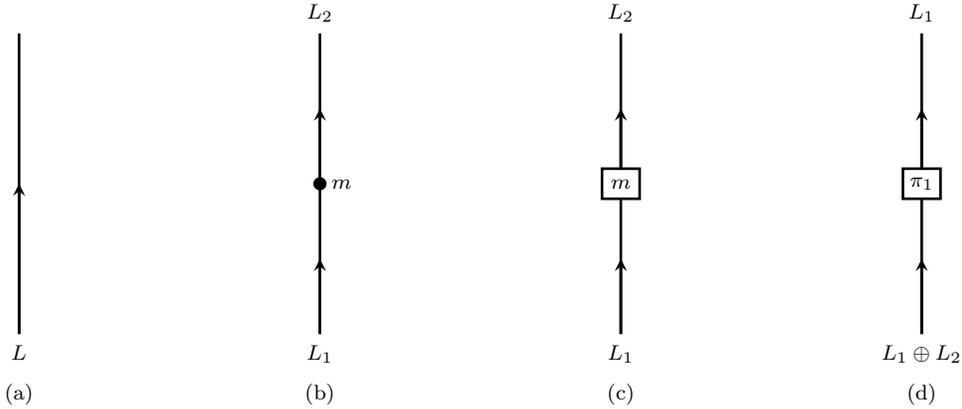

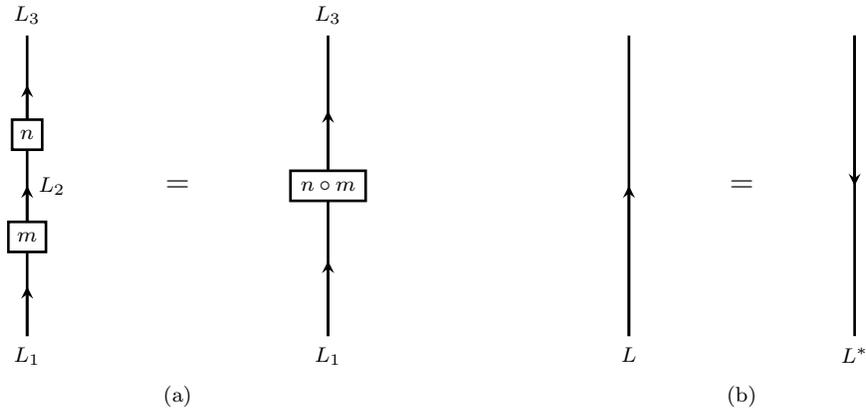
\begin{figure}
\centering
\begin{tikzpicture}[line width=1pt]
\begin{scope}[every node/.style={sloped,allow upside down}]
\coordinate (lowA1) at (0,0);
\coordinate (highA1) at ($(lowA1) + (0,4)$);
\coordinate (thirdA1) at ($(lowA1)!0.33!(highA1)$);
\coordinate (midA1) at ($(lowA1)!0.5!(highA1)$);
\coordinate (twothirdA1) at ($(lowA1)!0.67!(highA1)$);
\arrowpath{(lowA1)}{(thirdA1)}{0.5};
\arrowpath{(thirdA1)}{(twothirdA1)}{0.5};
\arrowpath{(twothirdA1)}{(highA1)}{0.5};
\coordinate (halfboxA11) at (0.25, 0.2);
\coordinate (halfboxA12) at (0.2, 0.2);
\draw[fill=white] ($(thirdA1) - (halfboxA11)$) rectangle ($(thirdA1) + (halfboxA11)$);
\draw[fill=white] ($(twothirdA1) - (halfboxA12)$) rectangle ($(twothirdA1) + (halfboxA12)$);
\node at (thirdA1) {\scriptsize{$m$}};
\node at (twothirdA1) {\scriptsize{$n$}};
\node[below] at (lowA1) {\scriptsize{$L_1$}};
\node[right] at (midA1) {\scriptsize{$L_2$}};
\node[above] at (highA1) {\scriptsize{$L_3$}};

\coordinate (eqA) at ($(midA1)+(2,0)$);
\node at (eqA) {$=$};

\coordinate (lowA2) at ($(lowA1)+(4,0)$);
\coordinate (highA2) at ($(lowA2) + (0,4)$);
\coordinate (midA2) at ($(lowA2)!0.5!(highA2)$);
\arrowpath{(lowA2)}{(midA2)}{0.5};
\arrowpath{(midA2)}{(highA2)}{0.5};
\coordinate (halfboxA2) at (0.5, 0.2);
\draw[fill=white] ($(midA2) - (halfboxA2)$) rectangle ($(midA2) + (halfboxA2)$);
\node at (midA2) {\scriptsize{$n \circ m$}};
\node[below] at (lowA2) {\scriptsize{$L_1$}};
\node[above] at (highA2) {\scriptsize{$L_3$}};

\node[below] at ($(lowA1)!0.5!(lowA2) + (0,-0.5)$) {\scriptsize{(a)}};

\coordinate (lowB1) at ($(lowA2) + (4,0)$);
\coordinate (highB1) at ($(lowB1) + (0,4)$);
\arrowpath{(lowB1)}{(highB1)}{0.5};
\node[below] at (lowB1) {\scriptsize{$L$}};

\coordinate (eqB) at ($(lowB1)!0.5!(highB1)+(1.5,0)$);
\node at (eqB) {$=$};

\coordinate (lowB2) at ($(lowB1) + (3,0)$);
\coordinate (highB2) at ($(lowB2) + (0,4)$);
\arrowpath{(highB2)}{(lowB2)}{0.5};
\node[below] at (lowB2) {\scriptsize{$L^*$}};

\node[below] at ($(lowB1)!0.5!(lowB2) + (0,-0.5)$) {\scriptsize{(b)}};

\end{scope}
\end{tikzpicture}
\caption{Various operations: (a) Fusion of local operators gives rise to composition of morphisms. (b) Changing the orientation of a line defect gives rise to the operation of taking \emph{dual} of an object.}
\label{fig:lines2}
\end{figure}

\begin{figure}
\begin{tikzpicture}[line width=1pt]
\begin{scope}[every node/.style={sloped,allow upside down}]
\coordinate (lowA) at (0,0);
\coordinate (lowB) at ($(lowA)+(0.5,0)$);
\coordinate (highA) at ($(lowA)+(0,4)$);
\coordinate (highB) at ($(lowB)+(0,4)$);

\arrowpath{(lowA)}{(highA)}{0.5};
\arrowpath{(lowB)}{(highB)}{0.5};

\node[below] at (lowA) {\scriptsize{$A$}};
\node[below] at (lowB) {\scriptsize{$B$}};

\coordinate (eq) at ($(lowB)+(1,2)$);
\node at (eq) {$=$};

\coordinate (lowAB) at ($(lowB)+(2,0)$);
\coordinate (highAB) at ($(lowAB)+(0,4)$);

\arrowpath{(lowAB)}{(highAB)}{0.5};

\node[below] at (lowAB) {\scriptsize{$A\otimes B$}};

\node[below] at ($(eq)-(0,2.5)$) {\scriptsize{(a)}};

\coordinate (C) at ($(lowAB)+(2.2,0)$);
\coordinate (mid) at ($(C)+(0,2)$);
\coordinate (A) at ($(mid)+(-0.7,2)$);
\coordinate (B) at ($(mid)+(0.7,2)$);

\arrowpath{(C)}{(mid)}{0.5};
\arrowpath{(mid)}{(A)}{0.5};
\arrowpath{(mid)}{(B)}{0.5};
\draw[fill=black] (mid) circle(2pt);

\node[below] at (C) {\scriptsize{$C$}};
\node[above] at (A) {\scriptsize{$A$}};
\node[above] at (B) {\scriptsize{$B$}};

\coordinate (eq2) at ($(mid)+(1.25,0)$);
\node at (eq2) {$=$};

\coordinate (C2) at ($(C)+(2.5,0)$);
\coordinate (mid2) at ($(C2)+(0,2)$);
\coordinate (AB) at ($(mid2)+(0,2)$);

\arrowpath{(C2)}{(mid2)}{0.5};
\arrowpath{(mid2)}{(AB)}{0.5};
\coordinate (halfbox) at (0.25, 0.2);
\draw[fill=white] ($(mid2) - (halfbox)$) rectangle ($(mid2) + (halfbox)$);

\node[below] at (C2) {\scriptsize{$C$}};
\node[above] at (AB) {\scriptsize{$A\otimes B$}};

\node[below] at ($(eq2)-(0,2.5)$) {\scriptsize{(b)}};

\coordinate (lowC1) at ($(C2)+(2,0)$);
\coordinate (highC1) at ($(lowC1) + (0,4)$);
\coordinate (midC1) at ($(lowC1)!0.5!(highC1)$);
\arrowpath{(lowC1)}{(midC1)}{0.5};
\arrowpath{(midC1)}{(highC1)}{0.5};
\coordinate (halfboxC1) at (0.25, 0.2);
\draw[fill=white] ($(midC1) - (halfboxC1)$) rectangle ($(midC1) + (halfboxC1)$);
\node at (midC1) {\scriptsize{$m$}};
\node[below] at (lowC1) {\scriptsize{$L_1$}};
\node[above] at (highC1) {\scriptsize{$L_3$}};

\coordinate (lowC2) at ($(lowC1)+(1,0)$);
\coordinate (highC2) at ($(lowC2) + (0,4)$);
\coordinate (midC2) at ($(lowC2)!0.5!(highC2)$);
\arrowpath{(lowC2)}{(midC2)}{0.5};
\arrowpath{(midC2)}{(highC2)}{0.5};
\coordinate (halfboxC2) at (0.2, 0.2);
\draw[fill=white] ($(midC2) - (halfboxC2)$) rectangle ($(midC2) + (halfboxC2)$);
\node at (midC2) {\scriptsize{$n$}};
\node[below] at (lowC2) {\scriptsize{$L_2$}};
\node[above] at (highC2) {\scriptsize{$L_4$}};

\coordinate (eqC) at ($(lowC2)!0.5!(highC2)+(1,0)$);
\node at (eqC) {$=$};

\coordinate (lowC3) at ($(lowC2)+(2.1,0)$);
\coordinate (highC3) at ($(lowC3) + (0,4)$);
\coordinate (midC3) at ($(lowC3)!0.5!(highC3)$);
\arrowpath{(lowC3)}{(midC3)}{0.5};
\arrowpath{(midC3)}{(highC3)}{0.5};
\coordinate (halfboxC3) at (0.5, 0.2);
\draw[fill=white] ($(midC3) - (halfboxC3)$) rectangle ($(midC3) + (halfboxC3)$);
\node at (midC3) {\scriptsize{$m \otimes n$}};
\node[below] at (lowC3) {\scriptsize{$L_1 \otimes L_2$}};
\node[above] at (highC3) {\scriptsize{$L_3 \otimes L_4$}};

\node[below] at ($(lowC2)!0.5!(lowC3) - (0, 0.5)$) {\scriptsize{(c)}};
\end{scope}
\end{tikzpicture}
\caption{Fusion: (a) Fusion of line defects gives rise to the tensor product of objects. (b) Three line defects coming together with a local operator placed at the point of intersection can be interpreted as a morphism from one line defect to the tensor product of other two line defects.  (c) Local operators between lines can also be fused to give rise to tensor product of morphisms.}\label{fig:lines3}
\end{figure}
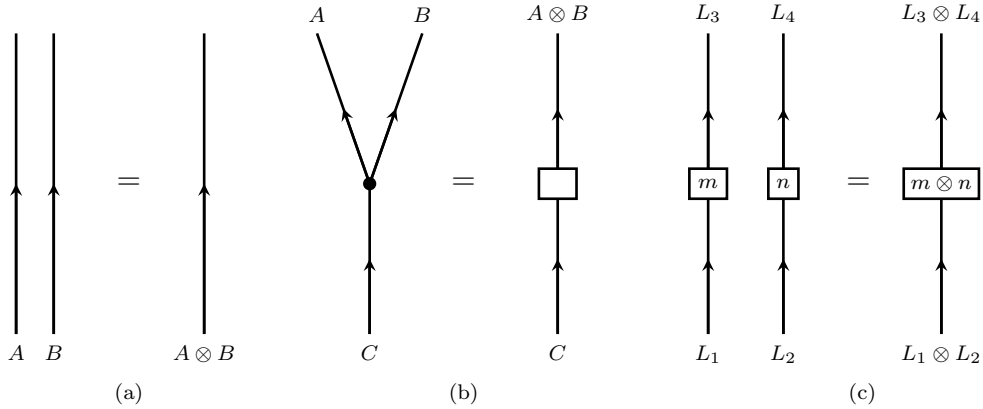

\begin{figure}
\centering
\begin{tikzpicture}[line width=1pt]
\begin{scope}[every node/.style={sloped,allow upside down}]
\coordinate (lowA) at (0,0);
\coordinate (highA) at ($(lowA)+(0,4)$);
\coordinate (lowB) at ($(lowA)+(0.2,0)$);
\coordinate (lowC) at ($(lowA)+(1.5,0)$);
\coordinate (highC) at ($(lowC)+(0,4)$);
\coordinate (highB) at ($(highC)-(0.2,0)$);
\coordinate (midBl) at ($(lowB)!0.5!(highB) - (0.1,0)$);
\coordinate (midBr) at ($(lowB)!0.5!(highB) + (0.1,0)$);
\arrowpath{(lowA)}{(highA)}{0.5};
\node at ($(lowA)-(0.2,0.2)$) {\scriptsize{$A$}};
\draw (lowB) .. controls (highA) and (lowC) .. (highB);
\node at ($(lowB)+(0.2,-0.2)$) {\scriptsize{$B$}};
\arrowpath{(midBl)}{(midBr)}{0.5};
\arrowpath{(lowC)}{(highC)}{0.5};
\node[below] at (lowC) {\scriptsize{$C$}};

\coordinate (eq) at ($(lowC)!0.5!(highC) + (0.75,0)$);
\node at (eq) {$=$};

\coordinate (lowT) at ($(lowC) + (1.5,0)$);
\coordinate (highT) at ($(lowT) + (0,4)$);
\coordinate (midT) at ($(lowT)!0.5!(highT)$);
\draw[fill=black] (midT) circle(2pt);
\arrowpath{(lowT)}{(highT)}{0.25};
\arrowpath{(lowT)}{(highT)}{0.75};
\node[below] at (lowT) {\scriptsize{$(A \otimes B)\otimes C$}};
\node[above] at (highT) {\scriptsize{$A \otimes (B\otimes C)$}};

\node[below] at ($(lowC) - (0,0.5)$) {\scriptsize{(a)}};

\coordinate (lowA1) at ($(lowT)+(2.5,0)$);
\coordinate (highA1) at ($(lowA1) + (0,2)$);
\coordinate (lowA2) at ($(lowA1)+(0.5,0)$);
\coordinate (highA2) at ($(lowA2) + (0,2)$);
\arrowpath{(lowA1)}{(highA1)}{0.5};
\arrowpath{(highA2)}{(lowA2)}{0.5};
\draw (highA1) arc[radius=0.25, start angle=180, end angle=0];
\node[below] at (lowA1) {\scriptsize{$L$}};

\coordinate (eqA) at ($(highA2)+(0.75,0)$);
\node at (eqA) {$=$};

\coordinate (lowA3) at ($(lowA2)+(1.75,0)$);
\coordinate (highA3) at ($(lowA3) + (0,4)$);
\coordinate (midA3) at ($(lowA3)!0.5!(highA3)$);
\arrowpath{(lowA3)}{(midA3)}{0.5};
\arrowpath{(midA3)}{(highA3)}{0.5};
\coordinate (halfboxA3) at (0.25, 0.2);
\draw[fill=white] ($(midA3) - (halfboxA3)$) rectangle ($(midA3) + (halfboxA3)$);
\node at (midA3) {\scriptsize{$e_L$}};
\node[below] at (lowA3) {\scriptsize{$L \otimes L^*$}};
\node[above] at (highA3) {\scriptsize{$1$}};

\node[below] at ($(lowA2) + (0.75,-0.5)$) {\scriptsize{(b)}};

\coordinate (lowB1) at ($(midA3)+(2,0)$);
\coordinate (highB1) at ($(lowB1) + (0,2)$);
\coordinate (lowB2) at ($(lowB1)+(0.5,0)$);
\coordinate (highB2) at ($(lowB2) + (0,2)$);
\arrowpath{(highB2)}{(lowB2)}{0.5};
\arrowpath{(lowB1)}{(highB1)}{0.5};
\draw (lowB1) arc[radius=0.25, start angle=180, end angle=360];
\node[above] at (highB1) {\scriptsize{$L$}};

\coordinate (eqB) at ($(lowB2)+(0.75,0)$);
\node at (eqB) {$=$};

\coordinate (lowB3) at ($(lowB2)+(1.75,-2)$);
\coordinate (highB3) at ($(lowB3) + (0,4)$);
\coordinate (midB3) at ($(lowB3)!0.5!(highB3)$);
\arrowpath{(lowB3)}{(midB3)}{0.5};
\arrowpath{(midB3)}{(highB3)}{0.5};
\coordinate (halfboxB3) at (0.25, 0.2);
\draw[fill=white] ($(midB3) - (halfboxB3)$) rectangle ($(midB3) + (halfboxB3)$);
\node at (midB3) {\scriptsize{$i_L$}};
\node[above] at (highB3) {\scriptsize{$L \otimes L^*$}};
\node[below] at (lowB3) {\scriptsize{$1$}};

\node[below] at ($(eqB) + (0,-2.5)$) {\scriptsize{(c)}};
\end{scope}
\end{tikzpicture}
\caption{Canonical maps: (a) Placing the lines as shown and fusing them gives rise to a canonical \emph{associator} map. (b) Folding a line as shown and fusing it with itself gives rise to a canonical \emph{evaluation} map. (c) Folding a line as shown and fusing it with itself gives rise to a canonical \emph{co-evaluation} map.}\label{fig:lines4}
\end{figure}
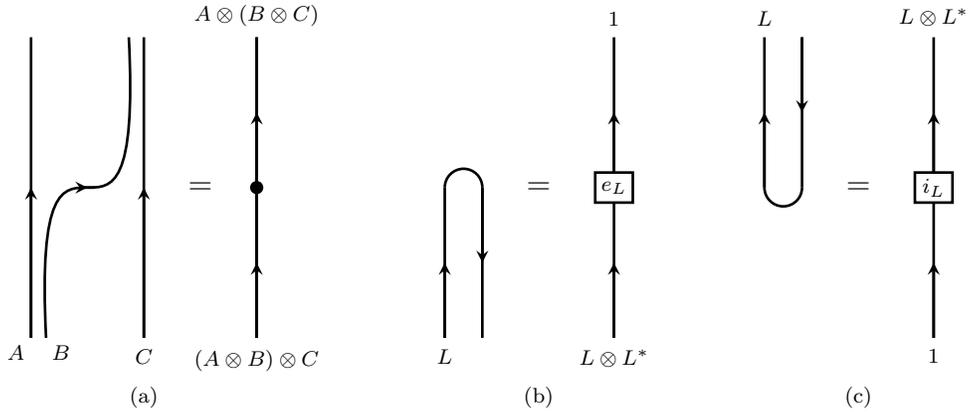

Depending on the dimension of space-time, the category of line defects will have further structures and constraints. 
Here we are interested in line defects which live on a gapped boundary condition. See Figures \ref{fig:lines1}, \ref{fig:lines2}, \ref{fig:lines3}, \ref{fig:lines4} for examples.

\begin{figure}
\centering
\begin{tikzpicture}[line width=1pt]
\begin{scope}[every node/.style={sloped,allow upside down}]
\coordinate (center) at (0,0);
\coordinate (frontup) at ($(center)+(0,0.25)$);
\coordinate (frontdown) at ($(center)+(0,-1)$);
\coordinate (topright) at ({2.5 * cos(45)}, {2.5 * sin(45)});
\coordinate (topleft) at ({2.5 * cos(135)}, {2.5 * sin(135)});
\coordinate (bottomleft) at ({2.5 * cos(225)}, {2.5 * sin(225)});
\coordinate (bottomright) at ({2.5 * cos(315)}, {2.5 * sin(315)});
\coordinate (backleft) at ($(center)+(-0.5,1.5)$);
\coordinate (backright) at ($(center)+(0.5,1.5)$);

\draw (center) circle (2.5 cm);
\arrowpathdashed{(backleft)}{(backright)}{0.6};
\arrowpathdashed{(backleft)}{(bottomleft)}{0.6};
\arrowpathdashed{(backright)}{(bottomright)}{0.6};
\arrowpath{(frontdown)}{(frontup)}{0.6};
\arrowpath{(frontup)}{($(frontup)!0.65!(topleft)+(0,-0.05)$)}{0.99};
\arrowpath{(frontup)}{($(frontup)!0.65!(topright)+(0,-0.05)$)}{0.99};
\arrowpath{($(frontdown)!0.5!(bottomleft)+(0.085,0)$)}{(frontdown)}{0.01};
\arrowpath{($(frontdown)!0.5!(bottomright)-(0.085,0)$)}{(frontdown)}{0.01};
\draw (frontup)..controls($(topright)+(-0.15,-0.2)$)..(topright);
\draw (frontup)..controls($(topleft)+(0.15,-0.2)$)..(topleft);
\draw (frontdown)..controls($(bottomright)+(0,-0.1)$)..(bottomright);
\draw (frontdown)..controls($(bottomleft)+(0,-0.1)$)..(bottomleft);
\draw (frontup) to (frontdown);
\draw[dashed] (topleft)..controls($(topleft)+(0,0.1)$)..(backleft);
\draw[dashed] (topright)..controls($(topright)+(0,0.1)$)..(backright);
\draw[dashed] (bottomleft) to (backleft);
\draw[dashed] (bottomright) to (backright);
\draw[dashed] (backleft) to (backright);

\draw[fill=black] (backleft) circle(2pt);
\draw[fill=black] (backright) circle(2pt);
\draw[fill=black] (frontup) circle(2pt);
\draw[fill=black] (frontdown) circle(2pt);

\node[below right] at (frontup) {\scriptsize{$m_1$}};
\node[above right] at (frontdown) {\scriptsize{$m_4$}};
\node[above] at (backleft) {\scriptsize{$m_2$}};
\node[above] at (backright) {\scriptsize{$m_3$}};

\node[left] at ($(frontup)!0.5!(frontdown)$) {\scriptsize{$A$}};
\node[below] at ($(backleft)!0.5!(backright)$) {\scriptsize{$E$}};
\node[left] at ($(backleft)!0.6!(bottomleft)$) {\scriptsize{$D$}};
\node[right] at ($(backright)!0.6!(bottomright)$) {\scriptsize{$F$}};
\node[below left] at ($(frontup)!0.65!(topleft)$) {\scriptsize{$B$}};
\node[below right] at ($(frontup)!0.65!(topright)$) {\scriptsize{$C$}};

\node[below] (a) at ($(center) + (0,-3)$) {\scriptsize{(a)}};

\coordinate (023) at ($(center)+(7,-2)$);
\coordinate (012) at ($(023)+(-1.5,1.2)$);
\coordinate (123) at ($(023)+(1.5,2.4)$);
\coordinate (013) at ($(023)+(0,3.6)$);
\coordinate (controlu) at ($(013)+(3,3)$);
\coordinate (controld) at ($(023)+(3,-3)$);
\coordinate (mid) at ($(controlu)!0.5!(controld)-(0.75,0)$);

\arrowpath{(023)}{(012)}{0.5};
\arrowpath{(023)}{(123)}{0.5};
\arrowpath{(012)}{(123)}{0.5};
\arrowpath{(012)}{(013)}{0.5};
\arrowpath{(123)}{(013)}{0.5};
\draw (013)..controls(controlu) and (controld)..(023);
\arrowpath{($(mid)+(0,0.1)$)}{($(mid)-(0,0.1)$)}{0.5};

\node[left] at (012) {\scriptsize{$m_2$}};
\node[below left] at (023) {\scriptsize{$m_1$}};
\node[above left] at (013) {\scriptsize{$m_4$}};
\node[right] at (123) {\scriptsize{$m_3$}};

\node[below left] at ($(023)!0.5!(012)$) {\scriptsize{$B$}};
\node[right] at ($(023)!0.5!(123)$) {\scriptsize{$C$}};
\node[above] at ($(012)!0.5!(123)$) {\scriptsize{$E$}};
\node[left] at ($(012)!0.5!(013)$) {\scriptsize{$D$}};
\node[above right] at ($(123)!0.5!(013)$) {\scriptsize{$F$}};
\node[right] at (mid) {\scriptsize{$A$}};

\draw[fill=black] (012) circle(2pt);
\draw[fill=black] (013) circle(2pt);
\draw[fill=black] (023) circle(2pt);
\draw[fill=black] (123) circle(2pt);

\node[below] (b) at ($(a) + (7.5,0)$) {\scriptsize{(b)}};
\end{scope}
\end{tikzpicture}
\caption{(a) A graph $\Gamma$ of boundary line defects drawn on a sphere. (b) The same graph drawn on a plane obtained after removing a point from the sphere.} \label{fig:sphere}
\end{figure}
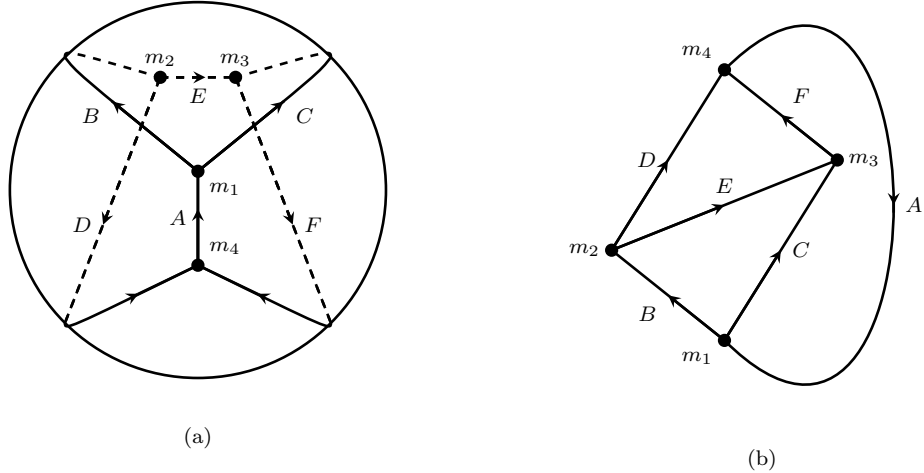

If the boundary condition itself has a single vacuum, the boundary line defects are expected to form a spherical fusion category $\cC$. 
The term spherical denotes a set of properties with a simple physical interpretation. Any graph $\Gamma$ of line defects 
drawn on a two-sphere, with a specific choice of local operators at the vertices, will produce a state in the two-sphere Hilbert space of the 
bulk theory. As the latter space is one-dimensional, the graph will effectively evaluate to a number $Z_\Gamma$, which can be interpreted 
as the partition function of the theory for a three-ball decorated by $\Gamma$, normalized by the partition function of the bare three-ball. See Figure \ref{fig:sphere}.

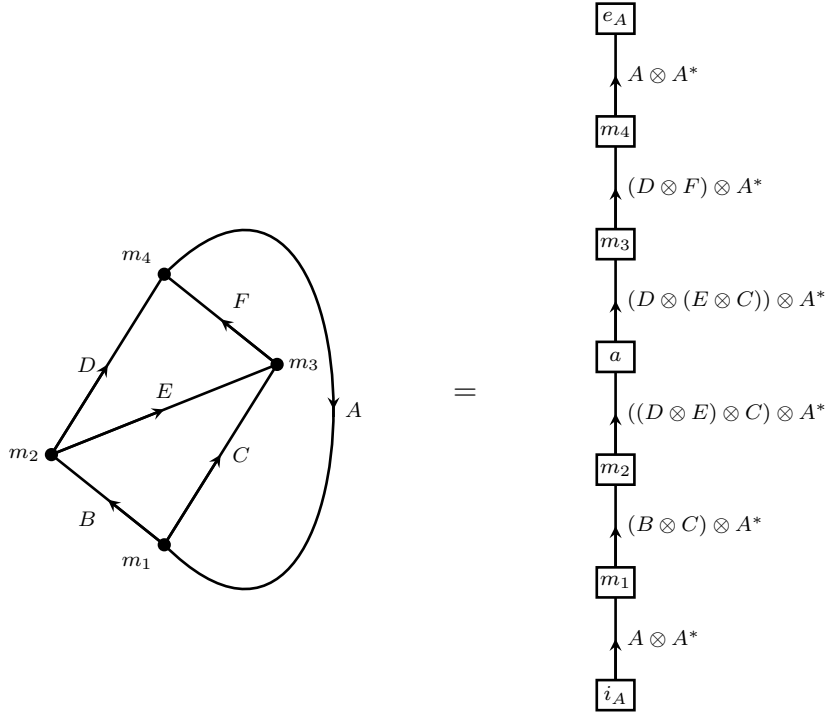
\begin{figure}
\centering
\begin{tikzpicture}[line width=1pt]
\begin{scope}[every node/.style={sloped,allow upside down}]
\coordinate (023) at (0,0);
\coordinate (012) at ($(023)+(-1.5,1.2)$);
\coordinate (123) at ($(023)+(1.5,2.4)$);
\coordinate (013) at ($(023)+(0,3.6)$);
\coordinate (controlu) at ($(013)+(3,3)$);
\coordinate (controld) at ($(023)+(3,-3)$);
\coordinate (mid) at ($(controlu)!0.5!(controld)-(0.75,0)$);

\arrowpath{(023)}{(012)}{0.5};
\arrowpath{(023)}{(123)}{0.5};
\arrowpath{(012)}{(123)}{0.5};
\arrowpath{(012)}{(013)}{0.5};
\arrowpath{(123)}{(013)}{0.5};
\draw (013)..controls(controlu) and (controld)..(023);
\arrowpath{($(mid)+(0,0.1)$)}{($(mid)-(0,0.1)$)}{0.5};

\node[left] at (012) {\scriptsize{$m_2$}};
\node[below left] at (023) {\scriptsize{$m_1$}};
\node[above left] at (013) {\scriptsize{$m_4$}};
\node[right] at (123) {\scriptsize{$m_3$}};

\node[below left] at ($(023)!0.5!(012)$) {\scriptsize{$B$}};
\node[right] at ($(023)!0.5!(123)$) {\scriptsize{$C$}};
\node[above] at ($(012)!0.5!(123)$) {\scriptsize{$E$}};
\node[left] at ($(012)!0.5!(013)$) {\scriptsize{$D$}};
\node[above right] at ($(123)!0.5!(013)$) {\scriptsize{$F$}};
\node[right] at (mid) {\scriptsize{$A$}};

\draw[fill=black] (012) circle(2pt);
\draw[fill=black] (013) circle(2pt);
\draw[fill=black] (023) circle(2pt);
\draw[fill=black] (123) circle(2pt);

\coordinate (eq) at ($(023)+(4,2)$);
\node at (eq) {$=$};

\coordinate (1) at ($(eq)+(2,-4)$);
\coordinate (2) at ($(1)+(0,1.5)$);
\coordinate (3) at ($(2)+(0,1.5)$);
\coordinate (4) at ($(3)+(0,1.5)$);
\coordinate (5) at ($(4)+(0,1.5)$);
\coordinate (6) at ($(5)+(0,1.5)$);
\coordinate (7) at ($(6)+(0,1.5)$);
\coordinate (halfbox) at (0.25, 0.2);

\arrowpath{(1)}{(2)}{0.5};
\arrowpath{(2)}{(3)}{0.5};
\arrowpath{(3)}{(4)}{0.5};
\arrowpath{(4)}{(5)}{0.5};
\arrowpath{(5)}{(6)}{0.5};
\arrowpath{(6)}{(7)}{0.5};

\draw[fill=white] ($(1) - (halfbox)$) rectangle ($(1) + (halfbox)$);
\draw[fill=white] ($(2) - (halfbox)$) rectangle ($(2) + (halfbox)$);
\draw[fill=white] ($(3) - (halfbox)$) rectangle ($(3) + (halfbox)$);
\draw[fill=white] ($(4) - (halfbox)$) rectangle ($(4) + (halfbox)$);
\draw[fill=white] ($(5) - (halfbox)$) rectangle ($(5) + (halfbox)$);
\draw[fill=white] ($(6) - (halfbox)$) rectangle ($(6) + (halfbox)$);
\draw[fill=white] ($(7) - (halfbox)$) rectangle ($(7) + (halfbox)$);

\node[right] at ($(1)!0.5!(2)$) {\scriptsize{$A\otimes A^*$}};
\node[right] at ($(2)!0.5!(3)$) {\scriptsize{$(B\otimes C)\otimes A^*$}};
\node[right] at ($(3)!0.5!(4)$) {\scriptsize{$((D\otimes E)\otimes C)\otimes A^*$}};
\node[right] at ($(4)!0.5!(5)$) {\scriptsize{$(D\otimes(E\otimes C))\otimes A^*$}};
\node[right] at ($(5)!0.5!(6)$) {\scriptsize{$(D\otimes F)\otimes A^*$}};
\node[right] at ($(6)!0.5!(7)$) {\scriptsize{$A\otimes A^*$}};

\node at (1) {\scriptsize{$i_A$}};
\node at (2) {\scriptsize{$m_1$}};
\node at (3) {\scriptsize{$m_2$}};
\node at (4) {\scriptsize{$a$}};
\node at (5) {\scriptsize{$m_3$}};
\node at (6) {\scriptsize{$m_4$}};
\node at (7) {\scriptsize{$e_A$}};
\end{scope}
\end{tikzpicture}
\caption{The computation of a graph on the plane involves the listed morphisms. Here $a$ denotes the associator tensored with identity morphism for $A^*$. The final result is the partition function $Z_\Gamma$ of the theory on a three-ball
decorated by the graph $\Gamma$.}\label{fig:sphere1}
\end{figure}

Mathematically, the graph is drawn on the plane as the evolution of a collection of lines, created, fused or annihilated 
at special points. The corresponding number is computed by Penrose calculus, 
as the composition of a sequence of maps associated to these individual processes,
which form the data of the spherical fusion category. See Figure \ref{fig:sphere1}. The axioms of the spherical fusion category guarantee that the answer is independent of how 
we draw the graph. This evaluation map for graphs on the two-sphere is the basic ingredient 
in state sum constructions.

If we are given two topological field theories $\fT$ and $\fT'$, with gapped boundary conditions $\fB$ and $\fB'$ 
associated to spherical fusion categories $\cC$ and $\cC'$, the product of the two theories with the product boundary condition 
is associated to the product $\cC \times \cC'$ of the fusion categories. 

\begin{figure}
\begin{tikzpicture}[line width=1pt]
\begin{scope}[every node/.style={sloped,allow upside down}]
\coordinate (lowO) at (0,0);
\coordinate (highO) at ($(lowO)+(1.5,4)$);
\coordinate (lowX) at ($(lowO)+(1.5,0)$);
\coordinate (highX) at ($(lowO)+(0,4)$);
\coordinate (OX) at (intersection of lowO--highO and lowX--highX);
\node [below] at (lowO) {\scriptsize{$O$}};
\node [below] at (lowX) {\scriptsize{$X$}};
\node [above] at (highO) {\scriptsize{$O$}};
\node [above] at (highX) {\scriptsize{$X$}};
\arrowpathdouble{(lowO)}{(highO)}{0.3};
\paddedline{(lowX)}{(highX)}{(0.1,0)};
\arrowpath{(lowX)}{(highX)}{0.3};

\coordinate (eq) at ( $(OX) + (1.5,0)$);
\node at (eq) {$=$};

\coordinate (lowOs) at ($(lowO) + (3,0)$);
\coordinate (highXs) at ($(lowOs) + (0,4)$);
\coordinate (lowXs) at ($(lowOs) + (1.5,0)$);
\coordinate (highOs) at ($(lowXs) + (0,4)$);
\coordinate (OXs) at (intersection of lowOs--highOs and lowXs--highXs);

\node [below] at (lowOs) {\scriptsize{$O$}};
\node [below] at (lowXs) {\scriptsize{$X$}};
\node [above] at (highOs) {\scriptsize{$O$}};
\node [above] at (highXs) {\scriptsize{$X$}};
\arrowpathdouble {(lowOs)}{(highOs)}{0.3};
\draw (lowXs)--(highXs);
\arrowpath{(lowXs)}{(highXs)}{0.3};
\coordinate (halfbox) at (0.3, 0.2);
\draw[fill=white] ($(OXs) - (halfbox)$) rectangle ($(OXs) + (halfbox)$);
\node at (OXs) {\scriptsize{$\beta_X$}};

\node[below] at ($(lowO)!0.5!(lowXs) - (0,0.5)$) {\scriptsize{(a)}};

\coordinate (lowO) at (7.5,0);
\coordinate (highO) at ($(lowO)+(1.5,4)$);
\coordinate (lowX) at ($(lowO)+(0.5,0)$);
\coordinate (highX) at ($(lowO)+(0.5,4)$);
\node [below] at (lowO) {\scriptsize{$O$}};
\node [below] at (lowX) {\scriptsize{$X$}};
\node [above] at (highO) {\scriptsize{$O$}};
\node [above] at (highX) {\scriptsize{$Y$}};
\arrowpathdouble{(lowO)}{(highO)}{0.7};
\paddedline{(lowX)}{(highX)}{(0.05,0)};
\draw[fill=black] ($(lowX)!0.5!(highX)$) circle(2pt);
\arrowpath{(lowX)}{(highX)}{0.2};
\arrowpath{(lowX)}{(highX)}{0.75};

\coordinate (eq) at ( $(lowX)!0.5!(highX) + (1.75,0)$);
\node at (eq) {$=$};

\coordinate (lowOs) at ($(lowO) + (3,0)$);
\coordinate (lowXs) at ($(lowOs) + (1,0)$);
\coordinate (highXs) at ($(lowXs) + (0,4)$);
\coordinate (highOs) at ($(highXs) + (0.5,0)$);
\coordinate (OXs) at (intersection of lowOs--highOs and lowXs--highXs);

\node [below] at (lowOs) {\scriptsize{$O$}};
\node [below] at (lowXs) {\scriptsize{$X$}};
\node [above] at (highOs) {\scriptsize{$O$}};
\node [above] at (highXs) {\scriptsize{$Y$}};
\arrowpathdouble {(lowOs)}{(highOs)}{0.3};
\paddedline{(lowXs)}{(highXs)}{(0.05,0)};
\draw[fill=black] ($(lowXs)!0.5!(highXs)$) circle(2pt);
\arrowpath{(lowXs)}{(highXs)}{0.25};
\arrowpath{(lowXs)}{(highXs)}{0.85};

\node[below] at ($(lowXs)!0.5!(lowX) - (0,0.5)$) {\scriptsize{(b)}};
\end{scope}
\end{tikzpicture}
\caption{Bulk lines and Drinfeld Center: (a) Bringing a bulk line $O$ to the boundary such that its image crosses a boundary line $X$ gives rise to a canonical \emph{half-braiding} given by morphism $\beta_X$. (b) Bringing $O$ to the boundary in two different ways as shown in the figure is equivalent and hence $\beta$ commutes with other morphisms.}\label{fig:drin}
\end{figure}
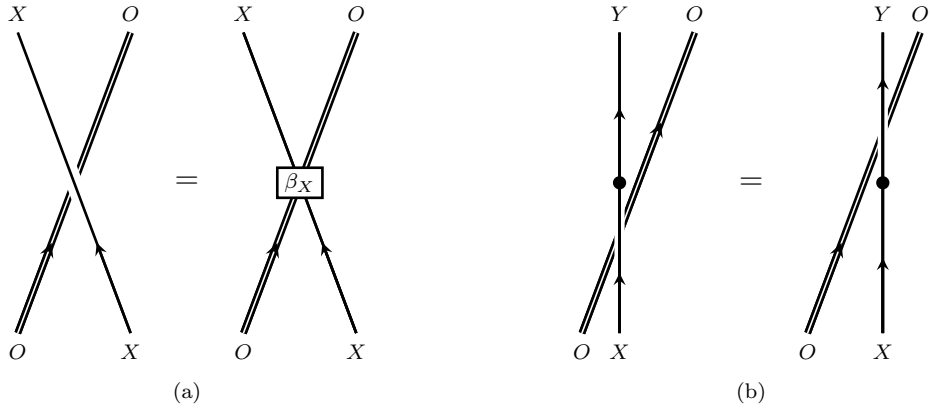

Bulk line defects can be fused with the boundary. If the boundary image crosses some pre-existing boundary line, 
the fusion produces some canonical local operator at the crossing. This physical process is encoded in the mathematical 
definition of Drinfeld center $Z[\cC]$. An element of the center is a pair $(O, \beta)$ of an object $O$ in $\cC$ together with a collection 
of crossing maps $\beta_X : O \otimes X \to X \otimes O$ for every other object $X$, satisfying certain axioms. See Figure \ref{fig:drin}.

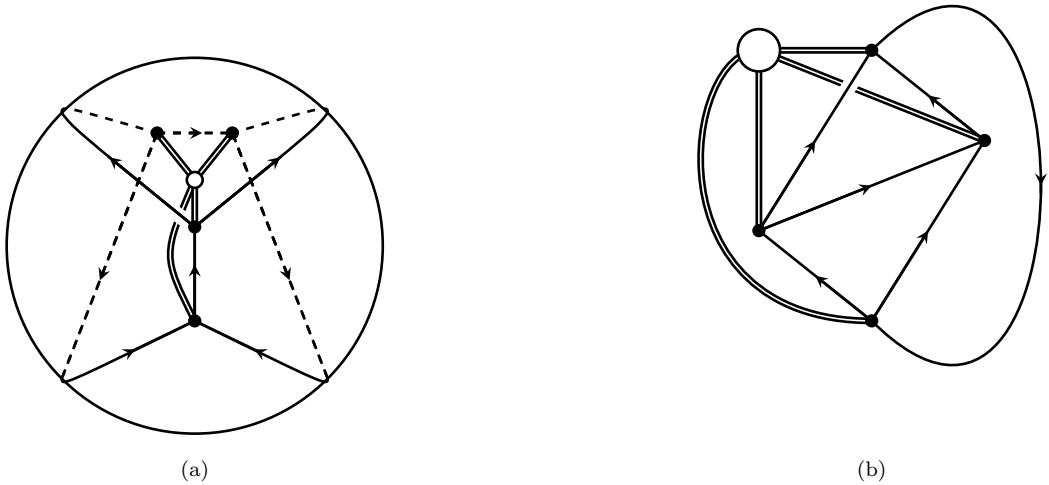
\begin{figure}
\begin{tikzpicture}[line width=1pt]
\begin{scope}[every node/.style={sloped,allow upside down}]

\coordinate (center) at (0,0);
\coordinate (frontup) at ($(center)+(0,0.25)$);
\coordinate (frontdown) at ($(center)+(0,-1)$);
\coordinate (topright) at ({2.5 * cos(45)}, {2.5 * sin(45)});
\coordinate (topleft) at ({2.5 * cos(135)}, {2.5 * sin(135)});
\coordinate (bottomleft) at ({2.5 * cos(225)}, {2.5 * sin(225)});
\coordinate (bottomright) at ({2.5 * cos(315)}, {2.5 * sin(315)});
\coordinate (backleft) at ($(center)+(-0.5,1.5)$);
\coordinate (backright) at ($(center)+(0.5,1.5)$);
\coordinate (backmid) at ($(backleft)!0.5!(backright)$);
\coordinate (blob) at ($(frontup)!0.5!(backmid)$);
\coordinate (control) at ($(backleft)!0.5!(bottomleft)+(0.7,0)$);
\coordinate (a) at ($(frontdown)+(0,-2)$);

\draw (center) circle (2.5 cm);
\draw[double] (backleft) to (blob);
\draw[double] (backright) to (blob);
\draw[double] (frontdown)..controls(control)..(blob);
\paddedline{(frontup)}{($(frontup)!0.2!(topleft)+(0,-0.05)$)}{(0,0.1)};
\draw[double] (frontup) to (blob);
\draw[fill=white] (blob) circle(3pt);
\arrowpathdashed{(backleft)}{(backright)}{0.6};
\arrowpathdashed{(backleft)}{(bottomleft)}{0.6};
\arrowpathdashed{(backright)}{(bottomright)}{0.6};
\arrowpath{(frontdown)}{(frontup)}{0.6};
\arrowpath{(frontup)}{($(frontup)!0.65!(topleft)+(0,-0.05)$)}{0.99};
\arrowpath{(frontup)}{($(frontup)!0.65!(topright)+(0,-0.05)$)}{0.99};
\arrowpath{($(frontdown)!0.5!(bottomleft)+(0.085,0)$)}{(frontdown)}{0.01};
\arrowpath{($(frontdown)!0.5!(bottomright)-(0.085,0)$)}{(frontdown)}{0.01};
\draw (frontup)..controls($(topright)+(-0.15,-0.2)$)..(topright);
\draw (frontup)..controls($(topleft)+(0.15,-0.2)$)..(topleft);
\draw (frontdown)..controls($(bottomright)+(0,-0.1)$)..(bottomright);
\draw (frontdown)..controls($(bottomleft)+(0,-0.1)$)..(bottomleft);
\draw (frontup) to (frontdown);
\draw[dashed] (topleft)..controls($(topleft)+(0,0.1)$)..(backleft);
\draw[dashed] (topright)..controls($(topright)+(0,0.1)$)..(backright);
\draw[dashed] (bottomleft) to (backleft);
\draw[dashed] (bottomright) to (backright);
\draw[dashed] (backleft) to (backright);

\draw[fill=black] (backleft) circle(2pt);
\draw[fill=black] (backright) circle(2pt);
\draw[fill=black] (frontup) circle(2pt);
\draw[fill=black] (frontdown) circle(2pt);

\node at (a) {\scriptsize{(a)}};

\coordinate (023) at ($(frontdown)+(9,0)$);
\coordinate (012) at ($(023)+(-1.5,1.2)$);
\coordinate (123) at ($(023)+(1.5,2.4)$);
\coordinate (013) at ($(023)+(0,3.6)$);
\coordinate (controlu) at ($(013)+(3,3)$);
\coordinate (controld) at ($(023)+(3,-3)$);
\coordinate (mid) at ($(controlu)!0.5!(controld)-(0.75,0)$);
\coordinate (blob) at ($(012)+(0,2.4)$);
\coordinate (controllu) at ($(blob)-(1,0)$);
\coordinate (controlld) at ($(023)-(3,0)$);

\draw[double] (123) to (blob);
\paddedline{(012)}{(013)}{(0.1,0)};
\draw[double] (013) to (blob);
\draw[double] (012) to (blob);
\arrowpath{(023)}{(012)}{0.5};
\arrowpath{(023)}{(123)}{0.5};
\arrowpath{(012)}{(123)}{0.5};
\arrowpath{(012)}{(013)}{0.5};
\arrowpath{(123)}{(013)}{0.5};
\draw (013)..controls(controlu) and (controld)..(023);
\arrowpath{($(mid)+(0,0.1)$)}{($(mid)-(0,0.1)$)}{0.5};
\draw[double] (023)..controls(controlld) and (controllu)..(blob);

\draw[fill=black] (012) circle(2pt);
\draw[fill=black] (013) circle(2pt);
\draw[fill=black] (023) circle(2pt);
\draw[fill=black] (123) circle(2pt);
\draw[fill=white] (blob) circle(8pt);

\node at ($(a)+(9,0)$) {\scriptsize{(b)}};
\end{scope}
\end{tikzpicture}
\caption{(a) A three-ball partition function decorated by a graph $\Gamma$ of bulk and boundary lines. (b) The graph is projected onto the sphere for evaluation. The different projections evaluate to the same result, thanks to the Drinfeld center axioms} \label{fig:drinsphere}
\end{figure}

These axioms have a simple interpretation. Consider a network of line defects in the three-ball, including boundary lines and 
bulk lines. If we project the network to a graph $\Gamma$ on the boundary and evaluate $Z_\Gamma$, the answer will not depend on the 
choice of projection.  See Figure \ref{fig:drinsphere}.
Every bulk line will thus map to an element of the center $Z[\cC]$. Conversely, the Turaev-Viro construction gives an explicit definition of 
a bulk line defect for every element of the center $Z[\cC]$. \footnote{From the point of view of the bulk theory, a gapped boundary condition can be characterized in terms 
the set of bulk lines which ``condense'' at the boundary, i.e. project to the trivial line on the boundary. 
They are a collection of mutually local bosons which is closed under fusion. }

\begin{figure}
\begin{tikzpicture}[line width=1pt]
\begin{scope}[every node/.style={sloped,allow upside down}]
\coordinate (lowb1) at (0,0);
\coordinate (lowb2) at ($(lowb1)+(1,0)$);
\coordinate (midb1) at ($(lowb1)+(0,1.5)$);
\coordinate (midb2) at ($(lowb2)+(0,1.5)$);
\coordinate (lowb3) at ($(lowb1)+(2,0)$);
\coordinate (highb3) at ($(lowb3)+(0,4)$);

\draw[double] (lowb1)--(midb1);
\draw[double] (lowb2)--(midb2);
\draw[double] (lowb3)--(highb3);
\draw[double] (midb1) arc[radius=0.5, start angle=180, end angle=0];

\node[below] at (lowb1) {\scriptsize{$L$}};
\node[below] at (lowb2) {\scriptsize{$L$}};
\node[below] at (lowb3) {\scriptsize{$L$}};

\coordinate (eq) at ($(lowb3)+(1,2)$);
\node at (eq) {$=$};

\coordinate (lowbb1) at ($(lowb3)+(2,0)$);
\coordinate (highbb1) at ($(lowbb1)+(0,4)$);
\coordinate (lowbb2) at ($(lowbb1)+(1,0)$);
\coordinate (lowbb3) at ($(lowbb2)+(1,0)$);
\coordinate (midbb2) at ($(lowbb2)+(0,1.5)$);
\coordinate (midbb3) at ($(lowbb3)+(0,1.5)$);

\draw[double] (lowbb2)--(midbb2);
\draw[double] (lowbb3)--(midbb3);
\draw[double] (lowbb1)--(highbb1);
\draw[double] (midbb2) arc[radius=0.5, start angle=180, end angle=0];

\node[below] at (lowbb1) {\scriptsize{$L$}};
\node[below] at (lowbb2) {\scriptsize{$L$}};
\node[below] at (lowbb3) {\scriptsize{$L$}};

\node[below] at ($(eq) + (0,-2.5)$) {\scriptsize{(a)}};

\coordinate (llowb1) at ($(lowbb3)+(1.75,0)$);
\coordinate (llowb2) at ($(llowb1)+(1.5,0)$);
\coordinate (hhighb2) at ($(llowb1)+(0,4)$);
\coordinate (hhighb1) at ($(llowb2)+(0,4)$);

\draw[double] (llowb1)--(hhighb1);
\paddedline{(llowb2)}{(hhighb2)}{(0.1,0)};
\draw[double] (llowb2)--(hhighb2);

\node[below] at (llowb1) {\scriptsize{$L$}};
\node[below] at (llowb2) {\scriptsize{$L$}};

\coordinate (eq2) at ($(llowb2)+(0.75,2)$);
\node at (eq2) {$=$};

\coordinate (pm) at ($(eq2)+(0.75,0)$);
\coordinate (llowbb1) at ($(llowb2)+(2,0)$);
\coordinate (hhighbb1) at ($(llowbb1)+(0,4)$);
\coordinate (llowbb2) at ($(llowbb1)+(1,0)$);
\coordinate (hhighbb2) at ($(llowbb2)+(0,4)$);

\draw[double] (llowbb2)--(hhighbb2);
\draw[double] (llowbb1)--(hhighbb1);

\node[below] at (llowbb1) {\scriptsize{$L$}};
\node[below] at (llowbb2) {\scriptsize{$L$}};
\node at (pm) {$\pm$};

\node[below] at ($(eq2) + (0,-2.5)$) {\scriptsize{(L)}};
\end{scope}
\end{tikzpicture}
\caption{$\mathbb{Z}_2$ 1-form symmetries: (a) There exists a bulk line $L$ with properties shown in the figure. (b) Half-braiding $L$ lines across each other gives a factor of $\pm 1$ when compared to $L$ lines without braiding. 
The factor of $+1$ arises for a bosonic 1-form symmetry generator $L \equiv B$ and $-1$ arises for a fermionic 1-form symmetry generator $L \equiv \Pi$. This minus sign implies that the symmetry is anomalous.}
\label{fig:oneform}
\end{figure}
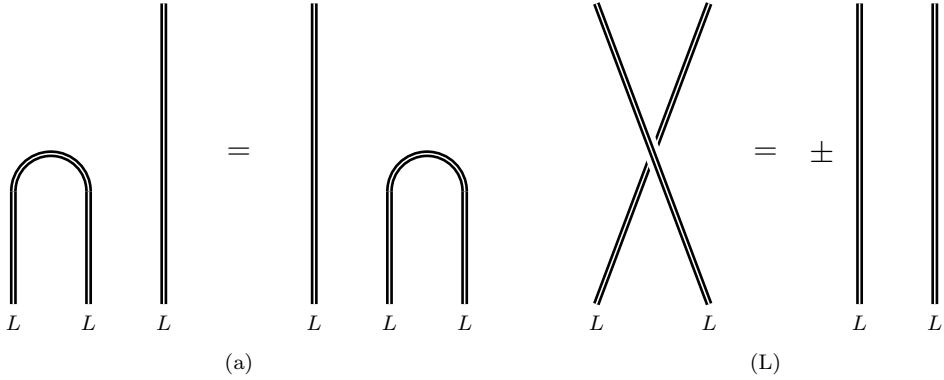

In particular, we can recognize the generators of bulk 1-form symmetries as special 
elements of the center. For example, a spherical fusion category $\cC_b$ represents a bulk theory equipped with a bosonic 
$\ZZ_2$ one-form symmetry if we can find a generator $B = (b, \beta)$, an element of the center $Z[\cC]$ such that $\beta_b =1_{b\otimes b}$ and 
such that there is an isomorphism $\xi_b:b \otimes b \to I$ with $\xi \otimes 1 = 1 \otimes \xi$ in $\Hom(b \otimes b \otimes b, b)$.
Essentially, this means that $B$ lines fuse to the identity and can be freely re-connected in pairs. See Figure \ref{fig:oneform}.

Similarly, a spherical fusion category $\cC_f$ represents a bulk theory equipped with a fermionic
$\ZZ_2$ one-form symmetry if we can find a generator $\Pi = (f, \beta)$, an element of the center $Z[\cC]$ such that $\beta_f =- 1_{f\otimes f}$ and 
such that there is an isomorphism $\xi_f:f \otimes f \to I$ with $\xi \otimes 1 = 1 \otimes \xi$ in $\Hom(f \otimes f \otimes f, f)$.
Essentially, this means that $f$ lines fuse to the identity and can be freely re-connected in pairs, at the price of a $-1$ sign for each 
crossing. See again Figure \ref{fig:oneform}. More generally, a monoidal category equipped with such a $\Pi$ is called  a ``monoidal $\Pi$-category'' in \cite{Brundan:2016aa}.

A couple variants to this setup may be useful. If the boundary condition has some accidental degeneracy, 
we should consider a spherical multi-fusion category. Local operators on the boundary 
are morphisms from the trivial line defect to itself, which is thus not simple. 
The category $\cC$ splits into multiple sub-categories $\cC_{a,b}$ representing 
line defects which interpolate between vacua $a$ and $b$. The objects in these categories fuse accordingly:
\begin{equation}
\cC_{a,b} \otimes \cC_{c,d} \in \delta_{b,c} \cC_{a,d}
\end{equation}

If the bulk theory and boundary condition have a non-anomalous discrete global symmetry $G$ (possibly broken at the boundary), 
we will have a $G$-graded spherical fusion category (see e.g. \cite{2012arXiv1208.5696T}), with sub-categories $\cC_g$ which fuse according to the group law:
\begin{equation}
\cC_{g} \otimes \cC_{g'} \in \cC_{g g'}
\end{equation}
The sector $\cC_e$ labelled by the group identity $e$ consists of standard boundary line defects while 
the other $\cC_g$ contain the boundary version of $g$-twist line defects. 

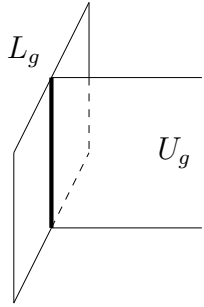
\begin{figure}
\begin{tikzpicture}[line width=0.2pt]
\begin{scope}[every node/.style={sloped,allow upside down}]
\node at (-5,0){};

\coordinate (A) at (0,0);
\coordinate (B) at ($(A)+(1,2)$);
\coordinate (C) at ($(B)-(0,2)$);
\coordinate (D) at ($(A)-(0,2)$);
\coordinate (AB) at ($(A)!0.5!(B)$);
\coordinate (CD) at ($(C)!0.5!(D)$);
\coordinate (rightup) at ($(AB)+(2,0)$);
\coordinate (rightdown) at ($(CD)+(2,0)$);
\coordinate (dash) at (intersection of B--C and AB--rightup);

\draw (A)--(B);
\draw (dash)--(B);
\draw (CD)--(D);
\draw (A)--(D);
\draw[ultra thick] (AB)--(CD);
\draw (AB)--(rightup);
\draw (CD)--(rightdown);
\draw[dashed] (dash)--(C);
\draw[dashed] (CD)--(C);

\node[above left] at (AB){$L_g$};
\node[left] at ($(rightup)!0.5!(rightdown)$){$U_g$};
\end{scope}
\end{tikzpicture}
\caption{Lines $L_g$ living at the intersection of a 0-form symmetry generator $U_g$ and the boundary form the sub-category $\mathcal{C}_g$.}
\end{figure}

Note that we can define a $G$-graded product of $G$-graded spherical fusion categories 
by letting $(\cC \times \cC')_g \equiv \cC_g \times \cC'_g$. Physically, this corresponds to taking the 
direct product of two theories $\fT$ and $\fT'$ and of their corresponding boundary conditions $\fB$ and $\fB'$.

If we gauge the symmetry $G$  (with Dirichlet boundary conditions for the gauge connection), 
all objects in $\cC$ become true boundary line defects. Bulk line defects are now associated to the center of the whole 
$\cC$. The center of $\cC$ includes Wilson loops, 
of the form $(I^{\oplus n}, \beta_{X_g} = T_g)$, $X_g$ is an arbitrary simple object of $\cC_g$ and the matrices $T_g$ define an $n$-dimensional representation of $G$.
\footnote{We are identifying here $\Hom_{\cC}(I^{\oplus n} \otimes X_g, X_g \otimes I^{\oplus n})$ 
with $n \times n$ matrices.} 
If $G$ is abelian, the Wilson loops are labelled by characters in the dual group $G^*$  and generate a non-anomalous  $G^*$  1-form symmetry. 

We can also gauge a subgroup $H$ of $G$. The resulting $H$ gauge theory should have a residual 
global symmetry given by the quotient $G_H=N_G(H)/H$ of the normalizer of $H$ by $H$. 
The corresponding $G_H$-graded category consists of 
\begin{equation}
\cC_{[g]} = \bigcup_{h \in H} \cC_{h g}
\end{equation}

Later in the paper, we will find it useful to build some interesting $G$-graded categories 
starting from SPT phases for a central extension $\hat G$ of $G$ by an Abelian group
and gauging the Abelian group as described above. 

Although a 1-form symmetry generator $B$ (or $\Pi$) for a $G$-graded theory is defined as a special element in $Z[\cC_e]$, 
we will often be interested in 1-form symmetries which are compatible with turning on $G$-flat connections 
or even gauging $G$. We will see that this is the case if $B$ (or $\Pi$) admits a lift to $Z[\cC]$. The lift may not be unique and
different lifts can be thought of as different ways to equip the theory with both $G$ symmetry and 1-form $\ZZ_2$ symmetry. 

\subsection{Example: toric code}
The simplest example of a category of boundary line defects occurs in the toric code, also known as topological $\ZZ_2$ gauge theory in 2+1d. 
Recall that the toric code has four quasi-particles, corresponding in the gauge theory to a trivial defect $1$, a Wilson 
loop $e$, a flux line $m$ and the fusion $\epsilon$ of the latter two. This topological field theory can be endowed with a $\ZZ_2$ global 
symmetry exchanging the $e$ and $m$ lines, which will be very important later on but which we ignore now. 

The $e$ and $m$ lines are bosons, while $\epsilon$ is a fermion. Indeed, $e$ generates a
non-anomalous $\ZZ_2^e$ one-form symmetry and in the language of the introduction
the toric code is the partner $\fT_{b}$ of a trivial $\fT_{\ZZ_2}$. Symmetrically, $m$ also generates a 
non-anomalous $\ZZ_2^m$ one-form symmetry (with a mixed anomaly with the $\ZZ_2^e$ symmetry).

On the other hand, $\epsilon$ generates precisely the sort of anomalous one-form $\ZZ_2^\epsilon$ 
symmetry we need for the shadow of a spin TFT. This will be an important example for us, especially after 
we make manifest the $\ZZ_2$ global symmetry exchanging $e$ and $m$. 

A $\ZZ_2$ gauge theory has two natural gapped boundary conditions: we can fix the flat connection at the boundary 
or let it free to fluctuate. The corresponding boundary conditions in the toric code, $\fB_e$ and $\fB_m$, 
condense either the $e$ or the $m$ particle. \footnote{In appendix \ref{app:toric} we describe a fermionic boundary condition $\fB_\epsilon$ 
at which $\epsilon$ condenses. }

In either case, the category of boundary line defects consists of two simple objects, $I$ and $P$, 
which fuse as 
\begin{equation}
I \otimes I = I \qquad I \otimes P = P \qquad P \otimes I = P \qquad P \otimes P = I
\end{equation}
All the associators and other data can be taken to be trivial. 

The four elements in the center, say for $\fB_e$, can be described as 
\begin{align}
1 &= (I; \beta_I =1, \beta_P = 1) \cr
e &= (I; \beta_I =1, \beta_P = -1) \cr
m &= (P; \beta_I =1, \beta_P = 1) \cr
\epsilon &= (P; \beta_I =1, \beta_P = -1) 
\end{align}
We recognize the required properties for generators of bosonic or fermionic 1-form symmetries. 

The toric code also offers a very simple example of gauging a $\ZZ_2$ symmetry at the level of spherical fusion categories: 
the trivial $\ZZ_2$ SPT phase is associated to a $\ZZ_2$-graded spherical fusion category, with $\cC_0$ consisting of the identity object $I$ 
and $\cC_1$ consisting of $P$. Dropping the grading gives us the $\ZZ_2$ gauge theory/toric code. 

\subsection{Example: bosonic SPT phases and group cohomology}

The group cohomology construction of bosonic SPT phases has precisely the form of a $G$-graded Turaev-Viro 
partition sum, based on a $G$-graded category $\cC$ with a single (equivalence class of) simple object $V_g$ in each  $\cC_g$ subcategory. 

The associator is a map from $V_{g g' g''}$ to itself, which can be written as 
$$
e^{ 2 \pi i \alpha_3(1,g,g g',g g' g'')} 1_{V_{g g' g''}}
$$
where $\alpha_3$ is a 3-cocycle on $BG$ with values in $U(1)$. The cocycle condition is 
equivalent to the pentagon axiom for the associator. Re-definitions of the isomorphisms $V_g \otimes V_{g'} \simeq V_{g g'}$ used in the definition will 
shift $\alpha_3$ by an exact cochain. 

\begin{figure}
\centering
\begin{tikzpicture}[line width=1pt]
\begin{scope}[every node/.style={sloped,allow upside down}]
\coordinate (2) at (0,0);
\coordinate (1) at ($(2)+(3,-0.5)$);
\coordinate (0) at ($(2)+(2.1,3.25)$);
\coordinate (3) at ($(1)+(1.5,1.5)$);

\arrowpath{(2)}{(3)}{0.5};
\paddedline{(0)}{(1)}{(0.1,0)};
\arrowpath{(0)}{(1)}{0.5};
\arrowpath{(0)}{(2)}{0.5};
\arrowpath{(1)}{(2)}{0.5};
\arrowpath{(0)}{(3)}{0.5};
\arrowpath{(1)}{(3)}{0.5};

\node[above] at (0) {\scriptsize{$1$}};
\node[below] at (1) {\scriptsize{$g$}};
\node[left] at (2) {\scriptsize{$gg'$}};
\node[right] at (3) {\scriptsize{$gg'g''$}};

\node[above right] at ($(0)!0.5!(1)$) {\scriptsize{$V_{g}$}};
\node[above left] at ($(0)!0.5!(2)$) {\scriptsize{$V_{gg'}$}};
\node[above right] at ($(0)!0.5!(3)$) {\scriptsize{$V_{gg'g''}$}};
\node[below] at ($(1)!0.5!(2)$) {\scriptsize{$V_{g'}$}};
\node[below right] at ($(1)!0.5!(3)$) {\scriptsize{$V_{g'g''}$}};
\node[above left] at ($(2)!0.5!(3)$) {\scriptsize{$V_{g''}$}};

\node (a) at ($(0)+(0.25,-4.5)$) {\scriptsize{(a)}};

\coordinate (023) at ($(2)+(9,0)$);
\coordinate (012) at ($(023)+(-1.5,1.2)$);
\coordinate (123) at ($(023)+(1.5,2.4)$);
\coordinate (013) at ($(023)+(0,3.6)$);
\coordinate (controlu) at ($(013)+(3,3)$);
\coordinate (controld) at ($(023)+(3,-3)$);
\coordinate (mid) at ($(controlu)!0.5!(controld)-(0.75,0)$);

\arrowpath{(023)}{(012)}{0.5};
\arrowpath{(023)}{(123)}{0.5};
\arrowpath{(012)}{(123)}{0.5};
\arrowpath{(012)}{(013)}{0.5};
\arrowpath{(123)}{(013)}{0.5};
\draw (013)..controls(controlu) and (controld)..(023);
\arrowpath{($(mid)+(0,0.1)$)}{($(mid)-(0,0.1)$)}{0.5};


\node[below left] at ($(023)!0.5!(012)$) {\scriptsize{$V_{gg'}$}};
\node[right] at ($(023)!0.5!(123)$) {\scriptsize{$V_{g''}$}};
\node[above] at ($(012)!0.5!(123)$) {\scriptsize{$V_{g'}$}};
\node[left] at ($(012)!0.5!(013)$) {\scriptsize{$V_{g}$}};
\node[above right] at ($(123)!0.5!(013)$) {\scriptsize{$V_{g'g''}$}};
\node[right] at (mid) {\scriptsize{$V_{gg'g''}$}};

\draw[fill=black] (012) circle(2pt);
\draw[fill=black] (013) circle(2pt);
\draw[fill=black] (023) circle(2pt);
\draw[fill=black] (123) circle(2pt);

\node at ($(a)+(7,0)$) {\scriptsize{(b)}};
\end{scope}
\end{tikzpicture}
\caption{Bosonic SPT phases: (a) A positively oriented tetrahedron with generic simple elements on edges. We label the vertices such that an edge going from $g$ to $h$ is assigned the element $V_{g^{-1}h}$. (b) The planar graph dual to the tetrahedron with all the morphisms as canonical identity morphisms. The graph evaluates to the associator $\alpha_3(1,g,gg',gg'g'')$. Notice that the faces of the dual graph correspond to vertices of the tetrahedron and are ordered correspondingly. 
Edges are oriented so that the face to the left is comes before the face to the right. For example, the outer face in the dual graph is the first.}
\label{fig:tetrag}
\end{figure}
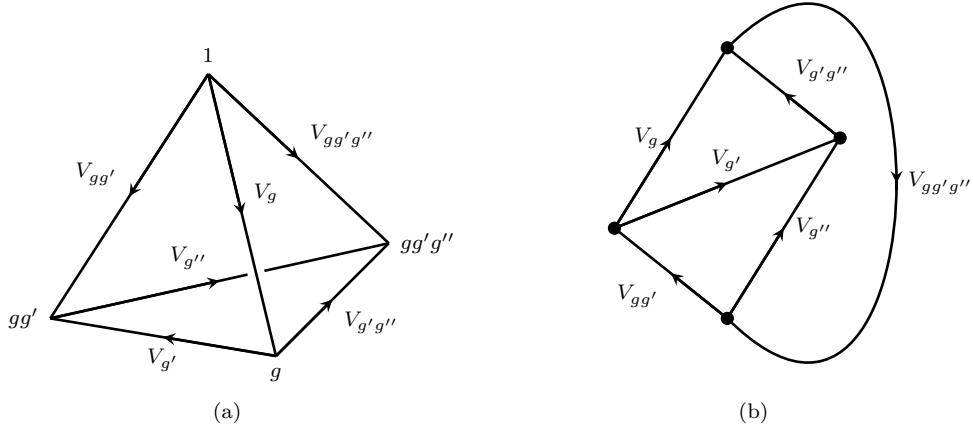

We refer the reader to Figure \ref{fig:tetrag} for a graphical explanation of the relation between associators 
and cocycle elements. An illustrative example is the non-trivial group cocycle for $G = \ZZ_2$: 
\begin{equation}
\alpha_3(0,\epsilon,\epsilon + \epsilon', \epsilon + \epsilon' + \epsilon'')=  \frac12 \epsilon \epsilon' \epsilon''
\end{equation} 
In terms of the cocycle $\epsilon_1$ defined by the group element on edges of the tetrahedron, $\alpha^{\ZZ_2}_3 = \frac12 \epsilon_1 \cup \epsilon_1 \cup \epsilon_1$.
\footnote{An alternative expression for the cocycle can be given via the Bockstein homomorphism: $\epsilon_1 \cup \epsilon_1$ is equivalent modulo $2$ to $\frac{1}{2} \delta \tilde \epsilon_1$, 
where $\tilde \epsilon_1$ is an integral lift of $\epsilon_1$. Thus we can write $\alpha^{\ZZ_2}_3 = \frac14 \tilde \epsilon_1 \cup \delta \tilde \epsilon_1$.}

\begin{figure}
\begin{tikzpicture}[line width=1pt]
\begin{scope}[every node/.style={sloped,allow upside down}]
\coordinate (2) at (0,0);
\coordinate (1) at ($(2)+(3,-0.5)$);
\coordinate (0) at ($(2)+(2.1,3.25)$);
\coordinate (3) at ($(1)+(1.5,1.5)$);

\arrowpath{(2)}{(3)}{0.5};
\paddedline{(0)}{(1)}{(0.1,0)};
\arrowpath{(0)}{(1)}{0.5};
\arrowpath{(0)}{(2)}{0.5};
\arrowpath{(1)}{(2)}{0.5};
\arrowpath{(0)}{(3)}{0.5};
\arrowpath{(1)}{(3)}{0.5};

\node[above] at (0) {\scriptsize{$0$}};
\node[below] at (1) {\scriptsize{$1$}};
\node[left] at (2) {\scriptsize{$2$}};
\node[right] at (3) {\scriptsize{$3$}};

\node (a) at ($(0)+(0.25,-4.5)$) {\scriptsize{(a)}};

\coordinate (1') at ($(2)+(7.5,0)$);
\coordinate (2') at ($(1')+(3,-0.5)$);
\coordinate (0') at ($(1')+(2.1,3.25)$);
\coordinate (3') at ($(2')+(1.5,1.5)$);

\arrowpath{(1')}{(3')}{0.5};
\paddedline{(0')}{(2')}{(0.1,0)};
\arrowpath{(0')}{(2')}{0.5};
\arrowpath{(0')}{(1')}{0.5};
\arrowpath{(1')}{(2')}{0.5};
\arrowpath{(0')}{(3')}{0.5};
\arrowpath{(2')}{(3')}{0.5};

\node[above] at (0') {\scriptsize{$0$}};
\node[left] at (1') {\scriptsize{$1$}};
\node[below] at (2') {\scriptsize{$2$}};
\node[right] at (3') {\scriptsize{$3$}};

\node (b) at ($(0')+(0.25,-4.5)$) {\scriptsize{(b)}};
\end{scope}
\end{tikzpicture}
\caption{Given a tetrahedron with a labeling of vertices by $i\in\{0,1,2,3\}$, we orient the edges such that vertex with label $i$ has $i$ incoming edges. This defines a \emph{local order} on the tetrahedron. Orientation is defined by using right-hand rule going from 0 to 1 to 2. If the thumb points inward, we say that the tetrahedron is positively oriented as shown in (a). If the thumb points outward, we say that the tetrahedron is negatively oriented as shown in (b).}
\end{figure}
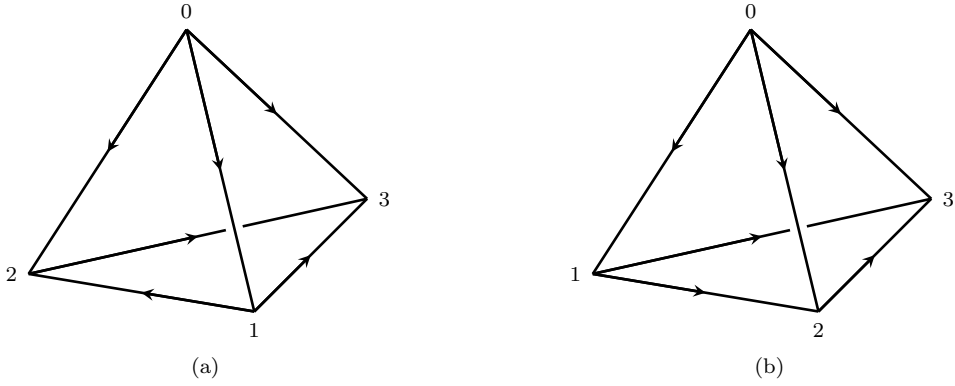

We can describe the corresponding $G$ gauge theory simply by ignoring the $G$ grading on $\cC$. 
For future reference, it is useful to describe objects in the Drinfeld center of $\cC$. The bulk defect lines (i.e. simple objects of the Drinfeld center)
turn out to be labelled by a pair $(g,\chi)$, where $g$ is an element of $G$ and 
$\chi$ an irreducible projective representation of the stabilizer $G_g$ of $g$ in $G$ \cite{Coste:aa}. 

The pair $(g,\chi)$ gives a center line of the form $(V_g^{\oplus n}, \beta_{g'} = \chi(g') )$. Notice that 
$\beta$ only needs to be specified if $g$ and $g'$ commute, in which case it is a matrix multiple of 
the basis element of $\Hom_{\cC}(V_g \otimes V_{g'}, V_{g'} \otimes V_g)\simeq \CC$. 
The definition of the Drinfeld center requires
\begin{equation}
\chi(g' g'') =e^{2 \pi i \alpha_3(1,g,gg',gg'g'') -2 \pi i \alpha_3(1,g',g' g,g' g g'') +2 \pi i \alpha_3(1,g',g' g'',g' g'' g) }\chi(g') \chi(g'')
\end{equation}
and fixes the group 2-cocycle associated to the projective representation in terms of $\alpha_3$ and $g$.  
Physically, this is a $g$-twist line dressed by a Wilson line.

\subsection{Example: $G$-equivariant $\ZZ_2$ gauge theory from a central extension}\label{sec:Gtoricfromcentral}

Consider a central extension 
\begin{equation}
0 \to \ZZ_2 \to \hat G \to G \to 0
\end{equation}
We can take a $\hat G$ SPT phase and gauge the $\ZZ_2$ subgroup. 

The result is a $G$-graded category with $\cC_g$ consisting of two objects. 
If we denote the pre-images of $g$ in $\hat G$ as $(g,0)$ and $(g,1)$, then 
$\cC_g$ consists of $V_{g,0}$ and $V_{g,1}$. The fusion rule is given by
\begin{equation}\label{GuWenfusion}
V_{g,\epsilon} \otimes V_{g',\epsilon'}\simeq V_{gg',\epsilon+\epsilon'+n_2(g,g')},
\end{equation}
where $n_2$ is the $\ZZ_2$-valued group 2-cocycle  corresponding to the central extension. 

We can now ask if the $\ZZ_2$ gauge theory has $\ZZ_2$ 1-form symmetry generators which are compatible with the $G$ 
global symmetry, i.e. map each $\cC_g$ to itself. That means we should look for objects of the center $Z[\cC]$ which project to 
either $V_{e,0}$ or $V_{e,1}$. The former case corresponds to the bare Wilson loop, which 
generates a bosonic $\ZZ_2$ 1-form symmetry. 

The latter case is more interesting, as the 2-cocycle for $(e,1)$-twist lines may 
be non-trivial. A $(1,1)$-twist line will be a bosonic (fermionic) $\ZZ_2$ generator 
if we can find a 1-dimensional projective representation of $\hat G$ 
with appropriate cocycle and $\chi((e,1)) = \pm 1$. 

This is a somewhat intricate constraint on the $\hat G$ 3-cocycle $\hat \alpha_3$ defining the 
initial SPT phase. Up to a gauge transformation, this constraint has a neat solution: 
$\hat \alpha_3$ must be given in terms of a group super-cohomology element 
$(\nu_3, n_2)$ as follows:
\begin{equation} \label{eq:hatalphap}
\hat \alpha_3 = \nu_3 +\frac12 n_2 \cup \epsilon_1.
\end{equation}
Here $\nu_3$ is an $\RR/\ZZ$-valued 3-cochain on $BG$ satisfying the Gu-Wen equation (\ref{GuWeneqs}), and 
where $\epsilon_1$ is the $\ZZ_2$-valued 1-cochain which sends $(g,\epsilon)$ to $\epsilon$. It is easy to see  that $\delta \epsilon_1 = n_2$, and 
thus the cocycle condition $\delta \hat \alpha_3=0$ follows from the Gu-Wen equations. 
The fermion $\Pi$ corresponds to the projective representation $\chi((g,\epsilon)) = (-1)^{\epsilon}$. 

Of course, the form given here for $\hat \alpha_3$ can be modified by gauge transformations. For example,
a transformation with parameter $\frac12 \epsilon_1 \cup_1 n_2$ would give another representative:
\begin{equation} \label{eq:hatalpha}
\hat \alpha'_3 = \nu'_3 +\frac12 \epsilon_1\cup n_2.
\end{equation}
with $\nu'_3 = \nu_3 + \frac12n_2 \cup_1 n_2$.

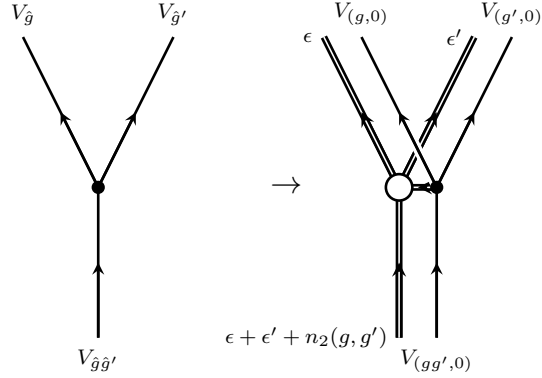
\begin{figure}
\centering
\begin{tikzpicture}[line width=1pt]

\begin{scope}[every node/.style={sloped,allow upside down}]
\coordinate (00) at (0,0);

\coordinate (C) at ($(00)+(1,0)$);
\coordinate (mid) at ($(C)+(0,2)$);
\coordinate (A) at ($(mid)+(-1,2)$);
\coordinate (B) at ($(mid)+(1,2)$);

\arrowpath{(C)}{(mid)}{0.5};
\arrowpath{(mid)}{(A)}{0.5};
\arrowpath{(mid)}{(B)}{0.5};
\draw[fill=black] (mid) circle(2pt);

\node[below] at (C) {\scriptsize{$V_{\hat g \hat g'}$}};
\node[above] at (A) {\scriptsize{$V_{\hat g }$}};
\node[above] at (B) {\scriptsize{$V_{\hat g'}$}};

\coordinate (eq2) at ($(mid)+(2.5,0)$);
\node at (eq2) {$\to$};

\coordinate (C) at ($(00)+(5,0)$);
\coordinate (mid) at ($(C)+(0,2)$);
\coordinate (A) at ($(mid)+(-1,2)$);
\coordinate (B) at ($(mid)+(1,2)$);

\arrowpathdouble{(C)}{(mid)}{0.5};
\arrowpathdouble{(mid)}{(A)}{0.5};
\arrowpathdouble{(mid)}{(B)}{0.5};

\coordinate (mid2) at (mid);

\node[left] at (C) {\scriptsize{$\epsilon + \epsilon' + n_2(g,g')$}};
\node[left] at (A) {\scriptsize{$\epsilon$}};
\node[left] at (B) {\scriptsize{$\epsilon'$}};

\coordinate (C) at ($(00)+(5.5,0)$);
\coordinate (mid) at ($(C)+(0,2)$);
\coordinate (A) at ($(mid)+(-1,2)$);
\coordinate (B) at ($(mid)+(1,2)$);

\arrowpathdouble{(mid)}{(mid2)}{0.5};
\draw[fill=white] (mid2) circle(5pt);

\arrowpath{(C)}{(mid)}{0.5};
\paddedline{(mid)}{(A)}{(0.05,0)};
\arrowpath{(mid)}{(A)}{0.5};
\arrowpath{(mid)}{(B)}{0.5};
\draw[fill=black] (mid) circle(2pt);

\node[below] at (C) {\scriptsize{$V_{(g g',0)}$}};
\node[above] at (A) {\scriptsize{$V_{(g,0)}$}};
\node[above] at (B) {\scriptsize{$V_{(g',0)}$}};

\end{scope}
\end{tikzpicture}
\caption{Gauge-fixing: A graphical representation of the partial gauge-fixing procedure used in computing the $\hat G$ group cocycle. 
Left: A choice of gauge is the same as a choice of basis vector in the space of junctions between line defects in the full $\hat G$ category. 
Right: we identify $V_{(g,\epsilon)}\simeq V_{(1,\epsilon)} \otimes V_{g,0}$ and identify $V_{(1,\epsilon)}$ with the corresponding elements $I$ or $\Pi$ of the center. 
We then express a general junction canonically 
in terms of a choice of junction between line defects labelled by $G$ elements. The double lines denote the center elements. The empty circle represents any choice of how to connect the center lines in a planar way.}
\label{fig:hatcocycle1}
\end{figure}

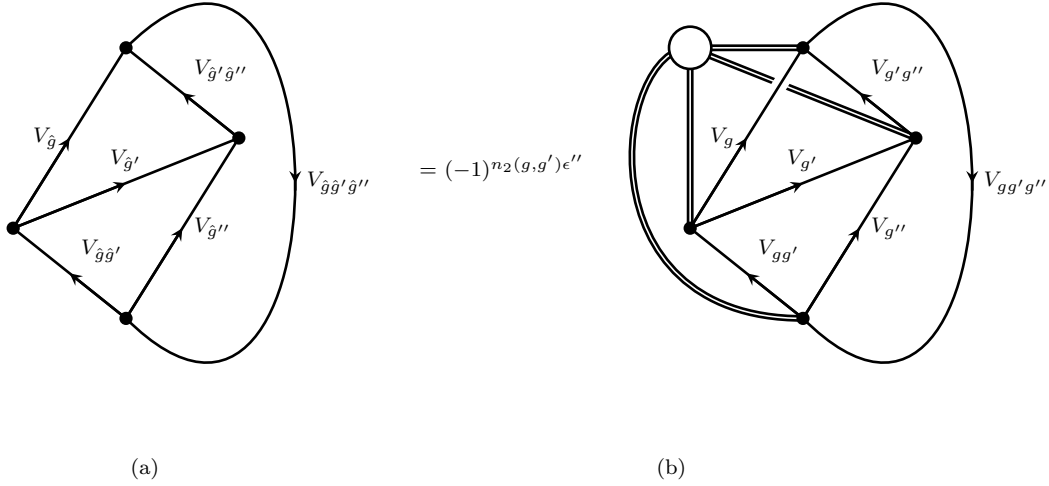
\begin{figure}
\centering
\begin{tikzpicture}[line width=1pt]
\begin{scope}[every node/.style={sloped,allow upside down}]
\coordinate (00) at (0,0);
\coordinate (023) at (00);
\coordinate (012) at ($(023)+(-1.5,1.2)$);
\coordinate (123) at ($(023)+(1.5,2.4)$);
\coordinate (013) at ($(023)+(0,3.6)$);
\coordinate (controlu) at ($(013)+(3,3)$);
\coordinate (controld) at ($(023)+(3,-3)$);
\coordinate (mid) at ($(controlu)!0.5!(controld)-(0.75,0)$);

\arrowpath{(023)}{(012)}{0.5};
\arrowpath{(023)}{(123)}{0.5};
\arrowpath{(012)}{(123)}{0.5};
\arrowpath{(012)}{(013)}{0.5};
\arrowpath{(123)}{(013)}{0.5};
\draw (013)..controls(controlu) and (controld)..(023);
\arrowpath{($(mid)+(0,0.1)$)}{($(mid)-(0,0.1)$)}{0.5};


\node[above right] at ($(023)!0.5!(012)$) {\scriptsize{$V_{\hat g \hat g'}$}};
\node[right] at ($(023)!0.5!(123)$) {\scriptsize{$V_{\hat g''}$}};
\node[above] at ($(012)!0.5!(123)$) {\scriptsize{$V_{\hat g'}$}};
\node[left] at ($(012)!0.5!(013)$) {\scriptsize{$V_{\hat g}$}};
\node[above right] at ($(123)!0.5!(013)$) {\scriptsize{$V_{\hat g' \hat g''}$}};
\node[right] at (mid) {\scriptsize{$V_{\hat g \hat g' \hat g''}$}};

\draw[fill=black] (012) circle(2pt);
\draw[fill=black] (013) circle(2pt);
\draw[fill=black] (023) circle(2pt);
\draw[fill=black] (123) circle(2pt);

\node (a) at ($(00)+(0.25,-2)$) {\scriptsize{(a)}};

\node (eq) at ($(00)+(5,2)$) {\scriptsize{= $(-1)^{n_2(g,g') \epsilon''}$}};

\coordinate (023) at ($(00)+(9,0)$);
\coordinate (012) at ($(023)+(-1.5,1.2)$);
\coordinate (123) at ($(023)+(1.5,2.4)$);
\coordinate (013) at ($(023)+(0,3.6)$);
\coordinate (controlu) at ($(013)+(3,3)$);
\coordinate (controld) at ($(023)+(3,-3)$);
\coordinate (mid) at ($(controlu)!0.5!(controld)-(0.75,0)$);
\coordinate (blob) at ($(012)+(0,2.4)$);
\coordinate (controllu) at ($(blob)-(1,0)$);
\coordinate (controlld) at ($(023)-(3,0)$);

\draw[double] (123) to (blob);
\paddedline{(012)}{(013)}{(0.1,0)};
\draw[double] (013) to (blob);
\draw[double] (012) to (blob);
\arrowpath{(023)}{(012)}{0.5};
\arrowpath{(023)}{(123)}{0.5};
\arrowpath{(012)}{(123)}{0.5};
\arrowpath{(012)}{(013)}{0.5};
\arrowpath{(123)}{(013)}{0.5};
\draw (013)..controls(controlu) and (controld)..(023);
\arrowpath{($(mid)+(0,0.1)$)}{($(mid)-(0,0.1)$)}{0.5};
\draw[double] (023)..controls(controlld) and (controllu)..(blob);

\node[above right] at ($(023)!0.5!(012)$) {\scriptsize{$V_{gg'}$}};
\node[right] at ($(023)!0.5!(123)$) {\scriptsize{$V_{g''}$}};
\node[above] at ($(012)!0.5!(123)$) {\scriptsize{$V_{g'}$}};
\node[left] at ($(012)!0.5!(013)$) {\scriptsize{$V_{g}$}};
\node[above right] at ($(123)!0.5!(013)$) {\scriptsize{$V_{g'g''}$}};
\node[right] at (mid) {\scriptsize{$V_{gg'g''}$}};

\draw[fill=black] (012) circle(2pt);
\draw[fill=black] (013) circle(2pt);
\draw[fill=black] (023) circle(2pt);
\draw[fill=black] (123) circle(2pt);
\draw[fill=white] (blob) circle(8pt);

\node at ($(a)+(7,0)$) {\scriptsize{(b)}};
\end{scope}
\end{tikzpicture}
\caption{Gu-Wen fermionic SPT phases: (a) The planar graph dual to the tetrahedron which computes $\hat \alpha$ in a gauge determined by the choice of morphism at the junctions. 
We partial gauge-fix as in the previous Figure. The resulting web of center lines can be simplified by bringing together all planar junctions and collapsing planar loops, up to resolving a single crossing (See next Figure {\protect \ref{fig:bis}}). 
Up to the corresponding sign, we obtain: (b) a graph which depends on $G$ elements only and defines $\nu_3$.}
\label{fig:hatcocycle2}
\end{figure}

There are two complementary ways to arrive at this solution. In Appendix \ref{app:twist} we give a derivation based on the analysis of anomalies in the $\ZZ_2$ gauge theory coupled to a $G$ gauge field. In Figures \ref{fig:hatcocycle1} and \ref{fig:hatcocycle2} we give a graphical/physical proof of \ref{eq:hatalpha} using the 
spherical fusion category associated to $\hat G$. Essentially, the existence of a Drinfeld center element 
of the form $(V_{e,1};\beta)$ allows certain topological manipulations of planar graphs, 
relating two graph which encode the left and right side of equation \ref{eq:hatalphap}.

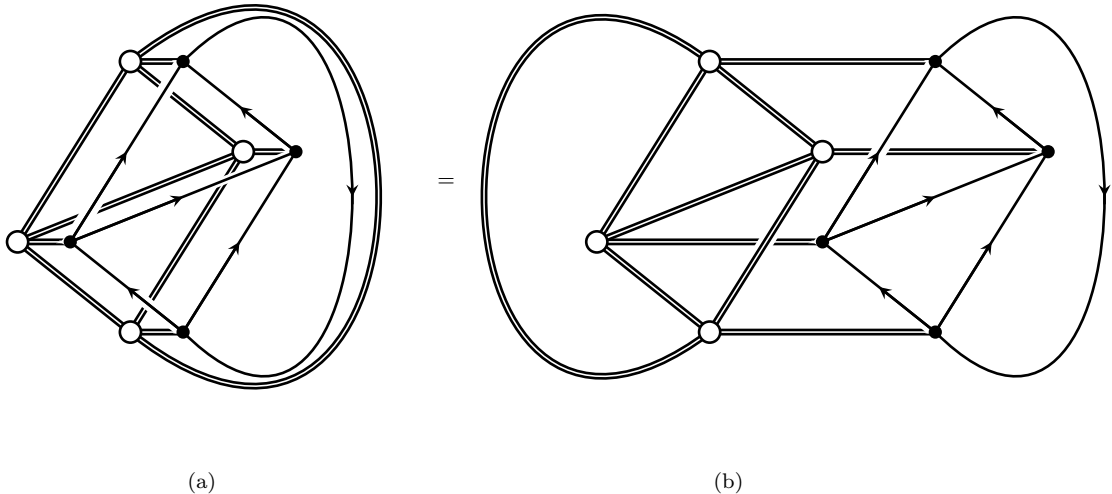
\begin{figure}
\centering
\begin{tikzpicture}[line width=1pt]
\begin{scope}[every node/.style={sloped,allow upside down}]
\coordinate (00) at (0,0);
\coordinate (023) at (00);
\coordinate (012) at ($(023)+(-1.5,1.2)$);
\coordinate (123) at ($(023)+(1.5,2.4)$);
\coordinate (013) at ($(023)+(0,3.6)$);
\coordinate (controlu) at ($(013)+(3,3)$);
\coordinate (controld) at ($(023)+(3,-3)$);
\coordinate (mid) at ($(controlu)!0.5!(controld)-(0.75,0)$);

\coordinate (023b) at ($(023)+ (-.7,0)$);
\coordinate (012b) at ($(012)+(-.7,0)$);
\coordinate (123b) at ($(123)+(-.7,0)$);
\coordinate (013b) at ($(013)+(-.7,0)$);
\coordinate (controlub) at ($(controlu)+(.7,.5)$);
\coordinate (controldb) at ($(controld)+(.7,-.5)$);

\draw[double] (023b) to (012b);
\draw[double] (023b) to (123b);
\draw[double] (012b) to (123b);
\draw[double] (012b) to (013b);
\draw[double] (123b) to (013b);
\draw[double] (013b)..controls(controlub) and (controldb)..(023b);

\draw[double] (023b) to (023);
\draw[double] (123b) to (123);
\draw[double] (012b) to (012);
\draw[double] (013b) to (013);

\draw[fill=white] (012b) circle(4pt);
\draw[fill=white] (013b) circle(4pt);
\draw[fill=white] (023b) circle(4pt);
\draw[fill=white] (123b) circle(4pt);

\paddedline{(023)}{(012)}{(.1,0)};
\paddedline{(023)}{(123)}{(.1,0)};
\paddedline{(012)}{(123)}{(.1,0)};
\paddedline{(012)}{(013)}{(.1,0)};
\paddedline{(123)}{(013)}{(.1,0)};

\arrowpath{(023)}{(012)}{0.5};
\arrowpath{(023)}{(123)}{0.5};
\arrowpath{(012)}{(123)}{0.5};
\arrowpath{(012)}{(013)}{0.5};
\arrowpath{(123)}{(013)}{0.5};
\draw (013)..controls(controlu) and (controld)..(023);
\arrowpath{($(mid)+(0,0.1)$)}{($(mid)-(0,0.1)$)}{0.5};

\draw[fill=black] (012) circle(2pt);
\draw[fill=black] (013) circle(2pt);
\draw[fill=black] (023) circle(2pt);
\draw[fill=black] (123) circle(2pt);

\node (a) at ($(00)+(0.25,-2)$) {\scriptsize{(a)}};

\node (eq) at ($(00)+(3.5,2)$) {\scriptsize{=}};

\coordinate (00) at (10,0);
\coordinate (023) at (00);
\coordinate (012) at ($(023)+(-1.5,1.2)$);
\coordinate (123) at ($(023)+(1.5,2.4)$);
\coordinate (013) at ($(023)+(0,3.6)$);
\coordinate (controlu) at ($(013)+(3,3)$);
\coordinate (controld) at ($(023)+(3,-3)$);
\coordinate (mid) at ($(controlu)!0.5!(controld)-(0.75,0)$);

\coordinate (023b) at ($(023)+ (-3,0)$);
\coordinate (012b) at ($(012)+(-3,0)$);
\coordinate (123b) at ($(123)+(-3,0)$);
\coordinate (013b) at ($(013)+(-3,0)$);
\coordinate (controlub) at ($(controlu)+(-10,0)$);
\coordinate (controldb) at ($(controld)+(-10,0)$);

\draw[double] (023b) to (023);
\draw[double] (123b) to (123);
\draw[double] (012b) to (012);
\draw[double] (013b) to (013);

\draw[double] (023b) to (012b);
\paddedline{(023b)}{(123b)}{(.1,0)};
\draw[double] (023b) to (123b);
\draw[double] (012b) to (123b);
\draw[double] (012b) to (013b);
\draw[double] (123b) to (013b);
\draw[double] (013b)..controls(controlub) and (controldb)..(023b);

\draw[fill=white] (012b) circle(4pt);
\draw[fill=white] (013b) circle(4pt);
\draw[fill=white] (023b) circle(4pt);
\draw[fill=white] (123b) circle(4pt);

\paddedline{(023)}{(012)}{(.1,0)};
\paddedline{(023)}{(123)}{(.1,0)};
\paddedline{(012)}{(123)}{(.1,0)};
\paddedline{(012)}{(013)}{(.1,0)};
\paddedline{(123)}{(013)}{(.1,0)};

\arrowpath{(023)}{(012)}{0.5};
\arrowpath{(023)}{(123)}{0.5};
\arrowpath{(012)}{(123)}{0.5};
\arrowpath{(012)}{(013)}{0.5};
\arrowpath{(123)}{(013)}{0.5};
\draw (013)..controls(controlu) and (controld)..(023);
\arrowpath{($(mid)+(0,0.1)$)}{($(mid)-(0,0.1)$)}{0.5};

\draw[fill=black] (012) circle(2pt);
\draw[fill=black] (013) circle(2pt);
\draw[fill=black] (023) circle(2pt);
\draw[fill=black] (123) circle(2pt);

\node at ($(a)+(7,0)$) {\scriptsize{(b)}};
\end{scope}
\end{tikzpicture}
\caption{Intermediate computational steps relating $\hat \alpha_3$ and $\nu_3$. The planar intersections (white circles) of center lines can be collapsed together safely, but the 
non-planar intersection has to be resolved first, at the price of a sign $(-1)^{n_2(g,g') \epsilon''}$. }
\label{fig:bis}
\end{figure}

In particular, we can define $\nu_3$ in terms of the spherical fusion category data as a tetrahedron graph of $(g,0)$ lines, 
with $n_2$ extra fermion lines at each vertex, exiting from the earliest face around the vertex and coming together to a common point where they are connected in a planar manner,
as in Figure \ref{fig:hatcocycle2} (b). 

In conclusion, we have a bijection between Gu-Wen fermionic SPT phases and potential shadows of $G$-symmetric 
spin-TFTs based on a $\ZZ_2$ theory.

Notice that the pair $(\nu_3,n_2)$ labels both the spherical fusion category and the choice of fermionic line, i.e. it labels the $\Pi$-category. 
The same spherical fusion category may admit multiple candidate fermionic lines. For example, if we are given a group homomorphism $\lambda_1$ 
from $G$ to $\ZZ_2$, we can dress $\Pi$ by a Wilson line for the corresponding representation, i.e. add a $(-1)^{\lambda_1}$ to $\beta$. Then the 
same choice of $\hat \alpha_3$ will give a $\nu_3$ which differs from the original by $\lambda_1 \cup n_2$. 

As an example of the construction, consider $\hat G = \ZZ_4$ as a $\ZZ_2$ central extension of $G = \ZZ_2$. Recall that $H^3(\ZZ_4,\RR/\ZZ)=\ZZ_4$. We claim that the generator of this group corresponds to a shadow of a Gu-Wen fermionic SPT. Indeed, if $[\eta_1]$ is the generator of $H^1(G,\ZZ_2)= \ZZ_2$, then the extension class corresponding to $\hat G$ can be written as $n_2 = [\frac12 \delta\tilde \eta_1]$, where $\tilde\eta_1$ is an integral lift of $\eta_1$. Concretely, $\eta_1$ is the $\ZZ_2$ cocycle defined by the $G$ elements on the edges of the triangulation and $[\frac12 \delta\tilde \eta_1]$
measures the failure of the group law for a $\hat G$ lift of the $G$ elements.
 
Therefore a possible solution of the equation (\ref{GuWeneqs}) is 
\begin{equation}\label{GuWenexample}
\nu_3=\frac{1}{8} \tilde\eta_1 \cup\delta \tilde\eta_1.
\end{equation}
The corresponding 3-cocycle on $\hat G$ is
\begin{equation}
\hat\alpha_3=\frac{1}{8}\tilde\eta_1 \cup\delta \tilde\eta_1 +\frac{1}{4} \delta\tilde\eta_1\cup \eps_1.
\end{equation}
Twice this cocycle is $\frac{1}{4} \tilde\eta_1 \cup\delta \tilde\eta_1\sim \frac12 \eps_1^3$, which is a pull-back of a 3-cocycle on $G=\ZZ_2$ generating $H^3(\ZZ_2,\ZZ_2)\simeq\ZZ_2$. 
Therefore this cocycle represents the generator of $H^4(\hat G,\RR/\ZZ)$. \footnote{Alternatively, we can re-write it directly in terms of the $\ZZ_4$ cocycle $\epsilon_1^{\ZZ_4} \equiv \tilde \eta_1 + 2 \epsilon_1$. 
It is easy to verify that $\hat\alpha_3$ is co-homologous to 
\begin{equation}\frac{1}{4} \epsilon_1^{\ZZ_4} \cup \epsilon_1^{\ZZ_4} \cup \epsilon_1^{\ZZ_4} = \frac{1}{4} \tilde \eta_1 \cup \tilde \eta_1 \cup \tilde \eta_1 + \frac{1}{2} \tilde \eta_1 \cup \tilde \eta_1 \cup \eps_1 + 
 \frac{1}{2} (\eps_1 \cup \tilde \eta_1+\tilde \eta_1 \cup \eps_1) \cup \tilde \eta_1,\end{equation}
 modulo 1. 
}
This is the shadow of a Gu-Wen phase with symmetry $\ZZ_2$. It is an abelian phase, in the sense that the fusion rules of the shadow TFT are abelian (based on an abelian group $\ZZ_4$).

Another solution of the Gu-Wen equations  with the same $n_2$ is 
\begin{equation}
\nu_3=-\frac{1}{8} \tilde\eta \cup\delta \tilde\eta .
\end{equation}
It differs from (\ref{GuWenexample}) by a closed 3-cochain $\frac{1}{4}\tilde\eta\cup\delta\tilde\eta$ whose class is the generator of $H^3(G,\RR/\ZZ)=\ZZ_2$. In physical terms, these two Gu-Wen phases (and their shadows) differ by tensoring with a bosonic SPT phase. Two more shadows of Gu-Wen phases are obtained by taking $\hat G=\ZZ_2\times\ZZ_2$. In this case $\hat\alpha_3$ is a pull-back of a 3-cocycle on $G=\ZZ_2$, which is otherwise unconstrained. Overall, we get four Gu-Wen phases with symmetry $G=\ZZ_2$. They are all abelian phases and are naturally labeled by elements of $\ZZ_4$.



\subsection{Example: $\ZZ_2$-equivariant toric code vs Ising}
The toric code has a $\ZZ_2$ symmetry which exchanges $e$ and $m$, which is not manifest
as an on-site symmetry in the standard microscopic formulation of the theory. 

The symmetry can be made manifest by extending the category of boundary line defects to a $\ZZ_2$-graded 
category which includes boundary twist lines for the $\ZZ_2$ symmetry and using the extended 
category as an input for a state sum or a string-net model. 

As the $\ZZ_2$ symmetry exchanges the $\fB_e$ and $\fB_m$ boundary conditions, the 
boundary twist lines interpolate between $\fB_e$ and $\fB_m$. 

Concretely the $\ZZ_2$-graded category can be identified with the Ising fusion category (see \cite{TY} or appendix B of \cite{DGNO} for a detailed discussion). 
There are three objects $I, S, P$ fusing as $P \otimes S = S \otimes P = S$ and $S \otimes S = I \oplus P$.
The object $S$ belongs to $\cC_1$, $I$ and $P$ to $\cC_0$. The nontrivial associators are
\begin{eqnarray}
a(P,S,P):& & (P\otimes S)\otimes P\ra P\otimes (S\otimes P),\\
a(S,P,S): & & (S\otimes P)\otimes S\ra S\otimes (P\otimes S),\\
a(S,S,S): & & (S\otimes S)\otimes S\ra S\otimes (S\otimes S).
\end{eqnarray}
The first one, regarded as an endomorphism of $S$, is $-1$. The second one, regarded as an endomorphism of $I\oplus P$,  is a vector $(1,-1)$.
The last associator is determined by the pentagon equation only up to an overall sign: the associator morphism regarded as an endomorphism of $S \oplus S$ is a matrix
\begin{equation}
\lambda^{-1} \begin{pmatrix} 1 & 1 \\ 1 & -1\end{pmatrix},
\end{equation}
where $\lambda=\pm \sqrt 2$.

The fusion rules can be explained as follows.  The fusion rules for $\cC_1$ are the usual fusion rules for the boundary lines on the $\fB_e$ boundary. Since $S$ is the termination of a $\ZZ_2$ domain wall which implements the particle-vortex symmetry transformation, we must have $S\otimes S\supset I$.: this means that a domain wall shaped as a hemisphere ending on a $\fB_e$ boundary can be shrunk away.  Finally, shrinking away the same hemispherical domain wall in the presence of a Wilson line $P$ shows that $S\otimes S\supset P$. The associators are fixed by the pentagon equation, up to an ambiguity in the sign of $\lambda$ \cite{TY}. 

This identification of the Ising category with the $\ZZ_2$ equivariant version of the toric code is consistent with the 
observation that gauging the $\ZZ_2$ symmetry of the toric code produces the
quantum double of the 3d Ising TFT, i.e. a TFT whose category of bulk like defects is the product of the Ising modular tensor category and its conjugate. 

The Ising modular tensor category has three simple objects $1,\sigma,\psi$ which fuse just as $I,S,P$ above. 
The quantum double (i.e. the Drinfeld center of the Ising fusion category) has bulk quasi-particles which are the product of $1, \sigma, \psi$ and $1, \bar \sigma, \bar \psi$. 
The $\psi \bar \psi$ particle is a boson to be identified with the Wilson loop. The $\psi$ and $\bar \psi$ fermions are two 
versions of the original $\epsilon$ particle. Thus, for a fixed $\lambda$, there is a two-fold ambiguity in the choice of the fermion $\Pi$ for the Ising fusion category. More precisely, crossing either $\psi$ or $\bar\psi$ with $P$ gived $-1$, while crossing a fermion with $S$ gives a phase $\xi^2$ satisfying \cite{DGNO}
\begin{equation}
\xi+\xi^{-1}=\lambda. 
\end{equation}
The two solutions of this equation correspond to taking $\Pi=\psi$ or $\Pi=\bar\psi$. It is easy to see that $\xi^4=-1$, so taking into account both the freedom in choosing $\lambda$ and the freedom in choosing $\Pi$ we get four $\ZZ_2$-equivariant versions of the toric code with a fermionic $\ZZ_2$ 1-form symmetry. They can be labeled by $\xi$, which is a fourth root of $-1$. The four versions of the theory are on equal footing, since none of the four roots is preferred.

Recall that fermionic SPT phases with a unitary $\ZZ_2$ symmetry have a $\ZZ_8$ classification \cite{Gu:2013aa}. Four of them correspond to Gu-Wen supercohomology phases. We will argue below that the shadows of the other four phases are given by the four versions of the Ising fusion category equipped with $\Pi$.  The latter phases are non-abelian, in the sense that the fusion rules of the shadow TFT are not group-like.

\subsection{Example: Ising pull-backs}\label{sec:Isingpullback}
If we are given a group $G$ with a group homomorphism $\pi_1: G \to \ZZ_2$, 
we can define a $G$-graded Ising-like category as follows.

If $\pi_1(g)=0$, we take $\cC_g$ to consist of two simple elements, $I_g$ and $P_g$.
If $\pi_1(g)=1$, we take $\cC_g$ to consist of a simple element $S_g$. We take the 
fusion rules to mimic the Ising category: 
\begin{align}
V_{g,\eps}\otimes V_{g,\eps'} &= V_{gg',\eps+\eps'},\\
V_{g,\eps}\otimes S_{g'} &= S_{gg'}\\
S_g\otimes V_{g',\eps'} &= S_{gg'},\\
S_g S_{g'} &= V_{gg',0}+V_{gg',1},
\end{align}
where we denoted $I_g=V_{g,0}$ and $P_g=V_{g,1}$. 
The associators can be taken from the Ising category.

The center particle with boundary image $P_1$ and $\beta$ taken from the fermion in the 
Ising category example equips this category with a fermionic 1-form symmetry.  We will call the corresponding $G$-equivariant  TFT an Ising pull-back and denote it $\cI^\xi_{\pi_1}$. It depends on a parameter $\xi$ satisfying $\xi^4=-1$ as well as $\pi_1:G\ra\ZZ_2$. We will see below that it is a shadow of a fermionic SPT phase with symmetry $G\times \ZZ_2^f$. 

A richer possibility is to consider a long exact sequence of groups
\begin{equation}
0 \to \ZZ_2 \to \hat G_0 \to G \to \ZZ_2 \to 0,
\end{equation}
where we denote the homomorphism from $G$ to $\ZZ_2$ by $\pi_1$. The kernel of $\pi_1$ will be denoted $G_0$, then $\hat G_0$ is a central extension of $G_0$ by $\ZZ_2$. Let $n_2$ be a 2-cocycle on $G_0$ corresponding to this central extension.

If $\pi_1(g)=0$, we take $\cC_g$ to have two simple objects, $V_{g, \eps}$, $\eps\in \ZZ_2$.
If $\pi_1(g)=1$, we take $\cC_g$ to have a single simple object $S_g$. We again take the 
fusion rules to still mimic the Ising category: $S_g S_{g'} = V_{g g',0} + V_{gg',1},$  etc.,
but now require the $V_{\hat g} \equiv V_{g, \epsilon}$ fusion to follow the $\hat G_0$ multiplication rules. 

It follows from the results of \cite{ENO} that for any such long exact sequence there exists a fusion category with these fusion rules, provided a certain obstruction $[O_4]\in H^4(G,\RR/\ZZ)$  constructed from $n_2$ and $\pi_1$ vanishes. Possible associators depend are parameterized by a 3-cochain $\nu_3\in C^3(G,\RR/\ZZ)$ such that $\delta\nu_3=O_4$. As argued in appendix \ref{app:twist},  such a category has a fermion if and only if $[n_2]$ is a restriction of a class $[\beta_2]$ in $H^2(G,\ZZ_2)$, in which case $\nu_3$ must satisfy the Gu-Wen equation (\ref{GuWeneqs}). This TFT is a candidate for a shadow of a fermionic SPT phase with symmetry $G\times \ZZ_2^f$. One can view this theory as a $G$-equivariant version of the toric code, where some elements of $G$ act by particle-vortex symmetry, and the fusion of $G$ domain walls is associative only up to $e$ and $m$ lines. This failure of strict associativity is controlled by the extension class $n_2\in H^2(G_0,\ZZ_2)$.

Thus we obtain categories labelled by a triple $(\nu_3, n_2, \pi_1)$ (and a choice of a fermion) which are shadows of fermionic TFTs with symmetry $G$. We will see below that all these TFTs are fermionic SPT phases, i.e. they are ``invertible''. On the other hand, one may argue that shadows of fermionic SPT phases with symmetry $G$ must be $G$-equivariant  versions of the toric code. Indeed, the component $\cC_1$ of such a category must contain the identity object, the fermion $\Pi$, and no other simple objects, since condensing the fermion must give an invertible fermionic TFT.  The fusion rules for $\cC_1$ must have the same form as in the toric code, because $\Pi$ generates a $\ZZ_2$ 1-form symmetry,
and the associator for $\cC_1$ must be trivial for $\Pi$ to be a fermion. Thus $\cC_1$ describes the toric code, and $\cC=\sum_g \cC_g$ is a $G$-equivariant extension of the toric code. 

\subsection{Gauging one-form symmetries in the presence of gapped boundary conditions}
Given a gapped boundary condition for $\fT_b$, we can derive in a simple manner 
a gapped boundary condition for $\fT_{\ZZ_2}$. Here we describe the process at the level of 
boundary line defects. In later sections we will test it at the level of partition sums and 
commuting projector Hamiltonians. \footnote{Although we specialize here to a $\ZZ_2$ one-form symmetry, 
the same procedure works for a general Abelian group}

 We start from a spherical fusion category $\cC_b$ equipped with a bosonic 
$\ZZ_2$ one-form symmetry generator $B = (b, \beta)$, an element of the center $Z[\cC]$ such that $\beta_b =1_{b\otimes b}$ and there is an
isomorphism $\xi_b:b \otimes b \to I$ such that $\xi \otimes 1 = 1 \otimes \xi$ in $\Hom(b \otimes b \otimes b, b)$.

In the condensed matter language, our objective is to condense the anyon $B$. The general mathematical formalism 
for anyon condensation is described in \cite{Kong:2013aa}. It should be applied to the commutative separable algebra $A = I + B$. 
We will use a somewhat simplified procedure for concrete calculations. 

\begin{figure}
\centering
\begin{tikzpicture}[line width=1pt]
\begin{scope}[every node/.style={sloped,allow upside down}]
\coordinate (lowb1) at (0,0);
\coordinate (lowb2) at ($(lowb1)+(1,0)$);
\coordinate (midb1) at ($(lowb1)+(0,1.5)$);
\coordinate (midb2) at ($(lowb2)+(0,1.5)$);
\coordinate (lowb3) at ($(lowb1)+(2,0)$);
\coordinate (highb3) at ($(lowb3)+(0,4)$);

\draw[double] (lowb1)--(midb1);
\draw[double] (lowb2)--(midb2);
\draw[double] (lowb3)--(highb3);
\draw[double] (midb1) arc[radius=0.5, start angle=180, end angle=0];

\node[below] at (lowb1) {\scriptsize{$b$}};
\node[below] at (lowb2) {\scriptsize{$b$}};
\node[below] at (lowb3) {\scriptsize{$b$}};

\coordinate (eq) at ($(lowb3)+(1,2)$);
\node at (eq) {$=$};

\coordinate (lowbb1) at ($(lowb3)+(2,0)$);
\coordinate (highbb1) at ($(lowbb1)+(0,4)$);
\coordinate (lowbb2) at ($(lowbb1)+(1,0)$);
\coordinate (lowbb3) at ($(lowbb2)+(1,0)$);
\coordinate (midbb2) at ($(lowbb2)+(0,1.5)$);
\coordinate (midbb3) at ($(lowbb3)+(0,1.5)$);

\draw[double] (lowbb2)--(midbb2);
\draw[double] (lowbb3)--(midbb3);
\draw[double] (lowbb1)--(highbb1);
\draw[double] (midbb2) arc[radius=0.5, start angle=180, end angle=0];

\node[below] at (lowbb1) {\scriptsize{$b$}};
\node[below] at (lowbb2) {\scriptsize{$b$}};
\node[below] at (lowbb3) {\scriptsize{$b$}};

\node[below] at ($(eq) + (0,-2.5)$) {\scriptsize{(a)}};

\coordinate (llowb1) at ($(lowbb3)+(1.75,0)$);
\coordinate (llowb2) at ($(llowb1)+(1.5,0)$);
\coordinate (hhighb2) at ($(llowb1)+(0,4)$);
\coordinate (hhighb1) at ($(llowb2)+(0,4)$);

\draw[double] (llowb1)--(hhighb1);
\paddedline{(llowb2)}{(hhighb2)}{(0.1,0)};
\draw[double] (llowb2)--(hhighb2);

\node[below] at (llowb1) {\scriptsize{$b$}};
\node[below] at (llowb2) {\scriptsize{$b$}};

\coordinate (eq2) at ($(llowb2)+(0.75,2)$);
\node at (eq2) {$=$};

\coordinate (pm) at ($(eq2)+(0.75,0)$);
\coordinate (llowbb1) at ($(llowb2)+(2,0)$);
\coordinate (hhighbb1) at ($(llowbb1)+(0,4)$);
\coordinate (llowbb2) at ($(llowbb1)+(1,0)$);
\coordinate (hhighbb2) at ($(llowbb2)+(0,4)$);

\draw[double] (llowbb2)--(hhighbb2);
\draw[double] (llowbb1)--(hhighbb1);

\node[below] at (llowbb1) {\scriptsize{$b$}};
\node[below] at (llowbb2) {\scriptsize{$b$}};
\node at (pm) {$\pm$};

\node[below] at ($(eq2) + (0,-2.5)$) {\scriptsize{(b)}};
\end{scope}
\end{tikzpicture}
\caption{$\mathbb{Z}_2$ 1-form symmetries: (a) There exists a bulk line $b$ with properties shown in the figure. (b) Half-braiding $b$ lines across each other gives a factor of $\pm 1$ when compared to $b$ lines without braiding. The factor of $+1$ arises for a bosonic 1-form symmetry and $-1$ arises for a fermionic 1-form symmetry. This minus sign implies that the symmetry is anomalous.}
\label{fig:condensing}
\end{figure}
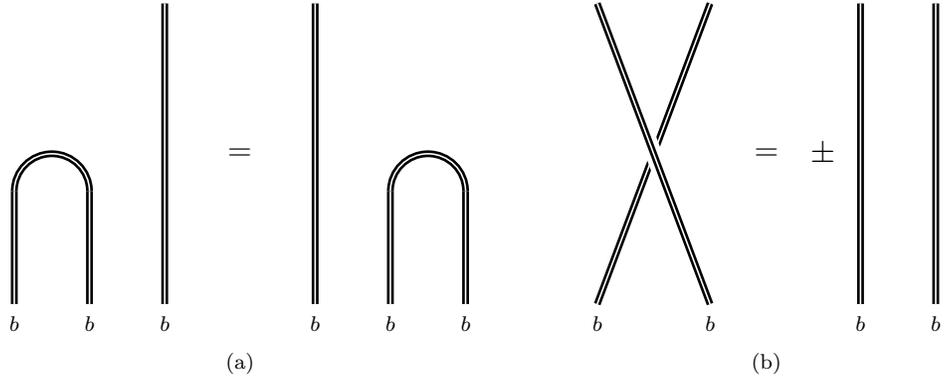

\begin{figure}
\begin{tikzpicture}[line width=1pt]
\begin{scope}[every node/.style={sloped,allow upside down}]
\coordinate (lowA1) at (0,0);
\coordinate (highA1) at ($(lowA1) + (0,4)$);
\coordinate (midA1) at ($(lowA1)!0.5!(highA1)$);
\arrowpath{(lowA1)}{(midA1)}{0.5};
\arrowpath{(midA1)}{(highA1)}{0.5};
\draw[fill=black] (midA1) circle(2pt);
\node[below] at (lowA1) {\scriptsize{$U$}};
\node[above] at (highA1) {\scriptsize{$V$}};

\coordinate (lowA2) at ($(lowA1)+(3,0)$);
\coordinate (highA2) at ($(lowA2) + (0,4)$);
\coordinate (midA2) at ($(lowA2)!0.5!(highA2)$);
\coordinate (b) at ($(highA2)-(1,0)$);
\arrowpath{(lowA2)}{(midA2)}{0.5};
\arrowpath{(midA2)}{(highA2)}{0.5};
\draw[double] (midA2)--(b);
\draw[fill=black] (midA2) circle(2pt);
\node[above] at (b) {\scriptsize{$b$}};
\node[below] at (lowA2) {\scriptsize{$U$}};
\node[above] at (highA2) {\scriptsize{$V$}};

\node[below] at ($(lowA1)!0.5!(lowA2) - (0, 0.5)$) {\scriptsize{(a)}};

\coordinate (lowB) at (7,0);
\coordinate (highB) at ($(lowB) + (0,4)$);
\coordinate (midB) at ($(lowB)!0.5!(highB)$);
\arrowpath{(lowB)}{(highB)}{0.17};
\arrowpath{(lowB)}{(highB)}{0.5};
\arrowpath{(lowB)}{(highB)}{0.83};
\draw[double] ($(lowB)!0.67!(highB)$) arc[radius=0.67, start angle=90, end angle=270];
\draw[fill=black] ($(lowB)!0.33!(highB)$) circle(2pt);
\draw[fill=black] ($(lowB)!0.67!(highB)$) circle(2pt);
\node[below] at (lowB) {\scriptsize{$U$}};
\node[above] at (highB) {\scriptsize{$V$}};

\node[below] at ($(lowB) - (0, 0.5)$) {\scriptsize{(b)}};

\coordinate (lowC1) at (11,0);
\coordinate (highC1) at ($(lowC1) + (0,4)$);
\coordinate (midC1) at ($(lowC1)!0.5!(highC1)$);
\coordinate (lowC2) at ($(lowC1)+(1,0)$);
\coordinate (highC2) at ($(lowC2) + (0,4)$);
\coordinate (midC2) at ($(lowC2)!0.5!(highC2)$);

\draw[double] (midC1) .. controls ($(midC1) + (-1,1.5)$) .. (midC2);

\paddedline{($(lowC1)!0.6!(highC1)$)}{(highC1)}{(0.05,0)};
\arrowpath{(lowC1)}{(highC1)}{0.25};
\arrowpath{(lowC1)}{(highC1)}{0.8};
\draw[fill=black] (midC1) circle(2pt);
\node[below] at (lowC1) {\scriptsize{$U_1$}};
\node[above] at (highC1) {\scriptsize{$V_1$}};

\arrowpath{(lowC2)}{(highC2)}{0.25};
\arrowpath{(lowC2)}{(highC2)}{0.8};
\draw[fill=black] (midC2) circle(2pt);
\node[below] at (lowC2) {\scriptsize{$U_2$}};
\node[above] at (highC2) {\scriptsize{$V_2$}};
\node[below] at ($(lowC1)!0.5!(lowC2) - (0, 0.5)$) {\scriptsize{(c)}};
\end{scope}
\end{tikzpicture}
\caption{Construction of $\mathcal{C}_0^{Z_2}$: (a) A morphism can involve a $b$ line or not. Notice that the direction of $b$ line is irrelevant as it is equal to its dual. (b) Composition of two morphisms involving a $b$ line is obtained by using the canonical map from $b\otimes b$ to identity to join the $b$ lines. (c) Tensor product of two morphisms involving a $b$ line is twisted by a half-braiding of $b$ across $V_1$.}
\label{fig:condensed}
\end{figure}
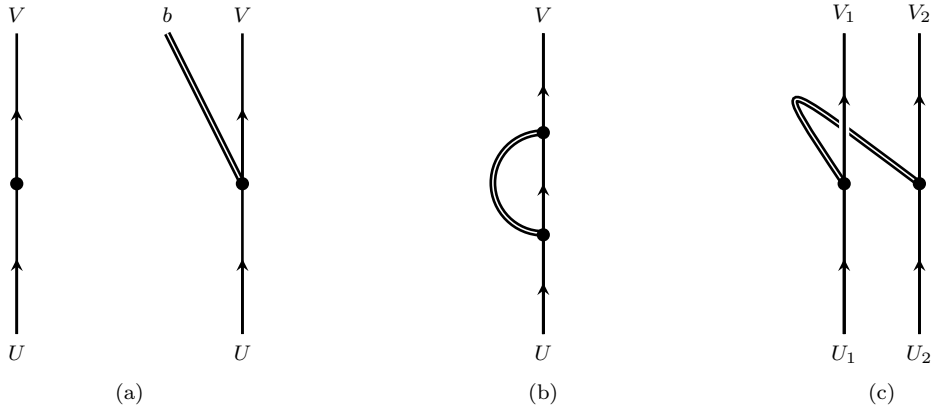

We then define a new category $\cC_0^{\ZZ_2}$ with the ``same'' objects and enlarged spaces of morphisms: 
\begin{equation}
\Hom_{\cC_0^{\ZZ_2}}(U_0,V_0) \equiv \Hom_{\cC_b}(U,V) \oplus \Hom_{\cC_{b}}(U,b \otimes V)
\end{equation}
The new morphisms should be thought of as $B$-twisted sectors. 
The morphisms are composed with the help of the $b \otimes b \to 1$ map and the 
tensor product is defined with the help of $\beta$, as in Figure \ref{fig:condensed}. 

The image of simple objects under this map may not be simple: if $X$ is simple, $b \otimes X$ 
is also simple and may or not coincide with $X$. In the former case, $\Hom_{\cC_0^{\ZZ_2}}(X_0,X_0)$
is two-dimensional and $X_0$ will split into two simples $X_0^\pm$. 

We then add to $\cC_0^{\ZZ_2}$ the simple summands of the objects inherited from $\cC_{b}$. Concretely, 
$X_0^\pm$ can be described as $X_0$ with the insertion of a projector $\pi^\pm_X$ along the line. 
The projectors will be linear combinations of the generator $1_X$ of $\Hom_{\cC_b}(X,X)$ and 
the generator $\xi_X$ of $\Hom_{\cC_{b}}(X,b \otimes X)$. We can compute $\xi_X^2 = \eta_X 1_X$ 
and define projectors 
\begin{equation}
\pi^\pm_X = \frac{1}{2}(1_X \pm (\eta_X)^{-1/2} \xi_X)
\end{equation}
Notice that if $b$ is the identity line in $\cC_b$, the identity in $\cC_0^{\ZZ_2}$ will itself split.

The final result will be a spherical (multi-)fusion category $\cC_0^{\ZZ_2}$. 
We can extend $\cC_0^{\ZZ_2}$ further to a $\ZZ_2$-graded category in the following manner. 
We extend the original category $\cC_b$ to a new graded category $\cC_b \times \ZZ_2$, a direct sum of $2$ copies of $\cC_b$.
We extend the $\ZZ_2$ 1-form symmetry to $\cC_b \times \ZZ_2$ by the center element $(b, \beta \times (-1)^\epsilon)$,
where $B$ it taken to lie in $\cC_b \times \{0\}$ and we twisted the original $\beta$ by a sign when 
crossing a line in $\cC \times \{1\}$. 

Finally, we proceed as before using the extended center element. Objects in $\cC_b \times \{\epsilon\}$ map to objects in 
$\cC_\epsilon^{\ZZ_2}$. Concretely, the only difference between objects in $\cC_0^{\ZZ_2}$ and $\cC_1^{\ZZ_2}$
is an extra sign in the tensor product of morphisms which appear when the $b$ line crosses a $\cC_1^{\ZZ_2}$ object. 

\subsection{Example: 1-form symmetries in the toric code}
Consider again the spherical fusion category $\cC$ modelled on $\ZZ_2$, with two objects $1$ and $P$ 
fusing as $P \otimes P \simeq 1$ and trivial associators. 
 
The Wilson loop in this $\ZZ_2$ gauge theory is the object $e = (1,\beta_P = -1)$ in the center of the category. 
It is a boson generating an ``electric'' $\ZZ_2^e$ 1-form symmetry. If we gauge this 1-form symmetry, we obtain a category $\cC_e$
with elements $I_0$, $P_0$ with a two-dimensional space of morphisms. We can denote the generators of these morphisms as
$1_1$, $\xi_1$, $1_P$, $\xi_P$. We have $\xi_1^2=1_1$ and $\xi_P^2=1_P$. 
 
 We can decompose $I_0= I^{++} + I^{--}$ and $P_0 = P^{+-} + P^{-+}$. Working out the fusion rules, 
 we find a multi-fusion category, with $P^{+-}$ and $P^{-+}$ being domain walls between the two vacua. 
 Each vacuum has a trivial category of line defects. 

Adding twisted sectors gives us two new objects, $I_1= I^{+-} + I^{-+}$ and $P_1 = P^{++} + P^{--}$. Hence our final graded multi-fusion category 
has has four sectors, $\cC_{\pm \pm}$, each consisting of an element of grading $0$ and an element of grading $1$. 
Physically, this is a boundary condition with two trivial vacua, each described by the trivial $\ZZ_2$-graded fusion category. 

This makes sense. We obtained the toric code by gauging the $\ZZ_2$ global symmetry of a trivial theory. 
In the absence of boundary conditions, gauging the dual 1-form symmetry effectively ungauges the $\ZZ_2$ gauge theory. 
In the presence of boundary conditions, gauging the standard $\ZZ_2$ symmetry with Dirichlet b.c. 
leaves us with a bulk $\ZZ_2$ gauge theory with a residual $\ZZ_2$ global symmetry at the boundary.  
This can be thought of as a $\ZZ_2$ gauge theory coupled to a boundary $\ZZ_2$-valued sigma model. 
After we gauge the 1-form symmetry, the boundary sigma model remains and the extra $\ZZ_2$ global symmetry
is spontaneously broken. 

On the other hand, gauging the 1-form symmetry generated by $m$ leads to a category $\cC_m$ with two isomorphic simple elements $I_{0}$, $P_{0}$. This is again a trivial category
of line defects. Adding twisted sectors, we find two more isomorphic objects, $I_1$ and $P_1$. We have obtained again the trivial $\ZZ_2$-graded fusion category. 

\subsection{The $\Pi$-product of shadows}
The product of two theories $\fT_f$ and $\fT'_f$ is equipped with a bosonic line $\Pi \Pi'$ which generates a standard 
bosonic $\ZZ_2$ 1-form symmetry. If we gauge $\Pi \Pi'$, we obtain a new theory which we can denote as $\fT_f \times_f \tilde \fT_f$.
This new theory still has a fermionic 1-form symmetry, generated by $\Pi$, or equivalently $\Pi'$ (the two coincide in the new theory).  
It is our candidate for the shadow of $\fT_s \times \tilde \fT_s$. 

The shadow product $\fT_f \times_f \fT'_f$ should be associative, as it corresponds to the product operation 
of the corresponding spin-TFTs. This is quite clear from the definition as well: the product of three shadows contain bosonic generators 
$\Pi \Pi'$, $\Pi' \Pi''$ and $\Pi \Pi''$ generating a $\ZZ_2\times \ZZ_2$ 1-form symmetry. Gauging the two $\ZZ_2$ in any order should be equivalent to 
gauging both. In the language of anyon condensation, we are condensing the algebra $A = I + \Pi \Pi'+\Pi' \Pi''+\Pi \Pi''$. 

We would like to explore the group structure of the candidate fermionic SPT phases we have encountered until now. Recall that 
we have introduced two basic classes of fermionic SPT phases: Ising pull-backs $\cI^\xi_{\pi_1}[G]$ and Gu-Wen phases $\cG_{\nu_3, n_2}[G]$. 

\subsubsection{Gu-Wen SPT phases}
As a simple example, consider two Gu-Wen phases $\cG_{\nu_3, n_2}[G]$ and $\cG_{\tilde \nu_3, \tilde n_2}[G]$.
We can take the $G$-graded product of the corresponding categories. The result 
is a $G$-graded category with objects $V_{g,\epsilon, \tilde \epsilon}$ 
which fuse according to a $\ZZ_2 \times \ZZ_2$ extension $G'$ of $G$, 
with cocycle $(n_2, \tilde n_2)$ and associators $\hat \alpha_3 \tilde {\hat \alpha}_3$. 

The bosonic symmetry generator is $V_{e,1,1}$, equipped with crossing $(-1)^{\epsilon+\tilde\epsilon}$.
As we gauge the symmetry, we will extend the morphisms so that $V^0_{g,\epsilon, \tilde \epsilon}$ and $V^0_{g,\epsilon+1, \tilde \epsilon+1}$
become isomorphic. Keeping this identification into account, the resulting objects will fuse according to 
the $\ZZ_2$  extension $\hat G$ of $G$ associated to the cocycle $n_2 + \tilde n_2$. 

Computing the associator of the new category takes a bit of effort. For concreteness, we can pick representative 
objects $V^0_{g, \epsilon,0}$. When we multiply them, we obtain, say, $V^0_{g g', \epsilon + \epsilon' + n_2(1,g,g g'),\tilde n_2(1,g,g g')}$
which has to be mapped back to $V^0_{g g', \epsilon + \epsilon' + n_2(1,g,g g')+\tilde n_2(1,g,g g'),0}$
by inserting $\tilde n_2(1,g,g g')$ extra intersections with $\Pi \tilde \Pi$ lines. 

We can gauge fix and then compute the associator via the tetrahedron graph. We obtain $\nu_3\tilde\nu_3(-1)^{\epsilon_1 \cup \tilde n_2+n_2 \cup \epsilon_1}$ 
where $\epsilon_1$ encodes the first $\ZZ_2$ grading of the elements placed on the edges. This differs from $\nu_3\tilde\nu_3(-1)^{(n_2+\tilde n_2)\cup\epsilon_1}$ by a sign
\begin{align}
&(-1)^{\epsilon_1\cup\tilde n_2+\tilde n_2\cup\epsilon_1}=(-1)^{\delta(\epsilon_1\cup_1\tilde n_2)+\epsilon_1\cup_1\delta\tilde n_2 +\delta\epsilon_1\cup_1\tilde n_2}
\end{align}
The second term above is zero and the first term can be absorbed into a gauge redefinition of the associator. Hence, we obtain a new Gu-Wen super-cohomology phase $(\nu_3',n_2+\tilde n_2)$ with
\begin{equation}
\nu_3'=\nu_3\tilde\nu_3(-1)^{n_2\cup_1\tilde n_2}
\end{equation}

This is indeed the expected group law for Gu-Wen fermionic SPT phases. 

\subsubsection{The squared equivariant toric code}
Another interesting example is the product of two equivariant toric codes. The resulting $\ZZ_2$-graded category 
has objects $II$, $PI$, $IP$, $PP$ in $\cC_0$ and $SS$ in $\cC_1$. The bosonic generator is then $\psi\psi$. 

We choose the two fourth roots $\xi_1$ and $\xi_2$ of $-1$ which identify specific Ising $\Pi$-categories $\cI^{\xi_1}$ and $\cI^{\xi_2}$. 
Recall that only $\xi^2$ affects the crossing phases. A flip $\xi \to - \xi$ changes the associator $SSS$ and thus effectively 
twists the category by a $\ZZ_2$ group cocycle, i.e. multiplies the theory by a bosonic $\ZZ_2$ SPT phase. 

Gauging the 1-form symmetry leads one to identify the pairs $II^0 \simeq PP^0$ and $PI^0 \simeq IP^0$, 
while $SS^0$ will split into some $S^0_+$ and $S^0_-$.

Fusion of $SS^0$ with $PI^0$ from the left involves crossing $\psi\psi$ across $PI$ and hence flips the sign of the non-trivial morphism of $SS^0$ to itself. 
On the other hand, fusion with $PI^0$ from the right flips the sign of the non-trivial morphism because of non-trivial $PSP$ associators for Ising category. 
We thus learn that $PI^0\otimes S_+^0=S_+^0\otimes PI^0=S_-^0$ and $PI^0\otimes S^0_-=S_-^0\otimes PI^0=S_+^0$. These fusion rules do not depend 
on $\xi_1$ or $\xi_2$. 

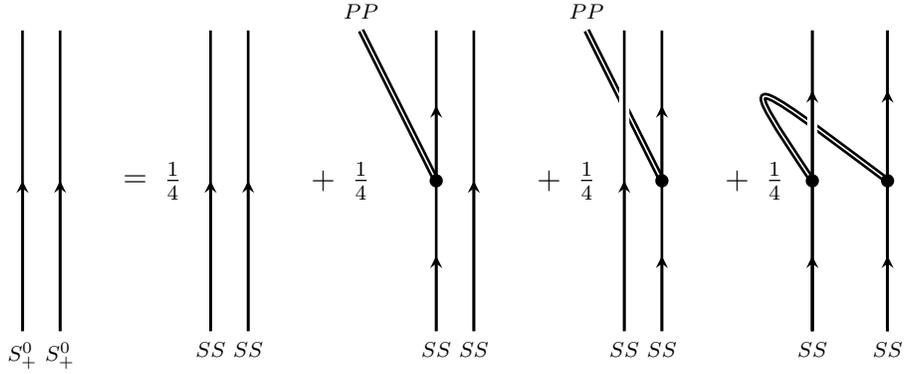
\begin{figure}
\centering
\begin{tikzpicture}[line width=1pt]
\begin{scope}[every node/.style={sloped,allow upside down}]
\coordinate (lowA) at (0,0);
\coordinate (lowB) at ($(lowA)+(0.5,0)$);
\coordinate (highA) at ($(lowA)+(0,4)$);
\coordinate (highB) at ($(lowB)+(0,4)$);

\arrowpath{(lowA)}{(highA)}{0.5};
\arrowpath{(lowB)}{(highB)}{0.5};

\node[below] at (lowA) {\scriptsize{$S_+^0$}};
\node[below] at (lowB) {\scriptsize{$S_+^0$}};

\coordinate (eq) at ($(lowB)+(1,2)$);
\node at (eq) {$=$};

\coordinate (lowS1) at ($(lowB)+(2,0)$);
\coordinate (lowS2) at ($(lowS1)+(0.5,0)$);
\coordinate (highS1) at ($(lowS1)+(0,4)$);
\coordinate (highS2) at ($(lowS2)+(0,4)$);

\arrowpath{(lowS1)}{(highS1)}{0.5};
\arrowpath{(lowS2)}{(highS2)}{0.5};

\node[below] at (lowS1) {\scriptsize{$SS$}};
\node[below] at (lowS2) {\scriptsize{$SS$}};

\coordinate (p1) at ($(lowS2)+(1,2)$);
\node at (p1) {$+$};

\coordinate (lowS3) at ($(lowS2)+(2.5,0)$);
\coordinate (lowS4) at ($(lowS3)+(0.5,0)$);
\coordinate (highS3) at ($(lowS3)+(0,4)$);
\coordinate (highS4) at ($(lowS4)+(0,4)$);

\coordinate (midS3) at ($(lowS3)!0.5!(highS3)$);
\coordinate (PP1) at ($(highS3)-(1,0)$);
\arrowpath{(lowS3)}{(midS3)}{0.5};
\arrowpath{(midS3)}{(highS3)}{0.5};
\draw[double] (midS3)--(PP1);
\draw[fill=black] (midS3) circle(2pt);
\node[above] at (PP1) {\scriptsize{$PP$}};

\arrowpath{(lowS4)}{(highS4)}{0.5};

\node[below] at (lowS3) {\scriptsize{$SS$}};
\node[below] at (lowS4) {\scriptsize{$SS$}};

\coordinate (p2) at ($(lowS4)+(1,2)$);
\node at (p2) {$+$};

\coordinate (lowS5) at ($(lowS4)+(2,0)$);
\coordinate (lowS6) at ($(lowS5)+(0.5,0)$);
\coordinate (highS5) at ($(lowS5)+(0,4)$);
\coordinate (highS6) at ($(lowS6)+(0,4)$);

\coordinate (midS6) at ($(lowS6)!0.5!(highS6)$);
\coordinate (PP2) at ($(highS6)-(1,0)$);
\arrowpath{(lowS6)}{(midS6)}{0.5};
\arrowpath{(midS6)}{(highS6)}{0.5};
\draw[double] (midS6)--(PP2);
\paddedline{($(lowS5)$)}{(highS5)}{(0.05,0)};
\draw[fill=black] (midS6) circle(2pt);
\node[above] at (PP2) {\scriptsize{$PP$}};

\arrowpath{(lowS5)}{(highS5)}{0.5};

\node[below] at (lowS5) {\scriptsize{$SS$}};
\node[below] at (lowS6) {\scriptsize{$SS$}};

\coordinate (p3) at ($(lowS6)+(1,2)$);
\node at (p3) {$+$};

\coordinate (lowS7) at ($(lowS6)+(2,0)$);
\coordinate (highS7) at ($(lowS7) + (0,4)$);
\coordinate (midS7) at ($(lowS7)!0.5!(highS7)$);
\coordinate (lowS8) at ($(lowS7)+(1,0)$);
\coordinate (highS8) at ($(lowS8) + (0,4)$);
\coordinate (midS8) at ($(lowS8)!0.5!(highS8)$);

\draw[double] (midS7) .. controls ($(midS7) + (-1,1.5)$) .. (midS8);

\paddedline{($(lowS7)!0.6!(highS7)$)}{(highS7)}{(0.05,0)};
\arrowpath{(lowS7)}{(highS7)}{0.25};
\arrowpath{(lowS7)}{(highS7)}{0.8};
\draw[fill=black] (midS7) circle(2pt);
\node[below] at (lowS7) {\scriptsize{$SS$}};

\arrowpath{(lowS8)}{(highS8)}{0.25};
\arrowpath{(lowS8)}{(highS8)}{0.8};
\draw[fill=black] (midS8) circle(2pt);
\node[below] at (lowS8) {\scriptsize{$SS$}};

\node at ($(eq)+(0.5,0)$) {$\frac{1}{4}$};
\node at ($(p1)+(0.5,0)$) {$\frac{1}{4}$};
\node at ($(p2)+(0.5,0)$) {$\frac{1}{4}$};
\node at ($(p3)+(0.5,0)$) {$\frac{1}{4}$};

\end{scope}
\end{tikzpicture}
\caption{A sample computation of fusion rules in shadow product of two equivariant toric codes with $\xi_1=\xi_2$: $S_+^0\otimes S_+^0$ is by definition a sum of four terms which involve associators and crossings. $II$ inside $SS\otimes SS$ is mapped to zero object as second term cancels against the third and the first term cancels against the fourth. $PI$ is mapped to $PI$ by the first and fourth terms and to $IP$ by the second and third terms. Hence, $S_+^0\otimes S_+^0\simeq PI^0$.} \label{fig:quad}
\end{figure}

The fusion rules involving $S_+^0$ and $S_-^0$, on the other hand, are affected by the $\beta_S$ crossing phases. We find that if $\xi_1 = \xi_2$, or more generally 
$\xi_1^2 \xi_2^2=-1$, we have $S_+^0\otimes S_+^0\simeq S_-^0\otimes S_-^0\simeq PI^0$ and $S_+^0\otimes S_-^0\simeq S_-^0\otimes S_+^0\simeq II^0$:
the objects in the new category fuse according to a $\hat G=\ZZ_4$ group law, generated, say, by $S^0_+$. We demonstrate an example of computation of fusion rules for this case in  fig. \ref{fig:quad}.

The $\hat G=\ZZ_4$ can be regarded as a $\ZZ_2$ central extension of $G=\ZZ_2$ 
with $V_{0,0}=II^0$, $V_{0,1}=PI^0$, $V_{1,0}=S_+^0$ and $V_{1,1}=S_-^0$. 
It can be easily checked that $PI^0$ equipped with crossing $(-1)^\epsilon$ is a fermionic bulk line. 
The result is the shadow of Gu-Wen phase for a $\ZZ_2$ global symmetry, with $\ZZ_4$ being the central extension. 

On the other hand, if $\xi_1^2 \xi_2^2=1$, we have $S_+^0\otimes S_+^0\simeq S_-^0\otimes S_-^0\simeq II^0$ and $S_+^0\otimes S_-^0\simeq S_-^0\otimes S_+^0\simeq PI^0$:
the objects in the new category fuse according to a $\hat G=\ZZ_2 \times \ZZ_2$ group law. We can set, say, $V_{0,0}=II^0$, $V_{0,1}=PI^0$, $V_{1,0}=S_+^0$ and $V_{1,1}=S_-^0$.
The result is the shadow of a Gu-Wen phase for a $\ZZ_2$ global symmetry, with trivial central extension, i.e. a bosonic SPT phase. 

We still need to compute the associator $\hat \alpha_3(\xi_1, \xi_2)$. We can compute the associativity phases for 
$S_\pm^0$ from the associator for $SS \otimes SS \otimes SS$ or by evaluating some tetrahedron planar graphs. 
The general calculation is somewhat tedious and we will omit it. It should be obvious that if $\xi_1 \xi_2 = 1$ all crossing or associator phases 
will cancel out among the two theories. Thus we expect to obtain a trivial associator as well as the trivial group extension. Thus we claim 
\begin{equation}
\cI^{\xi} \times_f \cI^{\xi^{-1}} \simeq I
\end{equation}
where $I$ denotes the trivial $\ZZ_2$ SPT phase. In particular, this proves the claim that the Ising $\Pi$-category is the shadow of an SPT phase!

On the other hand, $\cI^{\xi} \times_f \cI^{\xi}$ will be a root Gu-Wen $\ZZ_2$ SPT phase, which one of the two being determined by the value of 
$\xi^2$, as the sign of $\xi$ can be changed by adding a bosonic SPT phase. We can compute $\nu_3$ for that phase by looking at graphs 
involving $S_+^0$ and identity lines, with $\Pi$ lines emerging from junctions with two incoming $S_+$ lines. The only source of interesting 
 phases is the crossing phase of the fermion and $S_+$. We find that if $\xi^2 = \pm i$, then 
 $\nu_3 = \pm \frac{1}{4} \eta_1 \cup \eta_1 \cup \eta_1$. 

\subsubsection{Ising pull-back and Gu-Wen}

We can combine the Ising pull-back category with homomorphism $\pi_1$ with a Gu-Wen phase. 
The $G$-graded product has objects $I_{g,\epsilon}$, $P_{g,\epsilon}$ or $S_{g, \epsilon}$ depending 
on the value of $\pi_1(g)$. Gauging the bosonic 1-form symmetry identifies $P^0_{g,\epsilon}$ with $I^0_{g,\epsilon+1}$ 
and $S_{g,\epsilon}$ with $S_{g,\epsilon+1}$. 

We can restrict ourselves to objects $I^0_{g,\epsilon}$, or $S^0_{g, 0}$. Effectively, the $n_2$ cocycle has been 
restricted to a cocycle $n_2^0$ on $G_0=\ker\ \pi_1$.
The fusion rules of this category mimic our example based on a long exact sequence 
\begin{equation}
0 \to \ZZ_2 \to \hat G_0 \to G \to \ZZ_2 \to 0
\end{equation}
One might wonder what happens to the rest of the data of the $n_2$ cocycle which is not captured by $n_2^0$. This data goes into the associators for the new category. 
In particular, it is possible to extract the values of $n_2(g,g')$ from (relative) signs of certain associators. This is of course true in the Gu-Wen case 
as well, where $n_2(g,g')$ is also encoded, for example, by the sign in the associator $(V_{g,0}\otimes V_{g',0})\otimes V_{1,1}\to V_{g,0}\otimes(V_{g',0}\otimes V_{1,1})$.

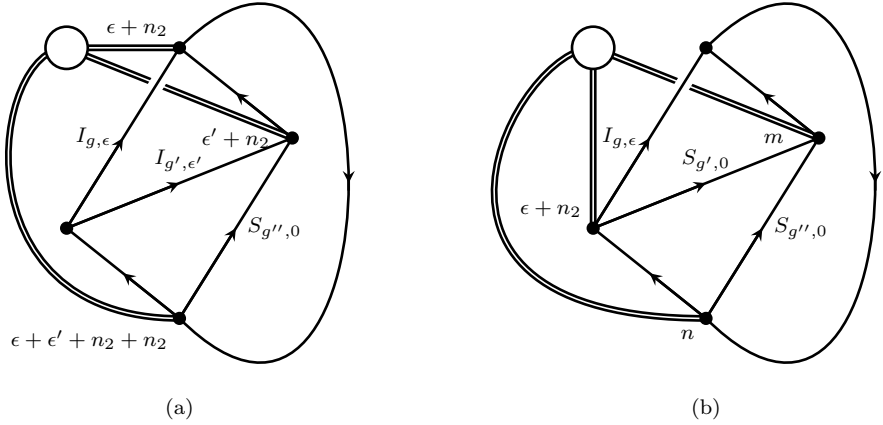
\begin{figure}
\begin{tikzpicture}[line width=1pt]
\begin{scope}[every node/.style={sloped,allow upside down}]
\coordinate (023) at (0,0);
\coordinate (012) at ($(023)+(-1.5,1.2)$);
\coordinate (123) at ($(023)+(1.5,2.4)$);
\coordinate (013) at ($(023)+(0,3.6)$);
\coordinate (controlu) at ($(013)+(3,3)$);
\coordinate (controld) at ($(023)+(3,-3)$);
\coordinate (mid) at ($(controlu)!0.5!(controld)-(0.75,0)$);
\coordinate (blob) at ($(012)+(0,2.4)$);
\coordinate (controllu) at ($(blob)-(1,0)$);
\coordinate (controlld) at ($(023)-(3,0)$);

\draw[double] (123) to (blob);
\paddedline{(012)}{(013)}{(0.1,0)};
\draw[double] (013) to (blob);
\arrowpath{(023)}{(012)}{0.5};
\arrowpath{(023)}{(123)}{0.5};
\arrowpath{(012)}{(123)}{0.5};
\arrowpath{(012)}{(013)}{0.5};
\arrowpath{(123)}{(013)}{0.5};
\draw (013)..controls(controlu) and (controld)..(023);
\arrowpath{($(mid)+(0,0.1)$)}{($(mid)-(0,0.1)$)}{0.5};
\draw[double] (023)..controls(controlld) and (controllu)..(blob);

\node[below left] at (023) {\scriptsize{$\epsilon+\epsilon'+n_2+n_2$}};
\node[above left] at (013) {\scriptsize{$\epsilon+n_2$}};
\node at ($(123)-(0.75,0)$) {\scriptsize{$\epsilon'+n_2$}};

\node[right] at ($(023)!0.5!(123)$) {\scriptsize{$S_{g'',0}$}};
\node[above] at ($(012)!0.5!(123)$) {\scriptsize{$I_{g',\epsilon'}$}};
\node[left] at ($(012)!0.5!(013)$) {\scriptsize{$I_{g,\epsilon}$}};

\draw[fill=black] (012) circle(2pt);
\draw[fill=black] (013) circle(2pt);
\draw[fill=black] (023) circle(2pt);
\draw[fill=black] (123) circle(2pt);
\draw[fill=white] (blob) circle(8pt);

\coordinate (023s) at ($(023)+(7,0)$);
\coordinate (012s) at ($(023s)+(-1.5,1.2)$);
\coordinate (123s) at ($(023s)+(1.5,2.4)$);
\coordinate (013s) at ($(023s)+(0,3.6)$);
\coordinate (controlus) at ($(013s)+(3,3)$);
\coordinate (controlds) at ($(023s)+(3,-3)$);
\coordinate (mids) at ($(controlus)!0.5!(controlds)-(0.75,0)$);
\coordinate (blobs) at ($(012s)+(0,2.4)$);
\coordinate (controllus) at ($(blobs)-(1,0)$);
\coordinate (controllds) at ($(023s)-(4.5,0)$);

\draw[double] (123s) to (blobs);
\paddedline{(012s)}{(013s)}{(0.1,0)};
\draw[double] (012s) to (blobs);
\arrowpath{(023s)}{(012s)}{0.5};
\arrowpath{(023s)}{(123s)}{0.5};
\arrowpath{(012s)}{(123s)}{0.5};
\arrowpath{(012s)}{(013s)}{0.5};
\arrowpath{(123s)}{(013s)}{0.5};
\draw (013s)..controls(controlus) and (controlds)..(023s);
\arrowpath{($(mids)+(0,0.1)$)}{($(mids)-(0,0.1)$)}{0.5};
\draw[double] (023s)..controls(controllds) and (controllus)..(blobs);

\node[above left] at (012s) {\scriptsize{$\epsilon+n_2$}};
\node[below left] at (023s) {\scriptsize{$n$}};
\node at ($(123s)-(0.6,0)$) {\scriptsize{$m$}};

\node[right] at ($(023s)!0.5!(123s)$) {\scriptsize{$S_{g'',0}$}};
\node[above] at ($(012s)!0.5!(123s)$) {\scriptsize{$S_{g',0}$}};
\node[left] at ($(012s)!0.5!(013s)$) {\scriptsize{$I_{g,\epsilon}$}};

\draw[fill=black] (012s) circle(2pt);
\draw[fill=black] (013s) circle(2pt);
\draw[fill=black] (023s) circle(2pt);
\draw[fill=black] (123s) circle(2pt);
\draw[fill=white] (blobs) circle(8pt);

\node (a) at ($(023)+(0,-1.2)$) {\scriptsize{(a)}};
\node (b) at ($(023s)+(0,-1.2)$) {\scriptsize{(b)}};
\end{scope}
\end{tikzpicture}
\caption{Two sample computations of associators for a phase corresponding to long exact sequence. The values at the starting of double lines encode the number of $P_{1,1}$ lines. We leave the argument of $n_2$ self-evident as it can be read from the diagram. $m$ and $n$ are numbers (defined modulo 2) associated to the choice of morphisms at the junctions where two $S$ lines converge and diverge respectively. $m$ is 1 if it corresponds to the morphism $S\otimes S\to P$ and 0 if it corresponds to $S\otimes S\to I$. $n$ is defined similarly. The graph in (b) evaluates to a non-zero number only if $n_2(g',g'')+n_2(g,g'g'')+n_2(gg',g'')+\epsilon+m+n=0$ which is the same as $n_2(g,g')+\epsilon+m+n=0$. As a result of this, the double lines always come in pairs. The graphs in (a) and (b) imply that the associators respectively are $(-1)^{\epsilon(\epsilon'+n_2(g',g''))}\nu_3$ and $\lambda^{(m+1)(n+1)}(-1)^{\epsilon m}\nu_3$ where $\lambda$ is a square root of 2.} \label{fig:les}
\end{figure}

The associators can be determined from tetrahedron graph by inserting the bosonic line $P_{1,1}$ at appropriate junctions. All of them can be written (modulo factors of square root of 2) as $\nu_3$ times a sign which depends on $n_2$, the choice of morphism $S\otimes S\to(I,P)$ and $\epsilon$ grading of lines. We show two sample associators and their results in Figure \ref{fig:les}. 

Choosing $\epsilon=1$, $\epsilon'=0$  and $g$ to be identity in Figure \ref{fig:les}(a) tells us that the associator equals $(-1)^{n_2(g',g'')}$. This means that sign of this associator determines $n_2(g',g'')$ for such that $\pi_1(g')=0$ and $\pi_1(g'')=1$. Similarly, we could compute the associator of $I_{g,\epsilon}$, $S_{g',0}$ and $I_{g'',\epsilon''}$ and choosing $g$ as identity, $\epsilon=1$ and $\epsilon''=0$ would determine $n_2(g',g'')$ such that $\pi_1(g')=1$ and $\pi_1(g'')=0$. Determining $n_2(g',g'')$ such that $\pi_1(g')=1$ and $\pi_1(g'')=1$ is a bit more non-trivial. It is determined by the associator in Figure \ref{fig:les}(b) when we choose $\epsilon=1$, $m=n_2(g',g'')$ and the particular $n$ for which the graph evaluates to a non-zero number. Notice that there is only one such $n$.

We will verify now that every long sequence example can be obtained in this manner.

\subsubsection{Ising pull-back and long exact sequence with the same $\pi_1$}
The product category has objects $V_{g_0, \epsilon}$, $PV_{g_0, \epsilon}$ and $SS_{g_1}$. 
The bosonic generator is associated to $P_{e,1}$. Condensation will identify $V_{g_0, \epsilon}$ and $PV_{g_0, \epsilon+1}$
and split $SS_{g_1}$ to $S_{g_1,\epsilon}$. 

It turns out that the consistency of fusion rules completely constrains them. First of all, we don't physically expect any of $S_{g,\epsilon}\otimes S_{g',\epsilon'}$ to be the zero object. This implies that they must fuse to a single object since the fusion of sums $(S_{g,0}\oplus S_{g,1})\otimes (S_{g',0}\oplus S_{g',1})$ is equal to sum of four objects $V_{gg',0}\oplus V_{gg',0}\oplus V_{gg',1}\oplus V_{gg',1}$. Using similar arguments, we find that the fusion of two simple objects must be a single simple obejct. Second, $V_{g,0}$ and $V_{g,1}$ must map $S_{g',\epsilon}$ to different objects. If, on the contrary $V_{g,0}\otimes S_{g',\epsilon}\simeq V_{g,1}\otimes S_{g',\epsilon}\simeq S_{gg',\epsilon'}$, then we could fuse by $S_{g'',\epsilon''}$ from the right to find that the elements in subcategory associated to $G_0$ do not fuse according to a cocycle, leading to a contradiction. Third, the fusion of $S$ elements with themselves must be captured by a cochain. This can be shown using a similar argument as above. This cochain can be combined with the cocycle for $G_0$ to give rise to a cochain for $G$ governing the fusion rules for the full category. Associativity of fusion then implies that this cochain must be a cocycle $n_2$.

Thus, we see that this is a Gu-Wen extension example, with objects $V_{g_0, \epsilon}$ and $V_{g_1, \epsilon} = S_{g_1,\epsilon}$. As $\cI^\xi_{\pi_1}$ and $\cI^{\xi^{-1}}_{\pi_1}$ are inverse to each other, we can express any long exact sequence example as the $\Pi$-product of 
$\cI^\xi_{\pi_1}$ and a Gu-Wen phase.

\subsubsection{Product of long exact sequence examples}
In a similar manner, we can verify that the $\times_f$ product of two long exact sequence examples 
is a new long exact sequence example. The product has a bosonic line $V_{e,1}\tilde V_{e,1}$. Let $\pi_1$ and $\tilde\pi_1$ be respectively the two homomorphisms. 

\begin{itemize}
\item In the $\pi_1(g)=\tilde\pi_1(g)=0$ sector, gauging the bosonic 1-form symmetry identifies $V_{g,\epsilon}\tilde V_{g,\epsilon'}$ with $V_{g,\epsilon+1}\tilde V_{g,\epsilon'+1}$ 
and we can choose representative objects as $V_{g,\epsilon}'=V_{g,\epsilon}\tilde V_{g,0}$. 

\item In the $\pi_1(g)=0, \tilde\pi_1(g)=1$ sector,$V_{g,\epsilon}\tilde S_{g}$ is identified with $V_{g,\epsilon+1}\tilde S_{g}$ and we choose representative object $S'_g=V_{g,0}\tilde S_g$. 

\item In the $\pi_1(g)=1, \tilde\pi_1(g)=0$ sector, $S_{g}\tilde V_{g,\epsilon'}$ is identified with $S_{g}\tilde V_{g,\epsilon'+1}$ and we choose representative object $S'_g=S_g \tilde V_{g,0}$. 

\item In the $\pi_1(g)=\tilde\pi_1(g)=1$ sector, $S_g \tilde S_g$ splits into two objects (as in the product of two equivariant toric codes above) which we denote as $V'_{g,0}$ and $V'_{g,1}$. 
\end{itemize}
The fusion rules of representative objects can be obtained analogously to the examples above. 

This can be identified with a long exact sequence 
\begin{equation}
0 \to \ZZ_2 \to G' \to G \to \ZZ_2 \to 0
\end{equation}
with $G\to\ZZ_2$ homomorphism $\pi_1'=\pi_1+\tilde\pi_1$. It is somewhat trickier to determine the $G'$ central extension: while the restriction to  $\pi_1(g_1)=\pi_1'(g_1)=\pi_1(g_2)=\pi_1'(g_2)=0$
coincides with $n_2(g_1,g_2)+\tilde n_2(g_1,g_2)$, the rest of it depends on the details of the associators of the two initial categories.  

We can attack the problem by specializing first to Ising pull-backs.

\subsection{Triple products and quaternions}
In consideration of our analysis, we expect some relation of the form 
\begin{equation}
\cI^\xi_{\pi_1}[G] \times_f \cI^\xi_{\pi'_1}[G] = \cG_{\nu_3(\pi_1, \pi'_1,\xi), n_2(\pi_1, \pi'_1,\xi)}[G] \times_f \cI^{\xi^{-1}}_{\pi_1+ \pi'_1}[G]
\end{equation}
We switched the $\xi$ phase for the Ising pull-back on the right hand side for future convenience. 

In order to extract the Gu-Wen phase which appears in this expression, we consider the triple product 
\begin{equation}
\cG_{\nu_3(\pi_1, \pi'_1), n_2(\pi_1, \pi'_1)}[G] = \cI^\xi_{\pi_1}[G]  \times_f \cI^{\xi}_{\pi_1+ \pi'_1}[G] \times_f \cI^\xi_{\pi'_1}[G] \label{triple}
\end{equation}

The details of the calculation only depend on the image of $G$ group elements under $\pi_1$ and $\pi'_1$. Without loss of generality, we can do our computation for
$G = \ZZ_2 \times \ZZ'_2$ with $\pi_1$ and $\pi'_1$ being the projections into the first and second factor respectively. The general answer will be obtained by pulling back 
the $\ZZ_2 \times \ZZ'_2$ answer by $\pi_1 \times \pi'_1$.

This is a rather non-trivial calculation, but it is somewhat simplified by the permutation symmetry acting on the triple $\pi_1$, $\pi'_1$, $\pi_1+ \pi'_1$, although
gauge-fixing choices may break the symmetry at intermediate stages of the calculation. The $[n_2]$ cocycle is actually independent from $\xi^2$: a shift of $\xi^2$ will be implemented by multiplying 
by the root Gu-Wen phase pulled back along $\pi_1$, $\pi'_1$ and $\pi_1+ \pi'_1$,
which shifts the cocycle by 
\begin{equation}
\pi_1 \cup \pi_1 + \pi'_1 \cup \pi'_1 +(\pi_1+ \pi'_1)\cup (\pi'_1+ \pi'_1) = \pi_1 \cup \pi'_1 + \pi'_1 \cup \pi_1
\end{equation}
which is exact. 

It turns out to be possible to pick a gauge-fixing in which $n_2$ is at least cyclically symmetric. We take triple product of elements of various Ising categories in the order mentioned in (\ref{triple}). For instance, $\pi_1=0,\pi'_1=1$ sector contains elements of the form $ISS$ and $PSS$. The lines $IPP$, $PIP$ and $PPI$ give rise to a $\ZZ_2\times\ZZ_2$ bosonic 1-form symmetry. There are two choices of junctions between these three lines. They correspond to canonical junctions between $IPP$, $PIP$ and $PPI$ lines taken in clockwise and counter-clockwise order respectively. Their product is clearly equal to 1 and their square is $-1$ as it involves a crossing. Hence, when we bring together these centre lines in calculations, we multiply the canonical junctions by $i$ and $-i$ respectively. \footnote{In the language of anyon condensation, 
this is the chocie of maps $A \otimes A \to A$ and $A \to A \otimes A$ with good properties.}  

When we condense, the three lines generate three non-trivial morphisms such that the product of two of these gives rise to the third. We choose $PPI$ to identify $ISS$ with $PSS$, $IPP$ to identify $SIS$ with $SPS$, and $PIP$ to identify $SSI$ with $SSP$ in a cyclic fashion. Similarly, we choose $IPP$ to split $ISS$ into $ISS_+$ and $ISS_-$ etc. in a cyclic manner. We summarize our choice of objects in the condensed category:
\begin{itemize}
\item In $\pi_1=\pi'_1=0$ sector, the objects are $III$ and $IPI$. $IPI$ equipped with an appropriate crossing is the generator of fermionic 1-form symmetry. We rename $III$ and $IPI$ as $V_1$ and $V_{-1}$ respectively.
\item In $\pi_1=0,\pi'_1=1$ sector, the objects are $ISS_+$ and $ISS_-$. We rename them as $V_{\pm i}$.
\item In $\pi_1=1,\pi'_1=0$ sector, the objects are $SSI_+$ and $SSI_-$. We rename them as $V_{\pm j}$.
\item In $\pi_1=\pi'_1=1$ sector, the objects are $SIS_+$ and $SPS_-$. We rename them as $V_{\pm k}$.
\end{itemize}

Some of the computations of fusion rules are completely analogous to the case of squared equivariant toric code. These are $(\pm q)\otimes(-1)\simeq(-1)\otimes (\pm q)\simeq \mp q$, $q\otimes q\simeq -1$ and $q\otimes (-q)\simeq 1$ where $q$ denotes either one of $i$, $j$ and $k$.

\begin{figure}
\centering
\begin{tikzpicture}[line width=1pt]
\begin{scope}[every node/.style={sloped,allow upside down}]
coordinate (lowS1) at (0,0);
\coordinate (lowS2) at ($(lowS1)+(0.5,0)$);
\coordinate (highS1) at ($(lowS1)+(0,4)$);
\coordinate (highS2) at ($(lowS2)+(0,4)$);

\coordinate (low1) at ($(lowS1)!0.5!(lowS2)$);
\coordinate (high1) at ($(highS1)!0.5!(highS2)$);

\arrowpath{(lowS1)}{(highS1)}{0.5};
\arrowpath{(lowS2)}{(highS2)}{0.5};

\draw($(low1)-(0,0.5)$)--(lowS1);
\draw($(low1)-(0,0.5)$)--(lowS2);
\draw($(high1)+(0,0.5)$)--(highS1);
\draw($(high1)+(0,0.5)$)--(highS2);
\draw($(low1)-(0,0.5)$)--($(low1)-(0,1)$);
\draw($(high1)+(0,0.5)$)--($(high1)+(0,1)$);
\draw[fill=black] ($(low1)-(0,0.5)$) circle(2pt);
\draw[fill=black] ($(high1)+(0,0.5)$) circle(2pt);

\node[left] at (lowS1) {\scriptsize{$ISS$}};
\node[right] at (lowS2) {\scriptsize{$SSI$}};

\node[below] at ($(low1)-(0,1)$) {\scriptsize{$SIS$}};
\node[above] at ($(high1)+(0,1)$) {\scriptsize{$SIS$}};

\coordinate (p1) at ($(lowS2)+(1,2)$);
\node at (p1) {$+$};

\coordinate (lowS3) at ($(lowS2)+(2.75,0)$);
\coordinate (lowS4) at ($(lowS3)+(0.5,0)$);
\coordinate (highS3) at ($(lowS3)+(0,4)$);
\coordinate (highS4) at ($(lowS4)+(0,4)$);

\coordinate (low2) at ($(lowS3)!0.5!(lowS4)$);
\coordinate (high2) at ($(highS3)!0.5!(highS4)$);

\coordinate (midS3) at ($(lowS3)!0.5!(highS3)$);
\coordinate (PP1) at ($(highS3)-(1,0)$);
\arrowpath{(lowS3)}{(midS3)}{0.5};
\arrowpath{(midS3)}{(highS3)}{0.5};
\draw[double] (midS3)..controls(PP1)..($(high2)+(0,1.5)$);
\draw[fill=black] (midS3) circle(2pt);
\node at ($(PP1)-(0.25,0)$) {\scriptsize{$IPP$}};

\draw($(low2)-(0,0.5)$)--(lowS3);
\draw($(low2)-(0,0.5)$)--(lowS4);
\draw($(high2)+(0,0.5)$)--(highS3);
\draw($(high2)+(0,0.5)$)--(highS4);
\draw($(low2)-(0,0.5)$)--($(low2)-(0,1)$);
\draw($(high2)+(0,0.5)$)--($(high2)+(0,1.5)$);
\draw($(high2)+(0,1.5)$)--($(high2)+(0,2)$);
\draw[fill=black] ($(low2)-(0,0.5)$) circle(2pt);
\draw[fill=black] ($(high2)+(0,0.5)$) circle(2pt);
\draw[fill=black] ($(high2)+(0,1.5)$) circle(2pt);

\arrowpath{(lowS4)}{(highS4)}{0.5};

\node[left] at (lowS3) {\scriptsize{$ISS$}};
\node[right] at (lowS4) {\scriptsize{$SSI$}};

\node[below] at ($(low2)-(0,1)$) {\scriptsize{$SIS$}};
\node[above] at ($(high2)+(0,2)$) {\scriptsize{$SIS$}};
\node[right] at ($(high2)+(0,1)$) {\scriptsize{$SPS$}};

\coordinate (p2) at ($(lowS4)+(1,2)$);
\node at (p2) {$+$};

\coordinate (lowS5) at ($(lowS4)+(2.75,0)$);
\coordinate (lowS6) at ($(lowS5)+(0.5,0)$);
\coordinate (highS5) at ($(lowS5)+(0,4)$);
\coordinate (highS6) at ($(lowS6)+(0,4)$);

\coordinate (low3) at ($(lowS5)!0.5!(lowS6)$);
\coordinate (high3) at ($(highS5)!0.5!(highS6)$);

\coordinate (midS6) at ($(lowS6)!0.5!(highS6)$);
\coordinate (PP2) at ($(highS6)-(1,0)$);
\arrowpath{(lowS6)}{(midS6)}{0.5};
\arrowpath{(midS6)}{(highS6)}{0.5};
\draw[double] (midS6)--($(PP2)+(0,0.75)$);
\paddedline{($(lowS5)$)}{(highS5)}{(0.05,0)};
\draw[fill=black] (midS6) circle(2pt);
\node at ($(PP2)-(0.2,0)$) {\scriptsize{$PPI$}};

\draw[double]($(high3)+(0,1.5)$)--($(PP2)+(0,0.75)$);
\draw[double]($(PP2)+(0,0.75)$)--($(PP2)+(-0.75,0.75)$);

\draw($(low3)-(0,0.5)$)--(lowS5);
\draw($(low3)-(0,0.5)$)--(lowS6);
\draw($(high3)+(0,0.5)$)--(highS5);
\draw($(high3)+(0,0.5)$)--(highS6);
\draw($(low3)-(0,0.5)$)--($(low3)-(0,1)$);
\draw($(high3)+(0,0.5)$)--($(high3)+(0,1.5)$);
\draw($(high3)+(0,1.5)$)--($(high3)+(0,2)$);
\draw[fill=black] ($(low3)-(0,0.5)$) circle(2pt);
\draw[fill=black] ($(high3)+(0,0.5)$) circle(2pt);
\draw[fill=black] ($(high3)+(0,1.5)$) circle(2pt);
\draw[fill=black] ($(PP2)+(0,0.75)$) circle(2pt);

\arrowpath{(lowS5)}{(highS5)}{0.5};

\node[left] at (lowS5) {\scriptsize{$ISS$}};
\node[right] at (lowS6) {\scriptsize{$SSI$}};

\node[below] at ($(low3)-(0,1)$) {\scriptsize{$SIS$}};
\node[above] at ($(high3)+(0,2)$) {\scriptsize{$SIS$}};
\node[right] at ($(high3)+(0,1)$) {\scriptsize{$SPS$}};

\node[above left] at ($(PP2)+(-0.25,0.75)$) {\scriptsize{$PIP$}};
\node at ($(high3)+(-0.6,1.4)$) {\scriptsize{$IPP$}};

\coordinate (p3) at ($(lowS6)+(1,2)$);
\node at (p3) {$+$};

\coordinate (lowS7) at ($(lowS6)+(3.25,0)$);
\coordinate (highS7) at ($(lowS7) + (0,4)$);
\coordinate (midS7) at ($(lowS7)!0.5!(highS7)$);
\coordinate (lowS8) at ($(lowS7)+(1,0)$);
\coordinate (highS8) at ($(lowS8) + (0,4)$);
\coordinate (midS8) at ($(lowS8)!0.5!(highS8)$);

\coordinate (low4) at ($(lowS7)!0.5!(lowS8)$);
\coordinate (high4) at ($(highS7)!0.5!(highS8)$);
\coordinate (meet) at ($(midS7)+(-1.25,2)$);

\draw[double](midS7)--(meet);
\draw[double](midS8)--(meet);
\draw[double](meet)--($(meet)+(-0.25,1)$);

\paddedline{($(lowS7)!0.6!(highS7)$)}{(highS7)}{(0.05,0)};
\arrowpath{(lowS7)}{(highS7)}{0.25};
\arrowpath{(lowS7)}{(highS7)}{0.8};
\draw[fill=black] (midS7) circle(2pt);
\node[left] at (lowS7) {\scriptsize{$ISS$}};

\draw[fill=black] (meet) circle(2pt);

\draw($(low4)-(0,0.5)$)--(lowS7);
\draw($(low4)-(0,0.5)$)--(lowS8);
\draw($(high4)+(0,0.5)$)--(highS7);
\draw($(high4)+(0,0.5)$)--(highS8);
\draw($(low4)-(0,0.5)$)--($(low4)-(0,1)$);
\draw($(high4)+(0,0.5)$)--($(high4)+(0,1)$);
\draw[fill=black] ($(low4)-(0,0.5)$) circle(2pt);
\draw[fill=black] ($(high4)+(0,0.5)$) circle(2pt);

\arrowpath{(lowS8)}{(highS8)}{0.25};
\arrowpath{(lowS8)}{(highS8)}{0.8};
\draw[fill=black] (midS8) circle(2pt);
\node[right] at (lowS8) {\scriptsize{$SSI$}};

\node at ($(meet)-(0.1,0.7)$) {\scriptsize{$IPP$}};
\node at ($(meet)+(0.8,-0.25)$) {\scriptsize{$PPI$}};
\node at ($(meet)+(0.25,0.75)$) {\scriptsize{$PIP$}};

\node[below] at ($(low4)-(0,1)$) {\scriptsize{$SIS$}};
\node[above] at ($(high4)+(0,1)$) {\scriptsize{$SIS$}};

\node at ($(eq)+(0.5,0)$) {$\frac{1}{4}$};
\node at ($(p1)+(0.5,0)$) {$\frac{1}{4}$};
\node at ($(p2)+(0.5,0)$) {$\frac{1}{4}$};
\node at ($(p3)+(0.5,0)$) {$\frac{1}{4}$};

\end{scope}
\end{tikzpicture}
\caption{The figure depicts various terms in the computation of $ISS_+\otimes SSI_+$. We need to convert $SPS$ into $SIS$ using the chosen isomorphisms. This results in junctions of the three bosonic lines with appropriate signs. After taking into account various factors from associators, crossings and junctions, we obtain $ISS_+\otimes SSI_+\simeq SIS_+$.}\label{fig:quat}
\end{figure}
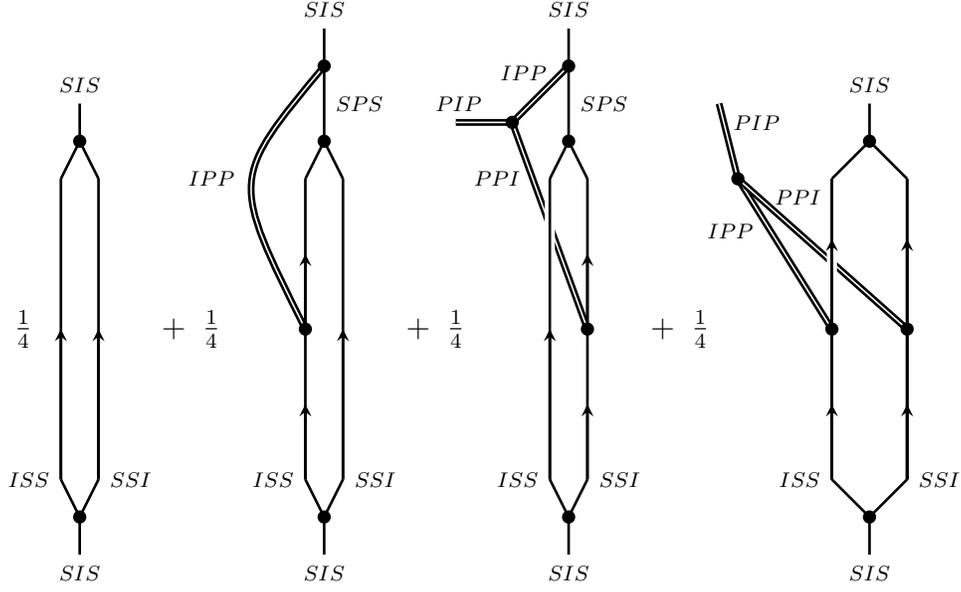

The other computations are analogous but we have to be careful about choosing correct sign for the junctions of three bosonic lines. We show how these junctions arise in a sample computation in \ref{fig:quat}. The final result is captured by the quaternion group:
\begin{align}
&i^2 = j^2 = k^2 = -1 \cr
&i j = - j i = k \cr
&j k = - k j = i \cr
&k i = - i k = j 
\end{align}
This corresponds to the cocycle
\begin{equation}
n_2(\pi_1, \pi'_1) = \pi_1 \cup \pi_1 + \pi'_1 \cup \pi'_1 +\pi_1 \cup \pi'_1 
\end{equation}
This describes the quaternion group as a $\ZZ_2$ central extension of $\ZZ_2 \times \ZZ'_2$!

The extension indeed enjoys $S_3$ permutation symmetry, up to a gauge transformation $i \to -i$, $j \to -j$, $k \to -k$ 
for odd permutations. 

Computing $\nu_3(\pi, \pi', \xi)$ is of course rather more cumbersome. We leave it as an exercise for the enthusiastic reader. 

\subsection{The group structure of fermionic SPT phases}
The dependence of $\cI^\xi[\ZZ_2]$ on $\xi$ is very mild: we can change $\xi$ to another fourth root of $-1$ by multiplying it 
by one of the four $\ZZ_2$ Gu-Wen phases. Correspondingly, we can change $\xi$ in $\cI^\xi_{\pi_1}[G]$ by multiplying it by 
the pull-back along $\pi_1$ of one of the four $\ZZ_2$ Gu-Wen phases. Consequently, we can just stick to a specific choice of $\xi$ in the following. 

We expect all fermionic SPT phases to take the form $\cG_{\nu_3, n_2}[G] \otimes \cI^\xi_{\pi_1}[G]$. We could label such a phase by a triple 
$(\nu_3, n_2, \pi_1)$. It is natural to ask what is the group law for such phases. 

We know that the product of two Gu-Wen phases is another Gu-Wen phase, with addition law
\begin{equation}
\cG_{\nu_3, n_2} \times \cG_{\nu'_3, n'_2} = \cG_{\nu_3+ \nu'_3+ \frac12 n_2 \cup_1 n_2', n_2+ n'_2}
\end{equation}
This can be expressed as the statement that the group $\cG[G]$ of Gu-Wen phases is a central extension 
\begin{equation}
0 \to H^3[BG,U(1)] \to \cG[G] \to H^2[BG,\ZZ_2] \to 0
\end{equation}
with cocycle $\frac12 n_2 \cup_1 n_2'$ valued in $H^3[BG,U(1)]$.

Similarly, when we add Ising pull-back phases the $\pi_1$ cocycles add up. Hence  the group of fermionic SPT phases $\cF[G]$ of the form $(\nu_3, n_2, \pi_1)$
will be a central extension 
\begin{equation}
0 \to \cG[G] \to \cF[G] \to H^1[BG,\ZZ_2] \to 0
\end{equation}
The $\cG[G]$-valued cocycle $\cG_2$ for this extension can be computed by the relation 
\begin{equation}
\cI^\xi_{\pi_1}[G] \times_f \cI^\xi_{\pi'_1}[G] = \cG_2(\pi_1, \pi'_1,\xi) \times_f \cI^{\xi}_{\pi_1+ \pi'_1}[G]
\end{equation}
Comparing with our previous computation, the change $\xi^{-1} \to \xi$ on the right hand side 
shifts $n_2$ by $(\pi_1+ \pi'_1)\cup (\pi_1+ \pi'_1)$. Thus $\cG_2(\pi_1, \pi'_1,\xi)$ 
has cocycle 
\begin{equation}
n_2(\pi_1, \pi'_1) = \pi'_1 \cup \pi_1 
\end{equation}
This corresponds to the dihedral group extension of $\ZZ_2 \times \ZZ'_2$.

A standard presentation of the dihedral group is given by elements $a$ and $b$ such that $a^4=b^2=1$ and $aba=b$. In our case, we can choose $V_a=SIS_+$ and $V_b=SSI_+$.

\subsection{$\Pi$-categories and $\Pi$-supercategories}
There is a known relationship between $\Pi$-categories and super-categories
which is analogous the the relation between $\cC_b$ and $\cC_{\ZZ_2}$ in the bosonic case \cite{Brundan:2016aa}. 

Given a $\Pi$-category $\cC_f$, we can build a super-category $\cC_s$ whose even morphisms 
are $\Hom_{\cC_f}(X,Y)$ and odd morphisms are $\Hom_{\cC_f}(X,\Pi \otimes Y)$.
This is a ``$\Pi$-supercategory'', i.e. a super-category equipped with an object $\Pi$ with is odd-isomorphic to 
$I$. Vice versa, we can go from a $\Pi$-supercategory to a $\Pi$-category by dropping the odd morphisms. 

In a previous work \cite{GaiottoKapustin} , we sketched a state-sum construction based on spherical super-fusion categories. For simplicity, 
we assumed the spherical super-fusion category had no Majorana objects, i.e. irreducible objects with an even and an odd endomorphisms. 
If we take the ``$\Pi$-envelope'' of such a super-category we will get a $\Pi$-supercategory with an even and an odd copy of each irreducible object.
Dropping odd morphisms we get  a $\Pi$-category.  The state-sum construction given in \cite{GaiottoKapustin}  builds up the 
spin TFT whose shadow is associated to this $\Pi$-category. In the next section, we will formulate the state-sum construction
for general $\Pi$-categories. It should be possible to re-formulate it in terms of the associated super-categories, with or without 
Majorana objects. 

\section{Spherical fusion categories and state sum constructions}\label{sec:swiss}

It is instructive to review the physical derivation of the Turaev-Viro construction for a 3d TFT with a single vacuum and a gapped boundary condition. 

\begin{figure}
\begin{center}
\includegraphics[width=15cm]{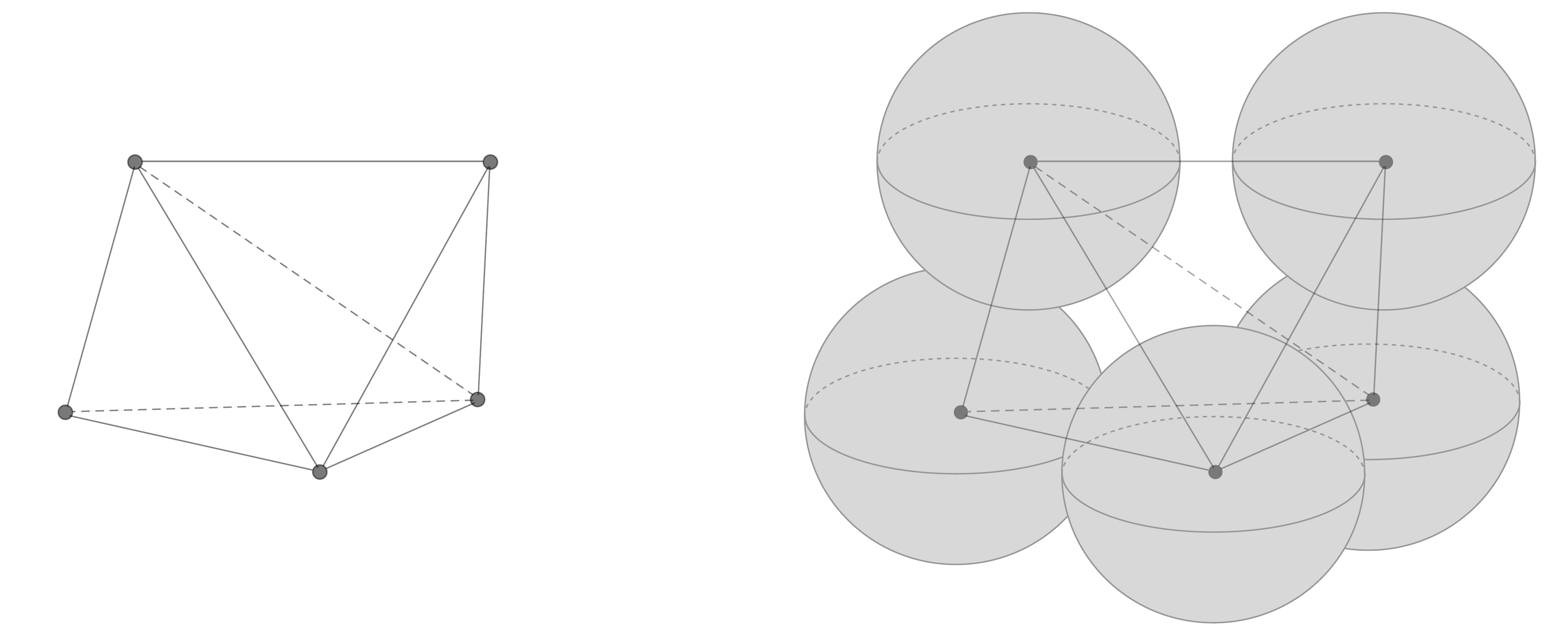}
\end{center}
\caption{The first step of a swiss cheese construction: the manifold is triangulated, and spherical holes are opened up at the vertices of the triangulation.}
\label{fig:foam1}
\end{figure}

\begin{figure}
\begin{center}
\includegraphics[width=15cm]{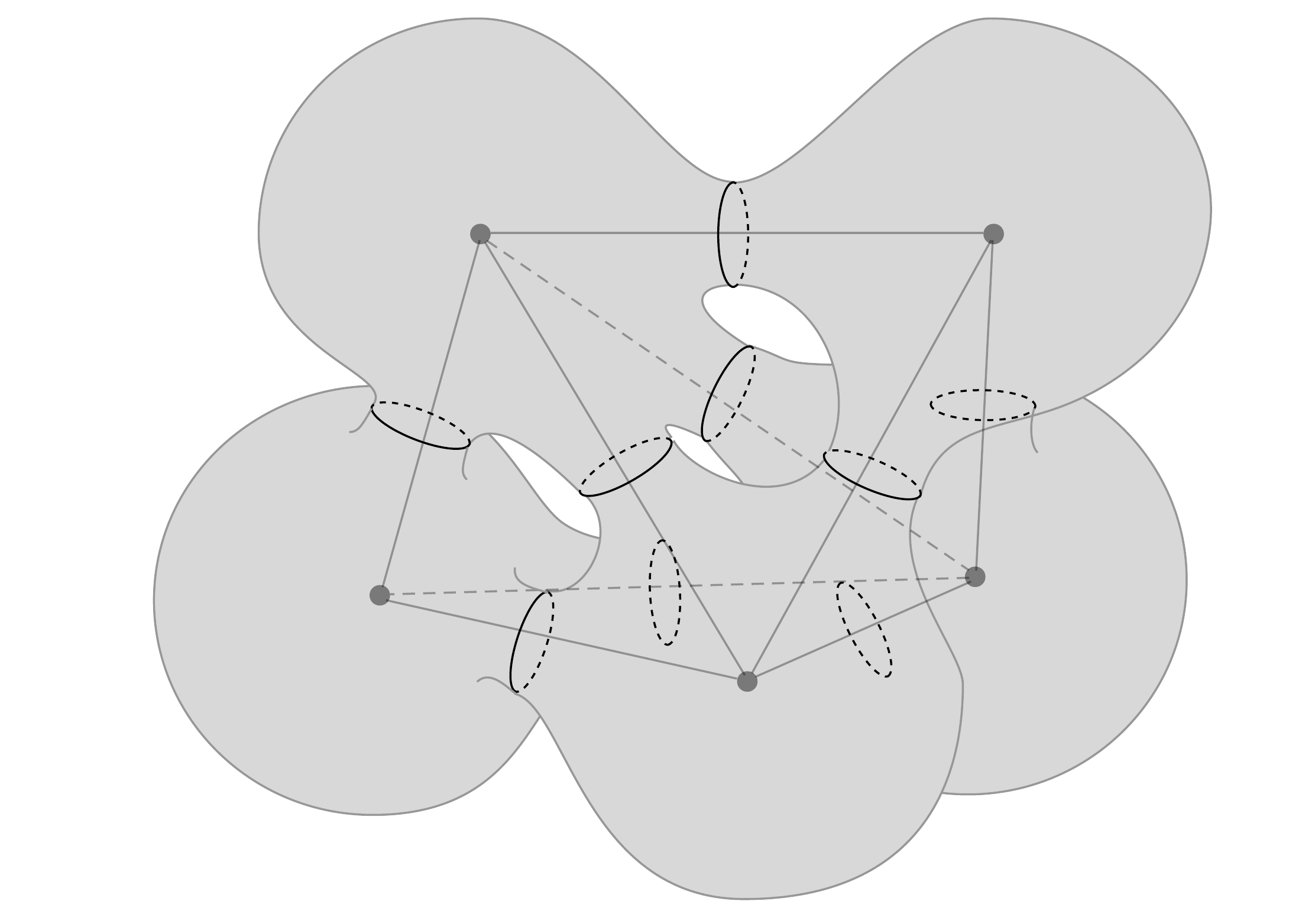}
\end{center}
\caption{The second step of a swiss cheese construction: we connect the holes by tubes running along the edges of the triangulation. A special line is added 
around the tubes to make them trivial.}
\label{fig:foam2}
\end{figure}

\begin{figure}
\begin{center}
\includegraphics[width=15cm]{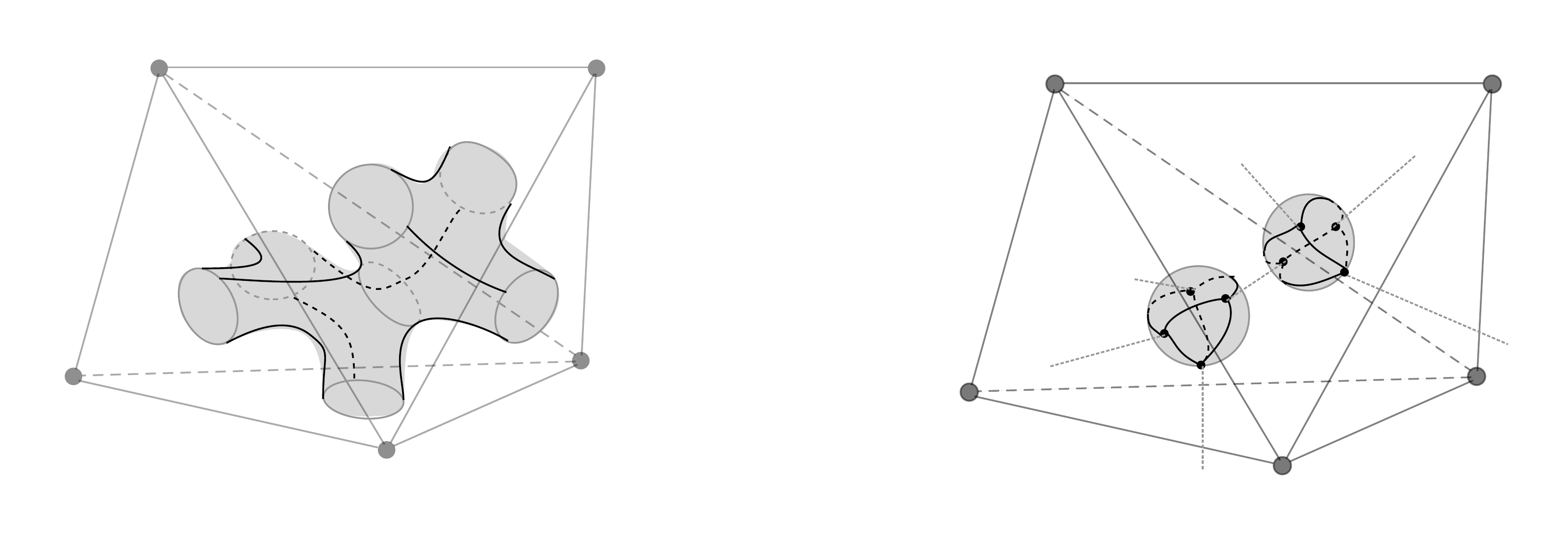}
\end{center}
\caption{The third step of a swiss cheese construction. The complement of the holes is a collection of solid cylinders running through 
faces of the triangulation, fused at tetra-valent vertices inside the tetrahedra.
The cross-section of the solid cylinders is a disk with three boundary punctures. We insert a complete basis of states along each cylinder.}
\label{fig:foam3}
\end{figure}

We begin with the observation that such a topological field theory $\fT$ associates a one-dimensional Hilbert space to 
a two-sphere. Thus a boundary $\fB$ with the topology of a two-sphere must create a state in that Hilbert space which is proportional to the 
state created by a three-ball, with some specific proportionality constant $C_\fB$ which depends on the theory and on the boundary condition.  

Consider a three-manifold $M$, say with no boundaries, for which we want to compute the partition function. 
Equip $M$ with some triangulation. Up to a factor of $C_\fB$ for each vertex, the partition function of $M$ will be 
the same as the partition function of a manifold $M'$ obtained from $M$ by removing a small ball around each vertex of the triangulation
and replacing it with a spherical boundary of type $\fB$. See Figure \ref{fig:foam1}.

We can enlarge the holes in $M'$ until the spherical boundaries collide with each other, so that each hole 
almost fills the corresponding 3-cell in the cell-decomposition of $M$ dual to the triangulation. 
The manifold $M'$ looks like a foam of empty bubbles. 

Next, we can ``pop'' the walls between bubbles. Concretely, this requires us to carve out parts of $M'$ with the topology of a cylinder 
with $\fB$ boundaries at each end, i.e. $[0,1]_{\fB,\fB} \times D^2$. The manifold is cut along the annulus $[0,1]_{\fB,\fB} \times \partial D^2$.
The path integral on the cylinder produces some state in the Hilbert space associated by the theory to the annulus with $\fB$ boundary conditions on the edge. 
We can replace the cylinder by some other geometry bounded by the same annulus, as long as they produce the 
same vector in the annulus Hilbert space. 

An example of such geometry is half a solid torus, bounded by that annulus and by an annulus with $\fB$ boundary condition, decorated by some
boundary line defect $L_i$ running along the annulus. It is natural to expect such geometries to produce a basis in the Hilbert space as the choice of $L_i$ is varied over 
all simple objects. \footnote{This should be analogous to the statement that solid tori with a bulk line defect give a basis of the Hilbert space associated to a torus.} 
Thus the cylinder path integral should produce a state which can be decomposed as a linear combination of these elements, with some coefficients $c_i$.
The correct choice of $c_i$ is known to coincide with the quantum dimensions $d_i$. 

We use the replacement of the cylinder geometry with the decorated half-solid torus to open holes in all the walls between bubbles, 
once for each 2-cell in the cell-decomposition of $M$ dual to the triangulation. The result is a sum over 
manifolds $M''_\ell$ labelled by the choice $\ell$ of lines for each 2-cell. See Figure \ref{fig:foam2}.

We can enlarge the holes in the walls until they almost fill the corresponding 2-cells in the cell-decomposition of $M$ dual to the triangulation. 
The manifold $M''_\ell$ looks like the 1-skeleton of the cell decomposition. Each 1-cell between 2-cells associated to lines $L_i$, $L_j$, $L_k$ 
corresponds to a component of the manifold with the cross-section of a disk with three punctures where the three lines $L_i$, $L_i$ and $L_k$ lie. 
See Figure \ref{fig:foam3}.

Finally, we can cut the 1-cells by using the Hilbert space associated to the disk with three boundary punctures. We can identify this Hilbert space with the 
space $V_{ijk}$ of local operators available at a junction between defects $L_i$, $L_j$, $L_k$ by the state-operator map. Inserting a complete basis of states across
each 1-cell we decompose the three-manifold to a collection of three-balls, with a tetrahedral graph of line defects drawn on the boundary. See again Figure \ref{fig:foam3}.
The partition function for each decorated three-ball can be evaluated using the data of $\cC$. 

These three steps express the original partition function as a state sum involving ingredients which can be computed fully in terms of the 
category $\cC$ of boundary line defects. If we take the basis of boundary line defects $L_i$ to consist of the simple objects in $\cC$ 
we obtain the  Turaev-Viro state sum.

The physical construction suggests that more general choices of collections of objects in $\cC$ should also reproduce the same 
partition sum, as long as one picks the correct $c_i$ coefficients to reproduce the correct sum of simples 
$\sum_i d_i L_i$. 

The construction can be extended to more general three manifolds, including a variety of extra topological defects. 
It is very simple to add boundaries with $\fB$ boundary conditions and arbitrary graphs of boundary line defects drawn on 
the boundary. This leads to the same state sum over a cell complex with boundary. 
\footnote{It is also possible to include other topological boundary conditions $\fB'$ (or interfaces) to the construction, but this requires extra data to be provided, 
in the form of the $\cC$-module ($(\cC,\cC)$-bimodule) category of domain lines between $\fB$ and $\fB'$.}

Another important example are bulk topological line defects, which turn out to be labelled by elements of the Drinfeld center $Z[\cC]$ of the spherical fusion category.
Concretely, an element of the Drinfeld center is an object in $\cC$ equipped with choice of canonical junction as it crosses any other element. The data encodes 
the image of the bulk line when brought to the boundary. 

The bulk lines $Y_\alpha$ ``run'' along the 1-skeleton of the construction, resulting into a modification of 
the vector spaces which appear along the 1-cells to the spaces $V_{\alpha;ijk}$ of local operators available at a junction between defects $Y_\alpha$,$L_i$, $L_j$, $L_k$.

\subsection{Symmetries}

The Turaev-Viro construction can be refined to deal with three manifolds equipped with a non-trivial flat connection 
for a discrete group $G$ \cite{Turaev:2012aa}. The starting point of such a construction is a $G$-graded spherical fusion category $\cC$,
which consists of a collection of sub-categories $\cC_g$ labelled by elements of $G$. Essentially, the flat connection 
is represented on the triangulation by group elements on the edges of the triangulation and the state sum decorates 
edges labelled by $g$ with objects in $\cC_g$.

The output of the construction is a topological field theory with a non-anomalous $G$ global symmetry. The theory is equipped 
with a topological boundary condition where the $G$ symmetry may be broken.

As before, we expect the converse to be true as well. A topological theory endowed with a non-anomalous
$G$ global symmetry and a topological boundary condition admits topological domain walls $U_g$ labelled by $G$ elements, 
which fuse according to the group law and admit canonical topological junctions. The boundary condition will be support 
categories $\cC_g$ of line defects at which a $U_g$ domain wall ends. Together, the $\cC_g$ form a $G$-graded spherical fusion category
which can be used to reconstruct the topological theory. 

In the absence of a flat connection, we can decorate all edges with the identity element $e$ and 
the state sum reduces to the Turaev-Viro construction for $\cC_e$. On the other hand, if we gauge the $G$ 
symmetry (with Dirichlet boundary conditions) the resulting theory is given by the Turaev-Viro construction for the whole $\cC$, forgetting about the grading. 

A topological field theory $T$ may also have a non-anomalous 1-form global symmetry. Concretely, that means that there is a set of 
bulk line defects $B_a$ which bosons, fuse according to a group law and braid trivially with each other. \footnote{Gauging a 1-form symmetry in $2+1$ dimensions 
should be a special case of the operation  of anyon condensation, which can be done to a theory which includes a topological line $A$ 
with sufficiently nice properties, generalizing the properties of $A = \oplus_a B_a$. }

A non-anomalous 1-form global symmetry allows one to couple the theory to a background 2-form flat connection. 
We will show how to include this coupling in the Turaev-Viro construction, by modifying the vector spaces 
attached to faces of the triangulation according to the value of the background 2-form. \footnote{Standard and 1-form global symmetries can be combined into 
the notion of 2-group. It would be interesting to integrate this possibility in our story.}

We will demonstrate that the anomaly of a fermionic $\ZZ_2$ 1-form symmetry can be
eliminated in a canonical way if the three-manifold is endowed with a spin structure. 
\subsection{Review of the Turaev-Viro construction}

We refer to \cite{2010arXiv1004.1533K,2010arXiv1010.1222B,2010arXiv1012.0560B} for a very clear discussion of the Turaev-Viro construction and of its relation to 
3d topological field theories $\fT$ equipped with a topological boundary condition $\fB$. 

We denote the spherical fusion category as $\cC$, with a finite set $I$ of (equivalence classes of) simple objects $V_i$. 
Remember that the space of local operators at a junction with outgoing line defects $V_1, \cdots V_n$ is $ \Hom_\cC\left(1, V_1 \otimes \cdots \otimes V_n \right)$. 

The building blocks of the Turaev-Viro construction are the spherical fusion category evaluation maps 
which assigns a complex number $Z(\Gamma)$ to a planar graph $\Gamma$ on a two sphere with 
edges labelled by objects and vertices labelled by morphism in $\cC$. 
More precisely, if we label the two ends of a segment $e$ by dual objects $V_e$ and $V^*_e$, 
a vertex $v$ joining edges $e_1, \cdots e_n$ is labelled by a morphism 
\begin{equation}
\varphi_v \in  \Hom_\cC\left(1, V_1 \otimes \cdots \otimes V_n \right)
\end{equation}
where $V_i$ are the objects associated to $v$ and $e_i$. 

The 3d partition function depends on a choice of manifold M, possibly decorated by bulk line defects $T_\alpha$ labelled by objects $Y_\alpha$ in the Drinfeld center $Z(\cC)$. 
The manifold may admit a boundary boundary $\partial M$, possibly decorated by boundary line defects $V_i$.

The first step in the calculation is to give a combinatorial description $\cM$ of $M$, which is essentially a decomposition of $M$ into convex polytopes, say tetrahedra. The partition function is computed as a sum over different ways to 
decorate the edges of $\cM$ by simple objects $l$ in $\cC$ (reversing the orientation of an edge conjugates the objects):
\begin{equation}
Z[\cM,\{Y_\alpha\}] = \sum_l \frac{\prod_i d_i^{\mathrm{edges}(\cM,i)}}{\cD^{2\mathrm{vertices}(\cM)}}Z[\cM,\{Y_\alpha\},l] 
\end{equation}
where we count bulk vertices with weight $1$ and boundary vertices with weight $1/2$ in $\mathrm{vertices}(\cM)$, and 
bulk edges with label $i$ with weight $1$ and boundary edges with label $i$ with weight $1/2$ in $\mathrm{edges}(\cM,i)$.
The $d_i$ and $\cD$ are quantum dimensions and total dimension.

The partial partition functions $Z[\cM,\{Y_\alpha\},l]$ are computed by gluing  
together contributions of the individual polytopes of $\cM$. Each face $C$ of the triangulation with counterclockwise edges $e_1, \cdots, e_n$ is associated to a vector space
\begin{equation}
H(C,l) = \Hom_\cC\left(1, l(e_1) \otimes \cdots \otimes l(e_n) \right)
\end{equation}
and the partial partition function is valued in $H(\partial \cM,l)  = \prod_{C \in \partial \cM} H(C,l)$. Pieces of a manifold are glued along 
faces $C$ and $\bar C$ by contracting the elements of dual vector spaces $H(C,l)$ and $H(\bar C,l)$. 

\begin{figure}
\centering
\begin{tikzpicture}[line width=1pt]
\begin{scope}[every node/.style={sloped,allow upside down}]
\coordinate (2) at (0,0);
\coordinate (1) at ($(2)+(3,-0.5)$);
\coordinate (0) at ($(2)+(2.1,3.25)$);
\coordinate (3) at ($(1)+(1.5,1.5)$);

\arrowpath{(2)}{(3)}{0.5};
\paddedline{(0)}{(1)}{(0.1,0)};
\arrowpath{(0)}{(1)}{0.5};
\arrowpath{(0)}{(2)}{0.5};
\arrowpath{(1)}{(2)}{0.5};
\arrowpath{(0)}{(3)}{0.5};
\arrowpath{(1)}{(3)}{0.5};

\node[above] at (0) {\scriptsize{$0$}};
\node[below] at (1) {\scriptsize{$1$}};
\node[left] at (2) {\scriptsize{$2$}};
\node[right] at (3) {\scriptsize{$3$}};

\node[above right] at ($(0)!0.5!(1)$) {\scriptsize{$a$}};
\node[above left] at ($(0)!0.5!(2)$) {\scriptsize{$b$}};
\node[above right] at ($(0)!0.5!(3)$) {\scriptsize{$c$}};
\node[below] at ($(1)!0.5!(2)$) {\scriptsize{$i$}};
\node[below right] at ($(1)!0.5!(3)$) {\scriptsize{$j$}};
\node[above left] at ($(2)!0.5!(3)$) {\scriptsize{$k$}};

\node (a) at ($(0)+(0.25,-4.5)$) {\scriptsize{(a)}};

\coordinate (023) at ($(2)+(9,0)$);
\coordinate (012) at ($(023)+(-1.5,1.2)$);
\coordinate (123) at ($(023)+(1.5,2.4)$);
\coordinate (013) at ($(023)+(0,3.6)$);
\node[draw,circle,inner sep=1pt] at ($(012)-(0.5,-0.25)$) {\scriptsize{$0$}};
\coordinate (controlu) at ($(013)+(3,3)$);
\coordinate (controld) at ($(023)+(3,-3)$);
\coordinate (mid) at ($(controlu)!0.5!(controld)-(0.75,0)$);
\coordinate (mid1) at ($0.33*(013)+0.33*(012)+0.33*(123)$);
\node[draw,circle,inner sep=1pt] at (mid1) {\scriptsize{$1$}};
\coordinate (mid2) at ($0.33*(023)+0.33*(012)+0.33*(123)$);
\node[draw,circle,inner sep=1pt] at (mid2) {\scriptsize{$2$}};
\node[draw,circle,inner sep=1pt] at ($(023)+(1,0.25)$) {\scriptsize{$3$}};

\arrowpath{(023)}{(012)}{0.5};
\arrowpath{(023)}{(123)}{0.5};
\arrowpath{(012)}{(123)}{0.5};
\arrowpath{(012)}{(013)}{0.5};
\arrowpath{(123)}{(013)}{0.5};
\draw (013)..controls(controlu) and (controld)..(023);
\arrowpath{($(mid)+(0,0.1)$)}{($(mid)-(0,0.1)$)}{0.5};


\node[below left] at ($(023)!0.5!(012)$) {\scriptsize{$b$}};
\node[right] at ($(023)!0.5!(123)$) {\scriptsize{$k$}};
\node[above] at ($(012)!0.5!(123)$) {\scriptsize{$i$}};
\node[left] at ($(012)!0.5!(013)$) {\scriptsize{$a$}};
\node[above right] at ($(123)!0.5!(013)$) {\scriptsize{$j$}};
\node[right] at (mid) {\scriptsize{$c$}};

\draw[fill=black] (012) circle(2pt);
\draw[fill=black] (013) circle(2pt);
\draw[fill=black] (023) circle(2pt);
\draw[fill=black] (123) circle(2pt);

\node at ($(a)+(7,0)$) {\scriptsize{(b)}};
\end{scope}
\end{tikzpicture}
\caption{Left: The basic ingredient of the state sum is a tetrahedron decorated with lines. Right: The dual planar graph in the spherical fusion category. For clarity we denoted with circled numbers the 
tetrahedron vertices dual to each face.} \label{fig:tetra}
\end{figure}
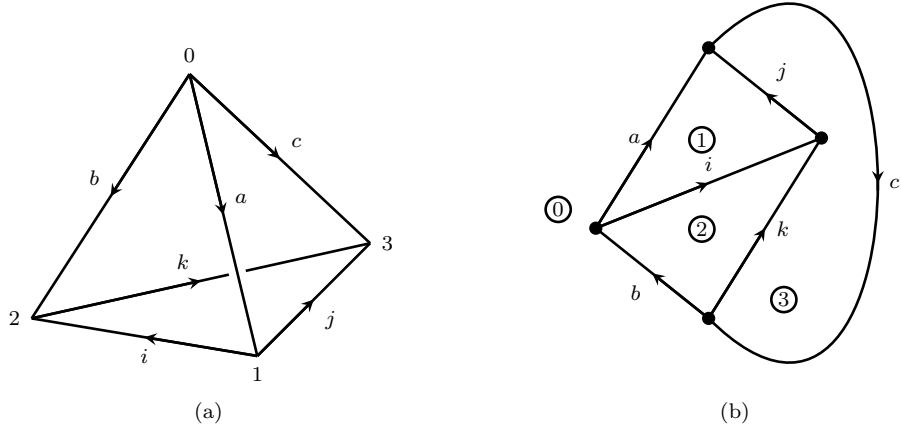

The contribution of an individual polytope is the output of the spherical fusion category evaluation map $Z_\Gamma$ for a spherical graph $\Gamma$ dual to the 
polytope. For example, a tetrahedron contribution is evaluated by the evaluation of a dual tetrahedral graph 
$\Gamma$, with vertices decorated by basis elements in $H(C,l)^*$. See Figure \ref{fig:tetra}.

\begin{figure}
\begin{tikzpicture}[line width=1pt,scale=1]
\begin{scope}[every node/.style={sloped,allow upside down}]
\coordinate (l1) at (0,0);
\coordinate (r1) at ($(l1) + (3,0)$);
\coordinate (cp1t) at ($(l1) + (0.5,1)$); 
\coordinate (cp2t) at ($(r1) + (-0.5,1)$); 
\coordinate (cp1b) at ($(l1) + (0.5,-1)$); 
\coordinate (cp2b) at ($(r1) + (-0.5,-1)$); 
\arrowpath{(l1)}{(r1)}{0.5};
\bezierarrowpath{(l1)}{(cp1t)}{(cp2t)}{(r1)}{0.5};
\bezierarrowpath{(l1)}{(cp1b)}{(cp2b)}{(r1)}{0.5};
\draw[fill=white] (l1) circle(0.7);
\draw[fill=white] (r1) circle(0.7);

\node at (l1) {$\Gamma_\ell$};
\node at (r1) {$\Gamma_r$};

\coordinate (eq) at ($(l1) + (5,0)$);
\node at (eq) {$=$};

\coordinate (l2) at ($(l1) + (7,0)$);
\coordinate (l2dot) at ($(l2) + (2,0)$);
\coordinate (cp1lt) at ($(l2) + (0,0.8)$); 
\coordinate (cp2lt) at ($(l2dot) + (-1,0.8)$); 
\coordinate (cp1lb) at ($(l2) + (0,-0.8)$); 
\coordinate (cp2lb) at ($(l2dot) + (-1,-0.8)$); 
\coordinate (r2dot) at ($(l2dot) + (2,0)$);
\coordinate (r2) at ($(r2dot) + (2,0)$);
\coordinate (cp1rt) at ($(r2dot) + (1,0.8)$); 
\coordinate (cp2rt) at ($(r2) + (0,0.8)$); 
\coordinate (cp1rb) at ($(r2dot) + (1,-0.8)$); 
\coordinate (cp2rb) at ($(r2) + (0,-0.8)$); 

\arrowpath{(l2)}{(l2dot)}{0.6};
\bezierarrowpath{(l2)}{(cp1lt)}{(cp2lt)}{(l2dot)}{0.7};
\bezierarrowpath{(l2)}{(cp1lb)}{(cp2lb)}{(l2dot)}{0.7};
\draw[fill=black] (l2dot) circle(2pt);
\draw[fill=white] (l2) circle(0.7);

\arrowpath{(r2dot)}{(r2)}{0.4};
\bezierarrowpath{(r2dot)}{(cp1rt)}{(cp2rt)}{(r2)}{0.3};
\bezierarrowpath{(r2dot)}{(cp1rb)}{(cp2rb)}{(r2)}{0.3};
\draw[fill=black] (r2dot) circle(2pt);
\draw[fill=white] (r2) circle(0.7);

\draw[dashed] (l2dot)--(r2dot);

\node at (l2) {$\Gamma_\ell$};
\node at (r2) {$\Gamma_r$};

\end{scope}
\end{tikzpicture}
\caption{A crucial identity for spherical fusion category evaluation maps: a graph $\Gamma$ (Left) can be split into two simpler graphs $\Gamma_1$ and $\Gamma_2$ (Right) 
with a sum over a complete set of dual local operators at the new junctions (dashed line).}\label{fig:cut}
\end{figure}
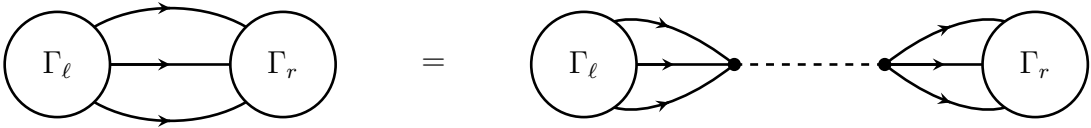

An important ingredient of the construction is a neat identity which holds for the evaluation maps. Consider a spherical graph $\Gamma$ 
and cut it along the equator of the sphere. We can obtain two simpler graphs $\Gamma_1$ and $\Gamma_2$ by taking either half of $\Gamma$ 
and bringing together the cut lines to a common junction. The two new junctions support dual spaces of local operators $V$ and $V^*$. 
Then 
\begin{equation}
Z_\Gamma = Z_{\Gamma_1} \cdot Z_{\Gamma_2}
\end{equation}
where the inner product denotes a sum over dual bases of local operators in $V$ and $V^*$. See Figure \ref{fig:cut}.

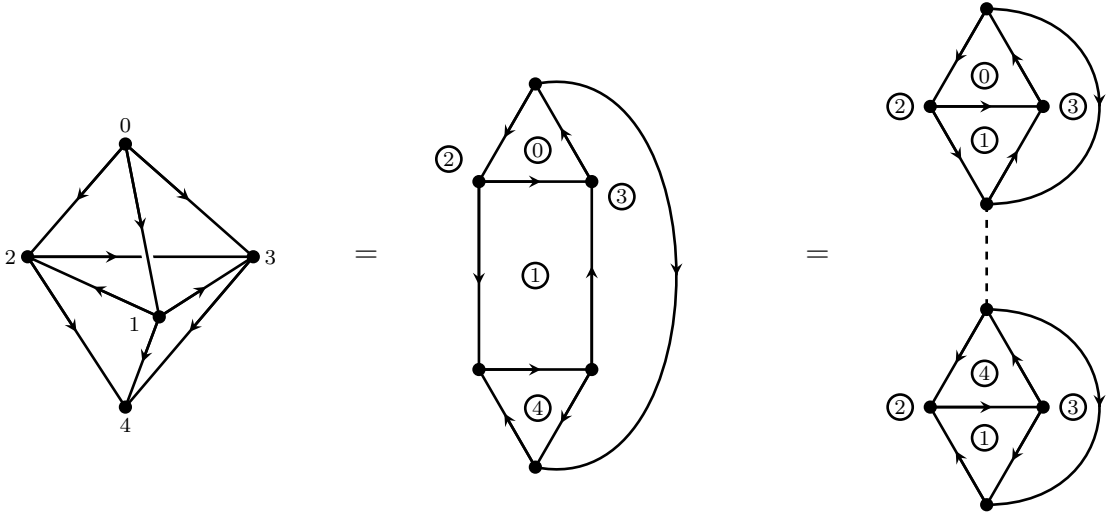
\begin{figure}
\begin{tikzpicture}[line width=1pt]
\coordinate (v2) at (0,0);
\coordinate (v3) at (3,0);
\coordinate (v0) at (1.3,1.5);
\coordinate (v1) at (1.75,-0.8);
\coordinate (v4) at (1.3,-2);

\arrowpath{(v2)}{(v3)}{0.4};
\draw[white,line width=1.5mm](v0)--(v1);
\arrowpath{(v0)}{(v1)}{0.5};
\arrowpath{(v0)}{(v2)}{0.5};
\arrowpath{(v0)}{(v3)}{0.5};
\arrowpath{(v1)}{(v2)}{0.5};
\arrowpath{(v1)}{(v3)}{0.5};
\arrowpath{(v1)}{(v4)}{0.5};
\arrowpath{(v2)}{(v4)}{0.5};
\arrowpath{(v3)}{(v4)}{0.5};
\draw[fill=black] (v0) circle(2pt); 
\draw[fill=black] (v1) circle(2pt); 
\draw[fill=black] (v2) circle(2pt); 
\draw[fill=black] (v3) circle(2pt); 
\draw[fill=black] (v4) circle(2pt); 
\node[above] at (v0) {\scriptsize{$0$}};
\node[left] at ($(v1)-(0.1,0.1)$) {\scriptsize{$1$}};
\node[left] at (v2) {\scriptsize{$2$}};
\node[right] at (v3) {\scriptsize{$3$}};
\node[below] at (v4) {\scriptsize{$4$}};

\coordinate (eq1) at ($(v3)+(1.5,0)$);
\node at (eq1) {$=$};

\coordinate (v02) at ($(eq1) + (1.5,1)$);
\coordinate (v03) at ($(v02) + (0:1.5)$);
\coordinate (v01) at ($(v02) + (60:1.5)$);
\coordinate (v42) at ($(eq1) + (1.5,-1.5)$);
\coordinate (v43) at ($(v42) + (0:1.5)$);
\coordinate (v41) at ($(v42) + (-60:1.5)$);

\arrowpath{(v01)}{(v02)}{0.55};
\arrowpath{(v02)}{(v03)}{0.55};
\arrowpath{(v03)}{(v01)}{0.55};
\arrowpath{(v41)}{(v42)}{0.55};
\arrowpath{(v42)}{(v43)}{0.55};
\arrowpath{(v43)}{(v41)}{0.55};
\arrowpath{(v02)}{(v42)}{0.55};
\arrowpath{(v43)}{(v03)}{0.55};
\bezierarrowpath{(v01)}{($(v01)+(2.5,0.5)$)}{($(v41)+(2.5,-0.5)$)}{(v41)}{0.5};

\draw[fill=black] (v01) circle(2pt); 
\draw[fill=black] (v02) circle(2pt); 
\draw[fill=black] (v03) circle(2pt); 
\draw[fill=black] (v41) circle(2pt); 
\draw[fill=black] (v42) circle(2pt); 
\draw[fill=black] (v43) circle(2pt); 

\node[draw,circle,inner sep=1pt] at ($0.33*(v01)+0.33*(v02)+0.33*(v03)+(0.1,0)$) {\scriptsize{$0$}};
\node[draw,circle,inner sep=1pt] at ($0.33*(v41)+0.33*(v42)+0.33*(v43)+(0.1,-0.1)$) {\scriptsize{$4$}};
\node[draw,circle,inner sep=1pt] at ($(v02)+(-0.4,0.3)$) {\scriptsize{$2$}};
\node[draw,circle,inner sep=1pt] at ($(v03)+(0.4,-0.2)$) {\scriptsize{$3$}};
\node[draw,circle,inner sep=1pt] at ($0.25*(v02)+0.25*(v03)+0.25*(v42)+0.25*(v43)$) {\scriptsize{$1$}};

\coordinate (eq2) at ($(eq1)+(6,0)$);
\node at (eq2) {$=$};

\coordinate (u02) at ($(eq2) + (1.5,2)$);
\coordinate (u03) at ($(u02) + (0:1.5)$);
\coordinate (u01) at ($(u02) + (60:1.5)$);
\coordinate (u04) at ($(u02) + (-60:1.5)$);
\coordinate (u42) at ($(eq2) + (1.5,-2)$);
\coordinate (u43) at ($(u42) + (0:1.5)$);
\coordinate (u41) at ($(u42) + (-60:1.5)$);
\coordinate (u40) at ($(u42) + (60:1.5)$);

\arrowpath{(u02)}{(u03)}{0.55};
\arrowpath{(u03)}{(u01)}{0.55};
\arrowpath{(u01)}{(u02)}{0.55};
\arrowpath{(u02)}{(u04)}{0.55};
\arrowpath{(u04)}{(u03)}{0.55};

\arrowpath{(u42)}{(u43)}{0.55};
\arrowpath{(u43)}{(u41)}{0.55};
\arrowpath{(u41)}{(u42)}{0.55};
\arrowpath{(u43)}{(u40)}{0.55};
\arrowpath{(u40)}{(u42)}{0.55};
\draw[dashed] (u04)--(u40);
\bezierarrowpath{(u01)}{($(u01)+(2,0)$)}{($(u04)+(2,0)$)}{(u04)}{0.5};
\bezierarrowpath{(u40)}{($(u40)+(2,0)$)}{($(u41)+(2,0)$)}{(u41)}{0.5};

\node[draw,circle,inner sep=1pt] at ($0.33*(u01)+0.33*(u02)+0.33*(u03)+(0.1,0)$) {\scriptsize{$0$}};
\node[draw,circle,inner sep=1pt] at ($0.33*(u04)+0.33*(u02)+0.33*(u03)+(0.1,0)$) {\scriptsize{$1$}};
\node[draw,circle,inner sep=1pt] at ($(u02)+(-0.4,0)$) {\scriptsize{$2$}};
\node[draw,circle,inner sep=1pt] at ($(u03)+(0.4,0)$) {\scriptsize{$3$}};

\node[draw,circle,inner sep=1pt] at ($0.33*(u41)+0.33*(u42)+0.33*(u43)+(0.1,0)$) {\scriptsize{$1$}};
\node[draw,circle,inner sep=1pt] at ($0.33*(u40)+0.33*(u42)+0.33*(u43)+(0.1,0)$) {\scriptsize{$4$}};
\node[draw,circle,inner sep=1pt] at ($(u42)+(-0.4,0)$) {\scriptsize{$2$}};
\node[draw,circle,inner sep=1pt] at ($(u43)+(0.4,0)$) {\scriptsize{$3$}};

\draw[fill=black] (u01) circle(2pt); 
\draw[fill=black] (u02) circle(2pt); 
\draw[fill=black] (u03) circle(2pt); 
\draw[fill=black] (u04) circle(2pt); 
\draw[fill=black] (u40) circle(2pt); 
\draw[fill=black] (u41) circle(2pt); 
\draw[fill=black] (u42) circle(2pt); 
\draw[fill=black] (u43) circle(2pt); 

\end{tikzpicture}
\caption{A triangular bi-pyramid (Right) can be obtained by gluing two tetrahedra. Correspondingly, the dual planar graph (Middle) can be obtained by fusing tetrahedral dual planar graphs along a pair of junctions (Right). For clarity, the faces dual to the original vertices 
are indicated by circled numbers.}\label{fig:bipi}
\end{figure}

This has a straightforward geometric interpretation: the polytope dual to $\Gamma$ can be decomposed into the polytopes dual to 
$\Gamma_1$ and $\Gamma_2$, glued along the faces dual to the new junctions. The partition functions are glued by contracting 
the dual vector spaces associated to these faces. See Figure \ref{fig:bipi}.

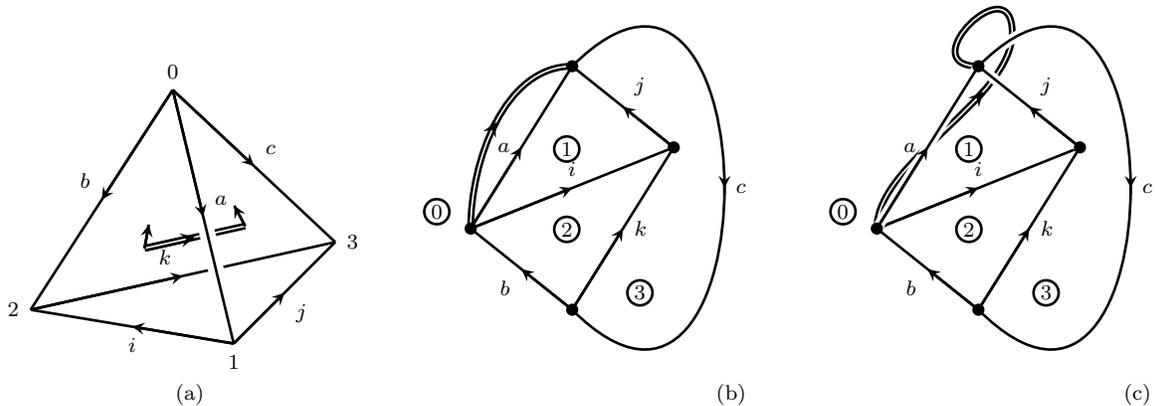
\begin{figure}
\centering
\begin{tikzpicture}[line width=1pt,scale=.9]
\begin{scope}[every node/.style={sloped,allow upside down}]
\coordinate (2) at (0,0);
\coordinate (1) at ($(2)+(3,-0.5)$);
\coordinate (0) at ($(2)+(2.1,3.25)$);
\coordinate (3) at ($(1)+(1.5,1.5)$);

\coordinate (mid012) at ($0.33*(0)+0.33*(1)+0.33*(2)$);
\coordinate (mid013) at ($0.33*(0)+0.33*(1)+0.33*(3)$);
\arrowpathdouble{(mid012)}{(mid013)}{.5};
\coordinate (mid0120) at  ($0.85*(mid012)+0.15*(0)$);
\coordinate (mid0130) at  ($0.85*(mid013)+0.15*(0)$);
\arrowpath{(mid012)}{(mid0120)}{1};
\arrowpath{(mid013)}{(mid0130)}{1};

\arrowpath{(2)}{(3)}{0.5};
\paddedline{(0)}{(1)}{(0.1,0)};
\arrowpath{(0)}{(1)}{0.5};
\arrowpath{(0)}{(2)}{0.5};
\arrowpath{(1)}{(2)}{0.5};
\arrowpath{(0)}{(3)}{0.5};
\arrowpath{(1)}{(3)}{0.5};

\node[above] at (0) {\scriptsize{$0$}};
\node[below] at (1) {\scriptsize{$1$}};
\node[left] at (2) {\scriptsize{$2$}};
\node[right] at (3) {\scriptsize{$3$}};

\node[above right] at ($(0)!0.5!(1)$) {\scriptsize{$a$}};
\node[above left] at ($(0)!0.5!(2)$) {\scriptsize{$b$}};
\node[above right] at ($(0)!0.5!(3)$) {\scriptsize{$c$}};
\node[below] at ($(1)!0.5!(2)$) {\scriptsize{$i$}};
\node[below right] at ($(1)!0.5!(3)$) {\scriptsize{$j$}};
\node[above left] at ($(2)!0.5!(3)$) {\scriptsize{$k$}};

\node (a) at ($(0)+(0.25,-4.5)$) {\scriptsize{(a)}};

\coordinate (023) at ($(2)+(8,0)$);
\coordinate (012) at ($(023)+(-1.5,1.2)$);
\coordinate (123) at ($(023)+(1.5,2.4)$);
\coordinate (013) at ($(023)+(0,3.6)$);
\node[draw,circle,inner sep=1pt] at ($(012)-(0.5,-0.25)$) {\scriptsize{$0$}};
\coordinate (controlu) at ($(013)+(3,3)$);
\coordinate (controld) at ($(023)+(3,-3)$);
\coordinate (mid) at ($(controlu)!0.5!(controld)-(0.75,0)$);
\coordinate (mid1) at ($0.33*(013)+0.33*(012)+0.33*(123)$);
\node[draw,circle,inner sep=1pt] at (mid1) {\scriptsize{$1$}};
\coordinate (mid2) at ($0.33*(023)+0.33*(012)+0.33*(123)$);
\node[draw,circle,inner sep=1pt] at (mid2) {\scriptsize{$2$}};
\node[draw,circle,inner sep=1pt] at ($(023)+(1,0.25)$) {\scriptsize{$3$}};

\coordinate (012l) at ($(012)+(0,1)$);
\coordinate (013l) at ($(013)+(-1,0)$);

\arrowpath{(023)}{(012)}{0.5};
\arrowpath{(023)}{(123)}{0.5};
\arrowpath{(012)}{(123)}{0.5};
\arrowpath{(012)}{(013)}{0.5};
\bezierarrowpathdouble{(012)}{(012l)}{(013l)}{(013)}{0.5};
\arrowpath{(123)}{(013)}{0.5};
\draw (013)..controls(controlu) and (controld)..(023);
\arrowpath{($(mid)+(0,0.1)$)}{($(mid)-(0,0.1)$)}{0.5};


\node[below left] at ($(023)!0.5!(012)$) {\scriptsize{$b$}};
\node[right] at ($(023)!0.5!(123)$) {\scriptsize{$k$}};
\node[above] at ($(012)!0.5!(123)$) {\scriptsize{$i$}};
\node[left] at ($(012)!0.5!(013)$) {\scriptsize{$a$}};
\node[above right] at ($(123)!0.5!(013)$) {\scriptsize{$j$}};
\node[right] at (mid) {\scriptsize{$c$}};

\draw[fill=black] (012) circle(2pt);
\draw[fill=black] (013) circle(2pt);
\draw[fill=black] (023) circle(2pt);
\draw[fill=black] (123) circle(2pt);

\node at ($(a)+(8,0)$) {\scriptsize{(b)}};

\coordinate (023) at ($(2)+(14,0)$);
\coordinate (012) at ($(023)+(-1.5,1.2)$);
\coordinate (123) at ($(023)+(1.5,2.4)$);
\coordinate (013) at ($(023)+(0,3.6)$);
\node[draw,circle,inner sep=1pt] at ($(012)-(0.5,-0.25)$) {\scriptsize{$0$}};
\coordinate (controlu) at ($(013)+(3,3)$);
\coordinate (controld) at ($(023)+(3,-3)$);
\coordinate (mid) at ($(controlu)!0.5!(controld)-(0.75,0)$);
\coordinate (mid1) at ($0.33*(013)+0.33*(012)+0.33*(123)$);
\node[draw,circle,inner sep=1pt] at (mid1) {\scriptsize{$1$}};
\coordinate (mid2) at ($0.33*(023)+0.33*(012)+0.33*(123)$);
\node[draw,circle,inner sep=1pt] at (mid2) {\scriptsize{$2$}};
\node[draw,circle,inner sep=1pt] at ($(023)+(1,0.25)$) {\scriptsize{$3$}};

\coordinate (012l) at ($(012)+(0,1)$);
\coordinate (013l) at ($(013)+(.5,-.5)$);
\coordinate (013r) at ($(013)+(.5,.5)$);
\coordinate (013ru) at ($(013)+(.5,1.5)$);
\coordinate (013ll) at ($(013)+(-1,0)$);

\bezierarrowpathdouble{(012)}{(012l)}{(013l)}{(013r)}{0.7};
\draw[double] (013r) .. controls (013ru) and (013ll) .. (013);
\arrowpath{(023)}{(012)}{0.5};
\arrowpath{(023)}{(123)}{0.5};
\arrowpath{(012)}{(123)}{0.5};
\draw[white,line width=4] (012) to (013);
\arrowpath{(012)}{(013)}{0.5};
\draw[white,line width=4] (123) to (013);
\arrowpath{(123)}{(013)}{0.5};
\draw[white,line width=4] (013)..controls(controlu) and (controld)..(023);
\draw (013)..controls(controlu) and (controld)..(023);
\arrowpath{($(mid)+(0,0.1)$)}{($(mid)-(0,0.1)$)}{0.5};


\node[below left] at ($(023)!0.5!(012)$) {\scriptsize{$b$}};
\node[right] at ($(023)!0.5!(123)$) {\scriptsize{$k$}};
\node[above] at ($(012)!0.5!(123)$) {\scriptsize{$i$}};
\node[left] at ($(012)!0.5!(013)$) {\scriptsize{$a$}};
\node[above right] at ($(123)!0.5!(013)$) {\scriptsize{$j$}};
\node[right] at (mid) {\scriptsize{$c$}};

\draw[fill=black] (012) circle(2pt);
\draw[fill=black] (013) circle(2pt);
\draw[fill=black] (023) circle(2pt);
\draw[fill=black] (123) circle(2pt);

\node at ($(a)+(14,0)$) {\scriptsize{(c)}};
\end{scope}
\end{tikzpicture}
\caption{a) A tetrahedron with an extra quasi-particle transversing two faces. We indicated the choice of framing at each face. 
b) The dual planar graph in the spherical fusion category. We drew the center line along the simplest choice of path. Alternative paths which self-intersect or wind around the endpoints (c)
would give answers which differ by framing phases}\label{fig:dritetra}
\end{figure}

The bulk line defects affect the partition sum by modifying the vector spaces associated to the faces crossed by the lines. Essentially, they replace faces $C$ 
with decorated faces $D_\alpha$ and $H(C,l)$ with 
\begin{equation}
H(D_\alpha,l) = \Hom_\cC\left(1, Y_\alpha \otimes l(e_1) \otimes \cdots \otimes l(e_n) \right)
\end{equation}
The computation of the contribution of a polyhedron with such modified faces involves adding an extra $Y_\alpha$ line attached to the appropriate vertices of $\Gamma$.
If $Y_\alpha$ crosses some other line in $\Gamma$ we can insert $\beta_\alpha$ there. The precise framed path followed by $Y_\alpha$ 
is immaterial because $Y_\alpha$ lies in the center. Changes of framing, though, change the answer appropriately. See Figure \ref{fig:dritetra}.

\subsection{Adding a flat connection}
Next, consider a $G$-graded spherical fusion category, a direct sum of sub-categories $\cC_g$ with the property that 
$\cC_g \otimes \cC_{g'} \in \cC_{g g'}$. We will denote the identity in $G$ as $1$.
The Turaev-Viro construction applied to $\cC_1$ gives a 3d TFT $\fT$ with boundary condition $\fB$, 
bulk lines in $Z[\cC_1]$ and boundary lines in $\cC_1$. 

We can extend this theory to manifolds equipped with a $G$ connection, simply representing the flat connection by edge 
elements $g_e$ and by prescribing that an edge $e$ of the triangulation is labelled by an object in $\cC_{g_e}$
and building the partition sum as before. This endows $\fT$ with a global symmetry $G$. 

Elements in $\cC_g$ can be interpreted as lines lying at the intersection of the boundary $\fB$ with a $U_g$ topological domain wall
implementing the $g$ symmetry. In general, there will not be any canonical choice of objects in $\cC_g$ with trivial associators,
meaning that the $G$ global symmetry is broken at the boundary. \footnote{Depending on the $\cC_g$ lines being dynamical or not in 
a UV completion of the theory, we can interpret the breaking as being spontaneous or explicit.}

Notice that bulk lines in the center $Z[\cC_1]$ are not equipped with a canonical crossing through lines in $\cC_g$. 
Physically, this corresponds to the fact that there is no canonical way for a bulk line to cross a $U_g$ domain wall.
\footnote{It is possible to define a $G$ action on the center $Z[\cC_1]$, corresponding to surrounding a bulk line with an $U_g$ 
domain wall, so that a canonical crossing morphism exists mapping $Y \otimes V_g$ to $V_g \otimes (g \circ Y)$. 
In the mathematical literature there is also the notion of $G$-center, corresponding to objects in $\cC$ with a canonical crossing through objects in $\cC_1$. 
These should correspond physically to bulk twist lines, at which $U_g$ defects may end.}

The group cohomology construction of bosonic SPT phases is the simplest example of a $G$-graded Turaev-Viro 
partition sum, based on a $G$-graded category with a single (equivalence class of) simple object $V_g$ in each  $\cC_g$ subcategory. 

The evaluation of a tetrahedron of positive (negative) orientation produces directly the associator $\alpha^\pm_3$ and the partition sum immediately reproduces the SPT partition function.

\subsection{Gauging standard global symmetries}
Gauging the $G$ global symmetry of $\fT$ should produce another topological field theory $\fT_G$. 
The partition function of $\fT_G$ should be obtained by summing the partition functions of $\fT$
over all possible choices of flat $G$ connections. We expect this to coincide with Turaev-Viro applied to the whole $\cC$,
disregarding the $G$ grading. This gives a sum over all flat connections rather than equivalence classes of flat connections, but 
the total quantum dimension should also change in such a way to compensate for that over-counting. 

Notice that in the presence of a $\fB$ boundary the Turaev-Viro construction applied to the whole $\cC$
will not sum over different choices of boundary lines. Correspondingly, the $G$ gauge theory has Dirichlet b.c.:
the flat connection is fixed at the boundary. 

Gauging a theory with a standard Abelian global symmetry $G$ should give a theory 
with a 1-form symmetry valued in the dual Abelian group $G^*$, generated by Wilson lines $B_a$. 
Using the definition of Wilson lines as center elements, we find that the 
insertion of a network of Wilson lines changes the sum over $G$ flat connections $\alpha_1$
by inserting a factor $e^{2 \pi i \int \alpha_1 \cup \beta_2}$, where we are contracting the 
group elements in the flat connection $\alpha_1$ with the characters in the background $G^*$ 2-form
 flat connection $\beta_2$. 
 
 In order to see that in a fully explicit manner, it is useful to put a local order on the vertices of the triangulation 
and pick the first vertex in every face as a framing for the $B_a$ lines. Then the decorated two-sphere graph associate 
to a tetrahedron has three $B_a$ lines coming out of vertices into the first face, and one coming out of the $234$ vertex towards the 
second face. In order to bring the $B_a$ lines together and define a consistent graph, we need to have the 
$B_a$ line from the $234$ vertex cross the line between the first and second faces. The twisting factor contributes 
$e^{2 \pi i (\hat \alpha_1)_{12} (\hat \beta_2)_{234}}$. This is precisely the contribution of a single tetrahedron to 
$e^{2 \pi i \int \hat \alpha_1 \cup \hat \beta_2}$. \footnote{In general, one can interpret the Fourier transform kernel $e^{2 \pi i \int \hat \alpha_1 \cup \hat \beta_2}$
as a very simple topological field theory with both a 1-form symmetry $G^*$ and a standard symmetry $G$,
generated by a spherical fusion category modelled on $G$ and $G^*$-valued lines $(1,e^{2 \pi i \chi \cdot h})$.}

\subsection{Adding a 2-form flat connection}
A theory endowed with a 1-form global symmetry $Z$ can be coupled to a 2-form flat connection, say described by a 2-cocycle 
$\beta_2$ valued in $Z$. Concretely, $\beta_2$ should tell us which symmetry generators $B_a= (b_a, \beta_a)$ run through each face of the triangulation. 
Without loss of generality, we can take the lines entering each tetrahedron to join together at some interior point, thanks to $\delta \beta_2=0$.
Gauge transformations on $\beta_2$ simply move around the lines or re-connect them. 

Thus we have a Turaev-Viro construction of the partition function $Z[\beta_2]$: we simply replace the vector spaces 
$\Hom_\cC\left(1, l(e_1) \otimes \cdots \otimes l(e_n) \right)$ with $\Hom_\cC\left(1, b_a l(e_1) \otimes \cdots \otimes l(e_n) \right)$
and project the $B_a$ lines for each tetrahedron to the surface, evaluating the corresponding graph as usual. 

On general grounds, gauging a theory with a 1-form symmetry $H$ should give a theory with a standard global symmetry valued in the dual 
Abelian group $H^*$. We have already described the process at the level of spherical fusion categories $\cC_b$ and $\cC^{\ZZ_2}$. 

The evaluation of tetrahedra in the $\cC^{\ZZ_2}_0$ category over objects inherited from $\cC_b$ will precisely match the evaluation of tetrahedra in 
 $\cC_b$ coupled to a general $\beta_2$. Adding a $\ZZ_2$ flat connection $\alpha_1$ simply adds the usual factor of $e^{\pi i \int \alpha_1 \cup \beta_2}$.
 
 The only non-trivial step in identifying the Turaev-Viro partition sum of $\cC^{\ZZ_2}$
 as the result of gauging the 1-form symmetry of the Turaev-Viro partition sum of $\cC_b$
 is to observe that the sum over simple object of $\cC_b$ of the images in $\cC^{\ZZ_2}_0$
 \begin{equation}
 \sum_i d_i^{\cC_b} (X_i)_0
 \end{equation}
reproduces the correct sum of simple objects in $\cC^{\ZZ_2}_0$.

\subsection{Example: toric code}
We can see now explicitly the equivalence between the toric code and $\ZZ_2$ gauge theory:
the decoration $I$ or $P$ of the edges of the triangulation encodes a $\ZZ_2$ cochain and all tetrahedra contributions equal to $1$. 
The total quantum dimension is $2$, and the partition function is 
\begin{equation}\label{toricpartitionfunction}
\frac{1}{2^v} \sum_{\epsilon_1 | \delta \epsilon_1=0} 1=\frac{1}{2} |H^1(M,\ZZ_2)|,
\end{equation}
where $v$ is the number of vertices. Remember that center of the category consists of objects $I = (I,\beta = 1)$, $e = (I,\beta_P = -1)$, $m = (P,\beta = 1)$, $\epsilon = (P,\beta_P = -1)$. The pre-factor $2^{-v}$ can be interpreted as the order of the $\ZZ_2$ gauge group .

We see explicitly that adding an $m$ line produces a vortex: the extra $P$ line passing through a face breaks the flatness condition there. 
If we couple the system to the corresponding flat connection $\beta_2^m$, the partition sum becomes 
\begin{equation}
\frac{1}{2^v} \sum_{\epsilon_1 | \delta \epsilon_1=\beta_2^m} 1.
\end{equation}
That is, it is equal to (\ref{toricpartitionfunction}) if $\beta_2^m$ is exact and equal to zero otherwise.
This is somewhat boring, but consistent. 

\begin{figure}
\centering
\begin{tikzpicture}[line width=1pt]
\begin{scope}[every node/.style={sloped,allow upside down}]

\coordinate (center) at (0,0);
\coordinate (frontup) at ($(center)+(0,0.25)$);
\coordinate (frontdown) at ($(center)+(0,-1)$);

\coordinate (023) at ($(frontdown)+(0,0)$);
\coordinate (012) at ($(023)+(-1.5,1.2)$);
\coordinate (123) at ($(023)+(1.5,2.4)$);
\coordinate (013) at ($(023)+(0,3.6)$);
\coordinate (controlu) at ($(013)+(3,3)$);
\coordinate (controld) at ($(023)+(3,-3)$);
\coordinate (mid) at ($(controlu)!0.5!(controld)-(0.75,0)$);
\coordinate (blob) at ($(012)+(0,2.4)$);
\coordinate (controllu) at ($(blob)-(1,0)$);
\coordinate (controlld) at ($(023)-(3,0)$);

\node[draw,circle,inner sep=1pt] at ($(023)+(-0.5,-0.5)$) {\scriptsize{$0$}};
\coordinate (mid) at ($(controlu)!0.5!(controld)-(0.75,0)$);
\coordinate (mid1) at ($0.33*(013)+0.33*(012)+0.33*(123)$);
\node[draw,circle,inner sep=1pt] at (mid1) {\scriptsize{$1$}};
\coordinate (mid2) at ($0.33*(023)+0.33*(012)+0.33*(123)$);
\node[draw,circle,inner sep=1pt] at (mid2) {\scriptsize{$2$}};
\node[draw,circle,inner sep=1pt] at ($(023)+(1,0.25)$) {\scriptsize{$3$}};

\draw[double] (123) to (blob);
\paddedline{(012)}{(013)}{(0.1,0)};
\draw[double] (013) to (blob);
\draw[double] (012) to (blob);
\arrowpath{(023)}{(012)}{0.5};
\arrowpath{(023)}{(123)}{0.5};
\arrowpath{(012)}{(123)}{0.5};
\arrowpath{(012)}{(013)}{0.5};
\arrowpath{(123)}{(013)}{0.5};
\draw (013)..controls(controlu) and (controld)..(023);
\arrowpath{($(mid)+(0,0.1)$)}{($(mid)-(0,0.1)$)}{0.5};
\draw[double] (023)..controls(controlld) and (controllu)..(blob);

\draw[fill=black] (012) circle(2pt);
\draw[fill=black] (013) circle(2pt);
\draw[fill=black] (023) circle(2pt);
\draw[fill=black] (123) circle(2pt);
\draw[fill=white] (blob) circle(8pt);

\end{scope}
\end{tikzpicture}
\caption{A canonical choice of framing for the dual tetrahedron graph dressed by 1-form symmetry generators.} \label{fig:dressed}
\end{figure}
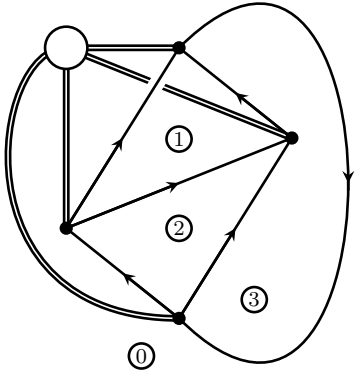

If we couple the system to a flat connection $\beta_2^e$, associated to the quasi-particle $e$, the partition sum becomes instead 
\begin{equation}
\frac{1}{2^v} \sum_{\epsilon_1 | \delta \epsilon_1=0} (-1)^{\int \epsilon_1 \cup \beta^e_2  }
\end{equation}
The cup product emerges as before from the evaluation of the tetrahedron: with a canonical choice of framing, a single $e$ line 
crosses a single edge as in Figure \ref{fig:dressed}. If $\beta^e_2$ is not exact, we can always find a dual 1-cocycle by which to shift $\epsilon_1$ 
in order to switch the sign of the cup product and thus cancel all terms in pairs. (This is equivalent to the statement that the mod-2 intersection pairing on cohomology is non-degenerate.) Thus the sum is not-vanishing only if $\beta_2^e$ is exact, in which case the integrand is a co-boundary and the sign drops out.  
This is consistent with the symmetry exchanging $e$ and $m$.

Next, we can try to couple the system to a flat connection $\beta_2^\epsilon$, associated to the fermion $\epsilon$.
The result should be anomalous, but yet instructive. The partition sum becomes 
\begin{equation}
\frac{1}{2^v} \sum_{\epsilon_1 | \delta \epsilon_1=\beta^\epsilon_2 } (-1)^{\int \epsilon_1 \cup \beta^\epsilon_2 }
\end{equation}
It is still true that if $\beta^\epsilon_2$ is not exact, we can always find a dual 1-cocycle by which to shift $\epsilon_1$ 
in order to switch the sign of the cup product and cancel all terms in pairs. Thus the sum 
is not-vanishing only if $\beta_2^\epsilon$ is exact. If we write $\beta_2^\epsilon = \delta \lambda_1$, we can 
absorb $\lambda$ into a shift of $\epsilon_1$ to get 
\begin{equation}
\frac{1}{2^v} \sum_{\epsilon_1 | \delta \epsilon_1=0 } (-1)^{\int \epsilon_1 \cup \delta \lambda_1 + \lambda_1 \cup \delta \lambda_1} = 
(-1)^{\lambda_1 \cup \delta \lambda_1} \frac{1}{2^v} \sum_{\epsilon_1 | \delta \epsilon_1=0 } 1=\frac12|H^1(M,\ZZ_2)||(-1)^{\lambda_1 \cup \delta \lambda_1}
\end{equation}
This is  the toric code partition function (\ref{toricpartitionfunction}) times $z_\Pi(\delta \lambda_1)$. In other words, the anomaly found in the toric code is precisely what we expected, at least for exact $\beta^\epsilon_2$ connections. 

\subsection{Gu-Wen $\Pi$-category}
The $\ZZ_2$ gauge theory based on the $\hat G$ SPT phase has a simple partition sum: lines are decorated by 
a fixed $G$ flat connection which is lifted to a $\hat G$ connection by some 1-cochain $\epsilon_1$. The fusion rules imply that 
$\delta \epsilon_1$ equals the value of $n_2$ on faces. The partition sum is 
\begin{equation}
\frac{1}{2^v} \sum_{\epsilon_1 | \delta \epsilon_1=n_2} \prod \hat \alpha_3.
\end{equation}
We can pull out the $\nu_3$ contribution and get 
\begin{equation}
\frac{1}{2^v} \prod \nu_3 \sum_{\epsilon_1 | \delta \epsilon_1=n_2} (-1)^{\int n_2 \cup \epsilon_1}.
\end{equation}

From the fact that the mod-2 intersection pairing on cohomology is non-degenerate, one again deduces that  the partition sum is non-zero only for $G$ flat connections for which the pull-back of $n_2$ is exact, 
i.e. $n_2 = \delta \lambda_1$. Then we have 
\begin{equation}
\frac{1}{2^v} (-1)^{\int \delta \lambda_1 \cup \lambda_1} \prod \nu_3 \sum_{\epsilon_1 | \delta \epsilon_1=0} 1=z_\Pi(\delta\lambda_1) 
\end{equation}

\begin{figure}
\centering
\begin{tikzpicture}[line width=1pt]

\begin{scope}[every node/.style={sloped,allow upside down}]
\coordinate (00) at (0,0);

\coordinate (C) at ($(00)+(1,0)$);
\coordinate (mid) at ($(C)+(0,2)$);
\coordinate (A) at ($(mid)+(-1,2)$);
\coordinate (B) at ($(mid)+(1,2)$);
\coordinate (D) at ($(mid)+(-1,0)$);

\arrowpath{(C)}{(mid)}{0.5};
\arrowpath{(mid)}{(A)}{0.5};
\arrowpath{(mid)}{(B)}{0.5};
\arrowpathdouble{(mid)}{(D)}{0.5};
\draw[fill=black] (mid) circle(2pt);

\node[below] at (C) {\scriptsize{$V_{\hat g \hat g'}$}};
\node[above] at (A) {\scriptsize{$V_{\hat g }$}};
\node[above] at (B) {\scriptsize{$V_{\hat g'}$}};

\coordinate (eq2) at ($(mid)+(2.5,0)$);
\node at (eq2) {$\to$};

\coordinate (C) at ($(00)+(5,0)$);
\coordinate (mid) at ($(C)+(0,2)$);
\coordinate (A) at ($(mid)+(-1,2)$);
\coordinate (B) at ($(mid)+(1,2)$);
\coordinate (D) at ($(mid)+(-1,0)$);

\arrowpathdouble{(C)}{(mid)}{0.5};
\arrowpathdouble{(mid)}{(A)}{0.5};
\arrowpathdouble{(mid)}{(B)}{0.5};
\arrowpathdouble{(mid)}{(D)}{0.5};

\coordinate (mid2) at (mid);

\node[left] at (C) {\scriptsize{$\epsilon + \epsilon' + n_2(g,g')$}};
\node[left] at (A) {\scriptsize{$\epsilon$}};
\node[left] at (B) {\scriptsize{$\epsilon'$}};

\coordinate (C) at ($(00)+(5.5,0)$);
\coordinate (mid) at ($(C)+(0,2)$);
\coordinate (A) at ($(mid)+(-1,2)$);
\coordinate (B) at ($(mid)+(1,2)$);

\arrowpathdouble{(mid)}{(mid2)}{0.5};
\draw[fill=white] (mid2) circle(5pt);

\arrowpath{(C)}{(mid)}{0.5};
\paddedline{(mid)}{(A)}{(0.05,0)};
\arrowpath{(mid)}{(A)}{0.5};
\arrowpath{(mid)}{(B)}{0.5};
\draw[fill=black] (mid) circle(2pt);

\node[below] at (C) {\scriptsize{$V_{(g g',0)}$}};
\node[above] at (A) {\scriptsize{$V_{(g,0)}$}};
\node[above] at (B) {\scriptsize{$V_{(g',0)}$}};

\end{scope}
\end{tikzpicture}
\caption{Gauge-fixing: A graphical representation of the partial gauge-fixing procedure used in computing the tetrahedron contribution of 
the $\hat G$ Gu-Wen phase decorated by an extra 2-form flat connection $\beta_2$.  
Left: A choice of gauge is the same as a choice of basis vector in the space of junctions between line defects in the full $\hat G$ category, possibly with an extra center line. 
Right: we identify $V_{(g,\epsilon)}\simeq V_{(1,\epsilon)} \otimes V_{g,0}$ and identify $V_{(1,\epsilon)}$ with the corresponding elements $I$ or $\Pi$ of the center. 
We then express a general junction canonically 
in terms of a choice of junction between line defects labelled by $G$ elements. The double lines denote the center elements. The empty circle represents any choice of how to connect the center lines in a planar way.}
\label{fig:hatcocycle4}
\end{figure}
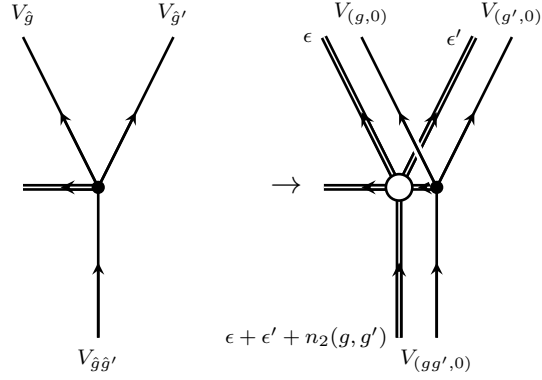

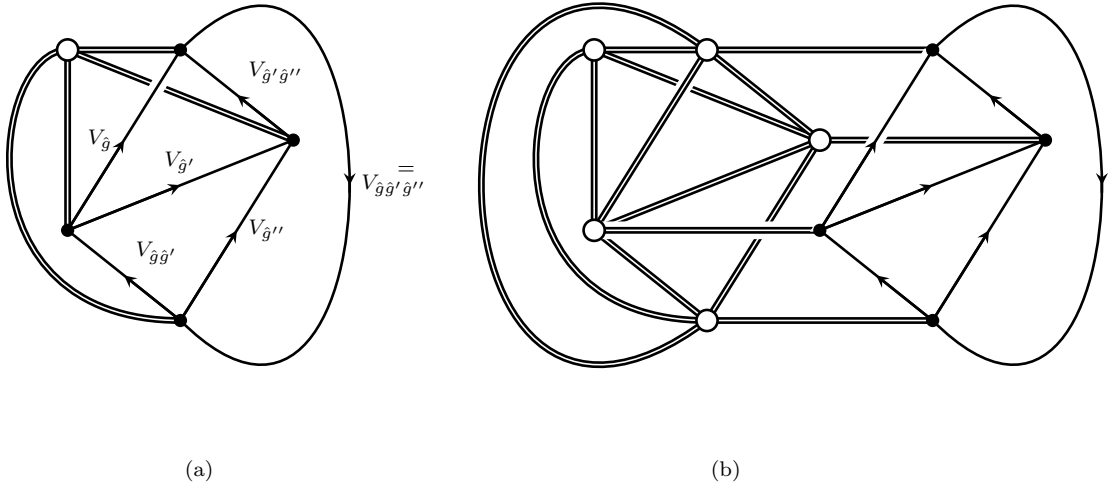
\begin{figure}
\centering
\begin{tikzpicture}[line width=1pt]
\begin{scope}[every node/.style={sloped,allow upside down}]
\coordinate (00) at (0,0);
\coordinate (023) at (00);

\coordinate (012) at ($(023)+(-1.5,1.2)$);
\coordinate (123) at ($(023)+(1.5,2.4)$);
\coordinate (013) at ($(023)+(0,3.6)$);
\coordinate (controlu) at ($(013)+(3,3)$);
\coordinate (controld) at ($(023)+(3,-3)$);
\coordinate (mid) at ($(controlu)!0.5!(controld)-(0.75,0)$);
\coordinate (blob) at ($(012)+(0,2.4)$);
\coordinate (controllu) at ($(blob)-(1,0)$);
\coordinate (controlld) at ($(023)-(3,0)$);

\draw[double] (123) to (blob);
\paddedline{(012)}{(013)}{(0.1,0)};
\draw[double] (013) to (blob);
\draw[double] (012) to (blob);
\arrowpath{(023)}{(012)}{0.5};
\arrowpath{(023)}{(123)}{0.5};
\arrowpath{(012)}{(123)}{0.5};
\arrowpath{(012)}{(013)}{0.5};
\arrowpath{(123)}{(013)}{0.5};
\draw (013)..controls(controlu) and (controld)..(023);
\arrowpath{($(mid)+(0,0.1)$)}{($(mid)-(0,0.1)$)}{0.5};
\draw[double] (023)..controls(controlld) and (controllu)..(blob);

\node[above right] at ($(023)!0.5!(012)$) {\scriptsize{$V_{\hat g \hat g'}$}};
\node[right] at ($(023)!0.5!(123)$) {\scriptsize{$V_{\hat g''}$}};
\node[above] at ($(012)!0.5!(123)$) {\scriptsize{$V_{\hat g'}$}};
\node[left] at ($(012)!0.5!(013)$) {\scriptsize{$V_{\hat g}$}};
\node[above right] at ($(123)!0.5!(013)$) {\scriptsize{$V_{\hat g' \hat g''}$}};
\node[right] at (mid) {\scriptsize{$V_{\hat g \hat g' \hat g''}$}};

\draw[fill=black] (012) circle(2pt);
\draw[fill=black] (013) circle(2pt);
\draw[fill=black] (023) circle(2pt);
\draw[fill=black] (123) circle(2pt);
\draw[fill=white] (blob) circle(4pt);

\node (a) at ($(00)+(0.25,-2)$) {\scriptsize{(a)}};

\node (eq) at ($(00)+(3,2)$) {\scriptsize{=}};

\coordinate (023) at ($(00)+(7,0)$);
\coordinate (012) at ($(023)+(-1.5,1.2)$);
\coordinate (123) at ($(023)+(1.5,2.4)$);
\coordinate (013) at ($(023)+(0,3.6)$);
\coordinate (controlu) at ($(013)+(3,3)$);
\coordinate (controld) at ($(023)+(3,-3)$);
\coordinate (mid) at ($(controlu)!0.5!(controld)-(0.75,0)$);
\coordinate (blob) at ($(012)+(0,2.4)$);
\coordinate (controllu) at ($(blob)-(1,0)$);
\coordinate (controlld) at ($(023)-(3,0)$);
\coordinate (controlu1) at ($(controlu)+(-7,0)$);
\coordinate (controld1) at ($(controld)+(-7,0)$);
\coordinate (mid1) at ($(controlu1)!0.5!(controld1)-(0.75,0)$);

\draw[double] (123) to (blob);
\paddedline{(012)}{(013)}{(0.1,0)};
\draw[double] (013) to (blob);
\draw[double] (012) to (blob);
\draw[double] (023) to (012);
\draw[double] (023) to (123);
\draw[double] (012) to (123);
\draw[double] (012) to (013);
\draw[double] (123)to (013);
\draw[double] (013)..controls(controlu1) and (controld1)..(023);
\draw[double] (023)..controls(controlld) and (controllu)..(blob);

\coordinate (023b) at ($(023)+ (3,0)$);
\coordinate (012b) at ($(012)+ (3,0)$);
\coordinate (123b) at ($(123)+ (3,0)$);
\coordinate (013b) at ($(013)+ (3,0)$);
\coordinate (controlub) at ($(controlu)+ (3,0)$);
\coordinate (controldb) at ($(controld)+ (3,0)$);
\coordinate (midb) at ($(mid)+ (3,0)$);
\coordinate (blobb) at ($(blob)+ (3,0)$);
\coordinate (controllub) at ($(controllu)+ (3,0)$);
\coordinate (controlldb) at ($(controlld)+ (3,0)$);

\paddedline{(012b)}{(012)}{(0,0.1)};
\draw[double] (023b) to (023);
\draw[double] (012b) to (012);
\draw[double] (123b) to (123);
\draw[double] (013b) to (013);

\draw[fill=white] (012) circle(4pt);
\draw[fill=white] (013) circle(4pt);
\draw[fill=white] (023) circle(4pt);
\draw[fill=white] (123) circle(4pt);
\draw[fill=white] (blob) circle(4pt);

\paddedline{(012b)}{(013b)}{(0.1,0)};

\arrowpath{(023b)}{(012b)}{0.5};
\arrowpath{(023b)}{(123b)}{0.5};
\arrowpath{(012b)}{(123b)}{0.5};
\arrowpath{(012b)}{(013b)}{0.5};
\arrowpath{(123b)}{(013b)}{0.5};
\draw (013b)..controls(controlub) and (controldb)..(023b);
\arrowpath{($(midb)+(0,0.1)$)}{($(midb)-(0,0.1)$)}{0.5};

\draw[fill=black] (012b) circle(2pt);
\draw[fill=black] (013b) circle(2pt);
\draw[fill=black] (023b) circle(2pt);
\draw[fill=black] (123b) circle(2pt);

\node at ($(a)+(7,0)$) {\scriptsize{(b)}};
\end{scope}
\end{tikzpicture}
\caption{The computation of the tetrahedron contribution for 
the $\hat G$ Gu-Wen phase decorated by an extra 2-form flat connection $\beta_2$.  }
\label{fig:hatcocycle3}
\end{figure}

Things become more interesting if we turn on the $\ZZ_2$ 2-form flat connection $\beta_2$ coupled to the 
fermionic 1-form symmetry generator. With appropriate gauge fixing, as in Figures \ref{fig:hatcocycle4} and \ref{fig:hatcocycle3}, the partition sum becomes 
\begin{equation}
\frac{1}{2^v} \prod \nu_3 \sum_{\epsilon_1 | \delta \epsilon_1=n_2+ \beta_2} (-1)^{\int n_2 \cup \epsilon_1 + \epsilon_1 \cup \beta_2}
\end{equation}
Now the partition sum is non-zero if the pull-back of $n_2$ is co-homologous to $\beta_2$. We can write 
$\beta_2 = n_2 + \delta \lambda_1$, shift $\epsilon_1$ and obtain 
\begin{equation}
\frac{1}{2^v} (-1)^{\int n_2 \cup \lambda_1 + \lambda_1 \cup n_2+ \lambda_1 \cup \delta \lambda_1}\prod \nu_3 \sum_{\epsilon_1 | \delta \epsilon_1=0} (-1)^{\int n_2 \cup \epsilon_1 + \epsilon_1 \cup n_2}
\end{equation}
The sign in the sum is actually a boundary and drops out. We get 
\begin{equation}
Z_f[\beta_2] = \frac{1}{2^v} (-1)^{\int n_2 \cup \lambda_1 + \lambda_1 \cup n_2+ \lambda_1 \cup \delta \lambda_1}\prod \nu_3 \sum_{\epsilon_1 | \delta \epsilon_1=0} 1
\end{equation}
Crucially, the answer transforms under gauge transformations precisely as $z_\Pi(\beta_2)$. Furthermore, the product 
$Z_f[\beta_2]z_\Pi(\beta_2)$ simply coincides with the (spin-structure corrected) Gu-Wen SPT phase partition function: 
the $\prod \nu_3$ combines with the Gu-Wen grassmann integral in $z_\Pi(\beta_2)$. We accomplished our main objective. 

\subsection{State sums and spin-TFTs}
We are ready to give our prescription for the Turaev-Viro partition sum of a spin TFT $\fT_s$ constructed from the 
spherical fusion category $\cC_f$ for its shadow $\fT_f$. 

Pick a spherical fusion $\Pi$-category $\cC_f$. Define the decorated Turaev-Viro 
partition sum $Z_f[\beta_2]$ by adding fermionic $\Pi$ lines through all faces where $\hat \beta_2=1$. 
The lines will be framed as in the Gu-Wen $\Pi$-category calculation, going out of each dual vertex in the direction of the 
earliest face in the order. See Figure \ref{fig:dressed}.

The amplitude for each tetrahedron is computed using the same projection on the two-sphere. The spin-TFT partition sum will be 
\begin{equation}
Z[M;\fT_s] =\frac{|H^0(M,\ZZ_2)|}{|H^1(M,\ZZ_2)|}\sum_{[\beta_2] \in H^2(M,\ZZ_2)} z_{\Pi}(\beta_2)Z[M;\fT_f;\beta_2]
\end{equation}
The spin-structure dependence is hidden in $z_{\Pi}(\beta_2)$.

Notice that the calculation of $z_{\Pi}(\beta_2)$ can be integrated into the Turaev-Viro calculation. The Gu-Wen Grassmann integral 
can be given a super-vector space interpretation: \begin{itemize}
\item We can assign fermion number $0$ to the $V_{ijk}$ spaces and fermion number $1$ to the $V_{\Pi;ijk}$ spaces. This mimics the assignment of 
Grassmann variables $\theta_f$, $\bar \theta_f$ to the faces of the triangulation. 
\item We can pick a specific order in the tensor product of face vector spaces which defined the Hilbert space associated to the boundary of a tetrahedron. 
The order mimics the choice of order for the Grassmann variables in the Gu-Wen integrand. 
\item When contracting pairs of dual vector spaces associated to each face, we keep track of the Koszul signs 
required to reorder the tensor product and bring the pair of spaces to be adjacent to each other. This 
mimics the choice of order for the Grassmann variables in the Gu-Wen integration measure
\end{itemize} 

As the combinatorics of super-vector space tensor products reconstruct the Gu-Wen Grassmann integral, all which is left is 
the linear coupling of $\beta_2$ to the chain $E$ of faces encoding the spin structure. 
We can write 
\begin{equation}
Z[M;\fT_s] =\frac{|H^0(M,\ZZ_2)|}{|H^1(M,\ZZ_2)|}\sum_{[\beta_2] \in H^2(M,\ZZ_2)} (-1)^{\int_E \beta_2} Z_{\mathrm{super}}[M;\fT_f;\beta_2]
\end{equation}

This has the form of a calculation in the spherical super-fusion category associated to the $\Pi$-category $\cC_f$.
It would be interesting to pursue this point further. 

\section{String net models}
The same data which goes into the Turaev-Viro construction can also be used to give a local lattice Hamiltonian construction 
of the theory. 

It is straightforward to give a physical motivation is analogous to the one we reviewed for the partition function.
The basic step is to relate the Hilbert space $\cH_\Sigma$ of the theory $\fT$ on some space manifold $\Sigma$ and the 
Hilbert space $\cH_{\Sigma'}$  on a manifold $\Sigma'$ with an extra circular hole with boundary condition $\fB$. 

In general, $\cH_{\Sigma'}$ is larger than $\cH_\Sigma$, but there will be maps $i, \pi$ embedding $\cH_\Sigma$ into $\cH_{\Sigma'}$
and projecting $\cH_{\Sigma'}$ to $\cH_\Sigma$, which can be described in terms of three-manifolds with the topology of $\Sigma \times [0,1]$ 
minus a half-sphere. It is easy to see that $\pi \circ i$ is a multiple of the identity map, as it corresponds to a three-manifolds with the topology of $\Sigma \times [0,1]$ 
minus a contractible sphere with boundary condition $\fB$. 

Thus we can describe $\cH_\Sigma$ as the image in $\cH_{\Sigma'}$ of the projector $i \circ \pi$, corresponding to a 
three-manifolds with the topology of $\Sigma \times [0,1]$ minus two half-spheres.
Furthermore, the projector $i \circ \pi$ can be given a simple description in terms of 
the combination $\sum_i d_i L_i$ we encountered in explaining the Turaev-Viro construction,
where the $L_i$ are interpreted as closed line defects going around the circumference of the hole,
acting on the Hilbert space. 

\begin{figure}
\begin{center}
\includegraphics[width=15cm]{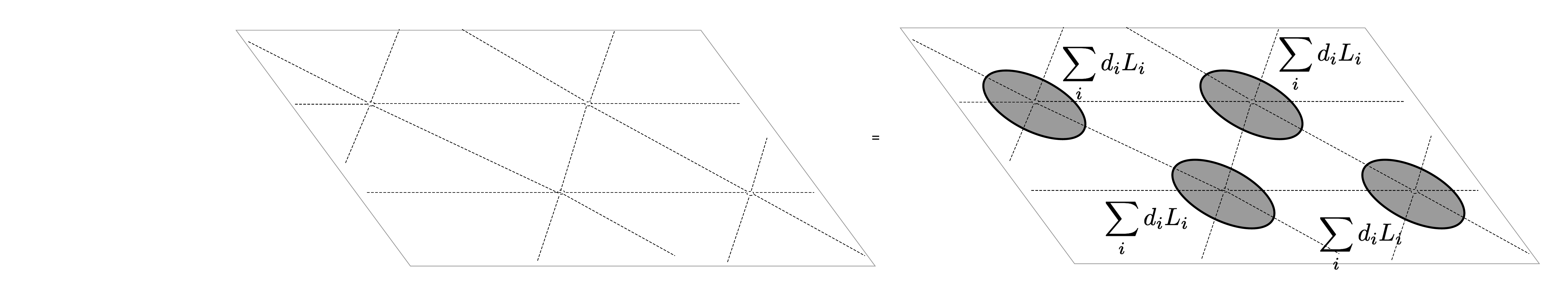}
\end{center}
\caption{The Hilbert space associated to a surface $\Sigma$ without holes (left) can be embedded in the Hilbert space associated to a surface $\Sigma'$ with holes (right), as the image of projectors
defined by the action of closed boundary lines $\sum_i d_i L_i$.}
\label{fig:hilb1}
\end{figure}

More generally, we can triangulate $\Sigma$ and carve out a circular hole at each vertex of the triangulation. Each hole will be associated to a separate 
projector $P_v = i_v \circ \pi_v$ and all projectors will commute. Thus $\cH_\Sigma$ is obtained from $\cH_{\Sigma'}$ as the ground state of a commuting projector Hamiltonian. See Figure \ref{fig:hilb1}.

We can readily give a local description of $\cH_{\Sigma'}$, by enlarging the holes until they almost fill the 2-cells dual to the vertices of the triangulation.
We can continue out decomposition as we did for the partition function. At the next step we cut at 1-cells and replace $\Sigma'$ with a collection $\Sigma''$ of 
disks associated to faces of the triangulation, with $\fB$ boundary conditions and three boundary lines for each disk. Each edge of the triangulation 
is associated with a pair of dual boundary lines in the disks corresponding to adjacent faces. 

\begin{figure}
\begin{center}
\includegraphics[width=15cm]{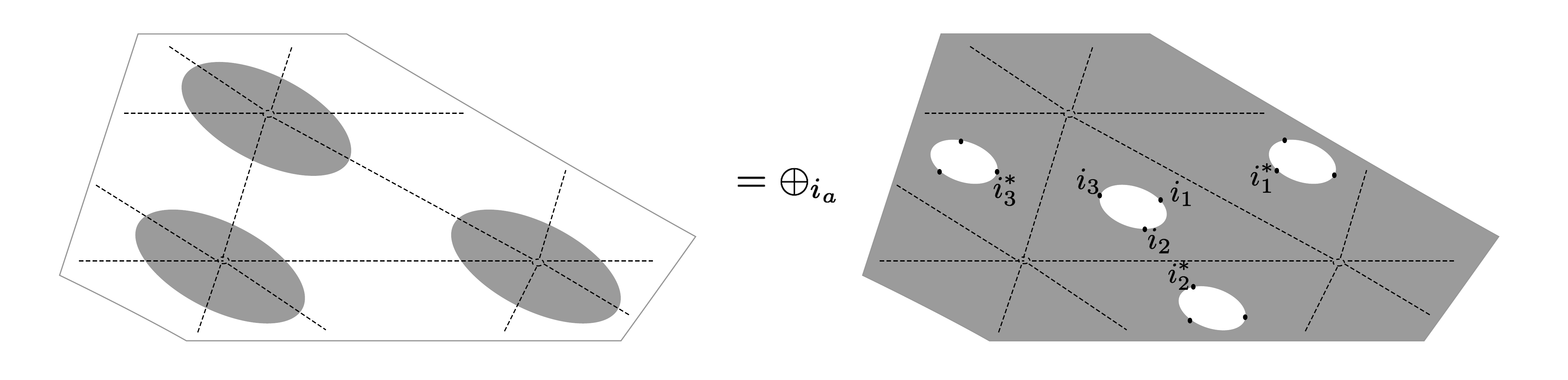}
\end{center}
\caption{The Hilbert space associated to a surface $\Sigma'$ with a regular arrangement of holes (left) can be identified with a direct sum of tensor product of Hilbert spaces associated to a collection of three-punctured disks. 
For clarity, we denote the boundary line defects $L_{i_a}$ and $L^*_{i_a}$ simply as ``$i_a$'' and ``$i_a^*$''. Disks are in correspondence to faces of the triangulation. Pairs of dual defects are in correspondences to the edges of the triangulation.}
\label{fig:hilb2}
\end{figure}

As long as the possible choices of lines run over all simples, or whatever other 
convenient basis of lines we employed in the state sum model, we can find maps $i_e$, $\pi_e$ embedding $\cH_{\Sigma'}$ in 
$\cH_{\Sigma''}$. See Figure \ref{fig:hilb2}.

If the lines we selected are simple objects, the embeddings are actual isomorphisms, as both $i_e \circ \pi_e$ and $\pi_e \circ i_e$ 
turn out to be multiples of the identity as long as the pairs of simple lines corresponding to an edge of a triangulation are dual to each other. 
The Hilbert space $\cH_{\Sigma''}$ is the tensor product of the corresponding morphism spaces 
$V_{ijk}$ for each disk.  
\footnote{If the lines $L_i$ are not simple, the $V_{ijk}$ are modules for the $\Hom(L_i,L_i)$ morphisms. Then $\pi_e \circ i_e$ is still a multiple of the identity but $i_e \circ \pi_e$ projects 
the naive tensor product of $V_{ijk}$ spaces to the tensor product over $\Hom(L_i,L_i)$.}  See again Figure \ref{fig:hilb2}.

The Hilbert space $\cH_{\Sigma''}$ is the microscopic Hilbert space for the string net model. The decoration of edges by line defects $L_i$ and the vector spaces $V_{ijk}$ 
decorating the faces are the microscopic degrees of freedom. This is also the Hilbert space associated in the state sum construction to a boundary with topology 
$\Sigma$, triangulated and decorated by all possible simple simple lines. 

\begin{figure}
\begin{center}
\includegraphics[width=15cm]{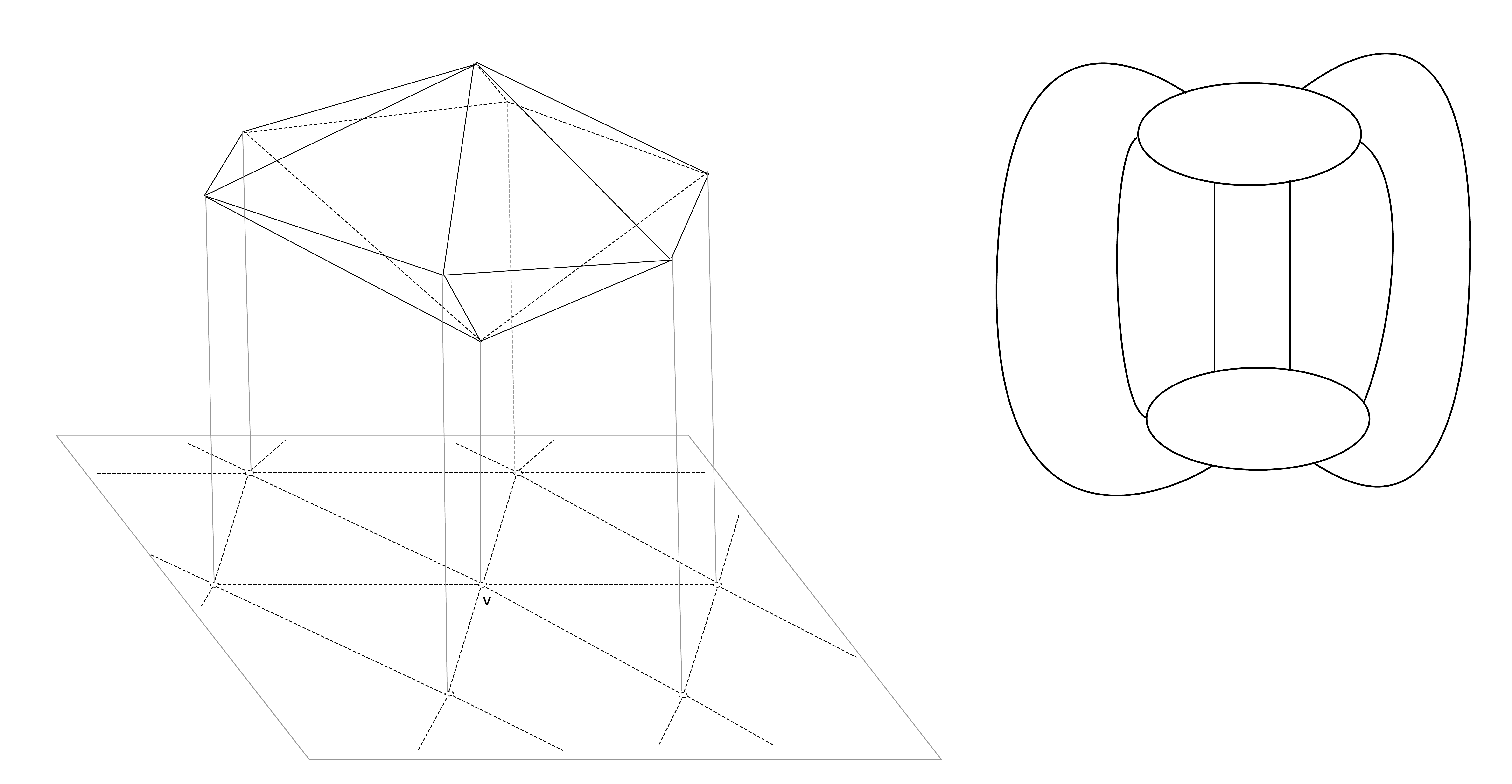}
\end{center}
\caption{Left: The projector $P_v$ can be computed as a state-sum model evaluation of a bi-pyramid. The bi-pyramid partition function is interpreted as a map from the dual of the vector space associated to the bottom faces
to the vector space associated to the top faces. The action of the projector on the microscopic Hilbert space of the string-net model corresponds to gluing the bi-pyramid 
on top of the vertex $v$. Right: The bi-pyramid is computed in the spherical fusion category by an appropriate planar graph dual to the bipyramid surface. The oval faces are dual to the top and bottom vertices of the bi-pyramid.}
\label{fig:proj1}
\end{figure}

The projectors $P_v$ can be computed as the state sum partition function for a geometry 
$M_v$, consisting of a bi-pyramid made of tetrahedra to be glued on top of the triangles adjacent to $v$. Of course, the bi-pyramid contribution 
can be computed directly by the evaluation of the dual graph in the spherical fusion category. See Figure \ref{fig:proj1}.

\begin{figure}
\begin{center}
\includegraphics[width=15cm]{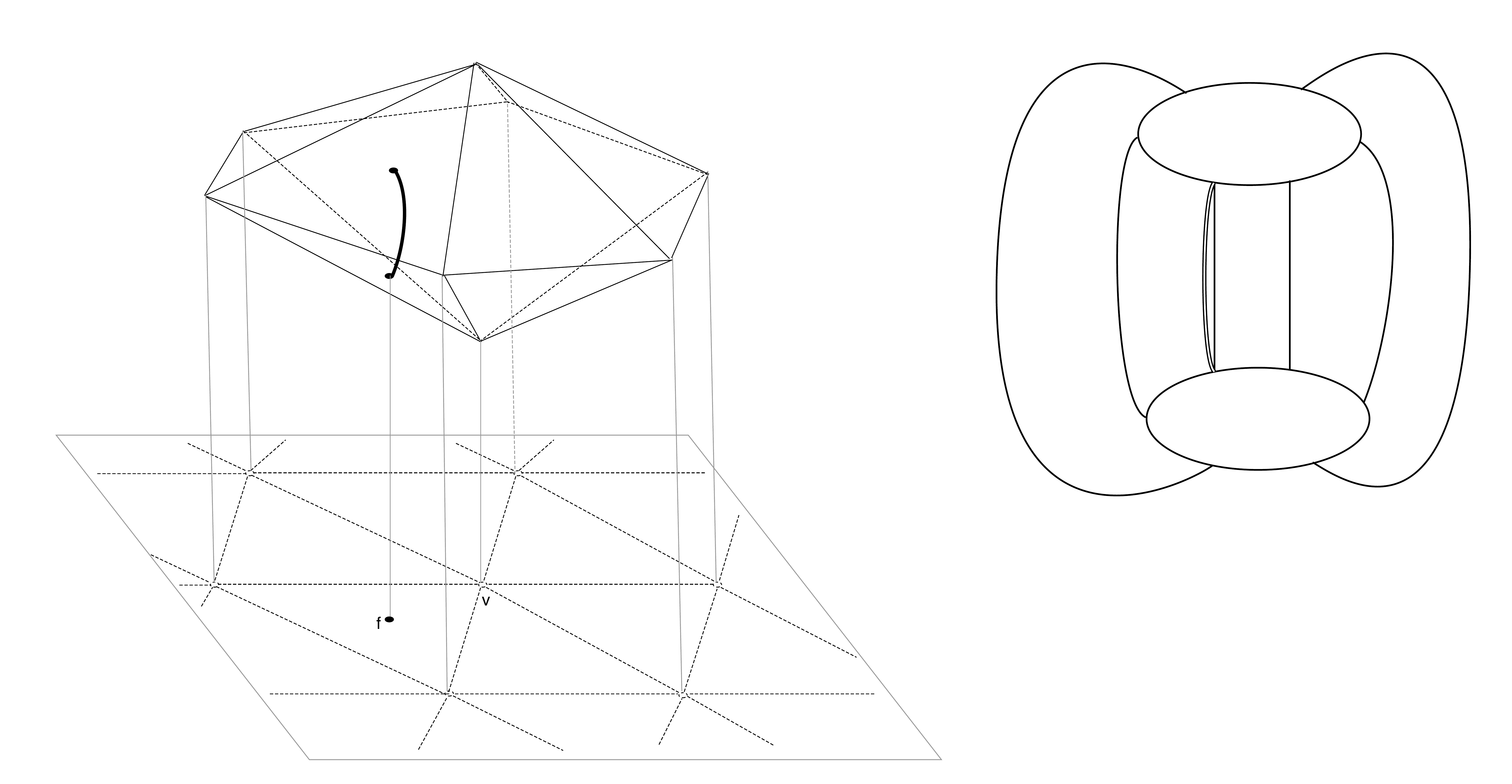}
\end{center}
\caption{Left: The projector $P_v$ in the presence of a quasi-particle at $f$ can be computed as a state-sum model evaluation of a bi-pyramid with an extra bulk line. 
Right: The decorated bi-pyramid is computed in the spherical fusion category by an appropriate planar graph including the appropriate center line corresponding to the quasi-particle. The quasi-particle 
joins the junctions dual to the bi-pyramid faces above $f$. We selected a specific framing for the quasi-particle (which direction it exits and enters the junctions) it and kept it constant}
\label{fig:proj2}
\end{figure}

If we want to add a bulk quasi-particle $Y$ as some point in $\Sigma$, say inside a face $f$ of the triangulation, 
we simply replace $V_{ijk}$ with the Hilbert space for a disk $V_{ijk;Y}$ with the extra bulk particle $Y$
in the middle, as in the state-sum model. The projectors for the vertices around $f$ 
are corrected by adding the quasi-particle to the state-sum calculation, going in and out 
the old and new $f$ face. See Figure \ref{fig:proj2}.

\begin{figure}
\begin{center}
\includegraphics[width=15cm]{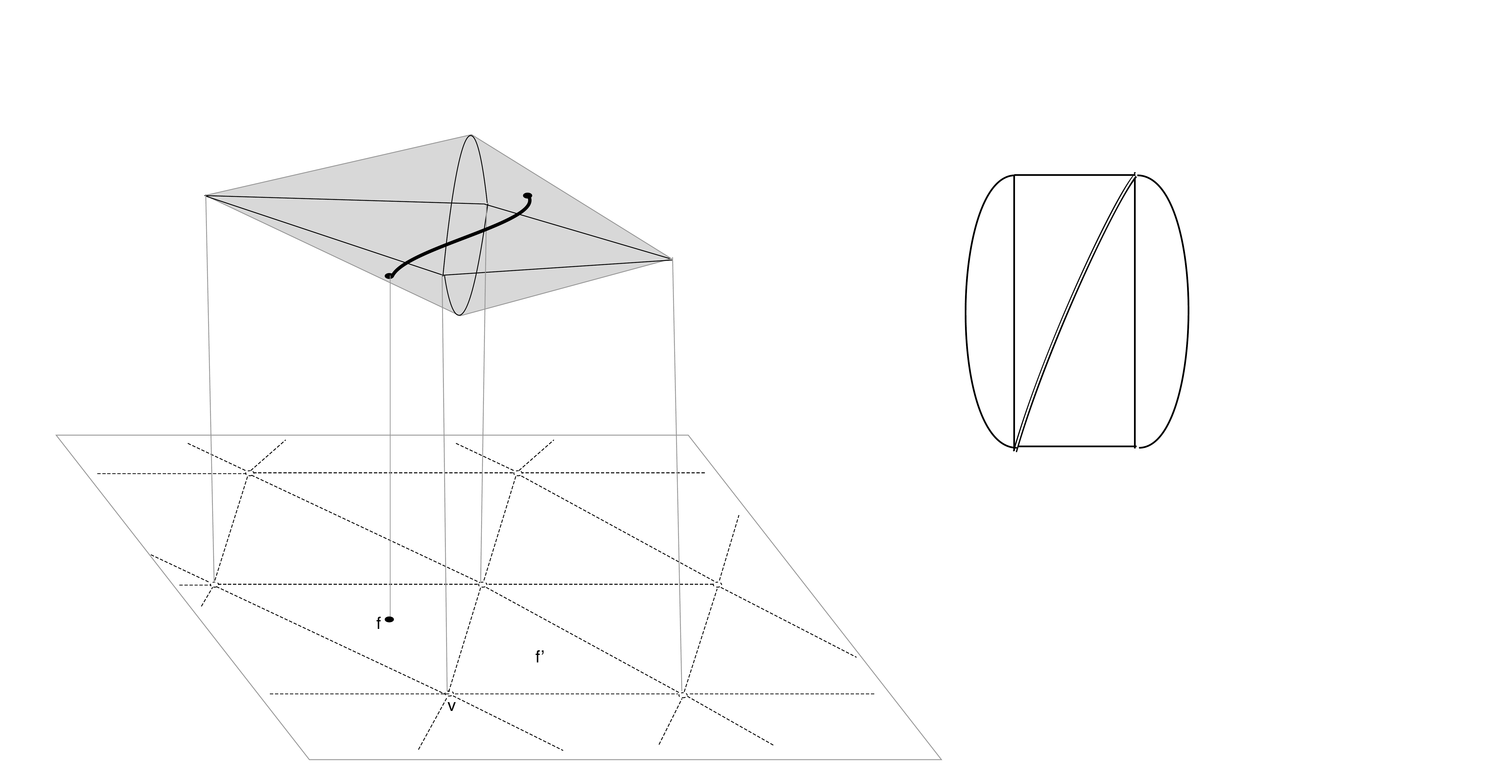}
\end{center}
\caption{Left: A very economical description of the operator $U^Y_{f,f'}[\ell]$ for adjacent faces $f$ and $f'$. The pillow-case geometry is the minimal way to interpolate between 
triangulations with quasi-particle insertions at $f$ and $f'$.  
Right: The decorated pillowcase is computed in the spherical fusion category by an appropriate planar graph including the appropriate center line corresponding to the quasi-particle. The quasi-particle 
joins the junctions dual to the bottom bi-pyramid face above $f$ and the top bi-pyramid face above $f'$. We selected a specific framing for the quasi-particle, pointing towards the vertex $v$.}
\label{fig:proj3}
\end{figure}

We can also consider operators $U^Y_{f,f'}[\ell]$ corresponding to state-sum geometries which interpolate between the original triangulation and a triangulation where the 
quasi-particle $Y$ has been moved to another face $f'$ along some framed path $\ell$ in $\Sigma \times [0,1]$. See Figure \ref{fig:proj3} for a crucial example.
Crucially, these operators will commute with the projectors in the Hamiltonian. Their algebra will mimic the 
topological properties of the corresponding quasi-particles. 

\subsection{Example: toric code}
The string net model for the toric code, based on the category with objects $I$ and $P$, 
is quite obviously a $\ZZ_2$ gauge theory: the configuration of edge decorations on the triangular lattice can be interpreted as 
a 1-cochain $\alpha_1$ with values in $\ZZ_2$. The fusion constraint requires $\alpha_1$ to satisfy $\delta \alpha_1=0$. Since all vector spaces $V_{ijk}$ are one-dimensional or zero-dimensional, each allowed edge decoration corresponds to a basis vector in $\cH_{\Sigma's}$ which will be denoted $|\alpha_1\rangle$. 
The projector $P_v$ simply acts as 
\begin{equation}
|\alpha_1\rangle \to \frac{1}{2}|\alpha_1\rangle + \frac12 |\alpha_1+ \delta \lambda^v_0\rangle
\end{equation}
where $\lambda^v_0(v') = \delta_{v,v'}$ is the 0-cochain supported on $v$. The product $\prod_v P_v$ projects to the subspace of gauge-invariant states. 

It is interesting to decorate this picture with quasi-particles. 
The $m$ quasi-particle at some face $f$ simply deforms the fusion constraint at $f$. 
More generally, a configuration $\beta_2^m$ of $m$ quasi-particles imposes the constraint 
$\delta \alpha_1=\beta_2^m$. The projectors are unchanged:
\begin{equation}
P_{\lambda_0}[\beta_2^m]   |\alpha_1\rangle = \frac{1}{2} |\alpha_1\rangle + \frac12|\alpha_1+ \delta \lambda_0\rangle
\end{equation}

Similarly, the operators $U^m_{\lambda_1}$ which change the locations of $m$ particles as $\beta^m_2 \to \beta^m_2 + \delta \lambda_1$ 
can be defined by combining individual $U^m_{e}$ which act on the two faces adjacent to an edge $e$, built from a pillowcase 
geometry. The map only changes $\alpha_1$ at the edge itself, and thus we have simply
\begin{equation}
U^m_{\lambda_1}  |\alpha_1\rangle = |\alpha_1+ \lambda_1\rangle .
\end{equation}
Clearly we have
\begin{equation}\label{Umalgebra}
U^m_{\lambda_1}U^m_{\lambda'_1}=U^m_{\lambda_1+\lambda'_1}.
\end{equation}
Note that if $\lambda_1$ is exact, $\lambda_1=\delta\mu_0$, we have
\begin{equation}\label{Uminv}
U^m_{\delta\mu_0}P_{\mu_0}=P_{\mu_0}.
\end{equation}
Therefore on the image of $\prod_v P_v$ the operator $U^m_{\lambda_1}$ is invariant under  $\lambda_1\mapsto \lambda_1+\delta{\mu_0}$. The ability to define operators $U^m_{\lambda_1}$ satisfying (\ref{Umalgebra}) and (\ref{Uminv}) indicates that the $\ZZ_2$ 1-form symmetry generated by the $m$ particle is non-anomalous. 

\begin{figure}
\begin{center}
\includegraphics[width=15cm]{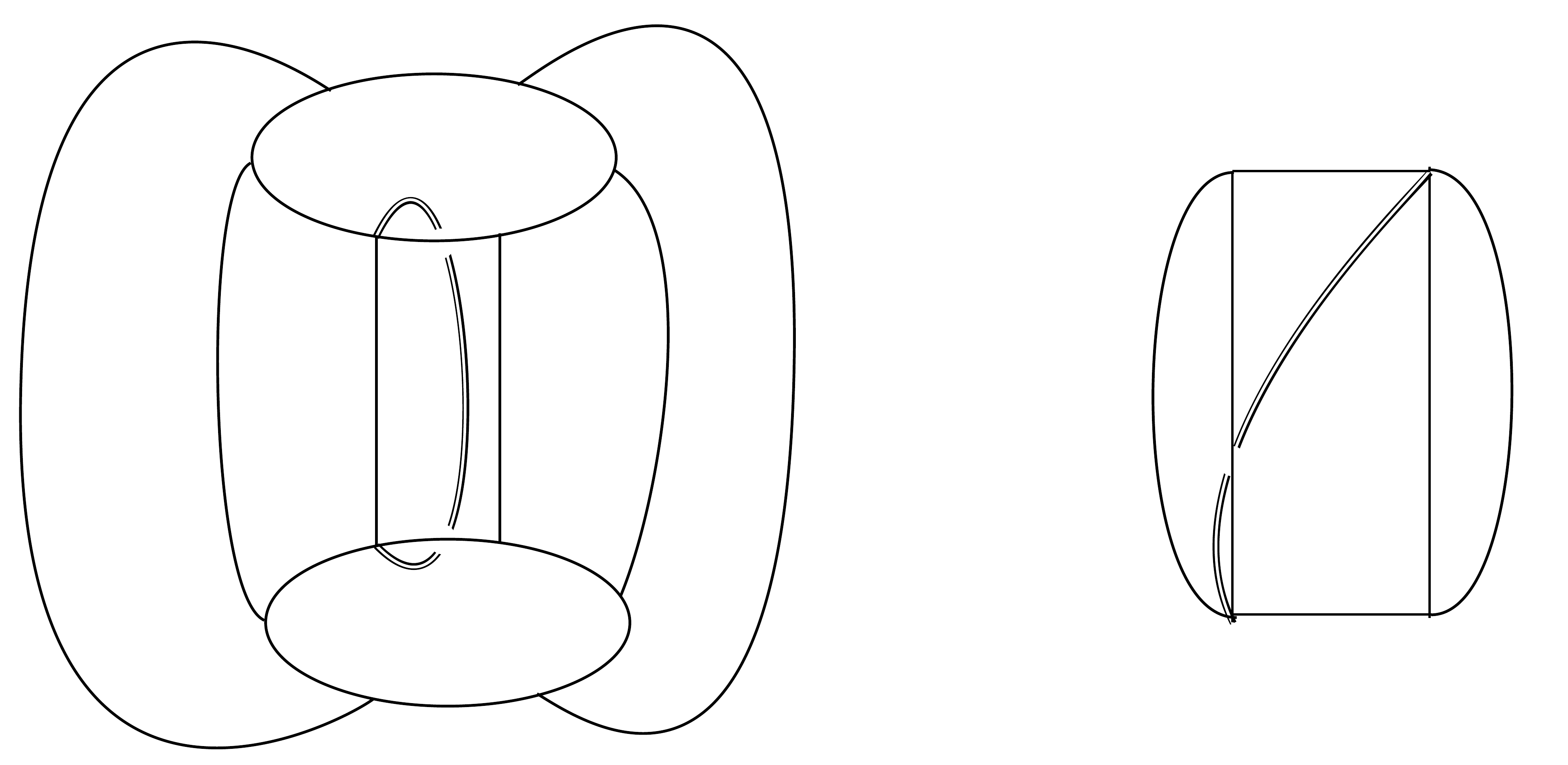}
\end{center}
\caption{Left: The only bi-pyramid contributing non-trivial signs to $P_v$ in the presence of an $e$ particle. The quasi-particle is framed towards $v$ and the decoration of the 
edges near $v$ must flip from $1$ to $P$ or viceversa. Right: An example of a pillowcase contributing a non-trivial sign to 
$U^e_e$. There is a potential sign whenever the quasi-particle is framed towards an oval face, i.e. the earliest vertex of a face (``0'') is opposite to the edge $e$. Then the sign measures the presence of $P$
along the $01$ edge of that face. This can be expressed as a cup product $\alpha_1 \cup \lambda^e_1$. }
\label{fig:signs}
\end{figure}

On the other hand, an $e$ quasi-particle at a face $f$ will not change the fusion constraint, but will change 
the form of the projectors for the vertices of $f$ by adding some signs. Inspection of the dual bi-pyramid graph 
shows that the $e$ center line only needs to cross other lines if it is framed towards $v$. In that case, we pick a sign 
$-1$ for each $P$ line it crosses. See Figure \ref{fig:signs}.

For definiteness, we should pick a canonical framing for quasi-particles. For example, we can add a branching structure (local order of vertices) on the 
triangulation  and frame quasi-particles towards the earliest vertex in each face. Then $P_v$ has a sign only if $v$ is the earliest vertex of $f$,
appearing in front of the $|\alpha_1+ \delta \lambda^v_0\rangle$ term. 

The projectors for a general configuration of $e$ particles $\beta^e_2$ become 
\begin{equation}
P_{\lambda_0}[\beta_2^e] |\alpha_1\rangle= \frac{1}{2} |\alpha_1\rangle +\frac12 (-1)^{\int \lambda_0 \cup \beta^e_2} |\alpha_1+ \delta \lambda_0\rangle
\end{equation}
Here we used the branching structure to define cup products. In a gauge theory language, the second term in the deformed projector inserts a Wilson line at the earliest vertex of the face $f$. 
The space of states in the presence of $e$ particles is the image of the projector $\prod_v P_v[\beta_2^e]$. Here $\alpha_1$ is closed because there are no $m$ quasi-particles present.

The operators $U^e_{\lambda_1}$ which change the locations of $e$ particles as $\beta^e_2 \to \beta^e_2 + \delta \lambda_1$ 
can be defined by combining individual $U^e_{e}$ which act on the two faces adjacent to an edge $e$, built from a pillowcase 
geometry. See again Figure \ref{fig:signs}. The operator is diagonal in the $|\alpha_1\rangle$ basis:
\begin{equation}
U^e_{\lambda_1}  |\alpha_1\rangle = (-1)^{\int \alpha_1 \cup \lambda_1} |\alpha_1\rangle
\end{equation}
Again we have
\begin{equation}\label{Uealgebra}
U^e_{\lambda_1}U^e_{\lambda'_1}=U^e_{\lambda_1+\lambda'_1},\quad U^e_{\delta\lambda_0} P^e_{\lambda_0}[\beta^e_2]=P^e_{\lambda_0}[\beta^e_2],
\end{equation}
indicating that the $\ZZ_2$ 1-form symmetry generated by the $e$ particle is non-anomalous. 

Notice that the $U^m$ and $U^e$ operators do not commute, as expected from the braiding phase of $e$ and $m$: 
\begin{equation}
U^m_{\lambda_1} U^e_{\lambda'_1} = (-1)^{\int \lambda_1 \cup \lambda'_1}U^e_{\lambda'_1}U^m_{\lambda_1}
\end{equation}
Neither is $U^e_{\lambda_1}$ invariant under $\lambda_1\mapsto\lambda_1+\delta\mu_0$ when $\delta\alpha_1=\beta^m_2$ is non-vanishing. This indicates a mixed anomaly for the two $\ZZ_2$ 1-form symmetries. 

Finally, we can insert $\epsilon$ particles. We both impose the constraint 
$\delta \alpha_1=\beta_2$ and use the projectors
\begin{equation}
P_{\lambda_0}[\beta_2] |\alpha_1\rangle= \frac{1}{2} |\alpha_1\rangle + \frac12(-1)^{\int \lambda_0 \cup \beta_2} |\alpha_1+ \delta \lambda_0\rangle
\end{equation}
These correspond to a specific choice of framing of the $\epsilon$ line in the pillowcase geometry: it joins the two junctions 
along the most direct path compatible with the framing of the junctions, crossing a minimum number of other edges in the 
planar graph.  

The operators which change the location of the $\epsilon$ particles around a single edge take the form 
\begin{equation}
U^\epsilon_e|\alpha_1\rangle\equiv U^\epsilon_{\lambda_1^e}  |\alpha_1\rangle = (-1)^{\int \alpha_1 \cup \lambda^e_1} |\alpha_1+ \lambda_1^e\rangle .
\end{equation}

We can tentatively define a general operator rearranging $\epsilon$ particles:
\begin{equation}
U^\epsilon_{\lambda_1}  |\alpha_1\rangle = (-1)^{\int \alpha_1 \cup \lambda_1} |\alpha_1+ \lambda_1\rangle
\end{equation}
but the anomaly pops out as expected: 
\begin{equation}
U^\epsilon_{\lambda_1} U^\epsilon_{\lambda'_1} = (-1)^{\int \lambda'_1 \cup \lambda_1}  U^\epsilon_{\lambda_1+\lambda'_1}  
\end{equation}

We can compute also 
\begin{equation}
U^\epsilon_{\lambda_1} P_{\lambda_0}[\beta_2] |\alpha_1\rangle= \frac{1}{2} (-1)^{\int \alpha_1 \cup \lambda_1} |\alpha_1+ \lambda_1\rangle +\frac12 (-1)^{\int \lambda_0 
\cup \beta_2}(-1)^{\int \delta \lambda_0 \cup \lambda_1} (-1)^{\int \alpha_1 \cup \lambda_1}|\alpha_1+ \delta \lambda_0+ \lambda_1\rangle
\end{equation}
and check that it coincides with $ P_{\lambda_0}[\beta_2+ \delta \lambda_1] U^\epsilon_{\lambda_1} |\alpha_1\rangle$, as expected.

Note that the pairing 
\begin{equation}
\omega_\Sigma (\lambda_1,\lambda'_1)= \int_\Sigma \lambda'_1\cup\lambda_1
\end{equation}
is symmetric modulo $2$ for closed cochains $\lambda_1,\lambda'_1$ but not in general. Rather, one has
\begin{equation}
\omega_\Sigma(\lambda_1,\lambda'_1)-\omega_\Sigma(\lambda'_1,\lambda_1)=\int_\Sigma (\delta\lambda\cup_1\lambda'_1+\lambda_1\cup_1\delta\lambda'_1).
\end{equation}
Thus some $U^\epsilon_{\lambda_1}$ and $U^\epsilon_{\lambda'_1}$ anti-commute rather than commute. Concretely, it is easy to check that $U^\epsilon_e$ and $U^\epsilon_{e'}$ anti-commute if the $e$ and $e'$ are adjacent to the same face and have the same orientation (induced by the branching structure) with respect to the face. They commute otherwise. Thus we cannot impose the constraint $U^\epsilon_e=1$ on the states for all $e$.\footnote{While it is true that the naive $U^\epsilon_e$ squares to $1$ for all $e$, it is not true that the naive $U^\epsilon_{\lambda_1}$ squares to $1$ for all $\lambda_1$. But this problem can be fixed by redefining $U^\epsilon_{\lambda_1}$ by suitable factors of $i$. The lack of commutativity for $U^\epsilon_e$ and $U^\epsilon_{e'}$ is a more serious problem.}

\subsection{A fermionic dressing operator}
To fix the sign problem in the algebra of the operators $U^\epsilon_{\lambda_1}$, let us place at each face $f$ of the triangulation a pair of Majorana fermions $\gamma_f$ and $\gamma'_f$. They are generators of a Clifford algebra $\Cl (2)$.
  
For any edge $e$, define 
\begin{equation}
S_e = i \gamma_{f_L[e]} \gamma'_{f_R[e]}
\end{equation}
where $f_{L,R}$ are the faces to the left and to the right of the edge (with respect to the branching structure orientation). We have the commutation relation
\begin{equation}\label{Sealgebra}
S_e S_{e'}=(-1)^{\int_\Sigma \delta_1(e)\cup \delta_1(e')} S_{e'} S_e,
\end{equation}
where $\delta_1(e)$ is a 1-cochain supported on the edge $e$. In words: $S_e$ operators commute unless the two edges share a face and have the same orientation with respect to the face, in which case they anti-commute.  We also have $S_e^2=1$ for all $e$.  The crucial point is that the combined operators $S_e U^\epsilon_e$ commute with each other for all $e$. 

Next we would like to define $S_{\lambda_1}$ for a general 1-cochain $\lambda_1$, so that
\begin{equation}\label{Salgebra}
S_{\lambda_1} S_{\lambda'_1}=(-1)^{\int_\Sigma \lambda'_1\cup\lambda_1} S_{\lambda_1+\lambda'_1}.
\end{equation}

We let
\begin{equation}\label{Slambda}
S_{\lambda_1}= (-1)^{\sum_{e<e' \in \eE} \int_\Sigma \delta_1(e) \cup \delta_1(e') }\prod_{e\in \eE} S_e.
\end{equation}
Here $\eE$ is the set of edges where $\lambda_1(e)=1$, ordered in some way, and the product is ordered from right to left. 
The sign factor in (\ref{Slambda}) can also be described as follows: we include $-1$ for every pair of edges in $\eE$ which share a face, have the same orientation (with respect to the branching structure),  and whose order along the face agrees with the ordering of $\eE$. It is easy to check that $S_{\lambda_1}$ does not depend on the choice or ordering of $\eE$ and that (\ref{Salgebra}) is satisfied. Thus, if we provisionally define $V_{\lambda_1}=S_{\lambda_1} U^\epsilon_{\lambda_1}$ on the tensor product of $\cH_{\Sigma'}$ and the fermionic Fock space, we will have the relations
\begin{equation}\label{Valgebra}
V_{\lambda_1}V_{\lambda'_1}=V_{\lambda_1+\lambda'_1}.
\end{equation}

A further issue which needs to be addressed is the behavior of $V_{\lambda_1}$ under transformations $\lambda_1\mapsto\lambda_1+\delta\mu_0$, where $\mu_0$ is a $\ZZ_2$-valued 0-cochain. A satisfactory generator of a $\ZZ_2$ 1-form symmetry must be invariant under such ``symmetries of symmetries''. Instead, in agreement with a general formula (\ref{Vnoninv}), we find
\begin{equation}
U^\epsilon_{\delta\mu_0}P_{\mu_0}[\beta_2]|\alpha_1\rangle=(-1)^{\int_\Sigma (\beta_2\cup\mu_0+\mu_0\cup\beta_2)} P_{\mu_0}[\beta_2]|\alpha_1\rangle. 
\end{equation}
Thus the operator $U^\epsilon_{\delta\mu_0}$ is nontrivial even after projection to the physical Hilbert space $\cH_\Sigma$. 

Similarly, we can compute $S_{\delta\mu_0}$. To write down the answer, note that basis elements in the fermionic Fock space are naturally labeled by $\ZZ_2$-valued 2-cochains $\nu_2$: for a given face $f$, the state $|\nu_2\rangle$ is an eigenstate of $i \gamma_f \gamma'_f$ with  eigenvalue $(-1)^{\nu_2(f)}$. Then we get
\begin{equation}
S_{\delta\mu_0}|\nu_2\rangle =(-1)^{\int_\Sigma (\nu_2\cup\mu_0+\mu_0\cup\nu_2+C_2\cup\mu_0)+\int_{{\tilde w}_2} \mu_0}|\nu_2\rangle.
\end{equation}
Here ${\tilde w}_2$ is a particular $\ZZ_2$-valued 0-chain representing the 2nd Stiefel-Whitney class of $\Sigma$, and $C_2$ is a $\ZZ_2$-valued 2-cochain which takes value $1$ on every face. 

We can cancel all $\nu_2$-dependent signs in $S_{\delta\mu_0}$ against $\beta_2$-dependent signs in $U^\epsilon_{\delta\mu_0}$ if we project to the subspace where $\nu_2=\beta_2$. To eliminate state-independent signs, we choose a 1-chain $E$ such that $\partial E={\tilde w}_2$. As discussed in \cite{GaiottoKapustin}, such a $E$ determines a spin structure on $\Sigma$. Then we define improved $E$-dependent Fock-space operators
$$
S^E_{\lambda_1}=(-1)^{\int_\Sigma C_1\cup \lambda_1+\int_E \lambda_1} S_{\lambda_1}. 
$$
Here $C_1$ is a $\ZZ_2$-valued 1-cochain taking value $1$ on every edge. It satisfies $\delta C_1=C_2$. We also define improved $E$-dependent dressed generators:
$$
V^E_{\lambda_1}=U^\epsilon_{\lambda_1} S^E_{\lambda_1}. 
$$
On the projected Hilbert space, they satisfy
$$
V^E_{\lambda_1} V^E_{\lambda'_1}=V^E_{\lambda_1+\lambda'_1},\quad V^E_{\lambda_1+\delta\mu_0}=V^E_{\lambda_1}
$$
for all 1-cochains $\lambda_1$ and all 0-cochains $\mu_0$. 

We can now define a commuting projector Hamiltonian for the phase $\fT_s$ on the projected Hilbert space as
$$
H^E=\sum_{e} \frac12 (1-V^E_e). 
$$

\subsection{Fermionic dressing for general $\Pi$-categories}
If the Drinfeld center of the fusion category $\cC$ contains a fermion $\Pi$, we can define $U^f_e$ operators in a manner completely analogous to the 
toric code example. The operator is evaluated by the pillowcase graphs framed as described above.  

\begin{figure}
\begin{center}
\includegraphics[width=15cm]{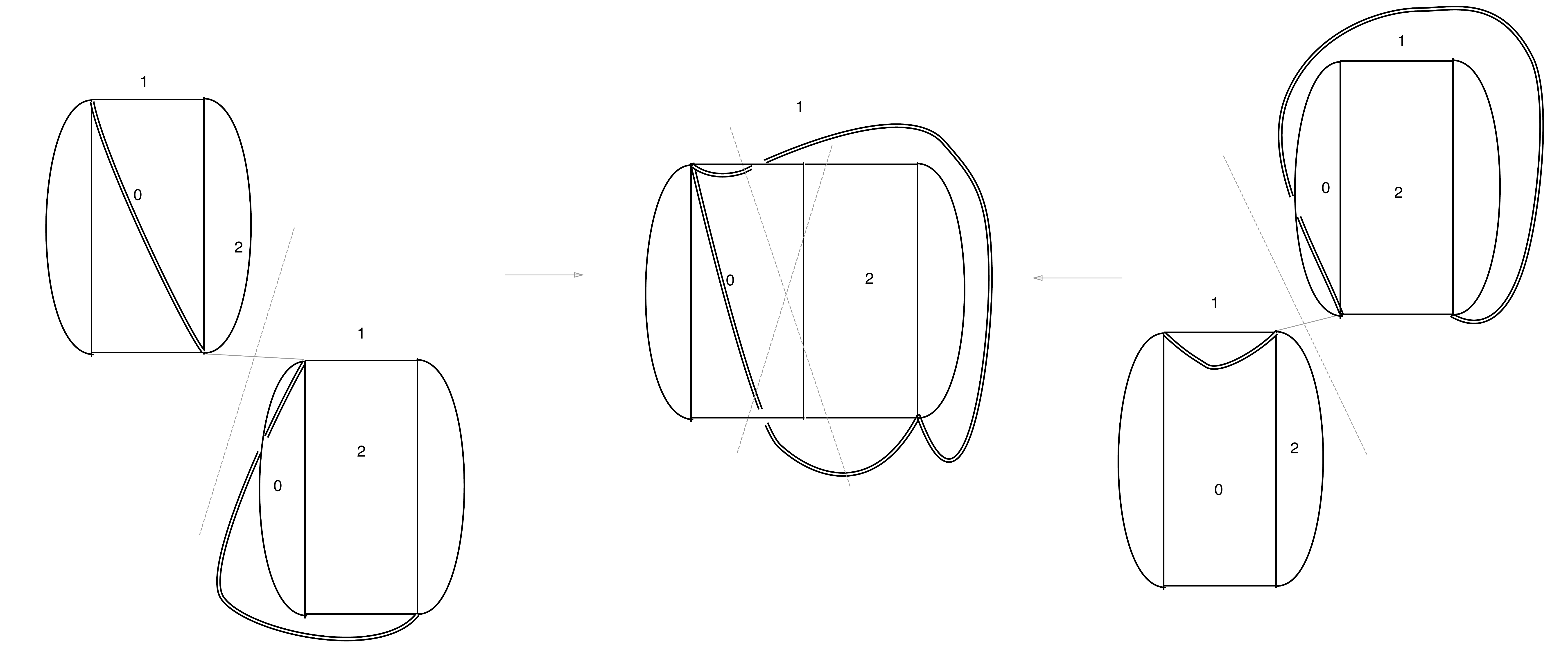}
\end{center}
\caption{A comparison of $U^f_e U^f_{e'}$ and $U^f_{e'} U^f_{e}$ for edges $e$, $e'$ adjacent to the same face $f$. Left and Right: The corresponding pairs of pillowcase geometries. Middle: The fused geometry. 
If the edges $e$ and $e'$ are the $01$ and $12$ edges of $f$, the center lines emanating from the fused junctions at $f$ reconnect as shown in the middle. 
The two center lines can be identifies only up to a change of one unit of framing. Similar pictures with the outer face labelled by $0$ or $2$ match with no change of framing. Hence associativity fails 
only if $e$ and $e'$ are the $01$ and $12$ edges or vice-versa.}
\label{fig:anticommute}
\end{figure}

Because of the universality of factors associated to changes of framing and recombination of $\Pi$ lines, we expect the same law as in the toric code.  
\begin{equation}
U^f_e U^f_{e'} = (-1)^{\int \delta_1(e')_1 \cup \delta_1(e)}  U^f_{e+e'}
\end{equation}
More directly, we can compare the geometries associated associated to $U^f_e U^f_{e'}$ and $U^f_{e'} U^f_{e}$. The corresponding pairs of pillowcase geometries 
can be glued together to give the same geometry, but the framing of the center lines in the new geometry may or not agree. A careful analysis of all cases reproduces 
the expected multiplication law. See Figure \ref{fig:anticommute}. Therefore the same dressing by Majorana fermions will give commuting operators $U^s_e \equiv S_e U_e^f$. 

From the toric code example, we also expect
\begin{equation}
U^f_{\delta\mu_0}=(-1)^{\int_\Sigma \left( \beta_2\cup\mu_0+\mu_0\cup\beta_2\right)}.
\end{equation}
This can be reproduced, with some effort, by counting the number of self-intersections of the $\Pi$ lines obtained by merging the 
chain of pillowcase graphs for the sequence of edges around a single vertex. 

Therefore the fermionic dressing ensures that $V^E_{\lambda_1}=U^f_{\lambda_1} S^E_{\lambda_1}$ is trivial when $\lambda_1$ is exact. 

It follows that we can gauge the 1-form symmetry generated by $\Pi$ by imposing the constraints $\beta_2(f)=\nu_2(f)$ for all faces $f$ and  $V^E_e=1$ for all edges $e$ 
in the tensor product of the Hilbert space of $\fT_f$ and the fermionic Fock space. 

This is our final prescription for a microscopic Hamiltonian for $\fT_s$, built from the data of $\cC_f$. 

\subsection{Including global symmetries}
It is easy to extend the string-net construction to models with global non-anomalous symmetry $G$. 
Such a model  is associated to a $G$-graded spherical fusion category $\cC=\oplus_g \cC_g$. In the Hamiltonian approach, $G$ acts on-site and commutes with the Hamiltonian. 

For example, we can model the lattice system on a discrete $G$-valued sigma model. That is, we put group variables $g_v$ 
at vertices.  Edges between vertices labeled by group elements $g$ and $g'$ are labeled by simple objects in 
$\cC_{g' g^{-1}}$. We have commuting projectors $P_v^{g,g'}$ which change the group element at $v$ from $g$ to $g'$,
built from a state-sum bi-pyramid with central edge decorated by $g' g^{-1}$ \cite{Chen:2011aa}.

Essentially by construction, adding $G$ gauge fields on the edges and gauge-fixing the vertex group elements 
reproduces the string net model for the theory where $G$ is gauged. 

Bosonic SPT phases provide an obvious, well understood example of this construction. In this case $\cC_g$ has a single simple object for all $g$, so the vertex variables $g_v$ are the only variables. 
For an explicit expression for the projectors $P_v^{g,g'}$ see \cite{Chen:2011aa}. 

\subsection{Example: the shadow of Gu-Wen phases}

Our next example is the $\ZZ_2$ gauge theory associated to a $\ZZ_2$ central extension $\hat G$ of a symmetry group $G$. 
In this case $\cC_g$ has two simple objects which we denote $V_{g,\epsilon}$, $\epsilon\in\ZZ_2$, as before. They fuse according to (\ref{GuWenfusion}).

We decorate the vertices of the triangulation with group elements in $G$ and the edges with $\ZZ_2$ variables $\epsilon_1$ so that the 
edge objects are $V_{g^{-1} g',\epsilon_1(e)}$. The fusion rules imply that $\delta \epsilon_1 = n_2$, where $n_2$ is 
evaluated on the $G$ group elements around each face. 

The projectors $P_v^{g,g'}$ involve two terms, each computed as a product of $\hat \alpha_3$. The two terms map a
state with given $\ZZ_2$ decoration to two states with $\ZZ_2$ decorations which differ by a gauge transformation at $v$. 
This is expected, as we are defining an equivariant version of $\ZZ_2$ gauge theory.

With a bit of patience, one can disentangle the contribution of  $\ZZ_2$ and $G$ variables to the bi-pyramid graph of $P_v^{g,g'}$.
For example, we can gauge-fix as in figure \ref{fig:hatcocycle4} every junction of the bi-pyramid graph. 
As we collapse the center line junctions to a single planar junction, we will get a factor of $-1$ from 
non-planar intersections. We can write these signs as $(-1)^{(n_2,\epsilon_1)}$, where the parenthesis 
indicates a certain bilinear pairing which is somewhat tedious to compute. This factor multiplies some 
expression $\tilde P_v^{g,g'}$ which depends on the $G$ variables only. 

We can populate the lattice with $\Pi$ particles along some cocycle $\beta_2$. Now $\delta \epsilon_1 = n_2+ \beta_2$.
We get deformed projectors $P_v^{g,g'}[\beta_2]$. Again, we can gauge-fix the junctions in a canonical way. 
The manipulation of $\Pi$ lines will give some new signs $(-1)^{(n_2,\epsilon_1)+ (\beta_2,\epsilon_1)'}$,
multiplying the same $\tilde P_v^{g,g'}$ expression as before. 

We can similarly compute the $U^f_e$ operators which change $\beta_2$ in the two faces adjacent to $e$. 
The calculation involves the same pillowcase graphs as before. 
The $G$ group elements do not change in the process, we only shift $\epsilon_1$ by $\lambda_1^e$. 

The contribution from $\Pi$ lines crossing other lines is again $(-1)^{\int \epsilon_1 \cup \lambda^e_1}$, as in the toric code.
Thus the fermionic dressing proceeds as before. 

\acknowledgments
We thank V. Ostrik for explaining the results of \cite{ENO} to one of us (A.K.). We are grateful to J. Brundan and A. Ellis for making their results \cite{Brundan:2016aa} available prior to publication. We also thank A. Karagiozova for helping LB and DG code the figures. A. K. is grateful to J. Morgan for communicating  to him the unpublished results of J. Morgan and G. Brumfiel  which helped to detect an error in the first version of the paper. 
The research of LB and DG was supported by the
Perimeter Institute for Theoretical Physics. Research at Perimeter Institute is supported
by the Government of Canada through Industry Canada and by the Province of Ontario
through the Ministry of Economic Development and Innovation. The work of A. K. was supported by the U.S. Department of Energy, Office of Science, Office of High Energy Physics, under Award Number DE-SC0011632
Opinions and conclusions expressed here are those of the authors and do not
necessarily reflect the views of funding agencies.

\appendix

\section{Spin-TFTs from Rational spin-CFTs} \label{app:RCFT}

Unitary rational conformal field theory is a rich source of examples of topological field theories in $2+1$ dimensions. 
The Hilbert space of the topological field theory is identified with the space of conformal blocks for the chiral algebra  
the conformal field theory.

If we endow the Riemann surface with a spin structure, we can consider conformal blocks for a chiral super-algebra $\cA = \cA_0 \oplus \cA_1$, 
which includes both bosonic currents of integral spin and fermionic currents of half-integral spin. The conformal blocks for $\cA$
can be naturally identified with the Hilbert space of a spin-TFT $\fT_s$.

We expect the shadow $\fT_f$ to be the TFT associated to the bosonic sub-algebra $\cA_0$. Notice that $\cA_1$ is a module for $\cA_0$ 
and thus gives a quasi-particle in $\fT_f$, a fermion. As $\cA_1$ currents fuse to $\cA_0$ currents, the fermionic quasi-particle fuses to the identity. 
We identify it with $\Pi$. Thus fermionic anyon condensation is related to fermionic current algebra extensions, in the same way as 
the standard anyon condensation is related to standard current algebra extensions.

It should be possible to pursue this analogy further and derive from spin-RCFTs appropriate axioms 
for ``super modular tensor categories''. 

We can give a few well-known examples of this construction.
\subsection{Ising model and a chiral fermion}
The Ising modular tensor category is naturally associated to the current algebra of a 
$c=\frac12$ Virasoro minimal model. 

The current algebra is generated by the stress tensor and can also be described as the coset 
$\frac{SU(2)_1 \times SU(2)_1}{SU(2)_2}$. It has three modules, which we can denote as $M_1 \equiv \cA_0$, $M_\sigma$ and $M_\psi$,
of conformal dimension $0$, $\frac{1}{16}$ and $\frac{1}{2}$. The latter is associated to the fermionc quasi-particle of the 
Ising modular tensor category. 

We can consider the super-algebra $\cA$ consisting of $\cA_0$ and $\cA_1 = M_\psi$. This is simply the algebra generated by a free chiral fermion $\psi(z)$. 
This algebra has a single Neveu-Schwarz (NS) module, $\cA$ itself. It corresponds to the identity quasi-particle in the bulk spin-TFT. 
On the other hand, $M\sigma$ is a Ramond module. It must lie at the end of a bulk defect with a 
non-bounding spin structure. 

The Ising 3d TFT is the shadow of the spin-TFT associated to a free chiral fermion. We could denote it as $\fT_s^\psi$. 

\subsection{$\Pi$-product of Ising models and multiple chiral fermions}
In order to find the shadow of the product of two free chiral fermions,
we are supposed to gauge the $\ZZ_2$ 1-form symmetry generated by $\Pi \Pi'$.
This is the same as a simple current extension. 

Consider two copies of the $c=\frac12$ Virasoro minimal model. The module corresponding to $\Pi \Pi'$ is 
$M_\psi \otimes M_{\psi'}$. We thus consider the chiral algebra $\cA_0 = M_1 \otimes M_{1'} \oplus M_\psi \otimes M_{\psi'}$. 
In other words, the chiral algebra generated by $\psi \partial \psi$, $\psi \psi'$ and $\psi' \partial \psi'$. 

By bosonization, we identify that with the algebra $\cA_0$ defined by a free boson current $\partial \phi$ and 
vertex operators $e^{2 n i \phi}$, of dimension $2 n^2$. In other words, this is the $U(1)_4$ current algebra. 
It has $4$ modules generated respectively by $1$, $e^{\frac{i \phi}{2}}$, $\psi = e^{i \phi}$ and $e^{-\frac{i \phi}{2}}$.

The chiral super-algebra generated by $\cA_0$ and by $\psi$ can be identified with $U(1)_1$ and is associated to the simplest
spin Chern-Simons TFT.  Again, it has a single quasi-particle, the identity, and an extra Ramond line defect associated to 
the module generated by $e^{\frac{i \phi}{2}}$. We can also identify it as the square of $\fT_s^\psi$.

More generally, a set of $N$ free chiral fermions has bosonic 1-form symmetry generators $\psi_i \psi_j$. 
Adjoining them to the identity module gives us the $SO(N)_1$ WZW current algebra, with a module $M_\psi$ 
generated by the $\psi_i$ and one or two modules generated by twist fields, associated to the spinor representation(s).
This is the shadow of the $N$-th power of $\fT_s^\psi$.

\subsection{$U(1)_{4k}$ Chern-Simons theories}
Consider the current algebra $\cA_0=U(1)_{4k}$ for odd $k$, generated by 
the bosonic current $\partial \phi$ and vertex operators $e^{2 \sqrt{k} n i \phi}$.
This is associated to an $U(1)_{4k}$ Chern-Simons theory.

The algebra $\cA_0$ has modules $M_m$ generated by $e^{\frac{m}{2 \sqrt{k}} i \phi}$, 
for $-2 k < m \leq 2 k$. In particular, $M_\Pi \equiv M_{2 k}$ is generated by $e^{\sqrt{k} i \phi}$,
which has half-integral dimension $k/2$. 

The chiral super-algebra $\cA$ generated by $\partial \phi$ and vertex operators $e^{\sqrt{k} n i \phi}$
is associated to an $U(1)_k$ spin Chern-simons theory. It has NS modules generated by $e^{\frac{m}{\sqrt{k}} i \phi}$
for $-k/2 < m < k/2$ and Ramond modules generated by $e^{\frac{2m+1}{2\sqrt{k}} i \phi}$.

In particular, $U(1)_{4k}$ is the shadow of $U(1)_k$.

\section{$G$-equivariant toric code}\label{app:twist}

\subsection{Symmetries of the toric code and their anomalies}

The toric code (also known as $\ZZ_2$ topological gauge theory in 2+1d with a trivial Dijkgraaf-Witten class) can be described by a Euclidean action
\begin{equation}\label{actiontoric}
S_{toric}=\pi i \int_M b\cup\delta a,
\end{equation}
where $a$ and $b$ are $\ZZ_2$-valued 1-cochains on a triangulation of an oriented 3-manifold $M$. One may call $a$ the gauge field, then $b$ is a Lagrange multiplier field imposing the constraint $\delta a=0$. The model has $\ZZ_2\times\ZZ_2$  0-form gauge symmetry:
\begin{equation}
a\mapsto a+\delta\lambda_a,\quad b\mapsto b+\delta\lambda_b,\quad \lambda_a,\lambda_b\in C^0(M,\ZZ_2).
\end{equation}
As for global symmetries, the toric code has a $\ZZ_2$ 0-form global symmetry $F_0$ exchanging $a$ and $b$. We will call it particle-vortex symmetry, since the Wilson line for $a$ represents an electrically charged particle, while the Wilson line for $b$ represents a vortex excitation. This symmetry is not manifest in the action since the cup product is not supercommutative on the cochain level. There is also $\ZZ_2\times\ZZ_2=F_1$ 1-form global symmetry
\begin{equation}\label{oneformtoric}
a\mapsto a+\alpha,\quad b\mapsto b+\beta,\quad \alpha,\beta\in Z^1(M,\ZZ_2). 
\end{equation}
Crucially, $F_0$ acts on $F_1$ by a nontrivial automorphism exchanging $\alpha$ and $\beta$. The combined symmetry is described by a 2-group (or equivalently a crossed module) \cite{KapustinThorngrenHigher}. In general, the equivalence class of a 2-group $\bbF$ is described by a pair of groups $F_0, F_1$,  where $F_1$ is abelian, an action $\pi$ of $F_0$ on $F_1$, and a Postnikov class $\gamma\in H^3_\pi(F_0, F_1)$. The Postnikov class describes the failure of the fusion of $F_0$ domain walls to be associative "on the nose". In the case of the toric code we can use Shapiro's lemma \cite{Evens} to compute $H^n_\pi(F_0, F_1)=H^n(\ker\ \pi, \ZZ_2)=0$ for $n>0$. Hence the Postnikov class is necessarily trivial. 

Global symmetries may have 't Hooft anomalies, i.e. obstructions to gauging. Such anomalies can always be canceled by coupling the theory to a topological gauge theory in one dimension higher. Thus anomalies are classified by topological actions for the gauge fields in one dimension higher. In the case of a 2-group symmetry, such actions have been classified in \cite{KapustinThorngrenHigher}. The gauge fields are a 1-form $F_0$ gauge field $\sA$ and a 2-form $F_1$ gauge field $\sB=(B_a,B_b)$. More precisely, $\sA$ is a 1-cocycle with values in $F_0$, while $\sB$ is a twisted 2-cocycle with values in a local system (i.e. flat bundle) with fiber $F_1$. The twist arises from the fact that $F_0$ acts nontrivially on $F_1$.
The most general action in 4d representing the anomaly for a 2-group symmetry is
\begin{equation}
S_{anomaly}=2\pi i \int_{M_4} \left(\frP_q(\sB)+\langle \sB,\cup c_2(\sA)\rangle+\omega_4(\sA)\right).
\end{equation}
The notation is as follows. We regard the pair $(\sA,\sB)$ as map from $M_4$ to the classifying space $B\bbF$ of the 2-group, which is a bundle over $BF_0$ with fiber $BF_1$. The action depends on a quadratic function $q: F_1\ra \RR/\ZZ$ invariant under the $F_0$ action, a class $c_2\in H^2_\pi(F_0,F_1^*)$, and a class $\omega_4\in H^4(F_0,\RR/\ZZ)$. $\frP$ denotes the Pontryagin square (a cohomological operation associated to the quadratic function $q$ which maps a twisted 2-cocycle $\sB$ to an $\RR/\ZZ$-valued 4-cocycle $\frP(\sB)$). In the case of interest to us, both $H^2_\pi(F_0,F_1^*)$ and $H^4(F_0,\RR/\ZZ)$ vanish (the former by Shapiro's lemma again), so there are no anomalies for $F_0$ or a mixed anomaly between $F_0$ and $F_1$. On the other hand, there exist $F_0$-invariant quadratic functions on $F_1$, so the 1-form symmetry $F_1$ could be anomalous. In fact, it is easy to see that the anomaly is nontrivial and corresponds to the quadratic function
\begin{equation}\label{quadratic}
q: \ZZ_2\times\ZZ_2\ra \RR/\ZZ,\quad q:(x,y)\mapsto\frac12 xy.
\end{equation}
Indeed, let the $F_0$ gauge field $\sA$ be trivial, so that $\sB$ is an ordinary 2-cocycle $(B_a,B_b)$ with values in $F_1=\ZZ_2\times\ZZ_2$.  If we perform the shifts (\ref{oneformtoric}) with not-necessarily-closed 1-cochains $\alpha$ and $\beta$, the action (\ref{actiontoric}) transforms as follows:
\begin{equation}
S_{toric}\mapsto S_{toric}+\pi i\int_M (b\cup\delta\alpha+\beta\cup\delta a+\beta\cup\delta\alpha).
\end{equation}
To cancel the terms which depend on $b$ and $a$ we couple the action to the 2-form gauge fields $B_a$ and $B_b$ which transform as $B_a\mapsto B_a+\delta\alpha$, $B_b\mapsto B_b+\delta\beta$ and define
\begin{equation}
S'_{toric}=\pi i \int_M (b\cup\delta a+b\cup B_a+ B_b\cup a).
\end{equation}
The new action transforms as
\begin{equation}
S'_{toric}\mapsto S'_{toric}+\pi i\int_M (\beta \cup B_a+B_b\cup\beta+\beta\cup\alpha),
\end{equation}
which is precisely the boundary term in the variation of 
\begin{equation}
\pi i \int_{M_4} B_b\cup B_a.
\end{equation}
This is nothing but the Pontryagin square of $\sB\in Z^2(M_4,F_1)$ for the quadratic function (\ref{quadratic}).\footnote{In general, if $\sA$ is nontrivial, $\sB$ is a twisted 2-cocycle, and the Pontryagin square for twisted cocycles is more difficult to write down.} 

To summarize, the anomaly action for the toric code is
\begin{equation}\label{toricanomalyaction}
S_{anomaly}=2\pi i \int_{M_4} \frP_q(\sB).
\end{equation}
In particular, the anomaly for the diagonal subgroup of $F_1$ is obtained by letting $B_a=B_b=B$. Note that this subgroup is $F_0$-invariant. The corresponding anomaly  action is 
\begin{equation}\label{squareanomaly}
S_{anomaly}=\pi i \int_{M_4} B\cup B,
\end{equation}
which means that the toric code is a shadow of a fermionic theory. The worldline ot the corresponding fermion $\Pi$ is represented by the Wilson line $\exp(i\pi \int(a+b))$. 

\subsection{$G$-equivariant toric code}

We are now ready to promote the toric code to a $G$-equivariant toric code, i.e. couple it to a $G$ gauge field $A$. Mathematically, this means embedding $G$ into the symmetry of the toric code. Since the ``target'' symmetry is a 2-group $\bbF$ rather than a group, this means specifying a homotopy class of maps $BG\ra B\bbF$. Physically, we make the fields $\sA$ and $\sB$ functions of $A$ so that under $G$ gauge transformations $\sA$ and $\sB$ transform by $F_0$ and $F_1$ gauge transformations, respectively. Such am embedding is characterized by a homomorphism $\pi:G\ra F_0$ and a cohomology class $[\Lambda_2]\in H^2_\pi(G,F_1)$. That is, we set
\begin{equation}
\sA=\pi(A),\quad \sB=\Lambda_2(A). 
\end{equation}
The corresponding anomaly action is obtained by substituting into (\ref{toricanomalyaction}):
\begin{equation}
S_{anomaly}=\int_{M_5} \frP_q(\Lambda_2(A)).
\end{equation}
This means that the symmetry $G$ is free of 't Hooft anomalies if and only if $\frP_q(\Lambda_2(A))$ is cohomologically trivial, i.e.  if and only if there exists a 3-cochain $\nu_3\in Z^3(G,\RR/\ZZ)$ such that
\begin{equation}\label{genGuWen}
\delta\nu_3=\frP_q(\Lambda_2).
\end{equation}
Then $G$ gauge-invariance can be restored by modifying the 3d action by a term $2\pi i \int_M \nu_3(A)$.

Let us specialize this to the case of trivial $\pi$. Then $\sA$ is trivial, and $\sB=\Lambda_2(A)$ is an ordinary (not twisted) 2-cocycle on $M$ with values in $F_1=\ZZ_2\times\ZZ_2$. We can write $\Lambda_2=(\beta_2^a,\beta_2^b)$, where $\beta_2^a,\beta_2^b\in Z^2(G,\ZZ_2)$. The condition (\ref{genGuWen}) simplifies to
\begin{equation}\label{trivialpigenGuWen}
\delta\nu_3=\frac12 \beta_2^b\cup\beta_2^a.
\end{equation}
The action of the equivariant toric code in this case is
\begin{equation}
2\pi i\int_M \left( \frac12 b\cup\delta a+ \frac12 b\cup \beta_2^a(A)+\frac12 \beta_2^b\cup a+\nu_3(A)\right).
\end{equation}
Regarding $b$ as a Lagrange multiplier, we see that it imposes a constraint $\delta a=\beta_2^a(A)$. This means that the pair $(A,a)$ is a 1-cocycle with values in $\hat G$, where $\hat G$ is a central extension of $G$ by $\ZZ_2$ whose extension class is $\beta_2^a$. The part of the action independent of $b$ can then be interpreted as an integral of a pull-back of a 3-cocycle $\hat\nu_3$, where $\hat\nu_3$ is given by
\begin{equation}
\hat\nu_3=\nu_3+\frac12 \beta_2^b\cup \eps,
\end{equation}
where $\eps$ is a $\ZZ_2$-valued 1-cochain on $\hat G$ trivializing the pull-back of $\beta_2^a$ and restricting to the identity on the central $\ZZ_2$ subgroup of $\hat G$. The corresponding fusion category is a twisted Drinfeld double of $\hat G$, with the twist given by $\hat\nu_3$. Essentially, we have shown that this is the most general $G$-equivariant extension of the toric code where $G$ acts trivially on the toric code quasi-particles (none of the elements of $G$ exchange $e$ and $m$).  Note that the model considered in section (\ref{sec:Gtoricfromcentral}) has this general form, but in addition has $\beta_2^a=\beta_2^b=\beta_2$. We will see shortly that this constraint arises if we require the model to contain a fermion. 

Now suppose $\pi$ is nontrivial. Let $G_0=\ker\ \pi$. It is easy to check that the action of $G$ on $F_1\simeq\ZZ_2\times\ZZ_2$ is induced from the trivial action of $G_0$ on
 $\ZZ_2$. .\footnote{We are grateful to V. Ostrik for pointing this out.} Therefore by Shapiro's lemma $H^2_\pi(G,F_1)\simeq H^2(G_0,\ZZ_2)$. Thus for nontrivial $\pi$ $G$-equivariant extensions of the toric code are labeled by a central extension of $G_0$ together with a trivialization $\nu_3$ of the corresponding Pontryagin square $\frP(\Lambda_2)$. The model considered in section (\ref{sec:Isingpullback}) is of this form. Below we will determine the condition on $\Lambda_2$ imposed by the existence of a fermion.

\subsection{One-form symmetries of the equivariant toric code}

Consider enlarging the symmetry group $G$ to a 2-group $\bbG$, such that the group of 1-form symmetries is $\ZZ_2$. Since $\ZZ_2$ has no nontrivial automorphisms, the 2-group structure of $\bbG$ is controlled by a Postnikov class $\Gamma\in H^3(G,\ZZ_2)$.  Enhancing the symmetry of the toric code  from $G$ to $\bbG$ involves  extending the map $BG\ra B\bbF$ to a map $B\bbG\ra B\bbF$. Since the Postnikov class $\gamma$ of $\bbF$ is trivial, this is only possible if $\Gamma$ is trivial. Physically, since the fusion of $F_0$ domain walls in the toric code is associative ``on the nose'',  this remains true even after we reinterpret them as $G$ domain walls via a homomorphism $\pi:G\ra F_0$. 

Specifying the homotopy class of a map $B\bbG\ra B\bbF$ is equivalent to specifying  $\sB$ and $\sA$ as functions of $B\in Z^2(M,\ZZ_2)$ and $A\in Z^1(M,G)$ in a way compatible with gauge transformations of $B$ and $A$.  This means
\begin{equation}
\sA=\pi(A),\quad \sB=\Lambda_2(A)+\rho(B),
\end{equation}
where $\pi$ and $\Lambda_2$ are as before, and $\rho$ is a nonzero homomorphism from $\ZZ_2$ to $F_1$ which is invariant with respect to the $G$ action on $F_1$ induced by $\pi:G\ra F_0$.


The anomaly for the 2-group $\bbG$ is obtained by substituting the expressions for $\sB$ and $\sA$ into (\ref{toricanomalyaction}). Using the properties of the Pontryagin square, we get
\begin{equation}\label{anomalytwogen}
S_{anomaly}=2\pi i \int_{M_4} \left(\frP_q(\Lambda_2(A))+b_q(\Lambda(A),\cup \rho(B)+\frac12 \rho(B)\cup \rho(B)\right),
\end{equation}
where $b_q$ is an $F_0$-invariant bilinear form on $F_1$ associated to the quadratic function $q$. Explicitly:
\begin{equation}
b_q(x_1,y_1;x_2,y_2)=\frac12 (x_1 y_2+y_1 x_2).
\end{equation}
We will assume as before that anomalies for $G$ are absent, i.e. $\frP(\Lambda_2)$ is cohomologically trivial. Then the first term in (\ref{anomalytwogen}) is exact. 
The second term describes the mixed anomaly between $\ZZ_2$ 1-form symmetry and $G$, so it must be exact for $\ZZ_2$ to be a global 1-form symmetry of the $G$-equivariant toric code. This means that the cohomology class of $\rho(1)$ must be orthogonal to the cohomology class of $\Lambda_2$ with respect to $b_q$. Finally, the last term describes the anomaly of the 1-form $\ZZ_2$ symmetry. 

Let us focus on fermionic $\ZZ_2$ 1-form symmetries. Such symmetries must have a nontrivial anomaly, so $q(\rho(1))\neq 0$. This uniquely fixes $\rho:$ it must send $1$ to the generator of $F_1^{F_0}$. That is, the fermion must be represented by a Wilson line $\exp(\pi i \int (a+b))$. The orthogonal complement of $F_1^{F_0}$ is $F_1^{F_0}$ itself, therefore the mixed anomaly is absent if and only if $[\Lambda_2]\in H^2_\pi(G,F_1)$ is in the image of the map $H^2(G,F_1^{F_0})\ra H^2_\pi(G,F_1)$. That is, we must have
\begin{equation}\label{psi}
\Lambda_2=\tilde\Lambda_2+\delta_\pi\tilde\psi_1,
\end{equation}
where $\tilde\Lambda\in Z^1(G,F_1^{F_0})$ and $\tilde\psi_1\in C^1_\pi(G,F_1)$. The anomaly action for symmetry $G$ is now exact, and if we let $M_4=\partial M_3$, it becomes
\begin{equation}\label{anomthreed}
S_{anomaly}=2\pi i \int_{M_3} \left(\nu_3(A)+b_q(\tilde\psi_1(A),\rho(B))\right)+\pi i \int_{M_4} B\cup B. 
\end{equation}
The first term is not invariant under 0-form and 1-form gauge symmetries, but it does not represent a true anomaly: it can be removed by modifying the action of the equivariant toric code by a local counterterm
\begin{equation}\label{Sct}
S_3^{ct}(A,B)=-2\pi i \int_{M_3} \left(\nu_3(A)+b_q(\tilde\psi_1(A),\rho(B))\right).
\end{equation}
The last term in (\ref{anomthreed}) is the correct anomaly for the 1-form $\ZZ_2$ symmetry to be fermionic.

The 1-cochain $\tilde\psi_1\in C^1(G,F_1)$ does not affect the cohomology class of the B-field and can be removed by a 1-form gauge transformation with a parameter $-\tilde\psi_1$. (This transformation also shifts $\nu_3$).  Then $\Lambda_2=\tilde\Lambda_2\in Z^2(G,\ZZ_2)$, and the constraint on $\nu_3$ simplifies:
\begin{equation}
\delta\nu_3=\frP_q(\tilde\Lambda_2)=\frac12\tilde \Lambda_2\cup\tilde\Lambda_2.
\end{equation}
This is nothing but the Gu-Wen equation. The counterterm action also takes a simple form:
\begin{equation}
S_3^{ct}(A,B)=-2\pi i \int_{M_3} \nu_3(A).
\end{equation}
We still retain the ability to perform 1-form symmetry transformations valued in $F_1^{F_0}\simeq\ZZ_2$. Indeed, while such transformations shift $\tilde\psi_1$, they do not affect $S_3^{ct}$ (\ref{Sct}), since $F_1^{F_0}$ is an isotropic subgroup of $F_1$. Under a transformation with a parameter $\lambda_1\in C^1(G,\ZZ_2)$ the data $(\nu_3,\tilde\Lambda_2)$ transform as follows:
\begin{equation}
\nu_3\mapsto \frac12 \lambda_1\cup\delta\lambda_1, \quad \tilde\Lambda_2\mapsto \tilde\Lambda_2+\delta\lambda_1.
\end{equation} 
Changing $\nu_3$ by exact cocycles also does not affect the action. 

We conclude that $G$-equivariant versions of the toric code with a fermionic $\ZZ_2$ 1-form symmetry are labeled by triples $(\pi,\tilde\Lambda_2,\nu_3)$, where $\pi$ is a homomorphism $G\ra\ZZ_2$ and $(\nu_3,\tilde\Lambda_2)\in C^3(G,U(1))\times Z^2(G,\ZZ_2)$ satisfy the Gu-Wen equations. We also described the identifications on this set which do not affect the model.

\section{Wen plaquette model} \label{app:toric}
The Wen placquette model for the toric code is defined on a square lattice with $\ZZ_2$ variables at each site.
The commuting projectors are $P_{i,j} = \sigma^x_{i,j}\sigma^y_{i+1,j}\sigma^y_{i,j+1}\sigma^x_{i+1,j+1}$,
associated to the plaquettes of the lattice. This is a realization of the toric code. 

The model realizes the $\ZZ_2$ symmetry of the equivariat toric code, but not in an on-site manner: 
the $\ZZ_2$ symmetry maps to translations of the lattice by one unit.

The $e$ and $m$ quasi-particles are described by switching the sign of a plaquette at even or 
odd locations on the lattice. The corresponding string operators can be taken to be, say, 
products of the form $\sigma^x_{i,j} \sigma^y_{i,j+1} \sigma^x_{i,j+2} \cdots$
which create a particle in the plaquette to the left and below the beginning of the string. 
A similar effect is archived by $\sigma^y_{i,j} \sigma^x_{i,j+1} \sigma^y_{i,j+2} \cdots$.

The combination $\sigma^z_{i,j} \sigma^z_{i,j+1} \sigma^z_{i,j+2} \cdots$
creates a pair of $e$ and $m$ particles at neighbouring plaquettes, i.e. an $\epsilon$ particle. 

\subsection{Fermionic ``boundary condition''}

Consider the model restricted to the upper half-plane. The bulk plaquettes do not gap the system completely. 
At the boundary, degrees of freedom survive which roughly correspond to a $\ZZ_2$ spin chain 
of twice the lattice spacing: the boundary operators $S_i =\sigma^y_{i,0}\sigma^x_{i+1,0}$ commute with all bulk plaquette 
operators. They anti-commute with nearest neighbours and commute with all others. 

The $\fB_e$ gapped boundary condition where $e$ condenses is easily described: we can add $\sum_i S_{2i}$ to the Hamiltonian. 
This commute with $e$ string operators ending on the boundary. 
Similarly, we obtain $\fB_m$ by adding $\sum_i S_{2i+1}$ to the Hamiltonian. This commute with $m$ string operators ending on the boundary. 
\footnote{In a gauge theory description, these boundary conditions are either Dirichlet, 
i.e. fix the connection at the boundary, or Neumann, i.e. leave the connection free to fluctuate at the boundary. }

Both choices break explicitly the translation symmetry along the boundary by one unit, which maps one into the other. 
This is compatible with the action of the $\ZZ_2$ symmetry of the toric code. 
A boundary condition defined by a bosonic boundary Hamiltonian which preserves the translation symmetry 
will flow to a $\ZZ_2$-symmetric boundary condition for the equivariant toric code. If gapped, it must 
coincide with a direct sum of $\fB_e$ and $\fB_m$, i.e. it must spontaneously break the $\ZZ_2$ symmetry. 

Simple choices of bosonic boundary conditions, such as adding $\sum_i S_{i}$ to the Hamiltonian, leave the $\ZZ_2$ symmetry unbroken
and give a gapless critical Ising model at the boundary. The Ising model is coupled to the bulk toric code in a straightforward manner: 
the bulk $\ZZ_2$ gauge theory couples to the Ising symmetry of the gapless theory. In particular, $e$ lines can end on $\sigma(z, \bar z)$ operators, 
$m$ lines on the dual $\mu(z,\bar z)$ operators and $\epsilon$ lines on the fermionic local operators $\psi(z)$ and $\bar \psi(\bar z)$. 

As the result of fermionic anyon condensation of $\epsilon$ is the root $\ZZ_2$ fermionic SPT phase, we should be able to produce a 
boundary condition $\fB_\epsilon$ at which $\epsilon$ ends and the $\ZZ_2$ symmetry is broken. 
The boundary condition should be related to a $\ZZ_2$-invariant interface between the toric code and the root $\ZZ_2$ fermionic SPT phase.

We can define $\fB_\epsilon$ as follows. First, we add a Majorana mode $\gamma_i$ at each boundary lattice site. 
Next, we add the following commuting projectors to the Hamiltonian: $\sum_i \gamma_i \gamma_{i+1} S_i$.
This Hamiltonian gaps the system. Indeed, if we fermionize locally the boundary Ising degrees of freedom in terms 
of new Majoranas $c_i$, the commuting projectors become $\sum_i \gamma_i \gamma_{i+1} c_i c_{i+1}$
and we have massive ground states  where all the $c_i \gamma_i$ have the same sign. 

The boundary Hamiltonian commutes with the $\epsilon$ line defects $c_i \sigma^z_{i,j} \sigma^z_{i,j+1} \sigma^z_{i,j+2} \cdots$
ending at the boundary. This motivates the identification with $\fB^\epsilon$. \footnote{In a gauge theory language, we expect $\fB^\epsilon$ to correspond to a deformed Neumann boundary condition, 
with extra boundary action given by the quadratic refinement of the intersection pairing associated to a spin structure on the boundary. }

Although the new Hamiltonian is naively invariant under translations along the boundary, in oder to 
define the actual Hilbert space we need to pair up the Majorana modes in some manner. This breaks the 
$\ZZ_2$ symmetry, as expected. 

There is a neat way to restore it: we can place on the lower half plane some gapped system which has 
Majorana boundary oscillators. An example is an infinite collection of Kitaev chains extended along the 
vertical direction. Concretely, we can put a pair of Majorana modes $\gamma_{i,j}$ and $\gamma'_{ij}$ 
at each site in the lower half plane and build the Hamiltonian with projectors $\gamma'_{i,j} \gamma_{i,j-1}$. 
These boundary plaquettes commute with the $\gamma_i \equiv \gamma_{i,-1}$ oscillators used in the boundary 
Hamiltonian. 

This is a microscopic description of the expected gapped interface between the root $\ZZ_2$ SPT phase and the equivariant toric code. 

\subsection{Anyon condensation}

It is straightforward to implement the bosonic anyon condensation in the plaquette model. 

We can populate the lattice with an arbitrary number of $e$ particles by 
removing from the Hamiltonian the odd plaquettes, using the Hamiltonian $- \sum_{i,j |i+j \mathrm{even}} P_{ij}$. 

The ``edge operators'' $U^b$ can be taken to be $\sigma^x_{ij}$ for even $i+j$ and 
$\sigma^y_{ij}$ for odd $ij$: they commute with the Hamiltonian and move or annihilate $e$ 
particles along diagonals in the lattice. They clearly all commute. 

Adding the $U^b$ edge operators to the Hamiltonian eliminates all the spin degrees of freedom 
and returns the trivial theory, as expected. 

Fermionic anyon condensation is a bit more subtle. We can populate the lattice with an arbitrary number 
of $\epsilon$ particles if we use the Hamiltonian 
\begin{equation}
H^\epsilon = \sum_{i,j} P_{2i,j} P_{2i+1,j} = \sum_{ij} \sigma^x_{2i,j}\sigma^z_{2i+1,j}\sigma^y_{2i+2,j}\sigma^y_{2i,j+1}\sigma^z_{2i+1,j+1}\sigma^x_{2i+2,j+1}
\end{equation}
We can visualize the $\epsilon$ particles as living in the middle of vertical edges at odd horizontal locations. 

The operator $\sigma^z_{2i+1,j}$ commute with the new Hamiltonian but anti-commute with the four  
$P$ operators around the vertex. It moves an $\epsilon$ particle vertically by one unit. 

The operator $\sigma^x_{2i-1,j} \sigma^z_{2i} \sigma^y_{2i+1}$ anti-commutes with 
the four $P$ operators above it. It moves an $\epsilon$ particle horizontally by one unit. 

As expected these $V^\epsilon$ operators do not commute: vertical operators commute with each other, 
but horizontal operators anti-commute with horizontal neighbours and vertical operators immediately below. 

We can add Majorana pairs. It is convenient to denote them as 
$\gamma_{2i+1,j}$ and $\gamma'_{2i+1,j}$. 

The dressed hopping operators $U^f$ take the form: $U^f_{2i+1,j}\equiv i \sigma^z_{2i+1,j}\gamma_{2i+1,j}\gamma'_{2i+1,j}$
and $U^f_{2i,j}\equiv i \sigma^x_{2i-1,j} \sigma^z_{2i,j} \sigma^y_{2i+1,j}\gamma_{2i-1,j}\gamma_{2i+1,j}$.

The $U^f$ operators square to $1$. The product of the operators around a closed path is 
\begin{equation}
P_{2i-1,j-1} P_{2i,j-1} \gamma'_{2i-1,j}  \gamma_{2i-1,j-1} \gamma'_{2i+1,j} \gamma_{2i+1,j-1}
\end{equation}

This will become $1$ as soon as we impose the Coulomb branch constraints 
\begin{equation}
C_{2i-1,j-1} \equiv P_{2i-1,j-1} \gamma'_{2i-1,j}  \gamma_{2i-1,j-1} = i (-1)^{i}
\end{equation}

Overall, we only need to impose the $C_{2i+1,j+1}$, $U^f_{2i+1,j}$ and $U^f_{2i,j}$ projectors,
as they imply the original $P_{2i,j} P_{2i+1,j}=1$ constraints. 

It is straightforward, if tedious, to show that we can use the $U^f_{2i+1,j}=1$ and $U^f_{2i,j}=1$
constraints to gauge-fix the spin variables. In terms of dressed fermionic operators 
\begin{equation}
\Gamma_{2j+1,j} = \gamma_{2j+1,j} \sigma^x_{2i+1,j} \sigma^y_{2i+2,j} \qquad \Gamma'_{2j+1,j} = \gamma'_{2j+1,j} \sigma^y_{2i+1,j} \sigma^x_{2i+2,j}
\end{equation}
commuting with the $U^f$ projectors, the Gauss law constraints involve $\Gamma'_{2j+1,j} \Gamma_{2j+1,j-1}$
and make the system into a collection of vertical Kitaev chains.   

\bibliography{References}{}
\bibliographystyle{JHEP_TD}
\end{document}